\documentclass[12pt]{iopart}
\usepackage{color}
\usepackage{graphicx}
\usepackage{epsf}
\usepackage{graphicx,epsfig}
 \def\neq {\not\equiv}

\def\dh{\dot{H}}
\def\doo{\dot{\phi}}
\def\ddo{\ddot{\phi}}
\def\cs2{c_{s}^{2}}

 \def\al{\alpha}
 \def\b{\beta}
 \def\ga{\gamma}
 \def\de{\delta}
 \def\De{\Delta}
 \def\ep{\varepsilon}
 \def\ze{\zeta}
 \def\th{\theta}

 \def\ph{\varphi}

 \def\df{\delta\phi}
 \def\p{\partial}
 \def\ddf{\delta\dot{\phi}}
 \def\alo{\alpha_{1}}
 \def\alt{\alpha_{2}}
 \def\tho{\theta_{1}}
 \def\tht{\theta_{2}}
 \def\bt{\bigtriangledown}
 \def\dP{\dot{\phi}}
 \def\epr{\eta^{'}}
 \def\eps{\eta^{''}}
 \def\Ga{\Gamma}
 \def\gap{{\gamma}^{'}}
 \def\dfp{{\delta \phi}^{'}}
 \def\Pig{\Pi^{\ga}}
 \def\Pif{\Pi^{\phi}}
\def\DB{\delta B}

 \def\be   {\begin{equation}}   \def\ee   {\end{equation}}
 \def\ba   {\begin{array}}      \def\ea   {\end{array}}
 \def\bea  {\begin{eqnarray}}   \def\eea  {\end{eqnarray}}
 \def\bean {\begin{eqnarray*}}  \def\eean {\end{eqnarray*}}
\begin{document}

\begin{titlepage}
\vspace{-.5cm}
\begin{center}
{\LARGE UNIVERSIT\`{A} DEGLI STUDI DI PADOVA} \\
\vskip 0.3cm
{\large DIPARTIMENTO DI FISICA ``G. GALILEI''} \\
\vskip 0.3cm
{\normalsize Scuola di dottorato di ricerca in fisica, ciclo XXII} 
\end{center}
\vspace{-0.5cm}
\hspace{-0.3cm}  \rule{14.5cm}{.2mm}
\vskip 5.5cm
\begin{center}
{\Large \bf Cosmological correlation functions in scalar and vector inflationary models}
\end{center}
\vskip 5.5cm
\parbox{15.cm} {\setlength{\baselineskip}{.5cm}Supervisori: \hspace{10cm} Dottoranda: \\
Prof. Sabino Matarrese \hspace{5.1cm} Emanuela Dimastrogiovanni\\Dr. Nicola Bartolo }
\vskip 1.5cm
\hspace{-0.7cm} \rule{14.5cm}{.2mm}
\end{titlepage}




\tableofcontents

\newpage

\section*{}

$\quad$$\quad$$\quad$$\quad$$\quad$$\quad$$\quad$$\quad$$\quad$$\quad$ \\
\bigskip
 \begin{center}\large{\textsl{to Matteo}}\end{center}

\newpage

\section{Introduction}
\setcounter{equation}{0}
\def\theequation{1.\arabic{equation}}

In the standard cosmological model, at very early times the Universe undergoes a quasi de Sitter exponential expansion driven by a scalar field, the inflaton, with an almost flat potential. The quantum fluctuations of this field are thought to be at the origin of both the Large Scale Structures and the Cosmic Microwave Background (CMB) fluctuations that we are able to observe at the present epoch \cite{lrreview}. CMB measurements indicate that the primordial density fluctuations are of order $10^{-5}$, have an almost scale-invariant power spectrum and are fairly consistent with Gaussianity and statistical isotropy \cite{smoot92,bennett96,gorski96,wmap3,Creminelli:2005hu,wmap5,Smith:2009jr,Senatore:2009gt}. All of these features find a convincing explanation within the inflationary paradigm. Nevertheless, deviations from the basic single-(scalar)field slow-roll model of inflation are allowed by experimental data. On one hand, it is then important to search for observational signatures that can help discriminate among all the possible scenarios; on the other hand, it is important to understand what the theoretical predictions are in this respect for the different models. \\

\noindent Non-Gaussianity and statistical anisotropy are two powerful signatures. A random field is defined ``Gaussian'' if it is entirely described by its two-point function, higher order connected correlators being equal to zero. Primordial non-Gaussianity \cite{review,prop} is theoretically predicted by inflation: it arises from the interactions of the inflaton with gravity and from self-interactions. However, it is observably too small in the single-field slow-roll scenario \cite{Acqua,Maldacena:2002vr,Seery:2006vu}. Alternatives to the latter have been proposed that predict higher levels of non-Gaussianity such as multifield scenarios \cite{Linde:1985yf,Kofman:1985zx,Polarski:1994rz,GarciaBellido:1995qq,Mukhanov:1997fw,Langlois:1999dw,Gordon:2000hv}, curvaton models \cite{Mollerach,Enqvist:2001zp,Lyth:2001nq,Lyth:2002my,Moroi:2001ct,Bartolo:2003jx} and models with non-canonical  Lagrangians \cite{Garriga:1999vw,ArmendarizPicon:1999rj,Ali,Chen:2006nt,ghostinfl}. Many efforts have been directed to the study of higher order (three and four-point) cosmological correlators in these models \cite{Bartolomulti,Barttrisp,SL3,VernizziWands,ChenWang,Chen:2006nt,Seery:2006vu,Byrnes,Seery:2008ax,SVW,Huang:2006eha,Arroja:2008ga,Arroja:2009pd,Chen:2009bc,Gao:2009gd,Mizuno:2009mv,Okamoto:2002ik,Kogo:2006kh} and towards improving the prediction for the two-point function, through quantum loop calculations \cite{Schwinger:1960qe,Calzetta:1986ey,Jordan:1986ug,Maldacena:2002vr,Weinberg:2005vy,Weinberg:2006ac,Seery:2007we,Seery:2007wf,Dimastrogiovanni:2008af}. From WMAP, the bounds on the bispectrum amplitude are given by $-4<f_{NL}^{loc}<80$ \cite{Smith:2009jr} and by $-125<f_{NL}^{equil}<435$ \cite{Senatore:2009gt} at $95\%$ CL, respectively in the local and in the equilateral configurations. For the trispectrum, WMAP provides $-5.6 \times 10^5<g_{NL}<6.4 \times 10^5$ \cite{Vielva:2009jz} ($g_{NL}$ is the ``local'' trispectrum amplitude from cubic contributions), whereas from Large-Scale-Structures data $-3.5\times 10^5<g_{NL}<8.2\times 10^5$ \cite{Desjacques:2009jb}, at $95\%$ CL. Planck \cite{http://planck.esa.int/} is expected to set further bounds on primordial non-Gaussianity.\\     
\noindent Statistical isotropy has always been considered one of the key features of the CMB fluctuations. The appearance of some ``anomalies'' \cite{de OliveiraCosta:2003pu,Vielva:2003et,Eriksen:2003db} in the observations though, after numerous and careful data analysis, suggests a possible a breaking of this symmetry that might have occurred at some point of the Universe history, possibly at very early times. This encouraged a series of attempts to model this event, preferably by incorporating it in theories of inflation. Let us shortly describe the above mentioned ``anomalies''. First of all, the large scale CMB quadrupole appears to be ``too low'' and the octupole ``too planar''; in addition to that, there seems to exist a preferred direction along which quadrupole and octupole are aligned \cite{Bennett:1996ce,Spergel:2003cb,de OliveiraCosta:2003pu,Efstathiou:2003tv,Land:2005ad}. Also, a ``cold spot'', i.e. a region of suppressed power, has been observed in the southern Galactic sky \cite{Vielva:2003et,Cruz:2006fy}. Finally, an indication of asymmetry in the large-scale power spectrum and in higher-order correlation functions between the northern and the southern ecliptic hemispheres was found \cite{Hansen:2004mj,Eriksen:2003db,Hansen:2004vq}. Possible explanations for these anomalies have been suggested such as improper foreground subtraction, WMAP systematics, statistical flukes; the possibilities of topological or cosmological origins for them have been proposed as well. Moreover, considering a power spectrum anisotropy due to the existence of a preferred spatial direction $\hat{n}$ and parametrized by a function $g(k)$ as 
\bea\label{disc}
P(\vec{k})=P(k)\left(1+g(k)(\hat{k}\cdot\hat{n})^2\right),
\eea
the five-year WMAP temperature data have been analyzed in order to find out what the magnitude and orientation of such an anisotropy could be. The magnitude has been found to be $g=0.29\pm 0.031$ and the orientation aligned nearly along the ecliptic poles \cite{Groeneboom:2009cb}. Similar results have been found in \cite{Hanson:2009gu}, where it is pointed out that the origin of such a signal is compatible with beam asymmetries (uncorrected in the maps) which should therefore be investigated before we can find out what the actual limits on the primordial $g$ are.  \\

\noindent Several fairly recent works have taken the direction of analysing the consequences, in terms of dynamics of the Universe and of cosmological fluctuations, of an anisotropic pre-inflationary or inflationary era. A cosmic no-hair conjecture exists according to which the presence of a cosmological constant at early times is expected to dilute any form of initial anisotropy \cite{Wald:1983ky}. This conjecture has been proven to be true for many (all Bianchi type cosmologies except for the the Bianchi type-IX, for which some restrictions are needed to ensure the applicability of the theorem), but not all kinds of metrics and counterexamples exist in the literature \cite{Barrow:2005qv,Barrow:2006xb,DiGrezia:2003ug}. Moreover, even in the event isotropization should occur, there is a chance that signatures from anisotropic inflation or from an anisotropic pre-inflationary era might still be visible today \cite{Pereira:2007yy,Pitrou:2008gk,Dimastrogiovanni:2008ua,Gumrukcuoglu:2008gi}. In the same context of searching for models of the early Universe that might produce some anisotropy signatures at late time, new theories have been proposed such as spinor models \cite{ArmendarizPicon:2003qk,Boehmer:2007ut,Watanabe:2009nc,deBerredoPeixoto:2009sb}, higher p-forms \cite{Germani:2009iq,Kobayashi:2009hj,Koivisto:2009ew,Germani:2009gg,Koivisto:2009fb,Koivisto:2009sd} and primordial vector field models. \\  
\noindent Within vector field models, higher order correlators had been computed in \cite{Yokoyama:2008xw,ValenzuelaToledo:2009af,Dimopoulos:2008yv,ValenzuelaToledo:2009nq,Karciauskas:2008bc} and, more recently, in \cite{Dimopoulos:2009am,Dimopoulos:2009vu} for $U(1)$ vector fields. We considered $SU(2)$ vector field models in \cite{Bartolo:2009pa,Bartolo:2009kg}. Non-Abelian theories offer a richer amount of predictions compared to the Abelian case. Indeed, self interactions provide extra contributions to the bispectrum and trispectrum of curvature fluctuations that are naturally absent in the Abelian case. We verified that these extra contributions can be equally important in a large subset of the parameter space of the theory and, in some case, can even become the dominant ones. \\   

\noindent The promising perspective of achieving more and more precise measurements for the cosmological observables thanks to Planck and future experiments and the search for signatures that may help identify the correct inflationary model, have also motivated studies of higher order corrections to cosmological correlation functions and to the power spectrum in particular. Indeed, loop corrections to the correlators arise from the interactions involving the fields during inflation and therefore carry some important information about the physics of the very early Universe. \\
\noindent Loop corrections may lead to interesting effects which scale like the power of the number of e-folds between horizon exit of a given mode $k$ and the end of 
inflation~\cite{Mukhanov:1996ak,Abramo:1997hu,boy1,boy2,olandesi}. The interest in loop corrections to the correlators of cosmological perturbations generated during an early epoch of inflation has been recently stimulated by two papers of Weinberg~\cite{weinberg1,weinberg2}. The reason is that one-loop corrections to the power spectrum of the 
curvature perturbation $\zeta$ seem to show some infra-red divergences which scale like $\ln(kL)$, where $L^{-1}$ is some infra-red comoving momentum cut-off \cite{sloth,sloth2,seery1,seery2}. However, it has been discussed in \cite{box4,box5} (see also~\cite{Enqvistetal}) that such potentially large corrections do not appear in quantities that are directly observable. \\
\noindent As to the power spectrum of curvature perturbations, one-loop corrections have been computed in single-field slow-roll inflation by D.~Seery \cite{Seery:2007we,Seery:2007wf} and by N.~Bartolo and myself \cite{Dimastrogiovanni:2008af}, in single-field slow-roll inflation. In \cite{Dimastrogiovanni:2008af} we completed the analysis carried out in \cite{Seery:2007we,Seery:2007wf}, where the metric tensor fluctuations had been neglected for simplicity, by including them in the calculations and proving that their contribution is as important as the one from the scalar perturbations. In the context of loop-calculations, we have also been working on corrections to the power spectrum in theories with non-canonical Lagrangians, which allow for higher and possibly observable corrections \cite{nuovo}. \\
\noindent It can be safely stated that in standard single-field slow-roll inflation, the perturbative expansion is well-behaved, in the sense that the agreement with observations found at tree-level for the power-spectrum is not spoiled by the radiative corrections and, on a more general basis, higher order loop corrections introduce smaller and smaller corrections as the perturbation series expansion progresses. This is not generically true in more general theories, such as for instance models with non-canonical Lagrangians, for which bounds need to be requested on the parameters of the theory in order to preserve the validity of the perturbative approach \cite{Leblond:2008gg,Shandera:2008ai}.  \\

\noindent This thesis collects the main results of our work on loop corrections to the power spectrum in theories of scalar inflation (Secs.~2 to 6) \cite{Dimastrogiovanni:2008af,nuovo}, on anisotropic pre-inflationary cosmologies (Sec.~7) \cite{Dimastrogiovanni:2008ua} and on primordial non-Gaussianity and anisotropy predictions from theories of inflation where vector fields can play a role in the production of the late time cosmological fluctuations (Secs.~8 to 12) \cite{Bartolo:2009pa,Bartolo:2009kg}. The $\delta$N and  the Schwinger-Keldysh formalisms are some of the main tools of our computation and will be briefly reviewed.

\newpage

\section{Schwinger-Keldysh formalism}
\setcounter{equation}{0}
\def\theequation{2.\arabic{equation}}

The temperature fluctuations in the CMB are rather small, of order $10^{-5}$. Theoretical predictions for the power-spectrum of curvature perturbations during inflation provide a very good match at tree level: this suggests that it is correct to use perturbation theory to evaluate cosmological correlators. A formalism conveniently employed to implement the perturbative approach is the Schwinger-Keldysh, also dubbed as ``in-in'', formalism. It was first formulated in \cite{Schwinger:1960qe,Calzetta:1986ey,Jordan:1986ug}, later applied by J.~Maldacena in \cite{Maldacena:2002vr} to the calculation of the bispectrum of curvature fluctuations and revived by S.~Weinberg in \cite{Weinberg:2005vy,Weinberg:2006ac}. In this formalism the expectation value of a field operator $\Theta(t)$ is given by

\be\label{skfirst}
\langle\Omega|\Theta(t)|\Omega\rangle=\left\langle 0\left|\left[\bar{T}\left(e^{i {\int}^{t}_{0}H_{I}(t')dt'}\right)\right]\Theta_{I}(t)\left[T \left(e^{-i {\int}^{t}_{0}H_{I}(t')dt'}\right)\right]\right|0\right\rangle,
\ee  
where $|\Omega\rangle$ represents the vacuum of the interacting theory, $T$ and $\bar{T}$ are time-ordering and anti-ordering operators, the subscript $I$ indicates the  fields in interaction picture and $H_{I}$ is the interaction Hamiltonian. The interaction picture has the advantage of allowing to deal with free fields only; the fields can be thus Fourier expanded in terms of quantum creation and annihilation operators
\bean
\df(\vec{x},t)=\int d^{3}k e^{i\vec{k}\vec{x}}\left[a_{\vec{k}} \df_{k}(t)+a^{+}_{-\vec{k}} \df_{k}^{*}(t)\right],
\eean
with commutation rules
\bean
\left[a_{\vec{k}},a^{+}_{\vec{k'}}\right]=(2 \pi)^{2}\de^{(3)}(\vec{k}-\vec{k'}).
\eean
\noindent The in-in formula has many similarities with the S-matrix in quantum field theory in terms of mathematical structure and perturbative approach, but they also have fundamental differences: the S-matrix corresponds to a transition amplitude between an initial and a final state; a cosmological correlation function is instead the expectation value of a given observable at a given time; moreover, asymptotic states in cosmology are only defined at very early times, when the same initial conditions as in Minkowsky spacetime apply for the free fields. \\
Using the positive and negative path technique of the in-in  formalism \cite{Weinberg:2005vy,Weinberg:2006ac}, the expectation value above can be recast in the form
\be\label{F}
\langle\Omega|\Theta(t)|\Omega\rangle = \left\langle 0\left|T\left(\Theta_{I}(t) e^{-i \int^{t}_{0}dt'\left(H_{I}^{+}(t')-H^{-}_{I}(t')\right)}\right)\right|0\right\rangle,
\ee
\noindent where the plus and minus signs indicate modified Feynman propagators, i.e. modified rules of contraction between interacting fields; schematically we have
\be
\langle T \left(\phi_{1}\phi_{2}...\phi_{n}\right)\rangle= \sum_{{ij,lm,...}}{[\widehat{\phi_{i}\phi_{j}},\widehat{\phi_{l}\phi_{m}},...]},
\ee
\noindent where the sum is taken over all of the possible sets of field contractions and
\bean
\widehat{\phi^{+}(\epr)\phi^{+}(\eps)} = G^{>}(\epr,\eps)\Theta(\epr-\eps)+G^{<}(\epr,\eps)\Theta(\eps-\epr),\\
\widehat{\phi^{+}(\epr)\phi^{-}(\eps)} = G^{<}(\epr,\eps),\\
\widehat{\phi^{-}(\epr)\phi^{+}(\eps)} = G^{>}(\epr,\eps),\\
\widehat{\phi^{-}(\epr)\phi^{-}(\eps)} = G^{<}(\epr,\eps)\Theta(\epr-\eps)+G^{>}(\epr,\eps)\Theta(\eps-\epr).
\eean
In momentum space we have
\bean
G^{>}_{k}(\epr,\eps)\equiv \delta \phi_{k}(\epr)\delta \phi_{k}^{*}(\eps),\\
G^{<}_{k}(\epr,\eps)\equiv \delta \phi_{k}^{*}(\epr)\delta \phi_{k}(\eps).
\eean
It is important to remember that, when we apply this formalism, the external fields are always supposed to be treated like $+$fields.

\newpage

\section{Scalar loop corrections to $P_{\zeta}$}
\setcounter{equation}{0}
\def\theequation{3.\arabic{equation}}

The power spectrum for the comoving curvature perturbation $\zeta$ is defined by
\be
\langle\ze_{\vec{k_{1}}}(t)\ze_{\vec{k_{2}}}(t)\rangle = (2 \pi)^{3} P_{\ze}(k) \delta^{(3)}(\vec{k_{1}}+\vec{k_{2}})\, ,
\ee
This and all other correlation functions presented in this thesis are computed using the $\delta N$ formula. $\zeta(\vec{x})$ at a given time $t$ can be interpreted as a geometrical quantity indicating the fluctuations in the local expansion of the universe; in fact, if $N(\vec{x},t^{*},t)$ is the number of e-foldings of expansion evaluated between times $t^{*}$ and $t$, where the initial hypersurface is chosen to be flat and the final one is uniform density, we have 
\bea\label{first}
\zeta(\vec{x},t)= N(\vec{x},t^{*},t)-N(t^{*},t)\equiv \delta N(\vec{x},t).
\eea
The number of e-foldings $N(\vec{x},t^{*},t)$ depends on all the fields and their perturbations on the initial slice. In principle, since the fields are governed by second order differential equations, it should also depend on their first time derivatives, but if we assume that slow-roll conditions apply, then the time derivatives will not count as independent quantities. \\

\noindent Let us apply Eq.~(\ref{first}) to the computation of $P_{\zeta}$ in single-field slow-roll inflation (the Lagrangian for the scalar field is given by $L_{\phi}=(1/2)g^{\mu\nu}\p_{\mu}\phi\p_{\nu}\phi-V(\phi)$)
\bea
\langle\ze_{\vec{k_{1}}}(t)\ze_{\vec{k_{2}}}(t)\rangle=\int \frac{d^{3}x_{1}}{(2 \pi)^{3}} \frac{d^{3}x_{2}}{(2 \pi)^{3}}e^{-i(\vec{k_{1}}\vec{x_{1}}+\vec{k_{2}}\vec{x_{2}})}\nonumber\\
\left\langle \left(\sum_{n}{\frac{N^{(n)}(t^{*},t)}{n!}\left(\df(\vec{x_{1}},t^{*})\right)^{n}}\right),\left(\sum_{m}{\frac{N^{(m)}(t^{*},t)}{m!}\left(\df(\vec{x_{2}},t^{*})\right)^{m}}\right)\right\rangle.
\eea
The sums can be expanded to the desired order. Up to one loop we have
\bea\label{ZZ}\fl
\langle\zeta_{\vec{k_{1}}}(t)\zeta_{\vec{k_{2}}}(t)\rangle&=&{N^{(1)}}^{2}\langle\df_{\vec{k_{1}}}\df_{\vec{k_{2}}}\rangle_{*}\nonumber\\\fl
&+&\frac{1}{2!}N^{(1)}N^{(2)}\int d^{3}q\langle\df_{\vec{k_{1}}}\df_{\vec{q}}\df_{\vec{k_{2}}-\vec{q}}\rangle_{*}+(\vec{k_{1}}\leftrightarrow \vec{k_{2}})\nonumber\\\fl
&+&\frac{1}{3!} N^{(1)}N^{(3)}\int d^{3}q d^{3}p\langle\df_{\vec{k_{1}}}\df_{\vec{q}}\df_{\vec{p}}\df_{\vec{q}+\vec{p}-\vec{k_{2}}}\rangle_{*} +(\vec{k_{1}}\leftrightarrow \vec{k_{2}})  \nonumber\\\fl
&+&\frac{1}{(2!)^2} \left(N^{(2)}\right)^{2}\int d^{3}q d^{3}p\langle\df_{\vec{q}}\df_{\vec{k_{1}}-\vec{q}}\df_{\vec{p}}\df_{\vec{k_{2}}-\vec{p}}\rangle_{*}. 
\eea
where a star indicates evaluation around the time of horizon crossing. Eq.~(\ref{ZZ}) can finally be rewritten as \cite{Byrnes:2007tm,seery2}
\bea\label{QQQ}\fl
\langle \zeta_{\vec{k_{1}}}(t)\zeta_{\vec{k_{2}}}(t) \rangle&=&(2 \pi)^{3}\delta^{(3)}(\vec{k_{1}}+\vec{k_{2}})\Big[\left( N^{(1)} \right)^{2}\Big(P_{{\rm tree}}(k_1)
+P_{{\rm one-loop}}(k_1)\Big) \nonumber\\\fl
&+&N^{(1)}N^{(2)}\int \frac{d^{3}q}{(2 \pi)^{3}} B_{\phi}(k_1,q,|\vec{k}_1-\vec{q}|)\nonumber\\
&+&\frac{1}{2}\left( N^{(2)} \right)^{2}\int \frac{d^{3}q}{(2 \pi)^{3}} P_{{\rm tree}}(q)P_{{\rm tree}}(|\vec{k}_1-\vec{q}|)
\nonumber\\\fl
&+&N^{(1)}N^{(3)}P_{{\rm tree}}(k)\int \frac{d^{3}q}{(2 \pi)^{3}} P_{{\rm tree}}(q)\Big]\, ,
\eea
$P_{{\rm tree}}(k)$ is the tree level power spectrum~(\ref{tree})
\bea\label{tree}
\langle\df_{\vec{k_{1}}}\df_{\vec{k_{2}}}\rangle_*&=& (2 \pi)^{3} P(k) \delta^{(3)}(\vec{k_{1}}+\vec{k_{2}})= (2 \pi)^{3}
\frac{H_*^{2}}{2k^{3}} \delta^{(3)}(\vec{k_{1}}+\vec{k_{2}})\, ,
\eea
where $H_*$ is the Hubble parameter evaluated at horizon exit (when $k=a H$). The variance per logarithmic interval in $k$ 
is given by ${\cal P}(k)=(k^3/ 2\pi^2) P(k)$. The one loop contribution to the power spectrum is given by
\begin{equation} 
P_{{\rm one-loop}}(k)=P_{{\rm scalar}}(k)+P_{{\rm tensor}}(k)\, ,
\end{equation}
where the first term on the right-hand side, $P_{{\rm scalar}}$, accounts for the contributions coming from the inflaton self-interactions and were computed by D.~Seery in ~\cite{seery1,seery2}
\bea
P_{{\rm scalar}}=\frac{H_{*}^{4}}{k^3}\left[g_{1}ln(k)+g_{2}\right],
\eea
where $g_{1}$ and $g_{2}$ are numerical factors. Their diagrammatic representation is given in Fig.$1$ for the leading order and in Fig.$2$ for the next-to-leading order corrections. The loop corrections $P_{{\rm tensor}}$, arising from interactions between the tensor (graviton) modes and the scalar field, were ignored for simplicity in \cite{seery1,sloth}, however they should be included since they are not slow-roll suppressed compared to loops of scalar modes. Their computation was presented for the first time in our paper \cite{Dimastrogiovanni:2008af} and will be reviewed in Secs.~4 to 6 of this thesis. \\

\begin{figure}
\begin{center}
\scalebox{0.5}{\includegraphics{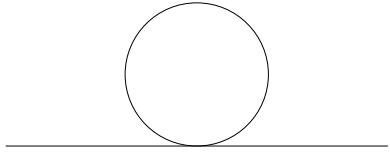}} 
\caption{Diagrammatic representation of the one loop corrections to the power spectrum of $\df$ from scalar modes to leading ($\sim {\epsilon}^{0}$) order in slow-roll.}
\label{1v-scal}
\end{center}
\end{figure}

\begin{figure}
\begin{center}
\scalebox{0.5}{\includegraphics{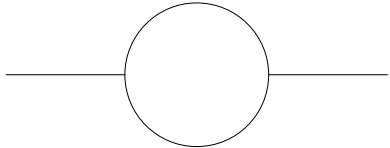}} 
\caption{Next-to-leading ($\sim \sqrt{\epsilon}$) order one loop corrections from scalar modes to the power spectrum of $\df$.}
\label{2v-scal}
\end{center}
\end{figure}

\noindent Both $P_{{\rm scalar}}$ and $P_{{\rm tensor}}$ are evaluated at around the time of horizon crossing and as such they are due to genuine quantum effects. \\

\noindent The contributions in the third and fourth lines of Eq.~(\ref{QQQ}), also dubbed as ``classical one-loop'', can be considered as classical loop contributions arising after the perturbation modes leave the horizon. The distinction between classical and quantum loops is intended as for example in \cite{seery2}: quantum loops find their origin in the Lagrangian interaction terms between the inflaton perturbations and the gravitational modes or from self-interaction of $\df$; classical loops are corrections merely coming from the expansion of $\zeta$ using the $\delta N$ formula and originate from zeroth order terms in the Schwinger-Keldysh formula. \\

\noindent Finally, the second line of (\ref{QQQ}) includes the integral of $B_{\phi}(k_{1},k_{2},k_{3})$, the bispectrum of the scalar field defined by
\bea
\langle {\df}_{\vec{k_{1}}}{\df}_{\vec{k_{2}}}{\df}_{\vec{k_{3}}} \rangle \equiv (2 \pi)^{3}\delta^{(3)}(\vec{k_{1}}+\vec{k_{2}}+\vec{k_{3}})B_{\phi}(k_{1},k_{2},k_{3})\, 
\eea
and from \cite{Maldacena:2002vr} we have
\bea\label{maldbis}
B_{\phi}\simeq\frac{\sqrt{\epsilon}H^{4}_{*}F(k_{i})}{m_{P}},
\eea 
where $m_{P}$ is the Planck mass, $\epsilon$ is the slow-roll parameter ($\epsilon\equiv-\dot{H}/H^2$) and $F$ is a function of the momenta moduli $k_{i}$ of dimension~$({\rm mass})^{-6}$.

\newpage

\section{Perturbative expansion of the Lagrangian in $P(X,\phi)$ theories}
\setcounter{equation}{0}
\def\theequation{4.\arabic{equation}}

\noindent In this and in the next two sections, we will review the computation of the tensor loop corrections to $P_{\zeta}$. For our purposes, the exponentials in Eq.~(\ref{F}) need to be expanded up to second order in the interaction Hamiltonian $H_{I}$
\bea \label{eq3}\fl
\langle\Omega|\Theta(\eta)|\Omega\rangle_{1L}&=&i\Big\langle 0\Big|T\left[\Theta \int^{\eta}_{- \infty}d \epr \left(H_{I}^{+}(\epr)-H_{I}^{-}(\epr)\right)\right]0\rangle\\
&+&\frac{(-i)^{2}}{2}\langle 0|T\Big[\Theta \int^{\eta}_{- \infty}d \epr \left(H_{I}^{+}(\epr)-H_{I}^{-}(\epr)\right)\int^{\eta}_{- \infty}d \eps \left(H_{I}^{+}(\eps)-H_{I}^{-}(\eps)\right)\Big]\Big|0\Big\rangle,\nonumber
\eea 
where $\Theta(t) \equiv \df_{\vec{k_{1}}}(\eta)\df_{\vec{k_{2}}}(\eta)$. One-loop power-spectrum diagrams require an expansion of the interaction Hamiltonian to third and fourth order in the field fluctuations, i.e. $H_{I}\equiv H_{I}^{(3)}+H_{I}^{(4)}$. We provide in Figs.~(3) and (4) the diagrammatic representation of the leading order corrections that we will find for the diagrams with tensor loops in single-field slow-roll inflation. The continuos lines represent scalar propagators, whereas the dotted lines indicate tensor propagators. In order to derive this result and the analytic expressions for these diagrams, we need to first calculate and expand $H_{I}$ up to fourth order in the field perturbations $\delta\phi$ and $\delta \gamma$. The starting point is the Lagrangian of the system. \\

\noindent We will begin with a more general Lagrangian for the scalar field than the usual $L_{\phi}=(1/2)g^{\mu\nu}\p_{\mu}\phi\p_{\nu}\phi-V(\phi)$, by introducing a non-conventional kinetic term, i.e.
\bea\label{beginning}
S=\frac{1}{2}\int d^{3}x dt \sqrt{-g}\left[M_{P}^{2}R+2P(X,\phi)\right],
\eea
where $P=P(X,\phi)$ is a generic function of the scalar field and of $X\equiv\frac{1}{2}g_{\mu \nu}\p^{\mu}\phi\p^{\nu}\phi$ and $R$ is the Ricci scalar in four dimensions. Notice that the action (\ref{beginning}) reduces to the standard case if $P=X-V$, where $V$ is the potential for the scalar field.\\
\noindent Theories of inflation where the Lagrangian kinetic term is a generic function of the scalar field and its first derivatives, like in Eq.~(\ref{beginning}), are string theory-inspired. They represent interesting alternatives to the basic inflationary scenario because of their non-Gaussianity predictions. The crucial quantity in this sense is represented by the speed of sound $c_{s}^{2}\equiv(\p_{X}P)/(\p_{X} P+2X \p_{XX}P)$, which is allowed to vary between $0$ and $1$. The perturbative expansion of the interaction Hamiltonian in this kind of models has coefficients proportional to inverse powers of the sound speed and therefore, for small values of $c_{s}$, allows both for non-negligible loop corrections to the power spectrum of the curvature fluctuations \cite{nuovo} and for large values for the amplitudes of three \cite{Chen:2006nt} and four \cite{Huang:2006eha,Arroja:2008ga,Arroja:2009pd,Chen:2009bc,Gao:2009gd,Mizuno:2009mv} point functions. In this thesis, we will carry out the calculations of the interaction Hamiltonian for these general theories up to a certain point and then, for simplicity in the presentation, focus on the canonical case (the remaining computations for more general Lagrangians will be found in \cite{nuovo}).\\

\noindent Let us list the background equations for the system
\bea\label{a}
2\dh+3H^2=-P ,\\
3H^2=2XP_{X}-P,\label{b}\\
\dot{X}\left(P_{X}+2XP_{XX}\right)&+&2\sqrt{3}{\left(2XP_{X}-P\right)}^{1/2}XP_{X}\nonumber\\&=&\sqrt{2X}\left(P_{\phi}-2XP_{X\phi}\right),\label{c}
\eea
where a dot indicates a derivative w.r.t. cosmic time and, to zeroth order, we have $X \equiv \frac{{\dot{\phi}}^{2}}{2}$.\\
\noindent The so called \textit{flow-parameters} are defined as
\bea
\ep\equiv -\frac{\dot{H}}{H^2}\label{d},\\
\eta \equiv \frac{\dot{\ep}}{\ep H}\label{e}.
\eea
These quantities reduce to the slow-roll parameters in the standard case, so it is natural to assume $|\ep| \ll 1$ and $|\eta|$. It is not correct to talk about \textsl{slow-roll} if $P$ is left as a generic function of $X$ and $\phi$, since the smallness of $\ep$ and $\eta$ does not necessarily indicate that ${\doo}^{2} \ll H^{2}$ and $|\ddo| \ll |3 H  \doo|$. It can be convenient to decompose $\ep$ as the sum $\ep=\ep_{\phi}+\ep_{X}$, where
\bea
\ep_{\phi} \equiv -\frac{\doo}{H^2}\frac{\p H}{\p \phi}\label{g},\\
\ep_{X} \equiv -\frac{\dot{X}}{H^2}\frac{\p H}{\p X}\label{h}.
\eea
The parameters that are expected to appear in the perturbative expansion of the Lagrangian are
\bea
\cs2=\frac{P_{X}}{P_{X}+2XP_{XX}}\label{ii},\\
s \equiv \frac{\dot{c_{s}}}{c_{s} H}\label{i},\\
u \equiv  1-\frac{1}{\cs2}\label{l},\\
\Sigma \equiv XP_{X}+2X^2P_{XX} \label{m},\\ 
\lambda \equiv  X^{2}P_{XX}+\frac{2}{3}X^{3}P_{XXX}\label{n},\\
\Pi \equiv X^{3}P_{XXX}+\frac{2}{5}X^4P_{XXXX}\label{o},
\eea
where $c_{s}$ is the sound speed. $c_{s}$ is allowed to vary between $0$ and $1$, so the quantity $|u|$ can freely range between $0$ and $\infty$. The only assumption we make is $s \ll 1$, from $c_{s}$ being constant in the standard case.

\begin{figure}
\begin{center}
\scalebox{0.5}{\includegraphics{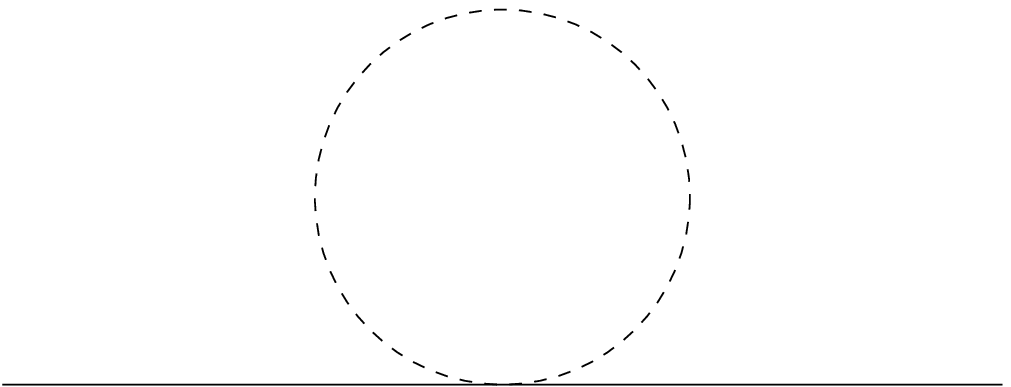}} 
\caption{Diagrammatic representation of the leading order (tensor mode) corrections from $H_{I}^{(4)}$ to the power spectrum of $\df$.}
\label{1v-tens}
\end{center}
\end{figure}

\begin{figure}
\begin{center}
\scalebox{0.5}{\includegraphics{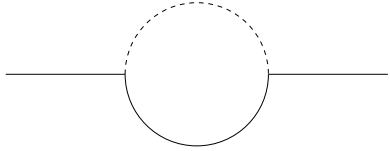}} 
\caption{Diagrammatic representation of the leading order (tensor mode) corrections from $H_{I}^{(3)}$ to the power spectrum of $\df$.}
\label{1v-tens}
\end{center}
\end{figure}

\subsection{Arnowitt-Deser-Misner (ADM) decomposition for $P(X,\phi)$ theories}

The Lagrangian in Eq.~(\ref{beginning}) will now undergo a perturbative expansion in terms of the field fluctuations $\delta \phi(\vec{x},t)\equiv \phi(\vec{x},t)-\phi_{0}(t)$ ($\phi_{0}$ is the homogeneous background value for the field) and of the metric fluctuations.\\
It is convenient to adopt the $3+1$ Arnowitt-Deser-Misner (ADM) splitting for the metric. In the spatially flat gauge the perturbed metric is
\be\label{eq2}
ds^2=-N^2dt^2+h_{ij}(dx^{i}+N^{i}dt)(dx^{j}+N^{j}dt),
\ee

\be
\label{smetric}
h_{ij}=a^{2}(t)(e^{\ga})_{ij},
\ee
where $a(t)$ is the scale factor, $\ga_{ij}$ is a tensor perturbation with $\p_{i}\ga_{ij}=\ga_{ii}=0$ (traceless and divergenceless) and 
det$(e^{\ga})_{ij}=1$. Notice that repeated lower indices are summed up with a Kronecker delta, so $\p_{i}\ga_{ij}$ stands for $\delta^{ik}\p_{i}\ga_{kj}$ and $\ga_{ij}=\delta^{ij}\ga_{ij}$.\\
In the ADM formalism, the action (\ref{beginning}) becomes \cite{Maldacena:2002vr}
\be\label{qua}
S=\frac{1}{2}\int{dtd^{3}x \sqrt{h}\left[NR^{(3)}+2NP+N^{-1}\left(E_{ij}E^{ij}-E^{2}\right)\right]},
\ee
where $R^{(3)}$ is the curvature scalar associated with the three dimensional metric $h_{ij}$ and
\bean
E_{ij}=\frac{1}{2}\left(\dot{h_{ij}}-\bt_{i}N_{j}-\bt_{j}N_{i}\right),\\
E=h^{ij}E_{ij}.
\eean
A dot indicates derivatives w.r.t. time $t$, all the spatial indices are raised and lowered with $h_{ij}$ and units of $M^{-2}_{Pl}\equiv 8 \pi G=1$ will be from now on employed. To $4th$ order we have
\be
R^{(3)}=-\frac{1}{4}\p_{i}\ga_{al}\p_{i}\ga_{al}.
\ee
The lapse and shift functions, $N$ and $N^{i}$, can be written as
\bean
N=1+\al \, ,\\
N_{j}=\p_{j}\th+\b_{j}\, ,
\eean
where $\al$, $\th$ and $\b$ are functions of time and space ($\b_j$ is divergenceless). We have exploited the gauge freedom to set two scalar and two vector modes to zero, thus leaving one scalar mode from $N$, one scalar and two vector modes from $N_{j}$ and two tensor modes (the two independent polarizations of the graviton) from $h_{ij}$, together with the inflaton field perturbation $\delta \phi$. $N$ and $N_{i}$ are non-dynamical degrees of freedom and can be expressed in terms of the other modes ($\delta \phi$ and $\gamma_{ij}$), once the Hamiltonian and the momentum constraints (we derive them in the next section) are solved.

\subsubsection{Solving Hamiltonian and momentum constraint equations}
${}$\\
Momentum and Hamiltonian constraints are derived from varying the action w.r.t. the shift and lapse functions respectively. It turns out that, in order to expand the action to a given order $n$, it is only necessary to perturb $N$ and $N_{i}$ up to order $n-2$ \cite{Maldacena:2002vr,Chen:2006nt}. Therefore we will solve the constraints to second order in the metric and scalar field fluctuations. \\
Let us begin with the expansions
\bean
\al=\alo+\alt,\\
\b_{i}=\b_{1i}+\b_{2i},\\
\th=\tho+\tht.
\eean
where $\alo$ and $\alt$ are respectively first and second order in the fields fluctuations (similarly for $\beta_{1i}$ and $\beta_{2i}$, and for $\tho$ and $\tht$). Let us then expand $P$ to second order. $P$ is a generic function of $X$ and $\phi$. We first need the expansion of $X$
\bea\fl
X&=&-g^{\mu\nu}\p_{\mu}\phi\p_{\nu}\phi=-\frac{1}{2}\left[g^{00}\dot{{\phi}^{2}}+2g^{0i}\p_{i}\phi \dot{\phi}+g^{ij}\p_{i}\phi\p_{j}\phi\right]\nonumber\\\fl
&=&-\frac{1}{2}\Big[-N^{-2}{\left(\dot{\phi}+\dot{\df}\right)}^{2}+2N^{-2}N^{i}\p_{i}\df\left(\dot{\phi}+\dot{\df}\right)+\left(h^{ij}-\frac{N^{i}N^{j}}{N^{2}}\right)\p_{i}\df\p_{j}\df\Big]\nonumber\\\fl&=&X_{0}+\Delta X
\eea
where $N^{i}\equiv h^{ij}N_{j}$, $X_{0}$ is the zeroth order part, i.e. $X_{0}=\frac{{\dot{\phi}}^{2}}{2}$ and $\Delta X$ is the perturbation to the desired order ($\Delta X=\Delta X_{1}+\Delta X_{2}+\Delta X_{3}+...$). Notice that  $\phi(t,\vec{x})=\phi_{0}(t)+\df(t,\vec{x})$, but for simplicity we will suppress the subscript '${0}$' in the background value of the field. \\
The expressions for the perturbations $\De X_{i}$ become
\bea\fl
\Delta X_{1}&=&2X_{0}\left[\frac{\dot{\df}}{\dot{\phi}}-\alo\right] \\\fl
\Delta X_{2}&=&X_{0}\Big[{\left(\frac{\ddf}{\dP}\right)}^{2}-4\alo\frac{\ddf}{\dP}-2\alt+3{\alo}^{2}-2N^{i}_{1}\p_{j}\frac{\ddf}{\dP}\Big]-\frac{1}{a^{2}{\dP}^{2}}\p_{i}\df\p_{i}\df
\eea
and so on for $\Delta X_{3}$ and higher order terms. The expansion of $P(X,\phi)$ up to second order becomes
\bea\fl
P(X,\phi)=P_{0}+P_{X}|_{0}\Delta X+P_{\phi}|_{0}\De \phi+\frac{1}{2!}P_{XX}|_{0}{(\Delta X)}^{2}+\frac{1}{2!}P_{\phi\phi}|_{0}{\De \phi}^{2}+P_{X\phi}|_{0}\Delta X \De \phi\nonumber\\
\eea
where as usual the subscript '${0}$' indicated the zeroth order, $P_{X}=\p_{X}P$, $P_{\phi}=\p_{\phi}P$ and similarly for the second order derivatives, $\Delta \phi=\df$ and $\Delta X$ needs to be expanded up to the needed order.\\

\noindent We are now ready to write the momentum and Hamiltonian contraints
\be
\bt_{i}\left[N^{-1}\left(E^{i}_{j}-\de^{i}_{j} E\right)\right]=N^{-1}P_{X}\left[\dot{\phi}-N^{l}\p_{l}\phi\right]\p_{j}\phi
\ee

\be
R^{(3)}+2P-4P_{X}X-N^{-2}\left(E_{ij}E^{ij}-E^{2}\right)-2P_{X}h^{ij}\p_{i}\phi\p_{j}\phi=0
\ee
The momentum constraint to first order reads
\be\label{der}
2H\p_{j}\alo-\frac{1}{2a^{2}}\p^{2}\beta_{1j}=P_{X}\dot{\phi}\p_{j}\df\, ,
\ee
where $H=\dot a/a$ is the Hubble parameter. Eq.~(\ref{der}) can be solved to derive $\alo$. Taking the derivative $\p^{j}$ of both sides of (\ref{der}) and using the divergenceless condition for $\b$, we have
\be\label{nonc1}
\alo=\frac{P_{X}\dP\df}{2H}.
\ee
Using the solution found for $\alo$, we find $\p^{2}\b_{1j}=0$, from which we can set $\b_{1j}=0$. Here $\p^{2}\equiv \delta^{ij}\p_{i}\p_{j}$, 
which we will indicate in the rest of the thesis also as $\p_{i}\p_{i}$, and from now on we define $\b_{i} \equiv \b_{2i}$ for simplicity.  \\

\noindent The momentum constraint to second order is
\bea
2H\p_{j}\alt-4H\alo\p_{j}\alo-\frac{1}{a^{2}}\p_{j}\alo\p^{2}\tho+\frac{1}{a^{2}}\p_{i}\alo\p_{i}\p_{j}\tho-\frac{1}{2}\p_{i}\alo\dot{\ga_{ij}}\nonumber\\-\frac{1}{2a^{2}}\p^{2}\b_{j}+\frac{1}{4}\dot{\ga_{ik}}\p_{i}\ga_{kj}-\frac{1}{4}\ga_{ik}\p_{i}\dot{\ga_{kj}}-\frac{1}{4}\dot{\ga_{ik}}\p_{j}\ga_{ik}+\frac{1}{2a^{2}}\p_{i}\tho\p^{2}\ga_{ij}\nonumber\\=P_{X}\p_{j}\df\ddf+2X P_{XX}\p_{j}\df\ddf-2X P_{XX}\dP\alo\p_{j}\df -P_{X}\dP\alo\p_{j}\df \nonumber \\
+P_{X\phi }\dP\df\p_{j}\df\, .
\eea
The solutions are
\bea
\alt&=&\frac{\alo^2}{2}+\frac{1}{2Ha^2}\p^{-2}\left[\p^{2}\alo\p^{2}\tho-\p_{i}\p_{j}\alo\p_{i}\p_{j}\tho\right]+\frac{P_{X}}{2H}\p^{-2}\Sigma \nonumber \\
&+&\frac{1}{4H}\p^{-2}\left[\dot{\ga_{ij}}\p_{i}\p_{j}\alo\right]-\frac{1}{4a^{2}H}\p^{-2}\left[\p_{i}\p_{j}\tho\p^{2}\ga_{ij}\right] \nonumber \\
&+&\frac{1}{8H}\p^{-2}\left[\p_{j}\dot{\ga_{ik}}\p_{j}\ga_{ik}\right]
+\frac{P_{X\phi }\dP}{2H}\p^{-2}\left[(\p_{j}\df)^2+\df\p^{2}\df\right] \nonumber \\
&+&\frac{X_{}P_{XX}}{H}\p^{-2}\left[\p^{2}\df\ddf+\p_{j}\df\p_{j}\ddf-\dP\left(\p_{j}\alo\p_{j}\df+\alo\p^{2}\df\right)\right]\, ,
\eea
where $\Sigma\equiv\p^{2}\df\ddf+\p_{j}\df\p_{j}\ddf$, and
\bea
\frac{1}{2a^{2}}\p^{2}\b_{j}&=&2H\p_{j}\alt-4H\alo\p_{j}\alo-\frac{1}{a^{2}}\p_{j}\alo\p^{2}\tho+\frac{1}{a^{2}}\p_{i}\alo\p_{i}\p_{j}\tho \nonumber \\
&-&\frac{1}{2}\p_{i}\alo\dot{\ga_{ij}} 
+\frac{1}{4}\dot{\ga_{ik}}\p_{i}\ga_{kj}-\frac{1}{4}\ga_{ik}\p_{i}\dot{\ga_{kj}}-\frac{1}{4}\dot{\ga_{ik}}\p_{j}\ga_{ik} \nonumber \\
&+&\frac{1}{2a^{2}}\p_{i}\tho\p^{2}\ga_{ij} -P_{X}\p_{j} \df\ddf 
-2X_{}P_{XX}\p_{j}\df\ddf\nonumber \\
&+&2X P_{XX}\dP\alo\p_{j}\df +P_{X}\dP\alo\p_{j}\df-P_{X\phi }\dP\df\p_{j}\df\, .
\eea

\noindent Let us now move to the Hamiltonian constraint which provides 
\bea\fl
\frac{4H}{a^2}\p^{2}\tho=-4XP_{X}\left(\frac{\ddf}{\dP}-\alo\right)+2P_{\phi}\df-8P_{XX}X^2\left(\frac{\ddf}{\dP}-\alo\right)-4 XP_{X\phi}\df
-12H^2\alo,\nonumber\\
\eea
\noindent to first order and
\begin{eqnarray}\label{nonc2}
\fl
-\frac{4H}{a^2}\p^{2}\tht =
(-2\alo)\Big[4XP_{X}\frac{\ddf}{\dP}+20P_{XX}X^{2}\frac{\ddf}{\dP}+2XP_{X\phi}\df+8P_{XXX}X^{3}\frac{\ddf}{\dP}\nonumber\\
+4P_{XX\phi}X^{2}\df+\frac{4H}{a^{2}}\p^{2}\tho\Big]
-\frac{4X\left(P_{X}+2XP_{XX}\right)}{a^{2}\dP}\p_{i}\tho\p_{i}\df-\frac{1}{a^{4}}\left(\p^{2}\tho\right)^{2}\nonumber\\
+\frac{1}{a^{2}}\Big[-\dot{\ga_{iq}}\p_{q}\p_{i}\tho+\frac{1}{a^{2}}\left(\p_{i}\p_{j}\tho\right)^{2}\Big]+
\left(-6H^{2}+2XP_{X}+4X^{2}P_{XX}\right)\Big[3\alo^{2}-2\alt\Big]\nonumber\\
+4\alo^{2}\left(3X^{2}P_{XX}+2X^{3}P_{XXX}\right)
+\frac{\ddf^{2}}{\dP^{2}}\Big[2XP_{X}+16X^{2}P_{XX}+8X^{3}P_{XXX}\Big]\nonumber\\
+\frac{\ddf\df}{\dP}\Big[4XP_{X\phi}+8X^{2}P_{XX\phi}\Big]
-\frac{{\left(\p_{i}\df\right)^{2}}}{a^{2}{\dP}^{2}}\Big[4X^{2}P_{XX}-2XP_{X}\Big]+{\df}^{2}\Big[-P_{\phi\phi}\nonumber\\
+2XP_{X\phi\phi}\Big]
+\frac{1}{4}\Big[\dot{\ga_{lj}}\dot{\ga_{lj}}+\frac{1}{a^{2}}\p_{a}\ga_{iq}\p_{a}\ga_{iq}\Big]-\frac{4H}{a^{2}}\ga_{ij}\p_{i}\p_{j}\tho,
\end{eqnarray}
\noindent to second order.\\

\subsubsection{Reduction to the canonical case}
${}$\\
In the canonical case, to zeroth order in perturbation theory $P=(\dot{\phi}^{2}/2)-V(\phi)$, so $P_{X}=1$ and $P_{{\phi}^{n}}=-\p^{n} V/ \p{\phi}^{n}$ with all other derivatives of $P$ being zero. The solutions above therefore reduce to \cite{Dimastrogiovanni:2008af}
\bea\label{una}
\alo=\frac{1}{2H}\dot{\phi}\df\label{K20},\\
\frac{4H}{a^{2}}\p^{2}\tho=-2V_{\phi}\df-2\dot{\phi}\ddf+2\alo\left(-6H^{2}+(\dot{\phi})^{2}\right),
\eea
\bea
\alt&=&\frac{{\alo}^{2}}{2}+\frac{1}{2H}\p^{-2}\Sigma+\frac{1}{2Ha^{2}}\p^{-2}\left[\p^{2}\alo\p^{2}\tho-\p_{i}\p_{j}\alo\p_{i}\p_{j}\tho\right]
 \\
&+&\frac{1}{4H}\p^{-2}\left[\dot{\ga_{ij}}\p_{i}\p_{j}\alo\right]
-\frac{1}{4Ha^{2}}\p^{-2}\left[\p_{i}\p_{j}\tho\p^{2}\ga_{ij}\right]+\frac{1}{8H}\p^{-2}\Big[\p_{j}\dot{\ga_{ik}}\p_{j}\ga_{ik}\Big]\nonumber \, ,
\eea
\bea
\frac{4H}{a^{2}}\p^{2}\tht&=&2\alo\left[2\dot{\phi}\ddf+\frac{4H}{a^{2}}\p^{2}\tho\right]+\frac{2}{a^{2}}\dot{\phi}\p_{i}\tho\p_{i}\df
-\frac{1}{a^{4}}\p_{i}\p_{j}\tho\p_{i}\p_{j}\tho \\
&+&\frac{1}{a^{4}}\left(\p^{2}\tho\right)^{2}
-\left(3\alo^{2}-2\alt\right)\left({\dot{\phi}}^{2}-6H^{2}\right)-\ddf^{2}-\frac{1}{a^{2}}\p_{i}\df\p_{i}\df\nonumber\\
&-&V_{\phi \phi}\df^{2}
-\frac{1}{4a^{2}}\p_{a}\ga_{iq}\p_{a}\ga_{iq}-\frac{1}{4}\dot{\ga_{lj}}\dot{\ga_{lj}}+\frac{1}{a^{2}}\dot{\ga_{iq}}\p_{i}\p_{q}\tho \nonumber \, ,
\eea

\bea\label{due}
\frac{1}{2a^{2}}\p^{4}\b_{j}&=&\frac{1}{a^{2}}\p^{2}\alo\p_{j}\p^{2}\tho
-\frac{1}{a^{2}}\p_{m}\p_{j}\alo\p_{m}\p^{2}\tho+\frac{1}{a^{2}}\p_{m}\alo\p_{m}\p_{j}\p^{2}\tho\nonumber\\
&-&\frac{1}{a^{2}}\p_{j}\alo\p^{4}\tho-\frac{1}{a^{2}}\p_{m}\p_{j}\p_{i}\alo\p_{i}\p_{m}\tho
+\frac{1}{a^{2}}\p^{2}\p_{i}\alo\p_{i}\p_{j}\tho\nonumber\\
&-&\frac{1}{a^{2}}\p_{i}\p_{j}\alo\p_{i}\p^{2}\tho+\frac{1}{a^{2}}\p_{m}\p_{i}\alo\p_{m}\p_{i}\p_{j}\tho 
+\p_{m}\p_{j}\ddf\p_{m}\df\nonumber\\
&-&\p^{2}\ddf\p_{j}\df+\p_{j}\ddf\p^{2}\df-\p_{m}\ddf\p_{m}\p_{j}\df
-\frac{1}{2}\p^{2}\left(\dot{\ga_{ij}}\p_{i}\alo\right)\nonumber\\
&-&\frac{1}{2a^{2}}\p^{2}\left(\p_{j}\ga_{bq}\p_{q}\p_{b}\tho\right)+\frac{1}{2a^{2}}\p^{2}\left(\p^{2}\ga_{jk}\p_{k}\tho\right)
-\frac{1}{4}\p^{2}\left(\ga_{il}\p_{i}\dot{\ga_{jl}}\right)\nonumber\\
&+&\frac{1}{4}\p^{2}\left(\dot{\ga_{ik}}\p_{i}\ga_{kj}\right).\label{K21}
\eea
\noindent where $V_{\phi \phi} \equiv \p^{2} V/\p {\phi}^{2} $ and $\p^{-2}$ is the inverse of the laplacian operator. Notice that the equations (\ref{K20}) through (\ref{K21}) agree with the results obtained in \cite{Seery:2006vu}, if we set $\ga_{ij}$ to zero.\\

\subsection{Fourth-order expansion of the Lagrangian in the canonical case}

The expansion of the action up to 4th order can be now derived by plugging the solutions (\ref{nonc1}) through (\ref{nonc2}) in (\ref{qua}). The final expression is quite lengthy and can be found in \cite{nuovo}. We will here only report the 4th order expansion of the action in the canonical case
\bea\fl
S_{4}&=&\int dt d^3x a^{3} \Bigg[- \frac{1}{24}V_{\phi\phi\phi\phi} \df^{4}+\frac{1}{2a^{2}}\p_{(i}\b_{j)}\p_{i}\b_{j}+\frac{1}{2a^{4}}\p_{j}\th_{1}\p_{j}\df \p_{m}\th_{1}\p_{m}\df\nonumber\\\fl
&-& \frac{1}{a^{2}}\ddf\left(\p_{j}\th_{2}+\b_{j}\right)\p_{j}\df+\left(\al^{2}_{1}\al_{2}-\frac{1}{2}\al_{2}^{2}\right)\left(-6H^{2}+\dot{\ph}^{2}\right)\nonumber\\\fl
&+&\frac{\al_{1}}{2}\Big[-\frac{1}{3}V_{\phi\phi\phi}\df^{3}-2V_{\phi}\al_{1}^{2}\df+\al_{1}\left(-\frac{1}{a^{2}}\p_{i}\df\p_{i}\df-V_{\phi\phi}\df^{2}\right)\nonumber\\\fl
&-& \frac{2}{a^{4}}\Big(\p_{i}\p_{j}\th_{2}\p_{i}\p_{j}\th_{1}-\p^{2}\th_{1}\p^{2}\th_{2}+\p_{i}\b_{j}\p_{i}\p_{j}\th_{1}\Big)
+\frac{2}{a^{2}}\Big(\dot{\phi}\Big(\p_{j}\th_{2}+\b_{j}\Big)\p_{j}\df\nonumber\\\fl&+&\ddf\p_{j}\th_{1}\p_{j}\df\Big)\Big]
+\alo^{2}\Big[\frac{1}{2a^{2}}\Big(\ga_{qi}\p_{a}\p_{i}\ga_{aq}-\frac{1}{2}\p_{a}\ga_{iq}\p_{a}\ga_{iq} \Big)-\frac{1}{4}\dot{\ga_{lj}}\dot{\ga_{lj}}+\dot{\ga_{iq}}\p_{i}\p_{q}\tho\Big]
\nonumber\\\fl&-&\frac{1}{a^{2}}\Big[\frac{1}{2}\ga_{ik}\ga_{kj}\p_{j}\df\p_{i}\df-\alo\ga_{ij}\p_{j}
\df\p_{i}\df+\alt\p_{i}\df\p_{i}\df\nonumber\\\fl
&-&\dot{\phi}\p_{j}\df\left(\ga_{ij}\p_{i}\tht+\ga_{ij}\b_{i}+\ga_{ij}\p_{i}\tho\right)
-\p_{k}\tht\dot{\ga_{ab}}\p_{b}\ga_{ak}-\b_{k}\dot{\ga_{ab}}\p_{b}\ga_{ak}\nonumber\\\fl&-&\frac{1}{2}\dot{\ga_{ab}}\b_{k}\p_{k}\ga_{ab}-\alo\Big(H\ga_{ab}\p_{a}\p_{b}\tht+\dot{\ga_{ab}}\p_{a}\p_{b}\tht+\dot{\ga_{ab}}\p_{a}\b_{b}\Big)+\frac{1}{2}\dot{\ga_{ab}}\p_{k}\ga_{ab}\p_{k}\tht\Big]\nonumber\\\fl&+&\frac{1}{2a^{4}}\Big(-8\ga_{ip}\p_{i}\p_{j}\tho\p_{p}\p_{j}\tht-4\ga_{ip}\p_{i}\p{j}\tho\p_{p}\b_{j}-4\ga_{ip}\p_{p}\p_{j}\tho\p_{j}\b_{i}\nonumber\\\fl&-&\p_{q}\tho\p_{i}\ga_{jq}\p_{i}\p_{j}\tht-\p_{q}\tht\p_{i}\ga_{jq}\p_{i}\p_{j}\tho-\b_{q}\p_{i}\ga_{jq}\p_{i}\p_{j}\tho\nonumber\\\fl
&-&\p_{q}\tho\p_{i}\ga_{jq}\p_{i}\b_{j}-\p_{q}\tho\p_{i}\ga_{jq}\p_{j}\b_{i}+\p_{q}\tho\p_{q}\ga_{ij}\p_{i}\p_{j}\tht\nonumber\\\fl
&+&2\p_{q}\tht\p_{q}\ga_{ij}\p_{i}\p{j}\tho+2\p_{q}\tho\p_{q}\ga_{ij}\p_{i}\b_{j}+2\b_{q}\p_{q}\ga_{ij}\p_{i}\p_{j}\tho\Big)\Bigg].
\eea
Similarly, the loop computation will be from now on performed considering this simpler case.


\newpage

\section{Tensor loop corrections to $P_{\zeta}$}

\setcounter{equation}{0}
\def\theequation{5.\arabic{equation}}

Let us then consider the terms in the interaction Hamiltonian $H_{I}$ that involve the tensor modes. The third order action in single-field slow-roll inflation with non-zero graviton fluctuations was calculated in \cite{Maldacena:2002vr}; we will focus on the leading order term in the slow-roll parameters, so we have    
\be
H_{I}^{(3)}(\eta)\equiv \frac{a^{2}(\eta)}{2}\int d^{3}x \ga_{ij}\p_{i}\df\p_{j}\df.
\ee
The fourth order action is given by Eq.~(\ref{Z}). Notice that some of the interaction terms involving the tensor modes in (\ref{Z}) appear with time derivatives, therefore the construction of the path integral formula requires additional care compared to the case where time derivatives only appear in the kinetic term of the Lagrangian. This issue will be discussed in Appendix~\ref{IHAM}. Also, it is possible to show that in Eq.~(\ref{Z}), of all the leading terms in the slow-roll expansion, only one will provide a non-zero contribution to the loop correction (see Appendix~\ref{Semplificazioni} for a detailed analysis) and contribute to the interaction Hamiltonian to fourth order which becomes
\bea
H_{I}^{(4)}(\eta)\equiv  \frac{a^{2}(\eta)}{4}\int d^{3}x\ga_{ik}\ga_{kj}\p_{i}\df\p_{j}\df
\eea
where the tensor fluctuations are
\bea
\ga_{ij}(\vec{x},t)=\int d^{3}k e^{i\vec{k}\vec{x}}\sum_{\lambda}{ \left[\ep_{ij}(\hat{k},\lambda)b_{\vec{k},\lambda}\ga_{k}(t)
+\ep^{*}_{ij}(-\hat{k},\lambda)b^{+}_{-\vec{k},\lambda}\ga^*_{k}(t)\right]},
\eea
with
\bea
\left[b_{\vec{k},\lambda},b^{+}_{\vec{k'},\lambda^{'}}\right]=(2 \pi)^{2}\de^{(3)}(\vec{k}-\vec{k'})\de_{\lambda,\lambda^{'}}.
\eea
\noindent The equation of motion for the eigenfunctions $\df_{k}(t)$ can be derived in the approximation of de-Sitter space from the second-order action 
\be
S_{2}=\int d\eta^{'}\frac{1}{(H \eta)^{2}}\left[\left(\df^{'}\right)^{2}-\left(\p_i \df \right)^{2}\right],
\ee
\noindent (where $d\eta = dt/a(t)$ is the conformal time) and they are given by the well-known expression
\be\label{uk}
u_{k}(\eta)=\frac{H}{\sqrt{2 k^{3}}}\left(1+i k \eta\right)e^{-ik\eta}.
\ee
In the same approximation, the eigenfunctions for the tensor modes $\gamma_k(\eta)$ are given by $u^{T}_{k}\equiv 2 u_{k}$.\\

\noindent Let us now begin with  the one-loop one-vertex part of the diagram (given in Fig.~$3$) which we label with the subscript $(1L,1v)$; this can be written as \cite{weinberg1}, \cite{weinberg2}
\be\fl
\langle\df_{\vec{k_{1}}}(\eta^{*})\df_{\vec{k_{2}}}(\eta^{*})\rangle_{(1L,1v)}=i\int d\epr \left\langle\left[H_{I}^{(4)}(\epr),\df_{\vec{k_{1}}}(\eta^{*})\df_{\vec{k_{2}}}(\eta^{*})\right]\right\rangle.
\ee 
We will study this in detail
\bea\fl
\langle\df_{\vec{k_{1}}}(\eta^{*})\df_{\vec{k_{2}}}(\eta^{*})\rangle_{(1L,1v)}&=&2i\int^{\eta^{*}}_{- \infty} d\epr a^{2}(\epr)\int \frac{d^{3}x}{(2 \pi)^{3}} 
\int d^{3}q_{1}d^{3}q_{2}d^{3}q_{3}d^{3}q_{4}\nonumber\\\fl
&\times&e^{-i \sum_{n}{\vec{q}}_{n}\cdot\vec{x}}P_{ij}(iq^{i}_{3})(iq^{j}_{4})u_{k_{1}}(\eta^{*})u^{*}_{q_{3}}(\epr)u_{k_{2}}(\eta^{*})u^{*}_{q_{4}}(\epr)u_{q_{1}}(\epr)u^{*}_{q_{2}}(\epr)\nonumber\\\fl
&\times &\delta^{(3)}(\vec{k_{1}}+\vec{q_{3}})\delta^{(3)}(\vec{k_{2}}+\vec{q_{4}})\delta^{(3)}(\vec{q_{1}}+\vec{q_{2}})+c.c.,
\eea
\noindent where the extra factor of $2$ accounts for the number of equivalent diagrams obtained by permuting the field contractions, $u_{k}(\eta)$ is given by 
Eq.~(\ref{uk}) and
\be P_{ij}=\sum_{\lambda,{\lambda}^{'}}\epsilon^{\lambda}_{ik}(\hat{q})\epsilon^{*{\lambda}^{'}}_{kj}(\hat{q})=2\sin^{2}\theta\delta_{ij}. 
\ee
Integration and the use of the delta function lead to a simpler form
\bea\label{K15}
\fl
\langle\df_{\vec{k_{1}}}(\eta^{*})\df_{\vec{k_{2}}}(\eta^{*})\rangle_{(1L,1v)}&=&-i{\delta}^{(3)}(\vec{k_{1}}+\vec{k_{2}})\frac{H_{*}^{4}}{2k^{4}}\int \frac{d^{3}q}{q^{3}}{\sin}^{2}\theta
\int^{\eta^{*}}_{- \infty}\frac{d {\eta}^{'}}{{\epr}^{2}}e^{2ik(\epr-\eta^{*})}{(1+ik\eta^{*})}^{2} \nonumber \\
&\times& {(1-ik\epr)}^{2}(1+iq\epr)(1-iq\epr)+c.c.,  
\eea
This equation is exact except for the approximation of using the de Sitter space formula for the scale factor, $a(\eta)=-(H\eta)^{-1}$, and evaluating the Hubble radius $H(\epr)$ at the time $\eta^{*}$. The reason why this is allowed is the following: the contribution to the integral w.r.t. time from regions well before horizon crossing is negligible compared to the contribution due to the region around horizon crossing \cite{Maldacena:2002vr,weinberg1,weinberg2}; in addition to that, we are choosing $\eta$ to be just a few e-folds after horizon crossing, so we can assume that the Hubble radius (as well as any of the slow-roll parameters of the theory) will not undergo a big variation during this interval of time. The same approximation will be applied to the two-vertex diagrams.\\
\\
We first solve the time integral. It is convenient to perform a change of variale like in \cite{sloth}, i.e. we set ${x}^{'}=-k\epr$ and $x^{*}=-k\eta^{*}$ 
so that
\bea\label{T}
\fl
\langle\df_{\vec{k_{1}}}(\eta^{*})\df_{\vec{k_{2}}}(\eta^{*})\rangle_{(1L,1v)}&=&{\delta}^{(3)}(\vec{k_{1}}+\vec{k_{2}})\frac{H_{*}^{4}}{2k^{4}}\int \frac{d^{3}q}{q^{3}}{\sin}^{2}\theta\, \, 
Im\Big[\int^{\infty}_{x^{*}}\frac{d{x}^{'}}{k}\frac{k^{2}}{{x}^{'2}}e^{2i({x}^{'}-x^{*})} \nonumber \\
&\times& {(1+ix^{*})}^{2} {(1-i{x}^{'})}^{2}(1+i\frac{q}{k}{x}^{'})(1-i\frac{q}{k}{x}^{'}) \Big]\, .  
\eea

\noindent After integrating the imaginary part, we end up with the following result
\bea
\label{onelooponev} \fl
\langle\df_{\vec{k_{1}}}(\eta^{*})\df_{\vec{k_{2}}}(\eta^{*})\rangle_{(1L,1v)}&=&
{\delta}^{(3)}(\vec{k_{1}}+\vec{k_{2}})\frac{H_{*}^{4}}{2k^{4}}\int \frac{d^{3}q}{q^{3}}{\sin}^{2}\theta
\frac{2k^{2}(3+x^{*2})+q^{2}(5+5x^{*2}+2x^{*4})}{4k}\nonumber\\\fl
&=&{\delta}^{(3)}(\vec{k_{1}}+\vec{k_{2}})\frac{H_{*}^{4}}{2k^{4}}2\pi\frac{4}{3}
\Big[\frac{k}{2}(3+x^{*2})\int\frac{dq}{q}
+\frac{1}{4k}(5+5x^{*2}+2x^{*4})\int dq q\Big]\, , \nonumber 
\eea
\noindent where the factor $4/3$ comes from integrating with respect to the azimuthal angle $\theta$ (notice that that the reference frame in momentum space has been chosen in such a way that the external wave vector $\vec{k}$ lies along the positive $z$ axis). We now solve the momentum integrals. Both the logarithmic and the quadratic one exhibit ultraviolet divergences and the logarithmic part diverges also at very low momenta. Ultraviolet divergences can be treated as in flat space; the infrared logarithmic divergence is fixed introducing a momentum lower cutoff ${\ell}^{-1}$ to be interpreted as 
a `box  size'~\cite{box4,box5,deltaN4,box2,box3} which can be fixed to be not much larger than the present horizon~\cite{box4,box5}. 
As an example, consider the first integral of Eq.~(\ref{onelooponev}) which is convenient to split as follows  
\be
\int_{{\ell}^{-1}}^{\Lambda}\frac{dq}{q}=\int_{{\ell}^{-1}}^{k}\frac{dq}{q}+\int_{k}^{\Lambda}\frac{dq}{q},
\ee
\noindent where we have introduced an upper cutoff $\Lambda$. The first integral gives $\ln(k \ell)$; the second integral can be renormalized introducing a counterterm $-\ln(\Lambda/k_{0})$, where $k_{0}$ is a renormalization constant. The final result for Eq.(\ref{onelooponev}) reads
\bea\label{FR1}
\fl
\langle\df_{\vec{k_{1}}}(\eta^{*})\df_{\vec{k_{2}}}(\eta^{*})\rangle_{(1L,1v)}=\pi{\delta}^{(3)}(\vec{k_{1}}+\vec{k_{2}})\frac{2H_{*}^{4}}{3k^{3}}(3+x^{*2})\Big[\ln(k \ell)-{\ln}(k)+\alpha\Big]\, ,
\nonumber 
\eea
where $\alpha$ is a left over constant from renormalization.\\

\noindent Let us now focus on the one-loop contribution from the $3rd$ order action with the gravitons (see Fig.~$4$ for its diagrammatic representation)
\bea \label{36}
\fl
\langle\df_{\vec{k_{1}}}(\eta^{*})\df_{\vec{k_{2}}}(\eta^{*})\rangle_{(1L,2v)}&=&\frac{(-i)^{2}}{2}\Big\langle T\Big[\df_{\vec{k_{1}}}(\eta^{*})\df_{\vec{k_{2}}}(\eta^{*}) \int^{\eta^{*}}_{- \infty}d \epr \Big(H_{I}^{+}(\epr)-H_{I}^{-}(\epr)\Big)\nonumber \\
&\times& \int^{\eta^{*}}_{- \infty}d \eps \left(H_{I}^{+}(\eps)-H_{I}^{-}(\eps)\right)\Big]\Big\rangle\nonumber\\
&=& \frac{(-i)^{2}}{2}\Big\langle T\Big[\df_{\vec{k_{1}}}(\eta^{*})\df_{\vec{k_{2}}}(\eta^{*})\Big(A+B+C+D\Big)\Big]\Big\rangle,
\eea 
\noindent where
\bea
A \equiv \int^{\eta^{*}}_{- \infty}d \epr H_{I}^{+}\int^{\eta^{*}}_{- \infty}d \eps H_{I}^{+},\\
B \equiv \int^{\eta^{*}}_{- \infty}d \epr H_{I}^{-}
\int^{\eta^{*}}_{- \infty}d \eps H_{I}^{-},\\
C \equiv -\int^{\eta^{*}}_{- \infty}d \epr H_{I}^{+}\int^{\eta^{*}}_{- \infty}d \eps H_{I}^{-},\\
D \equiv -\int^{\eta^{*}}_{- \infty}d \epr H_{I}^{-}
\int^{\eta^{*}}_{- \infty}d \eps H_{I}^{+}.
\eea
It is easy to check that $B=A^{*}$ and $C=C^{*}=D$. We can write Eq. (\ref{36}) as
\bea \fl
\langle\df_{\vec{k_{1}}}(\eta^{*})\df_{\vec{k_{2}}}(\eta^{*})\rangle_{(1L,2v)}&=&4 (-i)^{2}\delta^{(3)}(\vec{k_{1}}+\vec{k_{2}})k^{4}\int d^{3}q {\sin}^{4}\theta \\\fl
&\times& \int^{\eta^{*}}_{- \infty}\frac{d \epr}{{(H\epr)}^{2}}\int^{\eta^{*}}_{- \infty}\frac{d \eps}{{(H \eps)}^{2}}\left(w_{f}^{A}+w_{f}^{B}+w_{f}^{C}+w_{f}^{D}\right),\nonumber
\eea
\noindent where the factor $\sin^{4}\theta$ comes from contractions of the polarization tensors with external momenta \cite{polarization}
\be
\epsilon_{ij}(\vec{q})k^{i}k^{j}=\frac{k^{2}}{\sqrt{2}}\left[1-{\left(\frac{\vec{q}\cdot\vec{k}}{qk}\right)}^{2}\right]=\frac{k^{2}}{\sqrt{2}}\sin^{2}\theta,
\ee
\noindent and the wave fuctions $w_{f}$ are
\bea\fl
w_{f}^{A}(\epr,\eps)&=&u_{k}^{}(\eta^{*})u_{k}^{*}(\epr) u_{k}^{}(\eta^{*})u_{k}^{*}(\eps)[u_{|\vec{k}-\vec{q}|}^{}(\epr)u_{|\vec{k}-\vec{q}|}^{*}(\eps)u_{q}^{}(\epr)u_{q}^{*}(\eps)\nonumber\\ \fl
&\times&\Theta(\epr-\eps)+u_{|\vec{k}-\vec{q}|}^{*}(\epr)u_{|\vec{k}-\vec{q}|}^{}(\eps)u_{q}^{*}(\epr)u_{q}^{}(\eps)\Theta(\eps-\epr)]\, , \nonumber 
\eea
\bea\fl
w_{f}^{B}(\epr,\eps)&=&u_{k}^{*}(\eta^{*})u_{k}^{}(\epr) u_{k}^{*}(\eta^{*})u_{k}^{}(\eps)[u_{|\vec{k}-\vec{q}|}^{}(\epr)u_{|\vec{k}-\vec{q}|}^{*}(\eps)
u_{q}^{}(\epr)u_{q}^{*}(\eps) \nonumber\\\fl
&\times& \Theta(\eps-\epr)
+u_{|\vec{k}-\vec{q}|}^{*}(\epr)u_{|\vec{k}-\vec{q}|}^{}(\eps)u_{q}^{*}(\epr)u_{q}^{}(\eps)\Theta(\epr-\eps)]\, , \nonumber 
\eea
\bea\fl
w_{f}^{C}(\epr,\eps)&=&-u^{}_{k}(\eta^{*})u^{*}_{k}(\epr) u^{*}_{k}(\eta^{*})u^{}_{k}(\eps)u^{*}_{|\vec{k}-\vec{q}|}(\epr)
u^{}_{|\vec{k}-\vec{q}|}(\eps)u^{*}_{q}(\epr)u^{}_{q}(\eps)\, , \nonumber
\eea
\bea\fl
w_{f}^{D}(\epr,\eps)=-u^{*}_{k}(\eta^{*})u^{}_{k}(\epr) u^{}_{k}(\eta^{*})u^{*}_{k}(\eps)u^{}_{|\vec{k}-\vec{q}|}(\epr)
u^{*}_{|\vec{k}-\vec{q}|}(\eps)u^{}_{q}(\epr)u^{*}_{q}(\eps)\, , \nonumber
\eea
\noindent so $w_{f}^{C}(\epr,\eps)=w_{f}^{D}(\eps,\epr)$ and is a real number and $w_{f}^{B}(\epr,\eps)=w_{f}^{A*}(\eps,\epr)$. We will label the two contributions by $A$ and $C$, so that the one-loop contribution with two vertices to the two point function will be broken into two parts
\bea
\fl
\langle\df_{\vec{k_{1}}}(\eta^{*})\df_{\vec{k_{2}}}(\eta^{*})\rangle_{(1L,2v)}&=&
\langle\df_{\vec{k_{1}}}(\eta^{*})\df_{\vec{k_{2}}}(\eta^{*})\rangle_{(1L,2v)}^{A}
+\langle\df_{\vec{k_{1}}}(\eta^{*})\df_{\vec{k_{2}}}(\eta^{*})\rangle_{(1L,2v)}^{C}.
\eea
Let's look in details at the two parts. 
\bea
\fl
\langle\df_{\vec{k_{1}}}(\eta^{*})\df_{\vec{k_{2}}}(\eta^{*})\rangle_{(1L,2v)}^{A}&=&-\delta^{(3)}(\vec{k_{1}}+\vec{k_{2}})\frac{H_{*}^{4}}{2k^{2}}\int\frac{d^{3}q}{q^{3}}\frac{{\sin}^{4}\theta}{{|\vec{k}-\vec{q}|}^{3}}e^{-2ik\eta^{*}}{(1+ik\eta^{*})}^{2} \\
&\times& \int^{\eta^{*}}_{- \infty}\frac{d \epr}{{\epr}^{2}}e^{i\epr(k-q-|\vec{k}-\vec{q}|)}(1-ik\epr)(1+iq\epr)(1+i|\vec{k}-\vec{q}|\epr)
\nonumber\\
&\times& \int^{\epr}_{- \infty}\frac{d \eps}{{\eps}^{2}}e^{i(k+q+|\vec{k}-\vec{q}|)\eps}(1-ik\eps) (1-iq \eps)(1+i|\vec{k}-\vec{q}|\eps)+c.c.\nonumber 
\eea
The second time integral has $e^{i g\eps}\left[-\frac{1}{\eps}+\frac{c}{g}\eps-i\left(\frac{gb-c}{g^{2}}\right)\right]$ as its primitive function, where $g\equiv k+q+|\vec{k}-\vec{q}|$, $b \equiv -qk-(q+k)|\vec{k}-\vec{q}|$ and $c \equiv qk|\vec{k}-\vec{q}|$. This should be evaluated between $-\infty$ and $\epr$. It is soon evident that the lower bound represents a problem for this evaluation. We need to remind ourself, though, that the choice of the integration time contour needs to be deformed and to cross the complex plane to account for the right choice of the vacuum \cite{Maldacena:2002vr}. This is done by integrating in a slightly imaginary direction, i.e. taking $\eps\rightarrow \eps+i\epsilon|\eps|$, where $\epsilon$ is a fixed small real number; so for example
\be
\int_{-\infty}^{\eta} d\epr e^{ik\epr}=\frac{e^{ik\eta}}{i k}.
\ee    
With this contour prescription, our integral in $\eps$ vanishes at $- \infty$. Performing the same change of variables as in (\ref{T})
\bea\label{KKKK}
\fl
\langle\df_{\vec{k_{1}}}(\eta^{*})\df_{\vec{k_{2}}}(\eta^{*})\rangle_{(1L,2v)}^{A}&=&-\delta^{(3)}(\vec{k_{1}}+\vec{k_{2}})\frac{H_{*}^{4}}{k^{2}}\int\frac{d^{3}q}{q^{3}}\frac{{\sin}^{4}\theta}{{|\vec{k}-\vec{q}|}^{3}}
Re\Big[\int^{\infty}_{x^{*}}\frac{d{x}^{'}}{k}{\frac{k}{{x}^{'}}}{2}e^{2i({x}^{'}-x^{*})}{(1+ix^{*})}^{2} \nonumber \\
&\times& \left(\frac{k}{{x}^{'}}-\frac{c}{kg}{x}^{'}-i\left(\frac{gb-c}{g^{2}}\right)\right)
\left(1-i\frac{d}{k}{x}^{'}+\frac{s}{k^{2}}{{x}^{'}}^{2}-i\frac{c}{k^{3}}{{x}^{'}}^{3}\right)\Big]\, , \nonumber \\
\eea
\noindent where $d \equiv q-k+|\vec{k}-\vec{q}|$ and $s \equiv kq+(k-q)|\vec{k}-\vec{q}|$. The result of the integration w.r.t.time is a polynomial function of $\sin2x^{*}$, $\cos2x^{*}$, Si($2x^{*}$), Ci($2x^{*}$) and their products with coefficients which depend on $g,b,c,d,s,k$. Notice that in the large scale limit $x^{*}\rightarrow0$ a singularity similar to the one found in \cite{sloth} shows up in our result (see also \cite{Burgess:2009bs} for a recent discussion on these kind of singularities). However, by evaluating the 
power spectrum of $\delta \phi$ just a few e-folds after horizon crossing, we are safe from these kind of singular behaviour \cite{seery2}.\\  
The next step consist in performing the momentum integral. The integrals we need to evaluate are of the following kind 
\be
\int \frac{d^{3}q}{q^{3}} \frac{{\sin}^{4}\theta}{{|\vec{q}-\vec{k}|}^{3}}f(\vec{q}),
\ee
\noindent where $f(\vec{q})$ is a sum of functions of momentum. Let us begin for simplicity by considering the constant term of the sum, i.e. 
let us study 
\be\label{ex}
\int \frac{d^{3}q}{q^{3}} \frac{{\sin}^{4}\theta}{{|\vec{q}-\vec{k}|}^{3}}.
\ee
For the specific case of equation (\ref{ex}) the integrand function has singularities at $\vec{q}=0$ and at $\vec{q}=\vec{k}$ and shows no ultraviolet singularities. Based on an approximate evaluation performed considering a sphere of radius $\ell^{-1}$ around $\vec{q}=0$, where $\ell^{-1} \ll k$, the integral is proportional to a function $\ln(k\ell)$. The same result can be obtained working in a small sphere around  $\vec{q}=\vec{k}$ after a change of variables $\vec{q_{0}}=\vec{q}-\vec{k}$. The contribution from large values of q is negligible w.r.t. the ones from the singular points, so the integral over the whole momentum space is expected to be proportional to $\ln(k\ell)$. The exact value of the integral can been found 
after a change of variable from the $(q,\theta)$ to the $(q,p)$ space, where $p \equiv |\vec{q}-\vec{k}|$ and is equal to $(16 \pi/225k^{3})\left(1+30\ln(k\ell)\right) \sim k^{-3}\left(10^{-1}+10\ln(k\ell)\right)$.\\

\noindent Integrating Eq.~(\ref{KKKK}), we find ultaviolet power law and logaritmic singularities in addition to infrared logaritmic contributions. The final result of the integration is a function of $x^{*}=e^{-N_{*}}$, where $N_{*}=\ln(a_{*}/a_{k})$ is the number of e-foldings from horizon crossing
\bea\label{appC1}
\fl
\langle\df_{\vec{k_{1}}}(\eta^{*})\df_{\vec{k_{2}}}(\eta^{*})\rangle_{(1L,2v)}^{A}=\pi\delta^{(3)}(\vec{k_{1}}+\vec{k_{2}})\frac{H_{*}^{4}}{k^{3}}
\Big(a_{1}\ln(k)+a_{2}\ln(k \ell)+a_{3}\Big)\, , \\
\nonumber
\eea
where $a_{1}$, $a_{2}$ and $a_{3}$ are functions of $x^{*}$ (see Appendix~\ref{powerloops}). We are calculating the two point function for the scalar field a few e-foldings after horizon crossing, so $x^{*}$ may be chosen to range between $10^{-1}$ and $10^{-2}$. In this range $a_{1}
\sim {\cal O}(1)$ and negative, $a_{2}=-16/(15 x^{*2})+(8/15)(5-8 
{\rm Ci}(2x^{*}))$ and $a_{3}=-8/(225 x_*^2)+{\cal O}(1)+\rho$, where $\rho$ is  
a left-over scheme-dependent renormalization constant of the kind present in equation (\ref{FR1}). \\

\noindent Let us now move to part C of Eq. (\ref{36}) which we give below
\bea
\fl
\langle\df_{\vec{k_{1}}}(\eta^{*})\df_{\vec{k_{2}}}(\eta^{*})\rangle_{(1L,2v)}^{C}&=&\delta^{(3)}(\vec{k_{1}}+\vec{k_{2}})\frac{H_{*}^{4}}{2k^{2}}\int\frac{d^{3}q}{q^{3}}\frac{\sin^{4}\theta}{{|\vec{k}-\vec{q}|}^{3}}\left(1+{(k\eta^{*})^2}\right)
\nonumber\\
&\times& \int^{\eta^{*}}_{- \infty}\frac{d \epr}{{\epr}^{2}}e^{i g \epr}Q(\epr)
\int^{\eta_*}_{- \infty}\frac{d \eps}{{\eps}^{2}}e^{-i g \eps}Q^{*}(\eps)\nonumber\\
&=&\delta^{(3)}(\vec{k_{1}}+\vec{k_{2}})\frac{H_{*}^{4}}{2k^{2}}\int\frac{d^{3}q}{q^{3}}\frac{{\sin}^{4}\theta}{{|\vec{k}-\vec{q}|}^{3}}
\left(1+(k\eta^{*})^2\right)
\nonumber\\
&\times&  \Big[{\left(Re \int d \epr e^{ i g \epr }\frac{Q(\epr)}{{\epr}^{2}}\right)}^{2}+{\left(Im \int d \epr e^{i g \epr }\frac{Q(\epr)}{{\epr}^{2}}\right)}^{2}\Big],
\eea
\noindent where $Q(\epr) \equiv 1+ig\epr+b{\epr}^{2}-ic{\epr}^{3}$.
Let us integrate over conformal time
\bea \label{6}
\fl
\int^{\infty}_{x^{*}}\frac{d{x}^{'}}{k}\frac{k^{2}}{{{x}^{'}}^{2}} e^{-i\frac{g}{k} {x}^{'}}\left[1+i\frac{g}{k}{x}^{'}+\frac{b}{k^{2}}{{x}^{'}}^{2}-i\frac{c}{k^{3}}{{x}^{'}}^{3}\right] =\frac{e^{-i\frac{g}{k}}x^{*}}{k^{2}}\left(-\frac{k^{3}}{x}+\frac{ck}{g}x^{*}+i\frac{(gb-c)k^{2}}{g^{2}}\right)\, , \nonumber \\
\eea
\noindent where again the integration has been performed by continuing $\epr$ to the complex plane, i.e. $(\epr\rightarrow\epr+i\epsilon |\epr|)$, and then taking the limit $\epsilon\rightarrow 0$.\\ 
We are now ready to integrate over momentum
\bea
\fl
\langle\df_{\vec{k_{1}}}(\eta^{*})\df_{\vec{k_{2}}}(\eta^{*})\rangle_{(1L,2v)}^{C}&=&\delta^{(3)}(\vec{k_{1}}
+\vec{k_{2}})\frac{H_{*}^{4}}{2k^{2}}\int\frac{d^{3}q}{q^{3}}\frac{{\sin}^{4}\theta}{{|\vec{k}-\vec{q}|}^{3}}\left(1+{x^{*}}^{2}\right)
\nonumber\\
&\times& \left(\frac{k^{4}}{x^{*2}}-\frac{2k^{2}c}{g}-\frac{2k^{2}bc}{g^{3}}+\frac{k^{2}c^{2}}{g^{4}}+\frac{k^{2}b^{2}}{g^{2}}+\frac{x^{*2}c^{2}}{g^{2}}\right).
\eea
\noindent Similarly to what we have done in part A, one can check that there are no ultraviolet singularities in the remaining five integrals although some infrared logarithmic contributions are still present and the final result is
\bea\label{appC2}
\langle\df_{\vec{k_{1}}}(\eta^{*})\df_{\vec{k_{2}}}(\eta^{*})\rangle_{(1L,2v)}^{C}=\pi\delta^{(3)}(\vec{k_{1}}+\vec{k_{2}})\frac{H_{*}^{4}}{k^{3}}
\left(c_{1}+c_{2}\ln(k \ell)\right)\, , \\
\nonumber
\eea
where $c_{1}=(1/225)\left(8/x^{*2}+107+50x^{*2}\right)$ and $c_{2}=(16/15x^{*2})+(4/15)$. Notice that the $(x^{*})^{-2}$ coefficients in $c_1$ and $c_{2}$ exactly cancels the $(x^{*})^{-2}$ coefficients in $a_{2}$ and $a_3$. This is not surprising: based on \cite{sloth,seery2}, we expect we might observe a logarithmic singularity if we push $x^{*} \rightarrow 0$ in our results (which is indeed present in the Ci(2$x^{*}$) term of $a_{2}$), but no power-law singularities are actually expected.

\newpage

\section{Complete expression for $P_{\zeta}$ at one loop}

\setcounter{equation}{0}
\def\theequation{6.\arabic{equation}}

Let us now collect our results in the final formula for the power spectrum of the curvature perturbation $\zeta$ computed up to one-loop level. 
This can be derived from Eq.~(\ref{ZZ}), which follows from the $\delta N$ formula. Summing the main results of the previous section, Eqs. (\ref{FR1}), (\ref{appC1}) and (\ref{appC2}), we obtain the one-loop graviton correction to the inflaton power spectrum 

\be\label{FR}\fl
\langle\df_{\vec{k_{1}}}(\eta^{*})\df_{\vec{k_{2}}}(\eta^{*})\rangle_{1L}=\pi\delta^{(3)}(\vec{k_{1}}+\vec{k_{2}})\frac{H_{*}^{4}}{k^{3}}
\left[f_{1}\ln(k)+f_{2}\ln(k \ell)+f_{3}\right]\, ,
\ee
where 
\bea
f_{1}=-\frac{2}{15}\left(25+15x^{*2}+4x^{*4}\right)\, ,
\eea
\bea
f_{2} = 2+(2/3)x^{*2}+a_{2}+c_{2}\, ,
\eea
and $f_{3}$ is given by a left-over scheme-dependent renormalization constant plus contributions of order 
${\cal O}(1)$ (see Appendix~\ref{powerloops} for the complete expressions of $a_{2}$, $c_{2}$ and $f_3$). If we calculate the two point funtion of $\df$ a few e-foldings after horizon crossing, i.e. $x^{*}$ ranges for example between $10^{-1}$ and $10^{-2}$, $f_{1}$ reduces to a negative constant of order ${\cal O}(1)$ and $f_{2}\sim 4\left(1-{\rm Ci}(2x^{*})\right) \sim {\cal O}(10)$. In the limit where 
$x^{*} \rightarrow 1$ both $f_{1}$ and $f_2$ turn out to be of order unity. \\  
\noindent In order to understand which is the dominant contribution in Eq.~(\ref{QQQ}) and how big it is, one needs to 
(i) know the slow-roll order of the coefficients $N^{(i)}$: $N^{(1)} \sim {\epsilon}^{-1/2}$, $N^{(2)} \sim \epsilon^{0}$, 
$N^{(3)} \sim \epsilon^{1/2}$; 
(ii) compute the integrals involving the power spetrum $P(q)$. 
This is discussed in details in Ref.~\cite{seery2} (for the case of scalar perturbations only), see in particular Sec IV of~\cite{seery2}. 
It turns out that the crucial quantity is represented by the number of 
e-foldings of inflation between the times of 
horizon exit of the mode $\ell^{-1}$, which corresponds to the infrared cutoff, and the time of horizon exit of the mode $k$ we want to observe. 
However, to deal with observable quantitites one has to choose $\ell$ not much bigger than the present cosmological horizon $H_0^{-1}$~\cite{box4,box5}.\\
The relevant point about Eq. (\ref{FR}) is that it gives in Eq.~(\ref{QQQ}) a contribution which is of the same order of magnitude as 
those coming from loops which accounts for scalar perturbations only. Since in terms of the slow-roll parameters 
$\left( N^{(1)} \right)^2 \sim {\epsilon_*}^{-1}$ the magnitude of the one-loop graviton correction turns out to be
\begin{equation}
\label{fin}
\Delta P_{\zeta}^{1\rm{loop}}(k) \sim \frac{2 \pi^2}{k^3} \alpha(k) \frac{1}{\epsilon_*}{\cal P}^2_*(k)\, , 
\end{equation}  
where we have used Eq.~(\ref{tree}) for the power spectrum of the inflaton field. In Eq.~(\ref{fin}) 
$\alpha(k)$ includes the various coefficients of Eq.~(\ref{FR}), and it is ${\cal O}(1)$. 
Eq.~(\ref{fin}) 
allows a more direct comparison with the results of Ref.~\cite{seery2}, showing that the graviton contributions to the one-loop 
corrections are comparable to the ones computed only from scalar interactions. Notice that also for the tensor contributions we find terms of the form 
$\ln(k)$. \\

\noindent Summarizing our work: beyond linear order, the tensor perturbation modes produced during inflation unavoidably mix with scalar modes; this fact alone would require to include 
the tensor modes for a self-consistent computation. 
Most importantly, despite a naive expectation suggested by the fact that the power spectrum of the tensor modes is suppressed 
on large scales with respect to that of the curvature (scalar) perturbations, 
our results show explicitly that their inclusion is necessary since their contribution is not at all negligible with respect to the
loop corrections arising from interactions involving the inflaton field only.

\newpage

\section{Study of perturbations in anisotropic cosmologies}

\setcounter{equation}{0}
\def\theequation{7.\arabic{equation}}

The Bianchi models represent a classification of all homogeneous and anisotropic cosmologies. In the cosmic no-hair conjecture, any initial background, in the presence of a positive cosmological constant, eventually evolves in a de Sitter universe, where all the initial existing anisotropies are rapidly washed out. This conjecture was proven to be true for all of the Bianchi models except for the Bianchi-IX by Wald in \cite{Wald:1983ky}. For the Bianchi-IX this result is also true under the assumption that the cosmological constant overcomes the spatial curvature terms. More recently, there has been a revived interest in homogeneous but anisotropic models of the early Universe and several attempts to understand what kind of observational signatures these models might produce. A close look has in fact been given to the fluctuations in the metric and in the energy tensor developing during a hypotetical anisotropic stage either in a pre-inflationary epoch \cite{Dimastrogiovanni:2008ua,Gumrukcuoglu:2008gi} or during inflation itself \cite{Pereira:2007yy,Pitrou:2008gk,Watanabe:2009ct}. The Bianchi-I model is the simplest of all anisotropic and homogenous models and has often been the choice for such kind of studies. In the Bianchi-I model, the metric has a form
\bea\label{bianchi1}
ds^2=dt^2-e^{2\alpha}\left(e^{2\beta}\right)_{ij}dx^{i}dx^{j}
\eea
where $\alpha$ is a function of time as well as $\beta_{ij}$. The latter is a $3\times 3$ diagonal traceless matrix that anisotropizes the volume expansion
\bea
\left(e^{2\beta}\right)_{ij}=\delta_{ij}e^{2\beta_{i}},\quad\quad\quad\quad\sum_{i=1}^{3}\beta_{i}=0.
\eea
A metric described by Eq.~(\ref{bianchi1}) but with constant functions $\alpha$ and $\beta$ is often presented in the form
\bea\label{kasner}
ds^2=dt^2-t^{2p_{1}}dx^{2}_{1}+t^{2p_{2}}dx^{2}_{2}+t^{2p_{3}}dx^{2}_{3},
\eea
where the coefficients $p_{1}$, $p_{2}$, $p_{3}$ are numerical constants satisfying the relation
\bea
p_{1}+p_{2}+p_{3}=p_{1}^{2}+p_{2}^{2}+p_{3}^{2}=1.
\eea
The three parameters can be equal in pairs in the cases $(-\frac{1}{3},\frac{2}{3},\frac{2}{3},)$ and $(0,0,1)$. In all other cases they are distinct, one being negative and the other two being positive. The universe described by (\ref{kasner}) is spatially flat and, for any possible value of the coefficients $p_{i}$, its volume element is equal to $\sqrt{-g^{3}}d^3x=td^3x$, where $g^{3}$ is the determinant of the three-metric.\\
In \cite{Dimastrogiovanni:2008ua}, we consider a pre-inflationary epoch characterized by an initially expanding type-I Bianchi universe that evolves with an energy density not yet dominated by a cosmological constant. If by the time the cosmological constant has become the dominant form of energy the metric is still homogeneous, the Universe will eventually enter a de Sitter epoch; the fluctuations of this metric and their growth, however, can play an important role in determining the evolution of the Universe at this early stage. The question we ask concerns the kind of initial conditions that are needed for the Universe to remain homogeneous up to the time the cosmological constant eventually dominates the total energy density. We employ a metric as in Eq.~(\ref{bianchi1}) for the background and solve the first order Einstein equations for the perturbations considering a pressureless fluid. As expected, the anisotropy in the background is responsible for a coupling at the linear order of one of the tensor modes with the density contrast, so there is a correlation between the scalar and the gravitational perturbations. Moreover, the evolutions of the two tensor modes, the ``free'' one and the one that is coupled to the scalar, are very different from each other and very much depend on the scale of interest. It turns out that, for a reasonably large set of initial conditions, the growth of these perturbations could be fast enough so as to lead the Universe into an inhomogeneous state before the cosmological constant becomes the dominant form of energy. \\
A similar scenario was investigated in \cite{Gumrukcuoglu:2008gi}, where both expanding and contracting Bianchi-I Universes are studied during a pre-inflationary era. A scalar field dominates the matter content. In particular, the evolution of the gravitational perturbations of the metric (not to be confused with the gravitational waves generated from the vacuum fluctuation during inflation) is analysed, to point out that they may grow to significantly alter the geometry of spacetime before inflation begins and, in any case, they may leave an imprint on the observed CMB power spectrum. \\
Also in \cite{Pereira:2007yy,Pitrou:2008gk} a perturbative analysis was performed for Bianchi-I cosmologies leading to anisotropic inflation; in addition to that, predictions for the power spectrum of curvature fluctuations and gravity waves produced during inflation are provided for this models. Finally, a ``vector hair'' model and its observational consequences in terms of anisotropic signatures in the cosmological correlation functions and primordial gravitational waves, were recently discussed in \cite{Watanabe:2009ct}.\\

\newpage
\section{Inflation and primordial vector fields}\label{introvec}
\setcounter{equation}{0}
\def\theequation{8.\arabic{equation}}

The attempt to explain some of the CMB ``anomalous'' features as the indication of a break of statistical isotropy is the main reason behind ours and many of the existing inflationary models populated by vector fields, but not the only one. The first one of these models \cite{Ford:1989me} was formulated with the goal of producing inflation by the action of vector fields, without having to invoke the existence of a scalar field. The same motivations inspired the works that followed \cite{Golovnev:2008cf,Golovnev:2008hv,Golovnev:2009ks}. Lately, models where primordial vector fields can leave an imprint on the CMB have been formulated as an alternative to the basic inflationary scenario, in the search for interesting non-Gaussianity predictions \cite{Yokoyama:2008xw,ValenzuelaToledo:2009af,Dimopoulos:2008yv,ValenzuelaToledo:2009nq,Karciauskas:2008bc,Dimopoulos:2009am,Dimopoulos:2009vu,Bartolo:2009pa,Bartolo:2009kg}. Finally, vector fields models of dark energy have been proposed \cite{ArmendarizPicon:2004pm,Boehmer:2007qa,Koivisto:2007bp,Koivisto:2008xf,Jimenez:2009py,Jimenez:2009zza}. All this appears to us as a rich bag of motivations for investigating these scenarios.\\
Before we quickly sketch some of them and list the results so far achieved in this direction, it is important to briefly indicate and explain the main issues and difficulties that these models have been facing. We will also shortly discuss the mechanisms of production of the curvature fluctuations in these models. \\
 
\noindent Building a model where primordial vector fields can drive inflation and/or produce the observed spectrum of large scale fluctuations requires a more complex Lagrangian than the basic gauge invariant $\mathcal{L}_{vector}=-(\sqrt{-g}/4) F^{\mu\nu}F_{\mu\nu}$. In fact, for a conformally invariant theory as the one described by $\mathcal{L}_{vector}$, vector fields fluctuations are not excited on superhorizon scales. It is then necessary to modify the Lagrangian. For some of the existing models, these modifications have been done to the expense of destabilizing the theory, by ``switching on'' unphysical degrees of freedom. This was pointed out in \cite{Himmetoglu:2008zp,Himmetoglu:2008hx,Himmetoglu:2009qi}, where a large variety of vector field models was analyzed in which longitudinal polarization modes exist that are endowed with negative squared masses (the ``wrong'' signs of the masses are imposed for the theory to satisfy the constraints that allow a suitable background evolution). It turnes out that, in a range of interest of the theory, these fields acquire negative total energy, i.e. behave like ``ghosts'', the presence of which is known to be responsible for an unstable vacuum. A related problem for some of these theories is represented by the existence of instabilities affecting the equations of motion of the ghost fields \cite{Himmetoglu:2008zp,Himmetoglu:2008hx,Himmetoglu:2009qi}.\\
In the remaing part of this section, we are going to present some of these models together with some recent attempts to overcome their limits. \\

\noindent In all of the models we will consider, primordial vector fields fluctuations end up either being entirely responsible for or only partially contributing to the curvature fluctuations at late times. This can happen through different mechanisms. If the vector fields affects the universe expansion during inflation, its contribution $\zeta_{A}$ to the total $\zeta$ can be derived from combining the definition of the number of e-foldings ($N=\int H dt$) with the Einstein equation ($H^2=(8 \pi G/3)(\rho_{\phi}+\rho_{A})$, $\rho_{A}$ being the energy density of the vector field and $\rho_{\phi}$ the inflaton energy density) and using the $\delta$N expansion of the curvature fluctuation in terms of both the inflaton and the vector fields fluctuations (see Sec.~\ref{cf}). To lowest order we have \cite{Dimopoulos:2008yv}
\bea\label{direct}
\zeta_{A}=\frac{A_{i}}{2 m_{P}^2}\delta A_{i},
\eea
where a single vector field has been taken into account for simplicity ($m_{P}$ is the reduced Planck mass, $A$ is the background value of the field and $\delta A$ its perturbation). When calculating the amplitude of non-Gaussianity in Sec.~\ref{amplitude}, we will refer to this case as ``vector inflation'' for simplicity.\\
A different fluctuation production process is the curvaton mechanism which was initially formulated for scalar theories but it is also applicable to vectors \cite{Dimopoulos:2006ms,Dimopoulos:2008rf}. Specifically, inflation is driven by a scalar field, whereas the curvaton field(s) (now played by the vectors), has a very small (compared to the Hubble rate) mass during inflation. Towards the end of the inflationary epoch, the Hubble rate value starts decreasing until it equates the vector mass; when this eventually happens, the curvaton begins to oscillate and it will then dissipate its energy into radiation. The curvaton becomes responsible for a fraction of the total curvature fluctuation that is proportional to a parameter, $r$, related to the ratio between the curvaton energy density and the total energy density of the universe at the epoch of the curvaton decay~\cite{Dimopoulos:2008yv} 
\bea\label{indirect}
\zeta_{A}=\frac{r}{3}\frac{\delta \rho_{A}}{\rho_{A}},
\eea
where $r\equiv 3\rho_{A}/(3\rho_{A}+4\rho_{\phi})$. Anisotropy bounds on the power spectrum favour small values of $r$.\\
From Eqs.~(\ref{direct}) and (\ref{indirect}) we can see that, dependending on which one of these two mechanisms of production of the curvature fluctuations is considered, different coefficients will result in the $\delta$N expansion (see Eq.~(\ref{expansion})).\\

\noindent In this section we will describe both models where inflation is intended to be vector-field driven and those models in which, instead, the role of the inflaton is played by a scalar field, whereas the energy of the vector is a subdominant contribution to the total energy density of the universe during the entire inflationary phase.

\subsection{Self-coupled vector field models}

A pioneer work on vector field driven inflation was formulated by L.~H.~Ford \cite{Ford:1989me}, who considered a single self-coupled field $A_{\mu}$ with a Lagrangian
\bea
L_{vector}=-\frac{1}{4}F_{\mu\nu}F^{\mu\nu}+V(\psi)
\eea
where $F_{\mu\nu}\equiv \p_{\mu}B_{\nu}-\p_{\nu}B_{\mu}$ and the potential $V$ is a function of $\psi\equiv B_{\alpha}B^{\alpha}$. Different scenarios of expansion are analyzed by the author for different functions $V$. The universe expands anisotropically at the end of the inflationary era and this anisotropy either survives until late times or is damped out depending on the shape and the location of the minima of the potential.\\

\noindent The study of perturbations in a similar model was proposed by Dimopoulos in \cite{Dimopoulos:2006ms} where he showed that for a Lagrangian
\bea\label{dimopoulos}
L_{vector}=-\frac{1}{4}F_{\mu\nu}F^{\mu\nu}+\frac{1}{2}m^2 B_{\mu}B^{\mu}
\eea
and for $m^{2}\simeq -2H^2$, the transverse mode of the vector field is governed by the same equation of motion as a light scalar field in a de Sitter stage. A suitable superhorizon power spectrum of fluctuations could therefore arise. In order to prevent production of large scale anisotropy, in this model the vector field plays the role of the curvaton while inflation is driven by a scalar field. 

\subsection{Vector-field coupled to gravity}

The Lagrangian in Eq.~(\ref{dimopoulos}) may be also intended, at least during inflation, as including a non-minimal coupling of the vector field to gravity; indeed the mass term can be rewritten as
\bea\label{Lag}
L_{vector}\supset\frac{1}{2}\left(m^2_{0}+\xi R\right)B_{\mu}B^{\mu}
\eea
where, for the whole duration of the inflationary era, the bare mass $m_{0}$ is assumed to be much smaller than the Hubble rate and the Ricci scalar $R=-6\left[\frac{\ddot{a}}{a}+\left(\frac{\dot{a}}{a}\right)^2\right]$ can be approximated as $R\simeq-12 H^2$. For the specific value $\xi=1/6$, Eq.~(\ref{dimopoulos}) is retrieved.\\

\noindent For the Lagrangian just presented, Golovnev et al \cite{Golovnev:2008cf} proved that the problem of excessive anisotropy production in the case where inflation is driven by vector fields can be avoided if either a triplet of mutually orthogonal or a large number $N$ of randomly oriented vector fields is considered.\\

\noindent The Lagrangian (\ref{Lag}) with $\xi=1/6$ was also employed in \cite{Dimopoulos:2008rf}, where inflation is scalar-field-driven and a primordial vector field affects large-scale curvature fluctuations and, similarly, in \cite{Kanno:2008gn}, which includes a study of the backreaction of the vector field on the dynamics of expansion, by introducing a Bianchi type-I metric.

\subsection{Ackerman-Carroll-Wise (ACW) model}

A model was proposed in \cite{Ackerman:2007nb} where Lagrange multipliers ($\lambda$) are employed to determine a fixed norm primordial vector field $B_{\mu}B^{\mu}=m^2$
\bea\label{Acw}
L_{vector}\supset \lambda\left(B^{\mu}B_{\mu}-m^2\right)-\rho_{\Lambda}
\eea 
where $\rho_{\Lambda}$ is a vacuum energy. The expansion rate in this scenario is anisotropic: if we orient the $x$-axis of the spatial frame along the direction determined by the vector field, we find two different Hubble rates: along the $x$-direction it is equal to 
\bea
H_{b}^{2}=\frac{\rho_{\Lambda}}{m_{P}^{2}}\frac{1}{P(\mu)}, 
\eea
and it is given by $H_{a}=(1+c\mu^2)H_{b}$ along the orthogonal directions; $\mu\equiv m/m_{P}$, $P$ is a polynomial function of $\mu$ and $c$ is a parameter appearing in the kinetic part of the Lagrangian that we omitted in (\ref{Acw}) (see \cite{Ackerman:2007nb} for its complete expression). As expected, an isotropic expansion is recovered if the vev of the vector field is set to zero.

\subsection{Models with varying gauge coupling}

Most of the models mentioned so far successfully solve the problem of attaining a slow-roll regime for the vector-fields without imposing too many restrictions on the parameters of the theory and of avoiding excessive production of anisotropy at late times. None of them though escapes those instabilities related to the negative energy of the longitudinal modes (although a study of the instabilities for fixed-norm field models was done in \cite{Carroll:2009em} where some stable cases with non-canonical kinetic terms were found). As discussed in \cite{Himmetoglu:2008zp,Himmetoglu:2008hx,Himmetoglu:2009qi}, in the self-coupled model a ghost appears at small (compared to the horizon) wavelengths; in the non-minimally coupled and in the fixed-norm cases instead the instability concerns the region around horizon crossing.\\  

\noindent Models with varying gauge coupling can overcome the problem of instabilities and have recently attracted quite some attention. In \cite{Yokoyama:2008xw}, the authors consider a model of hybrid inflation \cite{Lyth:2005qk,Alabidi:2006wa,Salem:2005nd,Bernardeau:2004zz} with the introduction of a massless vector field
\bea
L\supset \frac{1}{2}\left(\p_{\mu}\phi\p^{\mu}\phi+\p_{\mu}\chi\p^{\mu}\chi\right)-\frac{1}{4}f^{2}(\phi)F_{\mu\nu}F_{\mu\nu}+V(\phi,\chi,B_{\mu})
\eea
where $\phi$ is the inflaton and $\chi$ is the so-called ``waterfall'' field. The potential $V$ is chosen in such a way as to preserve gauge invariance; this way the longitudinal mode disappears and instabilities are avoided.\\ 
Similarly, Kanno et al \cite{Watanabe:2009ct} consider a vector field Lagrangian of the type
\bea\label{}
L_{vector}=-\frac{1}{4}f^{2}(\phi)F^{\mu\nu}F_{\mu\nu},
\eea
but in a basic scalar field driven inflation model. Very recently, in \cite{Himmetoglu:2009mk,DG} the linear perturbations in these kind of models have been investigated.\\
Finally, in \cite{Dimopoulos:2008yv,Dimopoulos:2009vu} varying mass vector field models have been introduced
\bea\label{var}
L_{vector}=-\frac{1}{4}f^{2}(\phi)F^{\mu\nu}F_{\mu\nu}+\frac{1}{2}m^{2}B_{\mu}B^{\mu},
\eea
where $f\simeq a^{\alpha}$ and $m\simeq a$ ($a$ is the scale factor and $\alpha$ is a numerical coefficient). The special cases $\alpha=1$ and $\alpha=-2$ are of special interest. In fact, introducing the fields $\tilde{A}_{\mu}$ and $A_{\mu}$, related to one another by $\tilde{A}_{\mu}\equiv f B_{\mu}=a A_{\mu}$ ($\tilde{A}_{\mu}$ and $A_{\mu}$ are respectively the comoving and the physical vectors), it is possible to verify that the physical gauge fields are governed by the same equations of motion as a light scalar field in a de Sitter background. Vector fields in this theory can then generate the observed (almost) scale invariant primordial power spectrum.

\subsection{$SU(2)$ vector model}\label{ourmodel}

Let us consider some models where inflation is driven by a scalar field in the presence of an $SU(2)$ vector multiplet \cite{Bartolo:2009pa,Bartolo:2009kg}. A fairly general Lagrangian can be the following
\bea\label{ac}\fl
S=\int d^{4}x \sqrt{-g}\left[\frac{m_{P}^{2}R}{2}-\frac{f^{2}(\phi)}{4}g^{\mu\al}g^{\nu\b}\sum_{a=1,2,3}F_{\mu\nu}^{a}F_{\al\b}^{a}-\frac{M^2}{2}g^{\mu\nu}\sum_{a=1,2,3}B_{\mu}^{a}B_{\nu}^{a}+L_{\phi}\right],
\eea
where $L_{\phi}$ is the Lagrangian of the scalar field and $F_{\mu\nu}^{a}\equiv\p_{\mu}B^{a}_{\nu}-\p_{\nu}B^{a}_{\mu}+g_{c}\ep^{abc}B^{b}_{\mu}B^{c}_{\nu}$ ($g_{c}$ is the $SU(2)$ gauge coupling). Both $f$ and the effective mass $M$ can be viewed as generic functions of time. The fields $B_{\mu}^{a}$ are comoving and related to the physical fields by $A_{\mu}^{a}=\left(B_{0}^{a},B_{i}^{a}/a\right)$. The free field operators can be Fourier expanded in their creation and annihilation operators
\bea\label{ip}
\fl
\delta A_{i}^{a}(\vec{x},\eta)=\int \frac{d^{3}q}{(2\pi)^{3}}e^{i\vec{q}\cdot\vec{x}}\sum_{\lambda=L,R,long}\Big[e^{\lambda}_{i}(\hat{q})a_{\vec{q}}^{a,\lambda}\delta A_{\lambda}^{a}(q,\eta)+e^{*\lambda}_{i}(-\hat{q})\left(a_{-\vec{q}}^{a,\lambda}\right)^{\dagger}\delta A_{\lambda}^{*a}(q,\eta)\Big],
\eea
where the polarization index $\lambda$ runs over left ($L$), right ($R$) and longitudinal ($long$) modes and
\bea
\left[a^{a,\lambda}_{\vec{k}},(a_{\vec{k}^{'}}^{a^{'},\lambda^{'}})^{\dagger}\right]=(2 \pi)^3\delta_{a,a^{'}}\delta_{\lambda,\lambda^{'}}\delta^{(3)}(\vec{k}-\vec{k}^{'}). 
\eea
Here $\eta$ the conformal time ($d\eta=dt/a(t)$). Once the functional forms of $f$ and $M$ have been specified, the equations of motion for the vector bosons can be written. We compute cosmological correlation functions up to fourth order considering an action as in (\ref{ac}). The expression of the correlators that we derive, prior to explicitating the wavefunction for the gauge bosons, apply to any $SU(2)$ theory with an action as in (\ref{ac}), both for what we will call the ``Abelian'' and for the ``non-Abelian'' contributions. In particular, the structure of the interaction Hamiltonian is independent of the functional dependence of $f$ and $M$ and determines the general form of and the anisotropy coefficients appearing in the final ``non-Abelian'' expressions (see Sec.~\ref{cf}). When it comes to explicitate the wavefunctions, a choice that can help keeping the result as easy to generalize as possible is the following 
\bea\label{T}
\delta B^{T}=-\frac{\sqrt{\pi x}}{2\sqrt{k}}\left[J_{3/2}(x)+iJ_{-3/2}(x)\right],
\eea
for the transverse mode and
\bea\label{L}
\delta B^{||}=n(x)\delta B^{T}\, ,
\eea
for the longitudinal mode ($n$ is a unknown function of $x\equiv -k \eta$) \cite{Bartolo:2009pa,Bartolo:2009kg}. Let us see why. As previously stated, for $f \simeq a^{\alpha}$ and with $\alpha=0,1,-2$, it is possible to verify that the (physical) transverse mode behaves exactly like a light scalar field in a de Sitter background (see also Appendix~\ref{eomotion}). Considering the solution (\ref{T}) then takes into account at least these special cases. As to the longitudinal mode, a parametrization was adopted as in (\ref{L}) in order to keep the analysis more general and given that, because of the instability issues, introducing this degree of freedom into the theory requires special attention. We are going to keep the longitudinal mode ``alive'' in the calculations we present, by considering a nonzero function $n(x)$, and focus on the simplest case of $f=1$. This case is known to be affected by quantum instabilities in the longitudinal mode, anyway we choose $f=1$ for the sake of simplicity in our presentation. The results can be easily generalized to gauge invariant models (please refer to Sec.~\ref{gen} for a sample generalization of some of the calculations to massless $f\simeq a^{(1,-2)}$ models). 

\newpage

\section{Correlation functions of $\zeta$ in the $SU(2)$ model}
\label{cf}
\setcounter{equation}{0}
\def\theequation{9.\arabic{equation}}

We are now ready to compute the power spectrum, bispectrum and trispectrum for the curvature fluctuations $\zeta$ generated during inflation
\bea\label{ps}
\langle\zeta_{\vec{k}_{{1}}}\zeta_{\vec{k}_{{2}}}\rangle=(2 \pi)^3\delta^{(3)}(\vec{k_{1}}+\vec{k_{2}})P_{\zeta}(\vec{k}),
\\
\label{bisp}
\langle\zeta_{\vec{k}_{{1}}}\zeta_{\vec{k}_{{2}}}\zeta_{\vec{k}_{{3}}} \rangle=(2 \pi)^3\delta^{(3)}(\vec{k_{1}}+\vec{k_{2}}+\vec{k_{3}})B_{\zeta}(\vec{k_{1}},\vec{k_{2}},\vec{k_{3}})\\
\label{trisp}
\langle\zeta_{\vec{k}_{1}}\zeta_{\vec{k}_{2}}\zeta_{\vec{k}_{3}}\zeta_{\vec{k}_{4}} \rangle =(2 \pi)^3\delta^{(3)}(\vec{k}_{1}+\vec{k}_{2}+\vec{k}_{3}+\vec{k}_{4})T_{\zeta}(\vec{k}_{1},\vec{k}_{2},\vec{k}_{3},\vec{k}_{4}).
\eea
Notice that, on the right-hand side of (\ref{ps}) through (\ref{trisp}), we indicated a dependence from the direction of the wavevectors; in models of inflation where isotropy is preserved, the power spectrum and the bispectrum only depend on the moduli of the wave vectors. This will not be the case for the $SU(2)$ model.\\
\noindent The $\delta$N formula (\ref{first}) will be applied to our inflaton$+SU(2)$vector model
\bea
\label{expansion}\fl
\zeta(\vec{x},t)&=& N_{\phi}\df+N^{\mu}_{a}\delta A_{\mu}^{a}+\frac{1}{2}N_{\phi\phi}\left(\df\right)^2+\frac{1}{2}N^{\mu\nu}_{ab}\delta A_{\mu}^{a}\delta A_{\nu}^{b}+N_{\phi a}^{\mu}\df\delta A_{\mu}^{a}\nonumber\\\fl&+&\frac{1}{3!}N_{\phi\phi\phi}(\df)^3+\frac{1}{3!}N_{abc}^{\mu\nu\lambda}\de A_{\mu}^{a} \de A_{\nu}^{b} \de A_{\lambda}^{c}+\frac{1}{2}N_{\phi\phi a}^{\mu}(\df)^2\de A_{\mu}^{a}+\frac{1}{2}N_{\phi ab}^{\mu\nu}\df\de A_{\mu}^{a}\de A_{\nu}^{b}\nonumber\\\fl&+&\frac{1}{3!}N_{\phi\phi\phi\phi}(\df)^4+\frac{1}{3!}N_{abcd}^{\mu\nu\lambda\eta}\de A_{\mu}^{a}\de A_{\nu}^{b}\de A_{\lambda}^{c}\de A_{\eta}^{d}+...,
\eea
where now
\bea\label{NNN}
N_{\phi}\equiv \left(\frac{\p N}{\p \phi}\right)_{t^{*}},\,\,\,\,\,
N^{\mu}_{a}\equiv\left(\frac{\p N}{\p A^{a}_{\mu}}\right)_{t^{*}},\,\,\,\,\,N_{\phi a}^{\mu}\equiv\left(\frac{\p^{2} N}{\p \phi\p A^{a}_{\mu}}\right)_{t^{*}}
\eea
and so on for higher order derivatives.\\ Our plan is to show the derivation the correlation functions of $\zeta$ from the ones of $\df$ and $\delta A_{i}^{a}$, after a replacement of the $\delta$N expansion (\ref{expansion}) in Eqs.~(\ref{ps}) through (\ref{trisp}). \\
The correlation functions can be evaluated using the Schwinger-Keldysh formula (\ref{skfirst}), that we recall here for convenience
\bea\label{sk}
\langle\Omega|\Theta(t)|\Omega\rangle=\left\langle 0\left|\left[\bar{T}\left(e^{i {\int}^{t}_{0}H_{I}(t')dt'}\right)\right]\Theta_{I}(t)\left[T \left(e^{-i {\int}^{t}_{0}H_{I}(t')dt'}\right)\right]\right|0\right\rangle.
\eea

\noindent When calculating the spectra of $\zeta$, the perturbative expansions in Eq.~(\ref{expansion}) and (\ref{sk}) will be carried out to only include tree-level contributions, neglecting higher order ``loop'' terms, either classical, i.e. from the $\delta$N series, or of quantum origin, i.e. from the Schwinger-Keldysh series. Assuming that the $SU(2)$ coupling $g_{c}$ is ``small'' and that we are dealing with ``small'' fluctuations in the fields and given the fact that a slow-roll regime is being assumed, it turns out that it is indeed safe for the two expansions to be truncated at tree-level.\\ 

\noindent The correlation functions of $\zeta$ will then result as the sum of scalar, vector and (scalar and vector) mixed contributions. As to the vector part, this will be made up of terms that are merely generated by the $\delta$N expansion, i.e. they only include the zeroth order of the in-in formula (we call these terms ``Abelian'', being them retrievable in the $U(1)$ case), and by (``non-Abelian'') terms arising from the Schwinger-Keldysh operator expansion beyond zeroth order, i.e. from the gauge fields self-interactions.\\

\noindent Let us now discuss the level of generality of the results we will present in the next sections, w.r.t. the choice of a specific Lagrangian.\\ The expression for the Abelian contributions provided in Secs.~\ref{powerspectrum} and \ref{abelianc} apply to any $SU(2)$ model of gauge interactions with no direct coupling between scalar and vector fields (extra terms would be otherwise needed in Eqs.~(\ref{class1}) and (\ref{class2})). The next stage in the Abelian contributions computation would be to explicitate the derivatives of the e-foldings number and the wavefunctions of the fields: they both depend on the equations of motion of the system, therefore the fixing of a specific model is required at this point.\\
As to the non-Abelian contributions, the results in Eqs.~(\ref{class3}) and (\ref{class4}) are completely general except for assuming, again, that no direct vector-scalar field coupling exists. The structure of Eqs.~(\ref{fire}) and (\ref{fire1}) is instead due to the choice of a non-Abelian gauge group. The expressions of the anisotropy coefficients $I_{n}$ and $L_{n}$ in Eqs.~(\ref{fire}) and (\ref{fire1}) depend on the specific non-Abelian gauge group (for $SU(2)$ one of the $I_{n}$ is given in Eq.~(\ref{sample2})). Finally, the specific expressions of the isotropic functions $F_{n}$ (a sample of which is shown in Eq.~(\ref{sample1})) and $G_{n}$ were derived considering the Lagrangian (\ref{ac}) with $f=1$ and the eigenfunctions for the vector bosons provided in Eqs.~(\ref{T}) and (\ref{L}).

\subsection{The power spectrum}\label{powerspectrum}

The power spectrum of $\zeta$ can be straightforwardly derived at tree-level, using the $\delta$N expansion (\ref{expansion}), from the inflaton and the vector fields power spectra
\bea\label{power-zeta}
P_{\zeta}(\vec{k})&=&P^{iso}(k)\left[1+g^{ab}\left(\hat{k}\cdot\hat{N}_{a}\right)\left(\hat{k}\cdot\hat{N}_{b}\right)+is^{ab}\hat{k}\cdot\left(\hat{N}_{a}\times\hat{N}_{b}\right)\right].
\eea
The isotropic part of the previous expression has been factorized in 
\bea\label{piso}
P^{iso}(k)\equiv N^{2}_{\phi}P_{\phi}(k)+\left(\vec{N}_{c}\cdot\vec{N}_{d}\right)P^{cd}_{+},
\eea
where we have defined the following combinations
\bea
P_{\pm}^{ab}\equiv (1/2)(P_{R}^{ab}\pm P_{L}^{ab}),
\eea 
from the power spectra for the right, left and longitudinal polarization modes
\bea
P_{R}^{ab}&\equiv& \delta_{ab}\delta A_{R}^{a}(k,t^{*})\delta A_{R}^{b*}(k,t^{*}),\\
P_{L}^{ab}&\equiv& \delta_{ab}\delta A_{L}^{a}(k,t^{*})\delta A_{L}^{b*}(k,t^{*}),\\
P_{long}^{ab}&\equiv& \delta_{ab}\delta A_{long}^{a}(k,t^{*})\delta A_{long}^{b*}(k,t^{*}).
\eea
The anisotropic parts are weighted by the coefficients
\bea
g^{ab}\equiv \frac{N^{a}N^{b}\left(P^{ab}_{long}-P^{ab}_{+}\right)}{N^{2}_{\phi}P_{\phi}+\left(\vec{N}_{c}\cdot\vec{N}_{d}\right)P^{cd}_{+}},\\
s^{ab}\equiv \frac{N^{a}N^{b}P^{ab}_{-}}{N^{2}_{\phi}P_{\phi}+\left(\vec{N}_{c}\cdot\vec{N}_{d}\right)P^{cd}_{+}},
\eea
(where a sum is intended over indices $c$ and $d$ but not over $a$ and $b$). Eq.~(\ref{piso}) can also be written as
\bea
P^{iso}(k)=N^{2}_{\phi}P_{\phi}\left[1+\beta_{cd}\frac{P^{cd}_{+}}{P_{\phi}}\right],
\eea
after introducing the parameter
\bea\label{beta}
\beta_{cd}\equiv \frac{\vec{N}_{c}\cdot\vec{N}_{d}}{N^{2}_{\phi}}.
\eea
Notice that what when we say ``isotropic'', as far as the expression for the power spectrum is concerned, we simply mean ``independent'' of the direction of the wave vector. In this case instead, the vector bosons introduce three preferred spatial directions: the r.h.s. of Eq.~(\ref{power-zeta}) depends on their orientation w.r.t. the wave vector.\\

\noindent As expected, the coefficients $g^{ab}$ and $s^{ab}$ that weight the anisotropic part of the power spectrum are related to $\beta_{cd}$, i.e. to the parameters that quantify how much the expansion of the universe is affected by the vector bosons compared to the scalar field.\\
Assuming no parity violation in the model, we have $s^{ab}=0$; the parameters $g^{ab}$ and $\beta_{ab}$ are instead unconstrained. In the $U(1)$ case and for parity conserving theories, Eq.~(\ref{power-zeta}) reduces to \cite{Dimopoulos:2008yv}
\bea\label{pdisc}
P_{\zeta}(\vec{k})=P^{iso}(k)\left[1+g\left(\hat{k}\cdot\hat{n}\right)\right]
\eea
where $\hat{n}$ indicates the preferred spatial direction; also one can check that in this simple case, if $P_{+}\simeq P_{\phi}$ and $P_{long}=k P_{+}$ ($k\neq 1$), the relation $g=(k-1)\beta/(1+\beta)$ holds, where $\beta\equiv \left(N_{A}/N_{\phi}\right)^2$ (the anisotropy coefficient $g$ is not to be confused with the $SU(2)$ coupling constant $g_{c}$). If it is safe to assume $|g| \ll 1$ (see discussion following Eq.~(\ref{disc}) and references \cite{Groeneboom:2009cb,Hanson:2009gu}), a similar upper bound can also be placed on $\beta$.  \\
In the case where more than one special directions exists, as in the $SU(2)$ model, no such analysis on the anisotropy data has been so far carried out, the $g^{a}$ parameters cannot then be constrained, unless assuming that the three directions converge into a single one; in that case a constraint could be placed on the sum $|g|\equiv |\sum_{a}g^{a}|$, where $a=1,2,3$ and $P_{\zeta}(\vec{k})=P^{iso}_{\zeta}(k)\left[1+g^{a}\left(\hat{k}\cdot\hat{n}_{a}\right)\right]$.\\

\noindent In the next sections we will present the results for the tree-level contributions to the bispectrum and to the trispectrum of $\zeta$. \\
These can be classified in two cathegories, that we indicate as ``Abelian'' and ``non-Abelian''. The former are intended as terms that merely arise from the $\delta$N expansion and are thus retrievable in the Abelian case; the latter are derived from the linear and quadratic expansions (in terms of the gauge bosons interaction Hamiltonian) of the Schwinger-Keldysh formula and are therefore peculiar to the non-Abelian case.

\subsection{Bispectrum and trispectrum: Abelian contributions}\label{abelianc}

By plugging the $\delta$N expansion (\ref{expansion}) in Eqs.~(\ref{bisp}) and (\ref{trisp}), we have
\bea\label{class1}\fl
B_{\zeta}(\vec{k_{1}},\vec{k_{2}},\vec{k_{3}})&\supset&\frac{1}{2}N_{\phi}^{2}N_{\phi\phi}\left[P_{\phi}({k}_{1})P_{\phi}({k}_{2})+perms.\right]\nonumber\\\fl&+&\frac{1}{2}N_{a}^{\mu}N_{b}^{\nu}N_{cd}^{\rho\sigma}\left[\Pi^{ac}_{\mu\rho}(\vec{k}_{1})\Pi^{bd}_{\nu\sigma}(\vec{k}_{2})+perms.\right]
\nonumber\\\fl&+&\frac{1}{2}N_{\phi}N_{a}^{\mu}N_{\phi b}^{\nu}\left[P_{\phi}({k}_{1})\Pi^{ab}_{\mu\nu}(\vec{k}_{2})+perms.\right]\nonumber\\\fl&+&N_{\phi}N_{\phi}^{2}B_{\phi}(k_{1},k_{2},k_{3}),
\eea
for the bispectrum and
\bea\label{class2}\fl
T_{\zeta}(\vec{k_{1}},\vec{k_{2}},\vec{k_{3}},\vec{k_{4}})&\supset&N_{\phi}^{4}T_{\phi}(\vec{k}_{1},\vec{k}_{2},\vec{k}_{3},\vec{k}_{4})\nonumber\\\fl&+&N_{\phi}^{3}N_{\phi\phi}\left[P_{\phi}(k_{1})B_{\phi}(|\vec{k}_{1}+\vec{k}_{2}|,k_{3},k_{4})+perms.\right]\nonumber\\&+&N_{\phi}^{2}N_{a}^{\mu}N_{\phi b}^{\nu}\left[P_{\mu\nu}^{ab}(\vec{k}_{3})B_{\phi}^{}(k_{1},k_{2},|\vec{k}_{3}+\vec{k}_{4}|)+perms.\right]\nonumber\\\fl&+&N_{\phi}^{2}N_{\phi\phi}^{2}\left[P_{\phi}(k_{1})P_{\phi}(k_{2})P_{\phi}(|\vec{k}_{1}+\vec{k}_{3}|)+perms.\right]\nonumber\\&+&N_{\phi}^{3}N_{\phi\phi\phi}\left[P_{\phi}({k}_{1})P_{\phi}(k_{2})P_{\phi}(k_{3})+perms.\right]\nonumber\\\fl
&+&N_{\phi}^{2}N_{\phi a}^{\mu}N_{\phi b}^{\nu}\left[P_{\mu\nu}^{ab}(\vec{k}_{1}+\vec{k}_{3})P_{\phi}^{}({k}_{1})P_{\phi}^{}({k}_{2})+perms.\right]\nonumber\\\fl&+&N_{a}^{\mu}N_{b}^{\nu}N_{\phi c}^{\rho}N_{\phi d}^{\sigma}\left[P_{\mu\rho}^{ac}(\vec{k}_{1})P_{\nu\sigma}^{bd}(\vec{k}_{2})P_{\phi}^{}(|\vec{k}_{1}+\vec{k}_{3}|)+perms.\right]\nonumber\\\fl&+&N_{\phi}^{2}N_{a}^{\mu}N_{\phi\phi b}^{\nu}\left[P_{\phi}^{}({k}_{1})P_{\phi}^{}({k}_{2})P_{\mu\nu}^{ab}(\vec{k}_{3})+perms.\right]\nonumber\\\fl&+&N_{\phi}^{}N_{a}^{\mu}N_{b}^{\nu}N_{\phi cd}^{\rho\sigma}\left[P_{\mu\rho}^{ac}(\vec{k}_{1})P_{\nu\sigma}^{bd}(\vec{k}_{2})P_{\phi}^{}({k}_{3})+perms.\right]\nonumber\\\fl&+&N_{\phi\phi}N_{\phi}N_{\phi a}^{\mu}N_{b}^{\nu}\left[P_{\phi}(k_{2})P_{\phi}(|\vec{k}_{1}+\vec{k}_{2}|)P_{\mu\nu}^{ab}(\vec{k}_{4})+perms.\right]\nonumber\\\fl&+&N_{ab}^{\mu\nu}N_{c}^{\rho}N_{\phi d}^{\sigma}N_{\phi}\left[P_{ac}^{\mu\rho}(\vec{k}_{2})P_{bd}^{\nu\sigma}(\vec{k}_{1}+\vec{k}_{2})P_{\phi}(k_{4})+perms.\right]\nonumber\\\fl&+&N_{a}^{\mu}N_{b}^{\nu}N_{cd}^{\rho\sigma}N_{ef}^{\delta\eta}\left[P_{\mu\rho}^{ac}(\vec{k}_{1})P_{\nu\delta}^{be}(\vec{k}_{2})P_{\sigma\eta}^{df}(\vec{k}_{1}+\vec{k}_{3})+perms.\right]\nonumber\\&+&N_{a}^{\mu}N_{b}^{\nu}N_{c}^{\rho}N_{def}^{\sigma\delta\eta}\left[P_{\mu\sigma}^{ad}(\vec{k}_{1})P_{\nu\delta}^{be}(\vec{k}_{2})P_{\rho\eta}^{cf}(\vec{k}_{3})+perms.\right], 
\eea
for the trispectrum.\\ Before we proceed with explicitating these quantities and for the rest of the thesis, the $N^{a}_{0}$ coefficients will be set to zero: it is possible to verify that the temporal mode $B_{0}^{a}=0$ is a solution to the equations of motion for the vector bosons, after slightly restricting the parameter space of the theory (see Appendix~\ref{eomotion}); the adoption of this kind of solutions, which is related to the assumption of a slow-roll regime for the vector fields, implies that the derivatives of $N$ w.r.t. the temporal mode can be set to zero.\\
\noindent Let us now provide some definition for the quantities introduced in (\ref{class1})-(\ref{class2}): we are going to switch from the greek indices $\mu$, $\nu$, ... to the latin ones, generally used for labelling the three spatial directions, in order to stress that all of the vector quantities will be from now on three-dimensional
\bea
\Pi_{ij}^{ab}(\vec{k}) \equiv T^{even}_{ij}(\vec{k})P_{+}^{ab}+i T^{odd}_{ij}(\vec{k})P_{-}^{ab}+T^{long}_{ij}(\vec{k})P_{long}^{ab},
\eea 
where
\bea
T^{even}_{ij}(\vec{k}) &\equiv& e_{i}^{L}(\hat{k})e_{j}^{*L}(\hat{k})+e_{i}^{R}(\hat{k})e_{j}^{*R}(\hat{k}),\\
T^{odd}_{ij}(\vec{k}) &\equiv& i \Big[e_{i}^{L}(\hat{k})e_{j}^{*L}(\hat{k})-e_{i}^{R}(\hat{k})e_{j}^{*R}(\hat{k})\Big],\\
T^{long}_{ij}(\vec{k}) &\equiv& e_{i}^{l}(\hat{k})e_{j}^{*l}(\hat{k}).
\eea
The polarization vectors are $e^{L}(\hat{k})\equiv\frac{1}{\sqrt{2}} (\cos\theta\cos\phi-i\sin\phi,\cos\theta\sin\phi+i\cos\phi,-\sin\theta)$, $e^{R}(\hat{k})=e^{*L}(\hat{k})$ and $e^{l}(\hat{k})=\hat{k}=(\sin\theta\cos\phi,\sin\theta\sin\phi,\cos\theta)$, from which we have
\bea
T^{even}_{ij}(\vec{k})&=&\delta_{ij}-\hat{k}_{i}\hat{k}_{j},\\
T^{odd}_{ij}(\vec{k})&=&\epsilon_{ijk}\hat{k}_{k},\\
T^{long}_{ij}(\vec{k})&=&\hat{k}_{i}\hat{k}_{j}.
\eea

\noindent The purely scalar terms in Eqs.~(\ref{class1})-(\ref{class2}) are already known from the literature \footnote{In single-field slow-roll inflation $P_{\phi}=H_{*}^{2}/2k^3$, where $H_{*}$ is the Hubble rate evaluated at horizon exit; the bispectrum and the trispectrum of the scalar field ($B_{\phi}$ and $T_{\phi}$) can be found in \cite{Acqua,Maldacena:2002vr,Seery:2005wm,Seery:2006vu,Seery:2008ax}. For the bispectrum see also Eq.~(\ref{maldbis})}.\\ 

\noindent As to the mixed (scalar-vector) terms, they can be ignored if one considers a Lagrangian where there is no direct coupling between the inflaton and the gauge bosons but the latter condition is not sufficient for concluding that the mixed derivatives are null. As an example, it is useful to refer to \cite{Vernizzi:2006ve} which, among other things, includes an analytic study for the case of a set of slowly rolling fields with a separable quadratic potential. The number of e-folding is written as a sum of integrals over the different fields, to be evaluated between their values at an initial (generally set at around horizon crossing) and a final times. For each field, the value at the final time depends on the total field configuration at the initial time, so the mixed derivative of $N$ can in principle be non-zero. Anyway, if the final time approaches the end of inflation, it is reasonable to assume that, by then, the fields have stabilized to their equilibrium value and no longer carry the memory of their evolution. If this happens, the sum of integrals which defines $N$ becomes independent of the final field configuration and its mixed derivatives can therefore be shown to be zero. It turns out that we are allowed the same kind of analytic study, if we work with the Lagrangian in Eq.~(\ref{ac}) and introduce some slow-roll assumptions for the vector fields (see Appendix~\ref{eomotion} for a discussion about these assumptions and Appendix~\ref{deltader} for the actual calculation of $N$ and its derivatives).\\

\noindent Let us then look at the (purely) vector part. Its anisotropy features can be stressed by rewriting them as follows 
\bea\label{b}\fl
B_{\zeta}(\vec{k_{1}},\vec{k_{2}},\vec{k_{3}})&\supset&\frac{1}{2}N_{a}^{i}N_{b}^{j}N_{cd}^{kl}\Pi^{ac}_{ik}(\vec{k}_{1})\Pi^{bd}_{jl}(\vec{k}_{2})=M^{c}_{k} N^{kl}_{cd} M^{d}_{l}\\\label{tr}\fl
T_{\zeta}(\vec{k_{1}},\vec{k_{2}},\vec{k_{3}},\vec{k_{4}})&\supset&N_{a}^{\mu}N_{b}^{\nu}N_{cd}^{\rho\sigma}N_{ef}^{\delta\eta}P_{\mu\rho}^{ac}(\vec{k}_{1})P_{\nu\delta}^{be}(\vec{k}_{2})P_{\sigma\eta}^{df}(\vec{k}_{1}+\vec{k}_{3})\nonumber\\\fl&+&N_{a}^{\mu}N_{b}^{\nu}N_{c}^{\rho}N_{def}^{\sigma\delta\eta}P_{\mu\sigma}^{ad}(\vec{k}_{1})P_{\nu\delta}^{be}(\vec{k}_{2})P_{\rho\eta}^{cf}(\vec{k}_{3})\nonumber\\\fl&=&M_{i}^{c}L_{ce}^{ij}M_{j}^{e}+M_{i}^{f}M_{j}^{e}M_{k}^{d}N_{fed}^{ijk},
\eea
where
\bea\label{nonzero2}\fl
M_{k}^{c}(\vec{k})&\equiv& N_{a}^{i}P_{ik}^{ac}(\vec{k})=P^{ac}_{+}(k)\left[\delta_{ik}{N}_{a}^{i}+p^{ac}(k)\hat{k}_{k}\left(\hat{k}\cdot\vec{N}_{a}\right)+iq^{ac}(k)\left(\hat{k}\times\vec{N}_{a}\right)_{k}\right]\\\fl
L_{ce}^{jl}(\vec{k})&\equiv& N_{cd}^{ji}P_{ik}^{df}(\vec{k})N_{fe}^{kl}\nonumber\\\fl&=&P_{+}^{df}(\vec{k})\big[\vec{N}_{cd}^{j}\cdot\vec{N}_{ef}^{l}+p^{df}(k)\left(\hat{k}\cdot\vec{N}_{cd}^{j}\right)\left(\hat{k}\cdot\vec{N}_{ef}^{l}\right)+iq^{df}(k)\hat{k}\cdot\vec{N}^{j}_{cd}\times \vec{N}_{ef}^{l}\big]. 
\eea
In the previous equations, we defined
\bea
p^{ac}(k)&\equiv& \frac{P^{ac}_{long}-P^{ac}_{+}}{P^{ac}_{+}},\\
q^{ac}(k) &\equiv& \frac{P^{ac}_{-}}{P^{ac}_{+}},
\eea
with $\vec{N}_{a}\equiv (N_{a}^{1},N_{a}^{2},N_{a}^{3})$ and $\vec{N}_{cd}^{j}\equiv (N_{cd}^{j1},N_{cd}^{j2},N_{cd}^{j3})$.\\
Notice that, as for the power spectrum (\ref{power-zeta}), also in Eqs.~(\ref{b})-(\ref{tr}) the anisotropic parts of the expressions are weighted by coefficients that are proportional either to $P_{-}$ or to $(P_{long}-P_{+})$. When these two quantities are equal to zero, the (Abelian) bispectrum and trispectrum are therefore isotropized. $P_{-}=0$ in parity conserving theories, like the ones we have been describing. According to the parametrization (\ref{L}) of the longitudinal mode, we have $P_{long}-P_{+}=(|n(x)|^2-1) P_{+}$.

\subsection{Bispectrum and trispectrum: Non-Abelian contributions}\label{nonabelianc}

We list the non-Abelian terms for the bispectrum 
\bea\label{class3}\fl
B_{\zeta}(\vec{k_{1}},\vec{k_{2}},\vec{k_{3}})&\supset& N_{a}^{i}N_{b}^{j}N_{c}^{k}B_{ijk}^{abc}(\vec{k}_{1},\vec{k}_{2},\vec{k}_{3})
\eea
and for the trispectrum
\bea\label{class4}\fl
T_{\zeta}(\vec{k_{1}},\vec{k_{2}},\vec{k_{3}},\vec{k_{4}})&\supset& N_{a}^{i}N_{b}^{j}N_{c}^{k}N_{d}^{l}T_{ijkl}^{abcd}(\vec{k}_{1},\vec{k}_{2},\vec{k}_{3},\vec{k}_{4})\nonumber\\&+&N_{a}^{i}N_{b}^{j}N_{\phi}N_{\phi c}^{k}\left[P_{\phi}^{}({k}_{3})B_{ijk}^{abc}(\vec{k}_{1},\vec{k}_{2},\vec{k}_{3}+\vec{k}_{4})+perms.\right]\nonumber\\&+&N_{a}^{i}N_{b}^{j}N_{c}^{k}N_{de}^{lm}\left[P_{il}^{ad}(\vec{k}_{1})B_{jkm}^{bce}(\vec{k}_{1}+\vec{k}_{2},\vec{k}_{3},\vec{k}_{4})+perms.\right].
\eea
The computation of the vector bosons spectra
\bea
\langle \de A^{a}_{i} \de A^{b}_{j} \de A^{c}_{k}\rangle=\delta^{(3)}(\vec{k}_{1}+\vec{k}_{2}+\vec{k}_{3})B_{ijk}^{abc},\\
\langle \de A^{a}_{i} \de A^{b}_{j} \de A^{c}_{k} \de A^{d}_{l}\rangle=\delta^{(3)}(\vec{k}_{1}+\vec{k}_{2}+\vec{k}_{3}+\vec{k}_{4})T^{abcd}_{ijkl}, 
\eea
will be reviewed in this section. This requires the expansion of the in-in formula up to second order in the interaction Hamiltonian
\bea\label{skr}\fl
\langle\Theta(\eta^{*})\rangle &\supset& i\langle T \Big[\Theta \int_{-\infty}^{\eta^{*}}d\eta^{'}\left(H^{+}_{int}(\eta^{'})-H^{-}_{int}(\eta^{'})\right)\Big]\rangle\\\fl&+&
\frac{(-i)^2}{2}\langle T \Big[\Theta \int_{-\infty}^{\eta^{*}}d\eta^{'}\left(H^{+}_{int}(\eta^{'})-H^{-}_{int}(\eta^{'})\right) \int_{-\infty}^{\eta^{*}}d\eta^{''}\left(H^{+}_{int}(\eta^{''})-H^{-}_{int}(\eta^{''})\right)\Big]\rangle.\nonumber
\eea

\noindent The interaction Hamiltonian needs to be expanded up to fourth order in the fields fluctuations, i.e. $H_{int}=H_{int}^{(3)}+H_{int}^{(4)}$, where
\bea
\label{int3}
H_{int}^{(3)}&=&g_{c}\ep^{abc}g^{ik}g^{jl}\left(\p_{i}\DB_{j}^{a}\right)\DB^{b}_{k}\DB_{l}^{c}+g_{c}^2 \ep^{eab}\ep^{ecd}g^{ik}g^{jl}B^{a}_{i}\de B^{b}_{j}\de B^{c}_{k}\de B^{d}_{l}\\\label{int4}
H_{int}^{(4)}&=&g_{c}^2 \ep^{eab}\ep^{ecd}g^{ij}g^{kl}\de B^{a}_{i}\de B^{b}_{k}\de B^{c}_{j}\de B^{d}_{l}.
\eea

\begin{figure}\centering
 \includegraphics[width=0.3\textwidth]{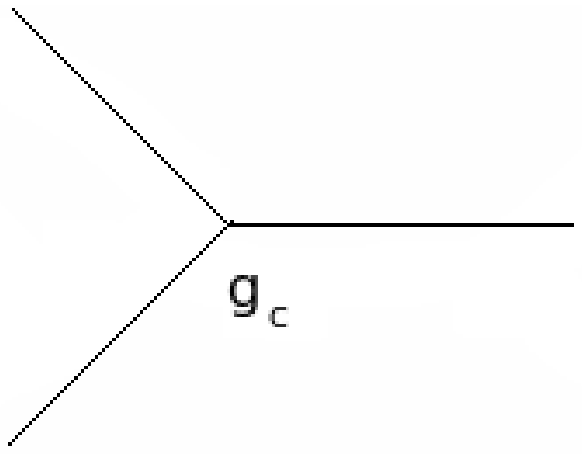}
\hspace{0.17\textwidth}
 \includegraphics[width=0.3\textwidth]{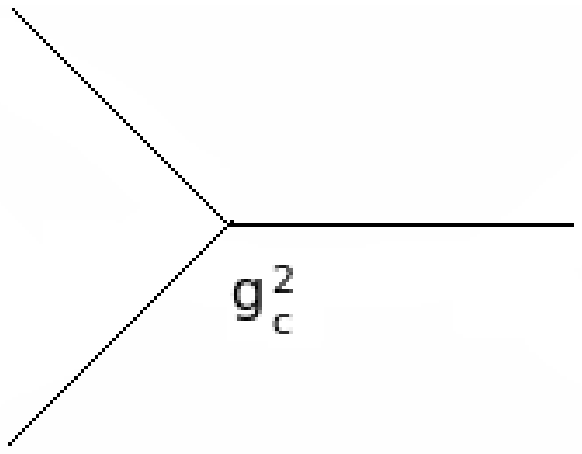}
\caption{ \label{Fig1} Diagrammatic representations of the tree-level contributions to the vector fields bispectrum.}
\end{figure}

\begin{figure}\centering
 \includegraphics[width=0.3\textwidth]{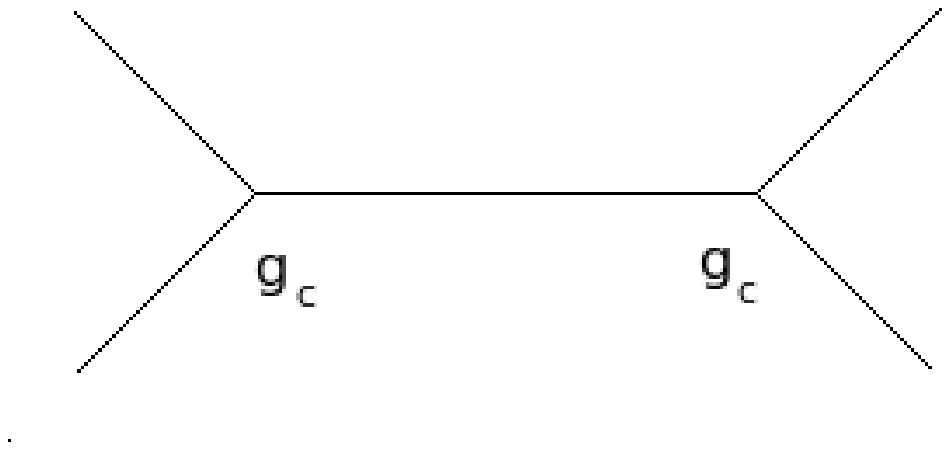}
\hspace{0.17\textwidth}
 \includegraphics[width=0.3\textwidth]{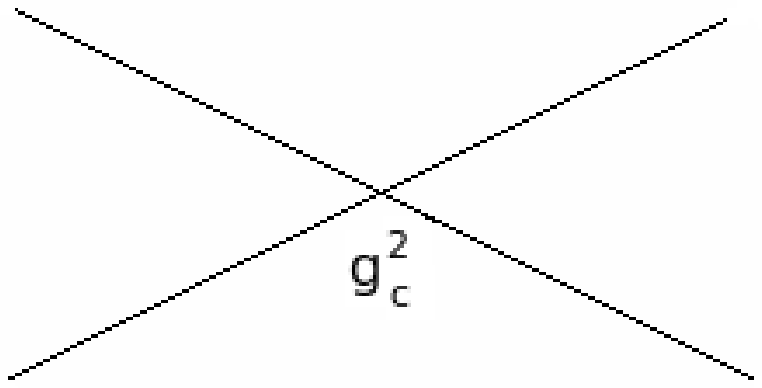}
\caption{ \label{Fig1} Diagrammatic representations of the tree-level contributions to the vector fields trispectrum: vector-exchange (on the left) and contact-interaction (on the right) diagrams.}
\end{figure}

\noindent To tree-level, the relevant diagrams are pictured in Figs.~5 and 6. By looking at Eqs.~(\ref{int3}) and (\ref{int4}), we can see that there is a bispectrum diagram that is lower in terms of power of the $SU(2)$ coupling ($\sim g_{c}$) compared to the trispectrum ($\sim g_{c}^2$); as a matter of fact, for symmetry reasons that we are going to discuss later in this section, $g_{c}^2$ interaction terms are needed to provide a non-zero contributions to the bispectrum. \\
\noindent The propagators for ``plus'' and ``minus'' fields are
\bea\fl
\widehat{\delta B^{a,+}_{i}(\epr)\delta B^{b,+}_{j}(\eps)} = \tilde{\Pi}_{ij}^{ab}(\epr,\eps)\Theta(\epr-\eps)+\bar{\Pi}_{ij}^{ab}(\epr,\eps)\Theta(\eps-\epr),\\\fl
\widehat{\delta B^{a,+}_{i}(\epr)\delta B^{b,-}_{j}(\eps)} = \bar{\Pi}_{ij}^{ab}(\epr,\eps),\\\fl
\widehat{\delta B^{a,-}_{i}(\epr)\delta B^{b,+}_{j}(\eps)} = \tilde{\Pi}_{ij}^{ab}(\epr,\eps),\\\fl
\widehat{\delta B^{a,-}_{i}(\epr)\delta B^{b,-}_{j}(\eps)} = \bar{\Pi}_{ij}^{ab}(\epr,\eps)\Theta(\epr-\eps)+\tilde{\Pi}_{ij}^{ab}(\epr,\eps)\Theta(\eps-\epr),
\eea
or
\bea\fl
\tilde{\Pi}_{ij}^{ab}(\vec{k})\equiv T_{ij}^{even}(\hat{k})\tilde{P}^{ab}_{+}+i T_{ij}^{odd}(\hat{k})\tilde{P}^{ab}_{ij}+T_{ij}^{long}(\hat{k})\tilde{P}^{ab}_{ij}\\\fl
\bar{\Pi}_{ij}^{ab}(\vec{k})\equiv T_{ij}^{even}(\hat{k})\bar{P}^{ab}_{+}+i T_{ij}^{odd}(\hat{k})\bar{P}^{ab}_{ij}+T_{ij}^{long}(\hat{k})\bar{P}^{ab}_{ij}
\eea
in Fourier space. In the previous equations we set $\tilde{P}^{ab}_{\pm}\equiv (1/2)(\tilde{P}^{ab}_{R}\pm\tilde{P}^{ab}_{L})$, $\tilde{P}^{ab}_{R}=\delta_{ab}\delta B^{ab}_{R}(k,\eta^{*})\delta B^{*ab}_{R}(k,\eta)$ and $\bar{P}^{ab}_{\pm}=\left(\tilde{P}^{ab}_{\pm}\right)^{*}$ (similar definitions apply for $\tilde{P}^{ab}_{L}$ and $\tilde{P}^{ab}_{long}$).\\

\noindent We are now ready to show the computation of the following contributions to the bispectrum and trispectrum of $\zeta$
\bea\label{bi}\fl
\langle \zeta_{\vec{k_{1}}}\zeta_{\vec{k_{2}}}\zeta_{\vec{k_{3}}}\rangle \supset N^{i}_{a}N^{j}_{b}N^{k}_{c}\langle \delta A^{a}_{i}(\vec{k_{1}})\delta A^{b}_{j}(\vec{k_{2}})\delta A^{c}_{k}(\vec{k_{3}})\rangle, \\\label{tri}\fl
\langle\zeta_{\vec{k}_{1}}\zeta_{\vec{k}_{2}}\zeta_{\vec{k}_{3}}\zeta_{\vec{k}_{4}} \rangle \supset N^{i}_{a}N^{j}_{b}N^{k}_{c}N^{l}_{d}\langle \de A^{a}_{i}(\vec{k_{1}}) \de A^{b}_{j}(\vec{k_{2}}) \de A^{c}_{k}(\vec{k_{3}}) \de A^{d}_{l}(\vec{k_{4}})\rangle.
\eea\\

\noindent Eq.~(\ref{bi}) becomes
\bea\label{bi1}\fl
\langle \zeta_{\vec{k_{1}}}\zeta_{\vec{k_{2}}}\zeta_{\vec{k_{3}}}\rangle &\supset&N_{a}^{i}N_{b}^{j}N_{c}^{k}\frac{\delta^{(3)}(\vec{k_{1}}+\vec{k_{2}}+\vec{k_{3}})}{a^{3}(\eta^{*})}\Big[\int d \eta a^{4}(\eta) \tilde{\Pi}_{im}(\vec{k_{1}})\tilde{\Pi}^{l}_{j}(\vec{k_{2}})\tilde{\Pi}^{m}_{k}(\vec{k_{3}})\nonumber\\\fl&\times&\Big(g_{c}\ep^{abc}k_{1l}+g_{c}^2\ep^{eda}\ep^{ebc}B^{d}_{l}\Big)\Big]+perms.+c.c.
\eea
Even before performing the time integration, one realizes that, because of the antisymmetric properties of the Levi-Civita tensor, the $\sim g_{c}$ contribution on the r.h.s. of Eq.~(\ref{bi1}) is equal to zero once the sum over all the possible permutations has been performed. The vector bosons bispectrum is therefore proportional to $g_{c}^{2}$. The final result from (\ref{bi1}) has the following form

\bea\label{fire}\fl
\langle \zeta_{\vec{k_{1}}}\zeta_{\vec{k_{2}}}\zeta_{\vec{k_{3}}}\rangle \supset (2 \pi)^3 \delta^{(3)}(\vec{k}_{1}+\vec{k}_{2}+\vec{k}_{3})g_{c}^{2}H_{*}^{2}\sum_{n}F_{n}(k_{i},\eta^{*})I_{n}(\hat{k}_{i}\cdot\hat{k}_{j},\vec{A}_{i}\cdot\vec{A}_{j},\hat{k}_{i}\cdot\vec{A}_{j})
\eea
where the $F_{n}$'s are isotropic functions of time and of the moduli of the wave vectors ($i=1,2,3$) and the $I_{n}$'s are anisotropic coefficients. The sum in the previous equation is taken over all possible combinations of products of three polarization indices, i.e. $n\in (EEE, EEl, ElE,..., lll)$, where $E$ stands for ``even'', $l$ for ``longitudinal''. The complete expressions for the terms appearing in the sum are quite lengthy and we list them in Appendix~\ref{bispquartic}. As an example, we report here one of these terms
\bea\label{sample1}\fl
F_{lll}&=&-n^{6}(x^{*})\frac{1}{24 k^{6}k_{1}^{2}k_{2}^{2}k_{3}^{2}x^{*2}}\left[A_{EEE}+\left(B_{EEE}\cos x^{*}+C_{EEE}\sin x^{*}\right)E_{i}x^{*}\right]\\\label{sample2}\fl
I_{lll}&=&\ep^{aa'b'}\ep^{ac'e}\Big[\Big(\left(\hat{k}_{1}\cdot\vec{N}^{a'}\right)\left(\hat{k}_{3}\cdot\vec{N}^{b'}\right)\left(\hat{k}_{2}\cdot\vec{N}^{c'}\right)\left(\hat{k}_{1}\cdot\hat{k}_{2}\right)\left(\hat{k}_{3}\cdot\hat{A}^{e}\right)\nonumber\\\fl&-&\left(\hat{k}_{3}\cdot\vec{N}^{a'}\right)\left(\hat{k}_{2}\cdot\vec{N}^{b'}\right)\left(\hat{k}_{1}\cdot\vec{N}^{c'}\right)\left(\hat{k}_{1}\cdot\hat{k}_{2}\right)\left(\hat{k}_{3}\cdot\hat{A}^{e}\right)\Big)+(1\leftrightarrow 3)+(2\leftrightarrow 3) \Big]
\eea
where $A_{EEE}$, $B_{EEE}$ and $C_{EEE}$ are functions of $x^{*}$ and of the momenta $k_{i}\equiv |\vec{k}_{i}|$ (they are all reported in Appendix~\ref{bispquartic}), $E_{i}$ is the exponential-integral function. As we will discuss in more details in Sec.~\ref{anisodiscuss}, one of the more interesting features of these models is that the bispectrum and the trispectrum turn out to have an amplitude that is modulated by the preferred directions that break statistical isotropy.
\\

\noindent Let us now move to the trispectrum. Again, we count two different kinds of contributions, the first from $\sim g_{c}$ and the second from $\sim g_{c}^{2}$ interaction terms, respectively in $H_{int}^{(3)}$ and $H_{int}^{(4)}$. The former produce vector-exchange diagrams, the latter are represented by contact-interaction diagrams (see Fig.~6). Their analytic expressions are different, but they both have a structure similar to (\ref{fire}) 
\bea\label{fire1}\fl
\langle \zeta_{\vec{k_{1}}}\zeta_{\vec{k_{2}}}\zeta_{\vec{k_{3}}}\zeta_{\vec{k_{4}}}\rangle &\supset& (2 \pi)^3 \delta^{(3)}(\vec{k}_{1}+\vec{k}_{2}+\vec{k}_{3}+\vec{k}_{4}) g_{c}^{2}H_{*}^{2}\fl\\&\times&\sum_{n}G_{n}(k_{i},k_{\hat{12}},k_{\hat{14}},\eta^{*})L_{n}(\hat{k}_{i}\cdot\hat{k}_{j},\vec{A}_{i}\cdot\vec{A}_{j},\hat{k}_{i}\cdot\vec{A}_{j})\nonumber
\eea
where we define $k_{\hat{12}}\equiv |\vec{k}_{1}+\vec{k}_{2}|$ and $k_{\hat{14}}\equiv |\vec{k}_{1}+\vec{k}_{4}|$. We will present the details of the computation of the gauge fields trispectrum and the explicit expression of the functions appearing in (\ref{fire1}) in the next section.


\subsubsection{Trispectrum from vector bosons: exchange diagram}
${}$\\
Let us begin with the two vertex diagram. Using the language of Eq.~(\ref{skr}), it can be put in the form 
\bea\label{2v}\fl
\langle\Theta(\eta^{*})\rangle \supset \frac{(-i)^2}{2}\langle T \Big[\Theta \int_{-\infty}^{\eta^{*}}d\eta^{'}\left(H^{+}(\eta^{'})-H^{-}(\eta^{'})\right) \int_{-\infty}^{\eta^{*}}d\eta^{''}\left(H^{+}(\eta^{''})-H^{-}(\eta^{''})\right)\Big]\rangle
\eea
where now $H\equiv H^{(3)}_{int}$, $\Theta\equiv \de A_{\mu}^{a}\de A_{\nu}^{b} \de A_{\rho}^{c}\de A_{\sigma}^{d}$ and the inclusion symbol as usual points out that what stands on the right-hand side is only one of the contributions to $\langle\Theta(\eta^{*})\rangle$. \\
\noindent Eq.~(\ref{2v}) can be rewritten as follows
\bea
\langle\Theta(\eta^{*})\rangle \supset\frac{(-i)^2}{2}\langle T\left[\Theta \left(\textit{A}+\textit{B}+\textit{C}+\textit{D}\right)\right]\rangle
\eea
where
\bea
\textit{A}\equiv \int_{-\infty}^{\eta^{*}}d\eta^{'}H^{+}(\eta^{'}) \int_{-\infty}^{\eta^{*}}d\eta^{''}H^{+}(\eta^{''})\nonumber\\
\textit{B}\equiv \int_{-\infty}^{\eta^{*}}d\eta^{'}H^{-}(\eta^{'}) \int_{-\infty}^{\eta^{*}}d\eta^{''}H^{-}(\eta^{''}) \nonumber\\
\textit{C}\equiv -\int_{-\infty}^{\eta^{*}}d\eta^{'}H^{+}(\eta^{'}) \int_{-\infty}^{\eta^{*}}d\eta^{''}H^{-}(\eta^{''}) \nonumber\\
\textit{D}\equiv -\int_{-\infty}^{\eta^{*}}d\eta^{'}H^{-}(\eta^{'}) \int_{-\infty}^{\eta^{*}}d\eta^{''}H^{+}(\eta^{''}) \nonumber
\eea
For each one of the integrals listed above, due to the presence of both the fields and their spatial derivatives in $H^{(3)}_{int}$, there are three different sets of contractions of the external with the vertex field-operators: for the first set, the field-operators with derivatives in the vertices are both contracted with external fields; for the second one, only one of the two field-operators with derivatives contracts with an external field (the other contracts with another internal field); for the third set, the field-operators with derivatives contract with each other.\\
A sample set of contractions of the first type is provided in the following equation
\bea\label{sample}\fl
T^{abcd}_{ijkl} &\supset&\frac{g_{c}^{2}}{2a^4(\eta^{*})}\ep^{a^{'}b^{'}c^{'}}\ep^{a^{''}b^{''}c^{''}}k_{m}k_{m^{'}}\nonumber\\\fl&\times&\int d\eta^{'}a^{4}(\eta^{'})\int d\eta^{''}a^{4}(\eta^{''})g^{mp}g^{m^{'}p^{'}}g^{nq}g^{n^{'}q^{'}}\tilde{\Pi}^{aa^{'}}_{in}\tilde{\Pi}^{bb^{'}}_{jp}\tilde{\Pi}_{kn^{'}}^{ca^{''}}\tilde{\Pi}^{db^{''}}_{lp^{'}}\tilde{\Pi}^{c^{'}c^{''}}_{qq^{'}}
\eea
where the first four $\tilde{\Pi}$s correspond to contractions between external and internal fields whereas the last one indicates the contraction between the two remaining internal fields. The $a^{-4}$ factor comes from expressing the external (physical) fields in terms of the comoving ones. As a reminder, we define $ (2\pi)^3\delta^{(3)}(\vec{k}_{1}+\vec{k}_{2}+\vec{k}_{3}+\vec{k}_{4})T^{abcd}_{ijkl}\equiv\langle \de A^{a}_{i} \de A^{b}_{j} \de A^{c}_{k} \de A^{d}_{l}\rangle$. The expression in (\ref{sample}) can be rewritten as follows

\bea\label{sec9}
T^{abcd}_{ijkl}&\supset&\frac{g_{c}^{2}}{2a^4(\eta^{*})}\ep^{a^{'}b^{'}c^{'}}\ep^{a^{''}b^{''}c^{''}}k_{m}k_{m^{'}}\delta^{aa^{'}}\delta^{bb^{'}}\delta^{ca^{''}}\delta^{db^{''}}\delta^{c^{'}c^{''}}\nonumber\\&\times&\sum_{\alpha,\beta,\gamma,\delta,\sigma}\left(\int d\eta^{'}\int d\eta^{''}\right)_{\alpha\beta\gamma\delta\sigma}T^{\alpha}_{in}T^{\beta}_{jm}T^{\gamma}_{kn^{'}}T^{\delta}_{lm^{'}}T^{\sigma}_{nn^{'}}
\eea

\noindent where the greek indices of the sum indicate either the transverse (E) or longitudinal (l) modes, the $\left(\int d\eta^{'}\int d\eta^{''}\right)$ stand for the integrals over the wave functions, the chosen time variable being $x\equiv -k\eta$ ($k\equiv \sum_{i=1,...,4}k_{i}$, $k_{i}\equiv |\vec{k}_{i}|$).\\
Let us define the coefficients $T_{ijkl}^{\alpha\beta\gamma\delta\sigma}\equiv k_{m}k_{m^{'}} T^{\alpha}_{in}T^{\beta}_{jm}T^{\gamma}_{kn^{'}}T^{\delta}_{lm^{'}}T^{\sigma}_{nn^{'}}$. They should be calculated for each one of the three different sets of contractions and for each permutation within the specific set. This is a straghtforward but rather lengthy and not particularly interesting calculation. A convenient way to proceed could be the following: we first compute the time integrals in order to find out which one among the combinations of longitudinal and transverse mode functions in the string $[\alpha,\beta,\gamma,\delta,\sigma]$ provides the highest amplitude for the trispectrum (in order to be able to perform this comparison we work, as it is usually done when trying to quantify the amplitude of a three or of a four-point function, in the so called ``equilateral configuration'', which for the trispectrum means taking $k_{1}=k_{2}=k_{3}=k_{4}$); for the combination with the highest amplitude, we then calculate the coefficients $T_{ijkl}^{\alpha\beta\gamma\delta\sigma}$ for all the different sets of contractions and sum over all the permutations.\\
Let us now perform our calculations. The wavefunctions we are going to adopt were introduced in Sec.~\ref{ourmodel} (see Eqs.~(\ref{T}) and (\ref{L})). It is possible to verify that 
$\textit{B}=\textit{A}^{*}$ and $\textit{D}=\textit{C}^{*}$ and that integrals of type \textit{A} are consistently smaller in amplitude than integrals of type \textit{C}. We therefore report the combined contribution $\textit{C}+\textit{D}=2 Re[\textit{C}]$ for one of the permutations

\bea
\fl\label{1a}
\left(\int d\eta^{'}\int d\eta^{''}\right)_{EEEEE}&=&\frac{1}{8k^{3}_{1}k_{2}^{3}k_{3}^{3}k_{4}^{3}k_{\hat{12}}^{3}(k_{\hat{12}}+k_{1}+k_{2})(k_{\hat{12}}+k_{3}+k_{4})x^{*8}}\\\fl&\times&\big[\big(M-2E\big)\big[\big(N-2F\big)\big(AB+CD\big)+\big(2H+L\big)\big(CB-AD\big)\big]\nonumber\\\fl&+&\big(2G+P\big)\big[\big(N-2F\big)\big(AD-CB\big)+\big(2H+L\big)\big(AB+CD\big)\big]\big]\nonumber\\\fl
\left(\int d\eta^{'}\int d\eta^{''}\right)_{EEEEl}&=&n^{2}(x^{*})\left(\int d\eta^{'}\int d\eta^{''}\right)_{EEEEE}\\\fl
\left(\int d\eta^{'}\int d\eta^{''}\right)_{EEEll}&=&n^{4}(x^{*})\left(\int d\eta^{'}\int d\eta^{''}\right)_{EEEEE}\\\fl
\left(\int d\eta^{'}\int d\eta^{''}\right)_{EElll}&=&n^{6}(x^{*})\left(\int d\eta^{'}\int d\eta^{''}\right)_{EEEEE}\\\fl
\left(\int d\eta^{'}\int d\eta^{''}\right)_{Ellll}&=&n^{8}(x^{*})\left(\int d\eta^{'}\int d\eta^{''}\right)_{EEEEE}\\\fl\label{2a}
\left(\int d\eta^{'}\int d\eta^{''}\right)_{lllll}&=&n^{10}(x^{*})\left(\int d\eta^{'}\int d\eta^{''}\right)_{EEEEE}
\eea
where $A$, $B$, $C$, $D$, $E$, $F$, $G$, $H$, $L$, $M$, $N$ and $P$ are functions of $x^{*}$ and of the momenta moduli to be provided in Appendix~\ref{exchangedetails} (see Eqs.~(\ref{eeeee1}) through (\ref{eeeee2})). Obviously, the value of the integrals does not change when permuting its labels $\alpha\beta\gamma\delta\sigma$, apart from a different power of the coefficient $n(x^{*})$. We need now to find out if there is one, among the integrals in Eqs.~(\ref{1a}) through (\ref{2a}), that has the largest amplitude, i.e. understand if something can be said about the order of magnitude of $n(x^{*})$. We could try to extrapolate some information about $n(x^{*})$ from what happens at very late times. In the models discussed in 
Ref.~\cite{Dimopoulos:2006ms,Dimopoulos:2008yv} it turns out that the longitudinal mode is $\delta B^{||}=\sqrt{2}\delta B^{T}$ \footnote{
As another example, in models with varying kinetic function and mass \cite{Dimopoulos:2009am,Dimopoulos:2009vu}, we have verified that $n(x)\gg 1$ at late times and for a vector field that is light until the end of inflation.}. If this is the correct asymptotic behaviour and we find it reasonable to extrapolate back until the horizon crossing epoch, it is then correct to conclude that in this case the amplitude is the largest for the integral among the ones listed in Eqs.~(\ref{1a}) through (\ref{2a}) containing the highest powers of $n$, i.e. for $\left(\int d\eta^{'}\int d\eta^{''}\right)_{lllll}$. The coefficients we intend to calculate are then of the kind $T_{ijkl}^{lllll}$ only. We list them below for the three different sets of contractions we introduced above and for one particular permutation (see again Appendix~\ref{exchangedetails} for more details)

\bea\label{tt}
T^{lllll(1)}_{ijkl}=k_{1}k_{3}k_{1234}\left(\hat{k}_{1}\cdot\hat{k}_{\hat{12}}\right)\left(\hat{k}_{1}\cdot\hat{k}_{2}\right)\left(\hat{k}_{3}\cdot\hat{k}_{4}\right)\left(\hat{k}_{3}\cdot\hat{k}_{\hat{12}}\right),\\
T^{lllll(2)}_{ijkl}=k_{3}k_{\hat{12}}k_{1234}\left(\hat{k}_{1}\cdot\hat{k}_{\hat{12}}\right)\left(\hat{k}_{2}\cdot\hat{k}_{\hat{12}}\right)\left(\hat{k}_{3}\cdot\hat{k}_{4}\right)\left(\hat{k}_{3}\cdot\hat{k}_{\hat{12}}\right),\\
T^{lllll(3)}_{ijkl}=k_{\hat{12}}k_{\hat{12}}k_{1234}\left(\hat{k}_{1}\cdot\hat{k}_{\hat{12}}\right)\left(\hat{k}_{2}\cdot\hat{k}_{\hat{12}}\right)\left(\hat{k}_{3}\cdot\hat{k}_{\hat{12}}\right)\left(\hat{k}_{4}\cdot\hat{k}_{\hat{12}}\right).
\eea 
We adopted the following notation: $k_{s}\equiv |\vec{k}_{s}|$, $\hat{k}_{s}\equiv\vec{k}_{s}/k_{s}$, the index $s$ running over the four external momenta; $k_{ss's''s'''}\equiv\hat{k}_{si}\hat{k}_{s'j}\hat{k}_{s''k}\hat{k}_{s'''l}$, with $s,s',s'',s'''=1,2,3,4$ and with the indices $i,j,k,l$ indicating the spatial components of the vectors; $\vec{k}_{\hat{ss'}}\equiv\vec{k}_{s}+\vec{k}_{s'}$, $k_{\hat{ss'}}\equiv|\vec{k}_{s}+\vec{k}_{s'}|$ and so $\hat{k}_{\hat{ss'}}\equiv\vec{k}_{\hat{ss'}}/k_{\hat{ss'}}$.\\

\noindent It is possible to prove that, once the Levi-Civita coefficients and the sum over the permutations are taken into account, only the first set of contractions is left. The final result after these cancellations can be written in the following form

\bea\label{final-2v}
\fl
\langle \de A^{a}_{i} \de A^{b}_{j} \de A^{c}_{k} \de A^{d}_{l}\rangle_{*}  &\supset&(2 \pi)^3\delta^{(3)}(\vec{k}_{1}+\vec{k}_{2}+\vec{k}_{3}+\vec{k}_{4}) g_{c}^{2}\left(\frac{H_{*}x^{*}}{k}\right)^4\ep^{abc^{''}}\ep^{cdc^{''}}\Big[I \times k_{1234}\times \left(\sum_{i=1}^{4}t_{i}\right)\nonumber\fl\\&+&II \times k_{1324}\times \left(\sum_{i=5}^{8}t_{i}\right)+III \times k_{1432}\times \left(\sum_{i=9}^{12}t_{i}\right)\Big].
\eea
All the possible permutations have been included in the previous equation and, as a reminder, the indices $i,j,k,l$ are hidden in $k_{ss's''s'''}$ on the right-hand side. We define
\bea\fl
I&\equiv& n^{10}\times\left(\frac{1}{8k^{3}_{1}k_{2}^{3}k_{3}^{3}k_{4}^{3}k_{\hat{12}}^{3}(k_{\hat{12}}+k_{1}+k_{2})(k_{\hat{12}}+k_{3}+k_{4})x^{*8}}\right)\nonumber\\\fl&\times&\big[\big(M-2E\big)\big[\big(N-2F\big)\big(AB+CD\big)+\big(2H+L\big)\big(CB-AD\big)\big]\nonumber\\\fl&+&\big(2G+P\big)\big[\big(N-2F\big)\big(AD-CB\big)+\big(2H+L\big)\big(AB+CD\big)\big]\big]
\eea
(from Eqs.~(\ref{1a})-(\ref{2a})). The function $II$ is defined from $I$ by exchanging $k_{2}$ with $k_{3}$ and $k_{\hat{12}}$ with $k_{\hat{13}}$; similarly, $III$ is defined from $I$ by exchanging $k_{2}$ with $k_{4}$ and $k_{\hat{12}}$ with $k_{\hat{14}}$, so they are all functions of the horizon crossing time $x^{*}\equiv -k \eta^{*}$ and of the moduli of the external momenta and of their sums. This amounts to seven independent variables, $x^{*}$, $k_{1}$, $k_{2}$, $k_{3}$, $k_{4}$, $k_{\hat{12}}$ and $k_{\hat{14}}$. The coefficients $t_{i}$ ($i=1,...,12$) come from $T_{ijkl}^{\alpha\beta\gamma\delta\sigma}$ and so they are also functions of the momenta moduli (see Eqs.~(\ref{t1}) through (\ref{t12}) for their expressions). Finally, the anisotropic part of Eq.~(\ref{final-2v}) is represented by the $k_{ss's''s'''}$ terms, which, in the final expression for the curvature perturbation trispectrum, have their spatial indices contracted with the derivatives $N_{i}^{a}$ of the number of e-foldings w.r.t. the vector fields as follows

\bea\label{IA}\fl
\langle\zeta_{\vec{k}_{1}}\zeta_{\vec{k}_{2}}\zeta_{\vec{k}_{3}}\zeta_{\vec{k}_{4}} \rangle \supset N_{i}^{a}N_{j}^{b}N_{k}^{c}N_{l}^{d}\langle \de A^{a}_{i} \de A^{b}_{j} \de A^{c}_{k} \de A^{d}_{l}\rangle_{*} &\supset& (2 \pi)^3\delta^{(3)}(\vec{k}_{1}+\vec{k}_{2}+\vec{k}_{3}+\vec{k}_{4}) g_{c}^{2}\left(\frac{H_{*}x^{*}}{k}\right)^4\nonumber\\\fl&\times&I\times\left(\sum_{i=1}^{4}t_{i}\right)\times \Delta_{I}+perms.
\eea
where the anisotropic term in the first permutation is
\bea
\Delta_{I}\equiv \ep^{abc^{''}}\ep^{cdc^{''}}N_{i}^{a}N_{j}^{b}N_{k}^{c}N_{l}^{d}k_{1234}.
\eea
It can be interesting to rewrite $\Delta_{I}$ in terms of all its variables
\bea\fl
\Delta_{I}=\sum_{(a<b)a,b=1}^{3}{\left[\left(N^{a}\right)^2\left(N^{b}\right)^2\times\prod_{[i,j]=[1,2],[3,4]}{det\left(M^{i,j,a,b}_{I}\right)}\right]}
\eea
The $M_{I}$'s are $2\times 2$ matrices whose entries are represented by the cosines of the angles between the wavevectors and the $\vec{N}^{a}$ 

\[
M_{I}^{i,j,a,b}\equiv \left| \begin{array}{cc}
\cos\theta_{ia} & \cos\theta_{ja}  \\
\cos\theta_{ib} & \cos\theta_{jb} 
\end{array} \right|.\]
i.e. $\cos\theta_{ia}\equiv\hat{k}_{i}\cdot\hat{N}^{a}$ and so on.\\ 
The two permutations in Eq.~(\ref{IA}) can be written in a similar fashion with anisotropic coefficients 

\bea\fl
\Delta_{II}\equiv \ep^{abc^{''}}\ep^{cdc^{''}}N_{i}^{a}N_{j}^{b}N_{k}^{c}N_{l}^{d}k_{1324}=\sum_{(a<b)a,b=1}^{3}{\left[\left(N^{a}\right)^2\left(N^{b}\right)^2\times\prod_{[i,j]=[1,3],[2,4]}{det\left(M^{i,j,a,b}_{II}\right)}\right]},\nonumber\\
\fl
\Delta_{III}\equiv \ep^{abc^{''}}\ep^{cdc^{''}}N_{i}^{a}N_{j}^{b}N_{k}^{c}N_{l}^{d}k_{1432}=\sum_{(a<b)a,b=1}^{3}{\left[\left(N^{a}\right)^2\left(N^{b}\right)^2\times\prod_{[i,j]=[1,4],[3,2]}{det\left(M^{i,j,a,b}_{III}\right)}\right]}\nonumber.
\eea
The number of angular variables is equal to $12$. These are to be added to the six scalar variables from the isotropic part of (\ref{IA}) ($k_{1}$, $k_{2}$, $k_{3}$, $k_{4}$, $k_{\hat{12}}$ and $k_{\hat{14}}$) and to three parameters represented by the lengths of the vectors $\vec{N}^{a}$ in $\Delta$. The anisotropy coefficients here become equal to zero in the event of an alignment of the gauge vectors along a unique direction.


\subsubsection{Trispectrum from vector bosons: point-interaction diagram}
${}$\\
Let us now move to the one-vertex diagrams
\bea
\langle\Theta(\eta^{*})\rangle &\supset& i\langle T \Big[\Theta \int_{-\infty}^{\eta^{*}}d\eta^{'}\left(H^{+}(\eta^{'})-H^{-}(\eta^{'})\right)\Big]\rangle, 
\eea
where now $H\equiv H_{int}^{(4)}$ and again $\Theta\equiv \de A_{\mu}^{a}\de A_{\nu}^{b} \de A_{\rho}^{c}\de A_{\sigma}^{d}$. After working out the Wick contractions, this becomes
\bea\label{trispec}\fl
N_{a}^{\mu}N_{b}^{\nu}N_{c}^{\rho}N_{d}^{\sigma}T_{\mu\nu\rho\sigma}^{abcd}(\vec{k}_{1},\vec{k}_{2},\vec{k}_{3},\vec{k}_{4}) &\supset&g_{c}^{2}\left(\frac{H_{*} x^{*}}{k}\right)^4\epsilon^{a'bc}\epsilon^{a'da}N^{m}_{a}N^{n}_{b}N^{o}_{c}N^{p}_{d}\nonumber\\\fl&\times&\sum_{\alpha\beta\gamma\delta}\left(\int dx\right)_{\alpha\beta\gamma\delta}T_{mi}^{\alpha}T_{nj}^{\beta}T_{oi}^{\gamma}T_{pj}^{\delta}+permutations.
\eea

\noindent Let us list the coefficients $T_{mnop}^{\alpha\beta\gamma\delta}\equiv  T_{mi}^{\alpha}T_{nj}^{\beta}T_{oi}^{\gamma}T_{pj}^{\delta}$ for one of the permutations

\bea\label{coeff1}\fl
T_{mnop}^{EEEE}&=& \de_{mo}\de_{np}-\de_{mo}\hat{k}_{p4}\hat{k}_{n4}-\de_{mo}\hat{k}_{n2}\hat{k}_{p2}+\de_{mo}\hat{k}_{n2}\hat{k}_{p4}\hat{k}_{2}\cdot\hat{k}_{4}-\de_{np}\hat{k}_{o3}\hat{k}_{m3}\nonumber\\\fl&+&k_{3434}+k_{3232}-\hat{k}_{2}\cdot\hat{k}_{4}\left(k_{3234}+k_{1214}\right)-\de_{np}\hat{k}_{m1}\hat{k}_{o1}+k_{1414}+k_{1212}\nonumber\\\fl&+&\de_{np}\hat{k}_{m1}\hat{k}_{o3}\hat{k}_{1}\cdot\hat{k}_{3}-\hat{k}_{1}\cdot\hat{k}_{3}\left(k_{1232}+k_{1434}\right)+\hat{k}_{1}\cdot\hat{k}_{3}\hat{k}_{2}\cdot\hat{k}_{4},\\\fl
T_{mnop}^{EEEl}&=& \de_{mo}\hat{k}_{p4}\hat{k}_{n4}-\de_{mo}\hat{k}_{n2}\hat{k}_{p4}\hat{k}_{2}\cdot\hat{k}_{4}-k_{3434}+\hat{k}_{2}\cdot\hat{k}_{4}\left(k_{3234}+k_{1214}\right)-k_{1414}\nonumber\\\fl&+&k_{1434}\hat{k}_{1}\cdot\hat{k}_{3}-k_{1234}\hat{k}_{1}\cdot\hat{k}_{3}\hat{k}_{2}\cdot\hat{k}_{4},\\\fl
T_{mnop}^{EElE}&=& \de_{np}\hat{k}_{o3}\hat{k}_{m3}-\de_{np}\hat{k}_{m1}\hat{k}_{o3}\hat{k}_{1}\cdot\hat{k}_{3}-k_{3434}+\hat{k}_{1}\cdot\hat{k}_{3}\left(k_{1434}+k_{1232}\right)-k_{3232}\nonumber\\\fl&+&k_{3234}\hat{k}_{2}\cdot\hat{k}_{4}-k_{1234}\hat{k}_{1}\cdot\hat{k}_{3}\hat{k}_{2}\cdot\hat{k}_{4},\\\fl
T_{mnop}^{ElEE}&=& \de_{mo}\hat{k}_{n2}\hat{k}_{p2}-\de_{mo}\hat{k}_{n2}\hat{k}_{p4}\hat{k}_{2}\cdot\hat{k}_{4}-k_{3232}+\hat{k}_{2}\cdot\hat{k}_{4}\left(k_{3234}+k_{1214}\right)-k_{1212}\nonumber\\\fl&+&k_{1232}\hat{k}_{1}\cdot\hat{k}_{3}-k_{1234}\hat{k}_{1}\cdot\hat{k}_{3}\hat{k}_{2}\cdot\hat{k}_{4},\\\fl
T_{mnop}^{lEEE}&=& \hat{k}_{m1}\hat{k}_{o1}\de_{np}-\hat{k}_{m1}\hat{k}_{o3}k_{13}\de_{np}-k_{1414}+\hat{k}_{1}\cdot\hat{k}_{3}\left(k_{1434}+k_{1232}\right)-k_{1212}\nonumber\\\fl&+&k_{1214}\hat{k}_{2}\cdot\hat{k}_{4}-k_{1234}\hat{k}_{1}\cdot\hat{k}_{3}\hat{k}_{2}\cdot\hat{k}_{4},\\\fl
T_{mnop}^{EEll}&=& k_{3434}-k_{3234}\hat{k}_{2}\cdot\hat{k}_{4}-k_{1434}\hat{k}_{1}\cdot\hat{k}_{3}+k_{1234}\hat{k}_{1}\cdot\hat{k}_{3}\hat{k}_{2}\cdot\hat{k}_{4},\\\fl
T_{mnop}^{ElEl}&=& \de_{mo}\hat{k}_{n2}\hat{k}_{p4}\hat{k}_{2}\cdot\hat{k}_{4}-\hat{k}_{2}\cdot\hat{k}_{4}\left(k_{3234}+k_{1214}\right)+k_{1234}\hat{k}_{1}\cdot\hat{k}_{3}\hat{k}_{2}\cdot\hat{k}_{4},\\\fl
T_{mnop}^{EllE}&=& k_{3232}-k_{3234}\hat{k}_{2}\cdot\hat{k}_{4}-k_{1232}\hat{k}_{1}\cdot\hat{k}_{3}+k_{1234}\hat{k}_{1}\cdot\hat{k}_{3}\hat{k}_{2}\cdot\hat{k}_{4},\\\fl
T_{mnop}^{llEE}&=& k_{1212}-k_{1214}\hat{k}_{2}\cdot\hat{k}_{4}-k_{1232}\hat{k}_{1}\cdot\hat{k}_{3}+k_{1234}\hat{k}_{1}\cdot\hat{k}_{3}\hat{k}_{2}\cdot\hat{k}_{4},\\\fl
T_{mnop}^{lEEl}&=& k_{1414}-k_{1214}\hat{k}_{2}\cdot\hat{k}_{4}-k_{1434}\hat{k}_{1}\cdot\hat{k}_{3}+k_{1234}\hat{k}_{1}\cdot\hat{k}_{3}\hat{k}_{2}\cdot\hat{k}_{4},\\\fl
T_{mnop}^{lElE}&=& \de_{np}\hat{k}_{m1}\hat{k}_{o3}\hat{k}_{1}\cdot\hat{k}_{3}-\hat{k}_{1}\cdot\hat{k}_{3}\left(k_{1434}+k_{1232}\right)+k_{1234}\hat{k}_{1}\cdot\hat{k}_{3}\hat{k}_{2}\cdot\hat{k}_{4},\\\fl
T_{mnop}^{lllE}&=& k_{1232}\hat{k}_{1}\cdot\hat{k}_{3}-k_{1234}\hat{k}_{1}\cdot\hat{k}_{3}\hat{k}_{2}\cdot\hat{k}_{4},\\\fl
T_{mnop}^{llEl}&=& k_{1214}\hat{k}_{2}\cdot\hat{k}_{4}-k_{1234}\hat{k}_{1}\cdot\hat{k}_{3}\hat{k}_{2}\cdot\hat{k}_{4},\\\fl
T_{mnop}^{lEll}&=& k_{1434}\hat{k}_{1}\cdot\hat{k}_{3}-k_{1234}\hat{k}_{1}\cdot\hat{k}_{3}\hat{k}_{2}\cdot\hat{k}_{4},\\\fl
T_{mnop}^{Elll}&=& k_{3234}\hat{k}_{2}\cdot\hat{k}_{4}-k_{1234}\hat{k}_{1}\cdot\hat{k}_{3}\hat{k}_{2}\cdot\hat{k}_{4},\\\label{coeff2}\fl
T_{mnop}^{llll}&=& k_{1234}\hat{k}_{1}\cdot\hat{k}_{3}\hat{k}_{2}\cdot\hat{k}_{4}.
\eea
When evaluating the integrals $\left(\int dx\right)_{\alpha\beta\gamma\delta}$, we use again the wavefunctions previously introduced in Eqs.~(\ref{T}) and (\ref{L}). The final result is

\bea\label{p-i}\fl
\left(\int dx\right)_{EEEE}&=& \frac{1}{24k^5k_{1}^{2}k_{2}^{2}k_{3}^{2}k_{4}^{2}x^{*7}}\big[Q_{EEEE}+A_{EEEE}\,\,\, ci x^{*}\left(B_{EEEE}\cos x^{*}+C_{EEEE}\sin x^{*}\right)\nonumber\\\fl&&+D_{EEEE}\,\,\, si x^{*}\left(E_{EEEE}\cos x^{*}+F_{EEEE} \sin x^{*} \right)\big] ,\\\fl 
\left(\int dx\right)_{EEEl}&=& n^{2}(x^{*})\left(\int dx\right)_{EEEE},\\\fl 
\left(\int dx\right)_{EEll}&=& n^{4}(x^{*})\left(\int dx\right)_{EEEE}  ,\\\fl 
\left(\int dx\right)_{lllE}&=&  n^{6}(x^{*})\left(\int dx\right)_{EEEE} ,\\\fl 
\left(\int dx\right)_{llll}&=& n^{8}(x^{*})\left(\int dx\right)_{EEEE} ,
\eea 
where $Q_{EEEE}$, $A_{EEEE}$, $B_{EEEE}$, $C_{EEEE}$, $D_{EEEE}$, $E_{EEEE}$ and $F_{EEEE}$ are functions of $x^{*}$ and of the momenta $k_{i}\equiv |\vec{k}_{i}|$, $ci$ and $si$ stand respectively for the CosIntegral and the SinIntegral functions. The expressions of these functions can be found in Appendix~\ref{pointdetails}.\\
It is again important noticing that the anisotropy coefficients become zero if the gauge fields are all aligned.\\
Finally, summing up the coefficients in Eqs.~(\ref{coeff1}) through (\ref{coeff2}), one realizes that if the longitudinal and the transverse mode evolve in the same way, the total contribution from the point-interaction diagram is isotropic

\bea\label{isot-trisp}\fl
\langle\zeta_{\vec{k}_{1}}\zeta_{\vec{k}_{2}}\zeta_{\vec{k}_{3}}\zeta_{\vec{k}_{4}} \rangle &\supset&(2 \pi)^3\delta^{3}\left(\vec{k}_{1}+\vec{k}_{2}+\vec{k}_{3}+\vec{k}_{4}+\right)g_{c}^{2}\left(\frac{H_{*} x^{*}}{k}\right)^4\epsilon^{a'bc}\epsilon^{a'da}\left(\vec{N}^{c}\cdot\vec{N}^{a}\right)\left(\vec{N}^{b}\cdot\vec{N}^{d}\right)\nonumber\\\fl&\times&\frac{1}{24k^5 k^{2}_{1}k^{2}_{2}k^{2}_{3}k^{2}_{4}x^{*7}}\big[Q_{EEEE}+A_{EEEE}ci x^{*}\left(B_{EEEE}\cos x^{*}+C_{EEEE}\sin x^{*}\right)\nonumber\\\fl&&+D_{EEEE}si x^{*}\left(E_{EEEE}\cos x^{*}+F_{EEEE} \sin x^{*} \right)\big]+permutations.
\eea

\newpage

\section{Amplitude of non-Gaussianity: $f_{NL}$ and $\tau_{NL}$}\label{amplitude}
\setcounter{equation}{0}
\def\theequation{10.\arabic{equation}}

Our definitions for the non-Gaussianity amplitudes are

\bea
\frac{6}{5}f_{NL}&=&\frac{B_{\zeta}(\vec{k}_{1},\vec{k}_{2},\vec{k}_{3})}{P^{iso}(k_{1})P^{iso}(k_{2})+perms.}\\
\tau_{NL}&=&\frac{2T_{\zeta}(\vec{k}_{1},\vec{k}_{2},\vec{k}_{3},\vec{k}_{4})}{P^{iso}(k_{1})P^{iso}(k_{2})P^{iso}(k_{\hat{14}})+23\,\,  {\rm perms.}}
\eea\\
\noindent The choice of normalizing the bispectrum and the trispectrum by the isotropic part of the power spectrum, instead of using its complete expression $P_{\zeta}$, is motivated by the fact that the latter would only introduce a correction to the previous equations proportional to the anisotropy parameter $g$, which is a small quantity.\\
The parameters $f_{NL}$ and $\tau_{NL}$ receive contributions both from scalar (``$s$'') and from vector (``$v$'') fields
\bea\label{uno}
f_{NL}=f_{NL}^{(s)}+f_{NL}^{(v)} ,\\
\tau_{NL}=\tau_{NL}^{(s)}+\tau_{NL}^{(v)} .
\eea
The latter can again be distinguished into Abelian ($A$) and non-Abelian ($NA$)
\bea
f_{NL}^{(v)}=f_{NL}^{(A)}+f_{NL}^{(NA)}  ,\\\label{due}
\tau_{NL}^{(v)}=\tau_{NL}^{(A_{1})}+\tau_{NL}^{(A_{2})}+\tau_{NL}^{({NA}_{1})}+\tau_{NL}^{({NA}_{2})}.
\eea 
The contribution $f_{NL}^{(A)}$ comes from Eq.~(\ref{b}), $f_{NL}^{(NA)}$ from (\ref{fire}), $\tau_{NL}^{(A_{1})}$ and $\tau_{NL}^{(A_{2})}$ from (\ref{tr}), finally $\tau_{NL}^{({NA}_{1})}$ from (\ref{fire1}) and $\tau_{NL}^{({NA}_{2})}$ from the last line of (\ref{class4}).\\ 

\noindent In order to keep the vector contributions manageable and simple in their structure, all gauge and vector indices will be purposely neglected in this section and so the angular functions appearing in the anisotropy coefficients will be left out of the final amplitude results. This is acceptable considering that these functions will in general introduce numerical corrections of order one. Nevertheless, it is important to keep in mind that the amplitudes also depend on the angular parameters of the theory. \\
We will now focus on the dependence of $f_{NL}$ and $\tau_{NL}$ from the non-angular parameters of the theory and quickly draw a comparison among the different contributions listed in Eqs.~(\ref{uno}) through (\ref{due}).\\ 

\begin{table}[t]\centering
\caption{Order of magnitude of $f_{NL}$ in different scenarios.\\}
\begin{tabular}{|c||c|c|c|}\hline 
 $$ & $f_{NL}^{s}$ & $f_{NL}^{A}$ & $f_{NL}^{NA}$ \\
\hline
\scriptsize{general case} &  \scriptsize{$\frac{1}{(1+\beta)^2}\frac{N_{\phi\phi}}{N_{\phi}^{2}}$} & \scriptsize{$\frac{\beta}{(1+\beta)^2}\frac{N_{AA}}{N_{\phi}^{2}}$} & \scriptsize{$\frac{\beta^2}{(1+\beta)^2}g_{c}^{2}\left(\frac{m}{H}\right)^2$} \\
\hline
\scriptsize{v.inflation} &  \scriptsize{$ \frac{\epsilon}{\left(1+\left(\frac{A}{m_{P}}\sqrt{\epsilon}\right)^2\right)^{2}}$} & \scriptsize{$\frac{\epsilon^2}{\left(1+\left(\frac{A}{m_{P}}\sqrt{\epsilon}\right)^2\right)^{2}} \left(\frac{A}{m_{P}}\right)^2 $} & \scriptsize{$\frac{\epsilon^2 g_{c}^{2}}{\left(1+\left(\frac{A}{m_{P}}\sqrt{\epsilon}\right)^2\right)^{2}} \left(\frac{A^2}{m_{P}H}\right)^2$} \\
\hline
\scriptsize{v.curvaton} & $\frac{\epsilon}{\left(1+\left(\frac{A m_{P}}{A_{tot}^2}\right)^{2}\epsilon r^2\right)^2}$ & $\frac{\epsilon^2 r^3}{\left(1+\left(\frac{A m_{P}}{A_{tot}^2}\right)^{2}\epsilon r^2\right)^2}\left(\frac{A m_{P}^{2}}{A_{tot}^{3}}\right)^2$ & $\frac{\epsilon^2 r^3 g_{c}^{2}}{\left(1+\left(\frac{A m_{P}}{A_{tot}^2}\right)^{2}\epsilon r^2\right)^2}\left(\frac{A^2 m_{P}^2}{A_{tot}^{3}H}\right)^2$ \\
\hline
\end{tabular}
\label{table1}
\end{table}

\begin{table}[t]\centering
\caption{Order of magnitude of the vector contributions to $\tau_{NL}$\\ in different scenarios.\\}
\begin{tabular}{|c||c|c|c|c|}\hline 
 $$ & \scriptsize{$\tau_{NL}^{{NA}_{1}}$} & \scriptsize{$\tau_{NL}^{{NA}_{2}}$} & \scriptsize{$\tau_{NL}^{A_{1}}$} & \scriptsize{$\tau_{NL}^{A_{2}}$}   \\
\hline 
\scriptsize{general case} & \scriptsize{$10^3\frac{\beta^2\epsilon g_{c}^{2}}{\left(1+\beta\right)^3}\left(\frac{m_{P}}{H}\right)^2$} & \scriptsize{$10^{-5}\frac{\beta^{3/2}\epsilon^{3/2}g_{c}^{2}}{\left(1+\beta\right)^3}\left(\frac{A}{H}\right)\left(\frac{m_{P}}{H}\right)m_{P}^2N_{AA}$} & \scriptsize{$\frac{\beta\epsilon^2}{\left(1+\beta\right)^3}m_{P}^4N_{AA}^2$} & \scriptsize{$\frac{\beta^{3/2}\epsilon^{3/2}}{\left(1+\beta\right)^3}m_{P}^3N_{AAA}$} \\
\hline
\scriptsize{v.inflation} & \scriptsize{same as above} & \scriptsize{$10^{-5}\frac{\beta^{3/2}\epsilon^{3/2}g_{c}^{2}}{\left(1+\beta\right)^3}\left(\frac{A}{H}\right)\left(\frac{m_{P}}{H}\right)$} & \scriptsize{$\frac{\beta\epsilon^2}{\left(1+\beta\right)^3}$} & $0$  \\
\hline
\scriptsize{v.curvaton} & \scriptsize{same as above} & \scriptsize{$10^{-5}\frac{r\beta^{3/2}\epsilon^{3/2}g_{c}^{2}}{\left(1+\beta\right)^3}\left(\frac{A}{H}\right)\left(\frac{m_{P}}{H}\right)\left(\frac{m_{P}}{A}\right)^2$} & \scriptsize{$\frac{r^2\beta\epsilon^2}{\left(1+\beta\right)^3}\left(\frac{m_{P}}{A}\right)^4$} & \scriptsize{$\frac{r\beta^{3/2}\epsilon^{3/2}}{\left(1+\beta\right)^3}\left(\frac{m_{P}}{A}\right)^3$} \\
\hline
\end{tabular}
\label{table1}
\end{table}

\noindent The expression of the number of e-foldings depends on the specific model and, in particular, on the mechanism of production of the fluctuations. Two possibilities have been described in Sec.~\ref{introvec}. For ``vector inflation'' we have 
\bea
N_{a}^{i}=\frac{A^{a}_{i}}{2m_{P}^{2}},\quad\quad\quad\quad\quad\quad\quad\quad N_{ab}^{ij}=\frac{\delta_{ab}\delta^{ij}}{2m_{P}^{2}} 
\eea
(see Appendix~\ref{deltader} for their derivation). In the vector curvaton model the same quantities become \cite{Dimopoulos:2008yv,Bartolo:2009pa}
\bea
N_{a}^{i}=\frac{2}{3}r\frac{A_{i}^{a}}{\sum_{b}|\vec{A}^{b}|^2} ,\quad\quad\quad\quad\quad N_{ab}^{ij}=\frac{1}{3}r\frac{\delta_{ab}\delta^{ij}}{\sum_{c}|\vec{A}^{c}|^2}.
\eea
Neglecting tensor and gauge indices, the expressions above can be simplified as $N_{A}\simeq A/m_{P}^2$ and $N_{AA}\simeq 1/m_{P}^{2}$ in vector inflation, $N_{A}\simeq r/A$ and $N_{AA}\simeq r/A^{2}$ in the vector curvaton model. Also we have $N_{AAA}=0$ in vector inflation and $N_{AAA}\simeq r/A^3$ in vector curvaton.\\

\noindent We are now ready to provide the final expressions for the amplitudes: in Table~1 we list all the contributions to $f_{NL}$, Table~2 includes the vector contributions to $\tau_{NL}$, the scalar contributions being given by
\bea
\tau_{NL}^{(s)}=\frac{\epsilon}{\left(1+\beta\right)^3}+\frac{\epsilon^2}{\left(1+\beta\right)^3}.
\eea

\begin{table}[t]\centering
\caption{Order of magnitude of the ratios $f_{NL}^{v}/f_{NL}^{s}$ in different scenarios.\\}
\begin{tabular}{|c||c|c|}\hline 
 $$ & \scriptsize{$f_{NL}^{A}/f_{NL}^{s}$} & \scriptsize{$f_{NL}^{NA}/f_{NL}^{s}$}   \\
\hline 
\scriptsize{general case} & \scriptsize{$\beta\frac{N_{AA}}{N_{\phi\phi}}$} & \scriptsize{$\beta^2 g_{c}^{2}\left(\frac{m}{H}\right)^2\frac{N_{\phi}^{2}}{N_{\phi\phi}}$}  \\
\hline
\scriptsize{v.inflation} & \scriptsize{$\beta$ } & \scriptsize{$\frac{\beta^2 g_{c}^{2}}{\epsilon}\left(\frac{m_{P}}{H}\right)^2$}   \\
\hline
\scriptsize{v.curvaton} & \scriptsize{$\beta r\left(\frac{m_{P}}{A}\right)^2$} & \scriptsize{$\frac{\beta^2 g_{c}^{2}}{\epsilon r}\left(\frac{A}{H}\right)^2$}  \\
\hline
\end{tabular}
\label{table2}
\end{table}

\begin{table}[t]\centering
\caption{Order of magnitude of the ratios $\tau_{NL}^{v}/\tau_{NL}^{s}$ in different scenarios.\\}
\begin{tabular}{|c||c|c|c|c|}\hline 
 $$ & \scriptsize{$\tau_{NL}^{{NA}_{1}}/\tau_{NL}^{s}$} & \scriptsize{$\tau_{NL}^{{NA}_{2}}/\tau_{NL}^{s}$} & \scriptsize{$\tau_{NL}^{A_{1}}/\tau_{NL}^{s}$} & \scriptsize{$\tau_{NL}^{A_{2}}/\tau_{NL}^{s}$}   \\
\hline 
\scriptsize{general case} & \scriptsize{$10^3\beta^2 g_{c}^{2}\left(\frac{m_{P}}{H}\right)^2$} & \scriptsize{$10^{-5}\beta^{3/2}\epsilon^{1/2}g_{c}^{2}\left(\frac{A}{H}\right)\left(\frac{m_{P}}{H}\right)m_{P}^2N_{AA}$} & \scriptsize{$\beta\epsilon m_{P}^4N_{AA}^2$} & \scriptsize{$\beta^{3/2}\epsilon^{1/2}m_{P}^3N_{AAA}$} \\
\hline
\scriptsize{v.inflation} & \scriptsize{same as above} & \scriptsize{$10^{-5}\beta^{3/2}\epsilon^{1/2}g_{c}^{2}\left(\frac{A}{H}\right)\left(\frac{m_{P}}{H}\right)$} & \scriptsize{$\beta\epsilon$} & $0$  \\
\hline
\scriptsize{v.curvaton} & \scriptsize{same as above} & \scriptsize{$10^{-5}r\beta^{3/2}\epsilon^{1/2}g_{c}^{2}\left(\frac{A}{H}\right)\left(\frac{m_{P}}{H}\right)\left(\frac{m_{P}}{A}\right)^2$} & \scriptsize{$r^2\beta\epsilon\left(\frac{m_{P}}{A}\right)^4$} & \scriptsize{$r\beta^{3/2}\epsilon^{1/2}\left(\frac{m_{P}}{A}\right)^3$} \\
\hline
\end{tabular}
\label{table2}
\end{table}

\noindent In the expressions appearing in the tables, numerical coefficients of order one have not been reported. Also, $m$ is by definition equal to $m_{P}$ in vector inflation and to $A/\sqrt{r}$ in the vector curvaton model; $N_{\phi}\simeq (m_{P}\sqrt{\epsilon})^{-1}$ and $N_{\phi\phi}\simeq m_{P}^{-2}$, with $\epsilon \equiv(\dot{\phi}^2)/(2 m_{P}^{2}H^{2})$.\\

\noindent The quantities involved in the amplitude expressions are $g$, $\beta$, $r$, $\epsilon$, $g_{c}$, $m_{P}/H$, $A/m_{P}$ and $A/H$. We already know that $g$ and $\beta$ are to be considered smaller than one (see discussion after Eq.~(\ref{pdisc})). Similarly, as mentioned after Eq.~(\ref{indirect}), $r$ has to remain small at least until inflation ends so as to attain an ``almost isotropic'' expansion. The slow-roll parameter $\epsilon$ and the $SU(2)$ coupling $g_{c}$ are small respectively to allow the inflaton to slowly roll down its potential and for perturbation theory to be valid. The ratio $m_{P}/H$ is of order $10^{5}$ (assuming $\epsilon \sim 10^{-1}$). Finally, $A/m_{P}$ and $A/H$ have no stringent bounds. A reasonable choice could be to assume that the expectation value of the gauge fields is no larger than the Planck mass, i.e. $A/m_{P}\leq 1$. As to the $A/H$ ratio, different possibilities are allowed, including the one where it is of order one (see Sec.~6 of \cite{Bartolo:2009pa} for a discussion on this).\\

\noindent Let us now compare the different amplitude contributions. The ratios between scalar and vector contributions are shown in Table~3 for the bispectrum and Table~4 for the trispectrum. We can observe that the dominance of a given contribution w.r.t. another one very much depends on the selected region of parameter space. It turns out that it is allowed for the vector contributions to be larger than the scalar ones and also for the non-Abelian contributions to be larger than the Abelian ones. This is discussed more in details in Sec.~6 of \cite{Bartolo:2009pa}. An interesting point is, for instance, the following: ignoring tensor and gauge indices, the ratio $g_{c}A/H$, that appears in many of the Tables entries, is a quantity smaller than one; if we consider the different configurations identified by gauge and vector indices, we realize that this is not always true, in fact the value of this ratio can be $\gg 1$ in some configurations (see also Appendix~\ref{eomotion} and Eqs.~(\ref{cond1}) through (\ref{cond3}) in particular). \\
Finally, it is interesting to compare bispectrum and trispectrum amplitudes (see Table~5). Again, it is allowed for the ratios appearing in Table~5 to be either large or small, depending on the specific location within the parameter space of the theory. For instance, the combination of a small bispectrum with a large trispectrum is permitted. The latter is an interesting possibility: if the bispectrum was observably small, we could still hope the information about non-Gaussianity to be accessible thanks to the trispectrum.\\ 
Another interesting feature of this model is that the bispectrum and the trispectrum depend on the same set of quantities. If these correlation functions were independently known, that information could then be used to test the theory and place some bounds on its parameters.

\begin{table}[t]\centering
\caption{Order of magnitude of the ratios $\tau_{NL}^{v}/\left(f_{NL}^{NA}\right)^2$ in different scenarios.\\}
\begin{tabular}{|c||c|c|c|c|}\hline 
 $$ & \scriptsize{$\tau_{NL}^{{NA}_{1}}/\left(f_{NL}^{NA}\right)^2$} & \scriptsize{$\tau_{NL}^{{NA}_{2}}/\left(f_{NL}^{NA}\right)^2$} & \scriptsize{$\tau_{NL}^{A_{1}}/\left(f_{NL}^{NA}\right)^2$} & \scriptsize{$\tau_{NL}^{A_{2}}/\left(f_{NL}^{NA}\right)^2$}   \\
\hline 
\scriptsize{v.i.} & \scriptsize{$10^9\frac{\epsilon\left(1+\beta\right)}{g_{c}^{2}\beta^2}\left(\frac{H}{m_{P}}\right)^2$} & \scriptsize{$10\frac{\epsilon^{3/2}\left(1+\beta\right)}{\beta^{5/2}g_{c}^{2}}\left(\frac{A}{H}\right)\left(\frac{H}{m_{P}}\right)^3$} & \scriptsize{$10^6\frac{\epsilon^2\left(1+\beta\right)}{\beta^3g_{c}^{4}}\left(\frac{H}{m_{P}}\right)^4$} & $0$  \\
\hline
\scriptsize{v.c.} & \scriptsize{$10^9\frac{r^2\epsilon\left(1+\beta\right)}{g_{c}^{2}\beta^2}\frac{m_{P}^{2}}{A^{2}}\frac{H^{2}}{A^{2}}$} & \scriptsize{$10\frac{r^5\epsilon^{3/2}\left(1+\beta\right)}{\beta^{5/2}g_{c}^{2}}\frac{H^{3}}{A^{3}}\frac{m_{P}}{H}\frac{m_{P}^{2}}{A^{2}}$} & \scriptsize{$10^6\frac{r^6\epsilon^2\left(1+\beta\right)}{\beta^3 g_{c}^{4}}\left(\frac{m_{P}}{A}\right)^4\left(\frac{H}{A}\right)^4$} & \scriptsize{$10^6\frac{r^3\epsilon^{3/2}\left(1+\beta\right)}{g_{c}^{2}\beta^{5/2}}\frac{m_{P}^{3}}{A^{3}}\frac{H^{4}}{A^{4}}$} \\
\hline
\end{tabular}
\label{table3}
\end{table}

\newpage

\section{Shape of non-Gaussianity and statistical anisotropy features}
\label{shapes}
\setcounter{equation}{0}
\def\theequation{11.\arabic{equation}}

Studying the shape of non-Gaussianity means understanding the features of momentum dependence of the bispectrum and higher order correlators (see e.g. \cite{Babich:2004gb}). If they also depend on variables other than momenta, it is important to determine how these other variables affect the profiles for any given momentum set-up. This is the case as far as the bispectrum and the trispectrum of the gauge fields are concerned, given the fact that they are functions, besides of momenta, also of a large set of angular variables (see Eqs.~(\ref{fire}) and (\ref{fire1})).

\subsection{Momentum dependence of the bispectrum and trispectrum}

\noindent We show the study of the momentum dependence of the $F_{n}$ and $G_{n}$ functions in Eqs.~(\ref{fire}) and (\ref{fire1}) first and then analyze the angular variables dependence of the spectra, once the momenta have been fixed in a given configuration. A natural choice would be to consider the configuration where the correlators are maximized. \\
The maxima can be easily determined for the bispectrum by plotting the isotropic functions $F_n$ and $G_n$ in terms of two of their momenta. These plots are provided in Fig.~7, where the variables are $x_{2}\equiv k_{2}/k_{1}$ and $x_{3}\equiv k_{3}/k_{1}$. Each one of the plots corresponds to one of the isotropic functions appearing in the sum in Eq.~(\ref{fire}). It is apparent that the maxima are mostly located in the in the so-called local region, i.e. for $k_{1}\sim k_{2}\gg k_{3}$; three out of the eight graphs do not have their peaks in this configuration but, at the same time, they show negligible amplitudes compared to the ``local'' peaked graphs.

\begin{figure}\centering
 \includegraphics[width=0.4\textwidth]{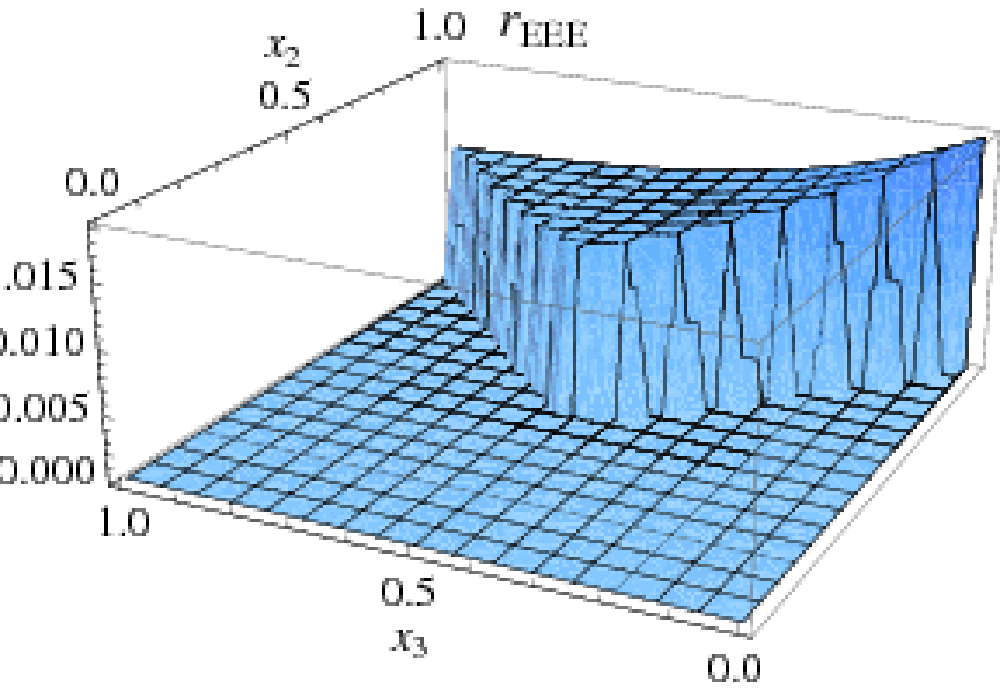}
\hspace{0.1\textwidth}
 \includegraphics[width=0.4\textwidth]{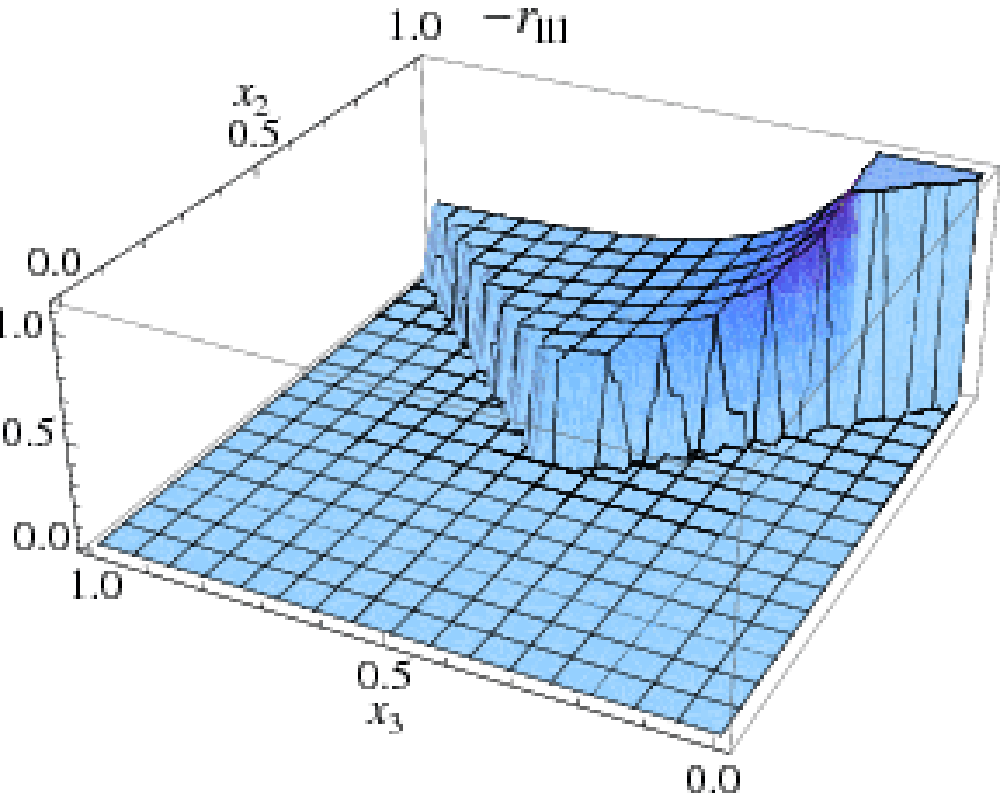}
\vspace{0.02\textwidth}
 \includegraphics[width=0.4\textwidth]{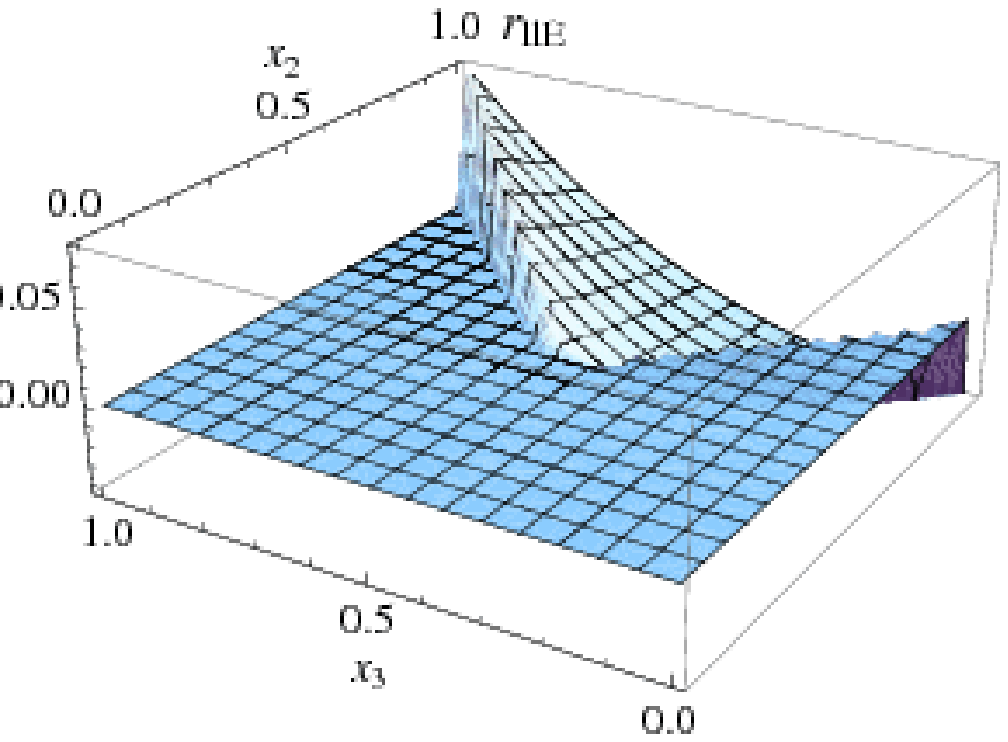}
\hspace{0.1\textwidth}
 \includegraphics[width=0.4\textwidth]{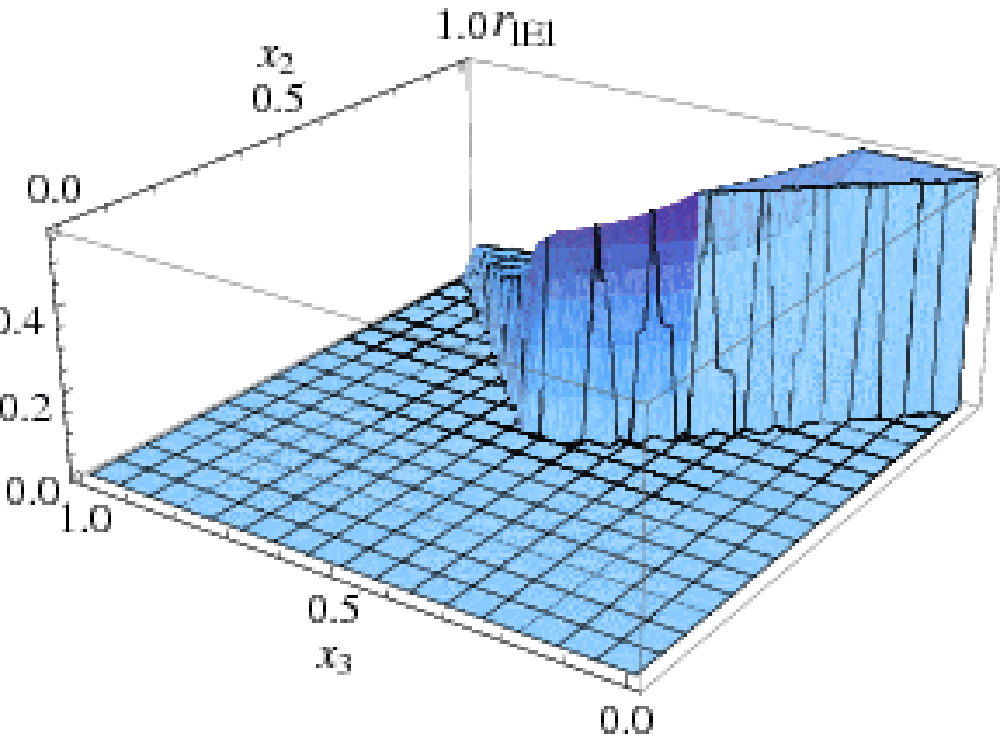}
\vspace{0.02\textwidth}
 \includegraphics[width=0.4\textwidth]{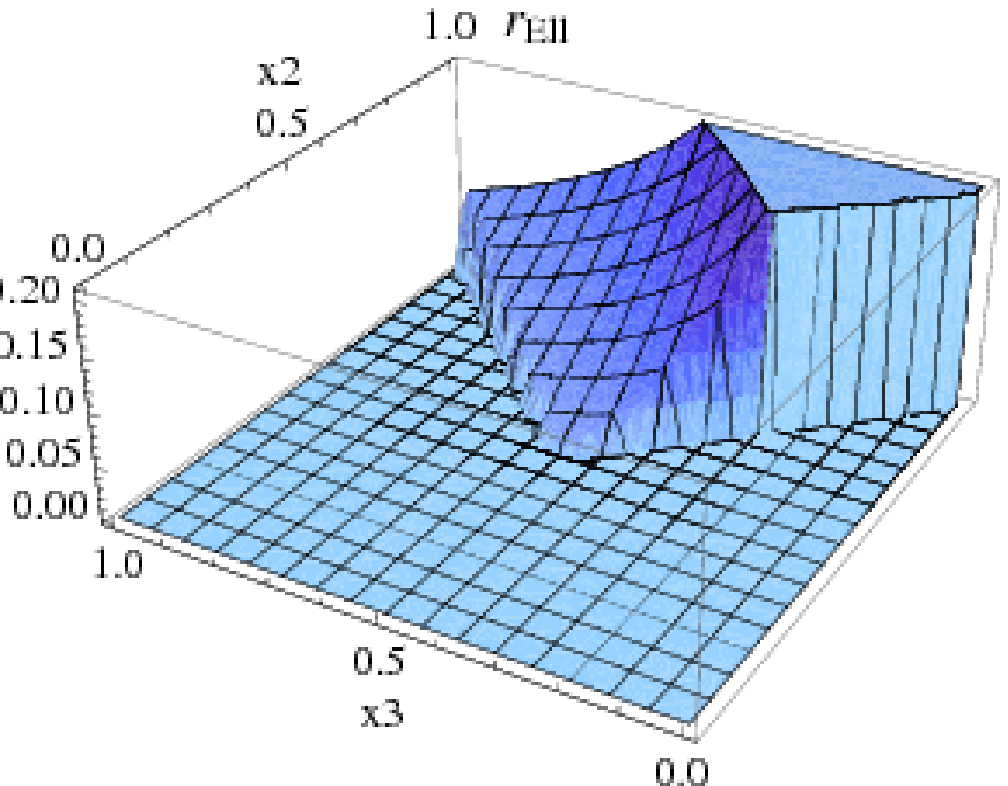}
\hspace{0.1\textwidth}
 \includegraphics[width=0.4\textwidth]{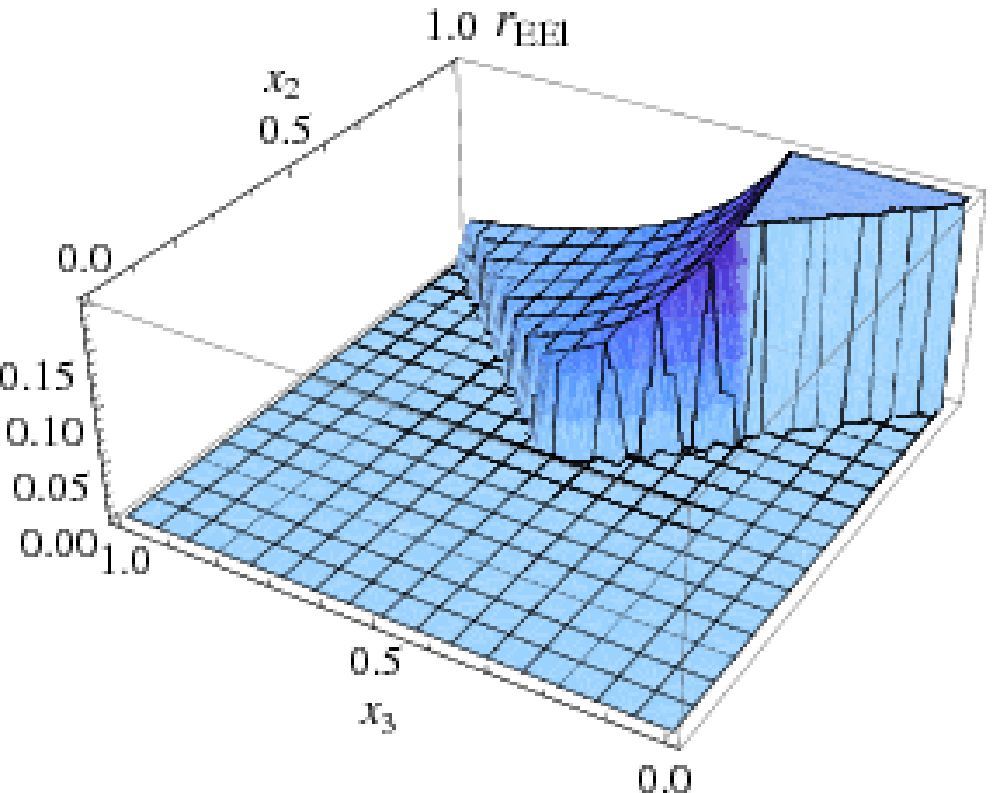}
\vspace{0.02\textwidth}
 \includegraphics[width=0.4\textwidth]{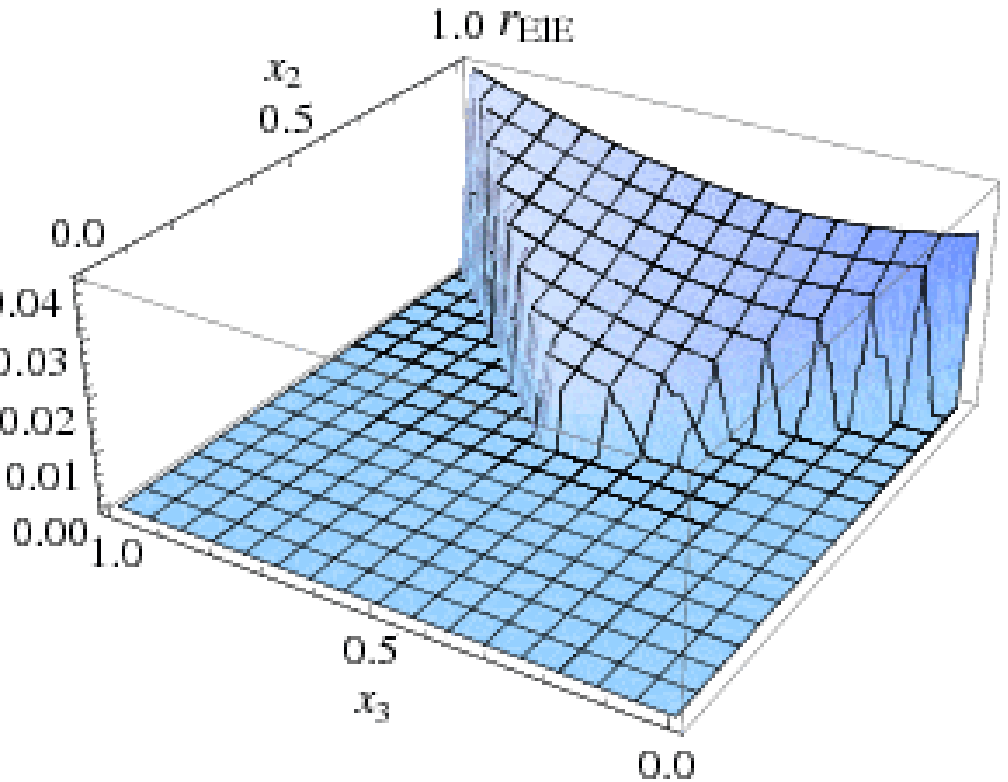}
\hspace{0.1\textwidth}
 \includegraphics[width=0.4\textwidth]{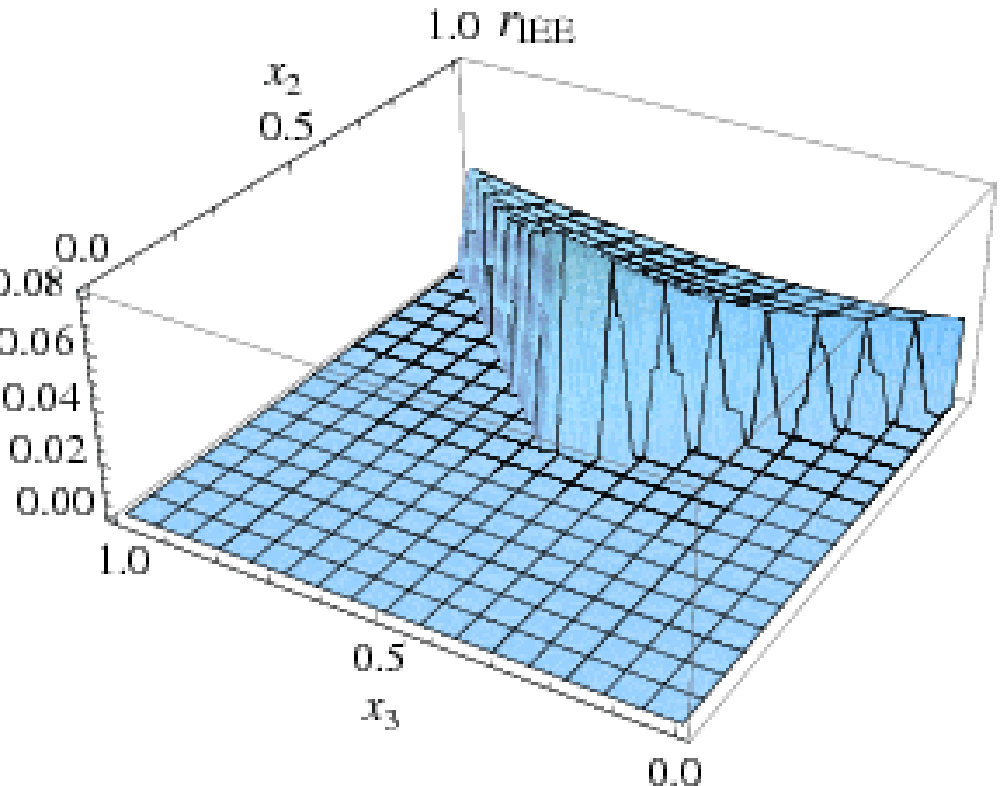}
\caption{ \label{Fig1}Plot of $r_{n}\equiv \Theta(x_{2}-x_
{3})\Theta(x_{3}-1+x_{2}) x_{2}^{2}x^{2}_{3}R_{n}(x_{2},x_{3})$, where\\we define $R_{n}=k_{1}^{6}F_{n}$. The Heaviside step functions $\Theta$ help
restricting\\the plot domain to the region $(x_{2},x_{3})$ that is allowed
for the triangle\\$\vec{k}_{1}+\vec{k}_{2}+\vec{k}_{3}=0$ (in particular, we
set $x_{3}<x_{2}$). We also set $x^{*}=1$.}
\end{figure}

\noindent The situation is much more complex for the trispectrum, being the number of momentum variables larger than three ($k_{1}$, $k_{2}$, $k_{3}$, $k_{4}$, $k_{\hat{12}}$ and $k_{\hat{14}}$). The momentum dependence of the isotropic functions can be studied by selecting different configurations for the tetrahedron made up by the four momentum vectors, in such a way as to narrow the number of independent momentum variables down to two. A list of possible configurations was presented in \cite{Chen:2009bc}. We consider two of them, the ``equilateral'' and the ``specialized planar''.

\begin{figure}\centering
\includegraphics[width=0.4\textwidth]{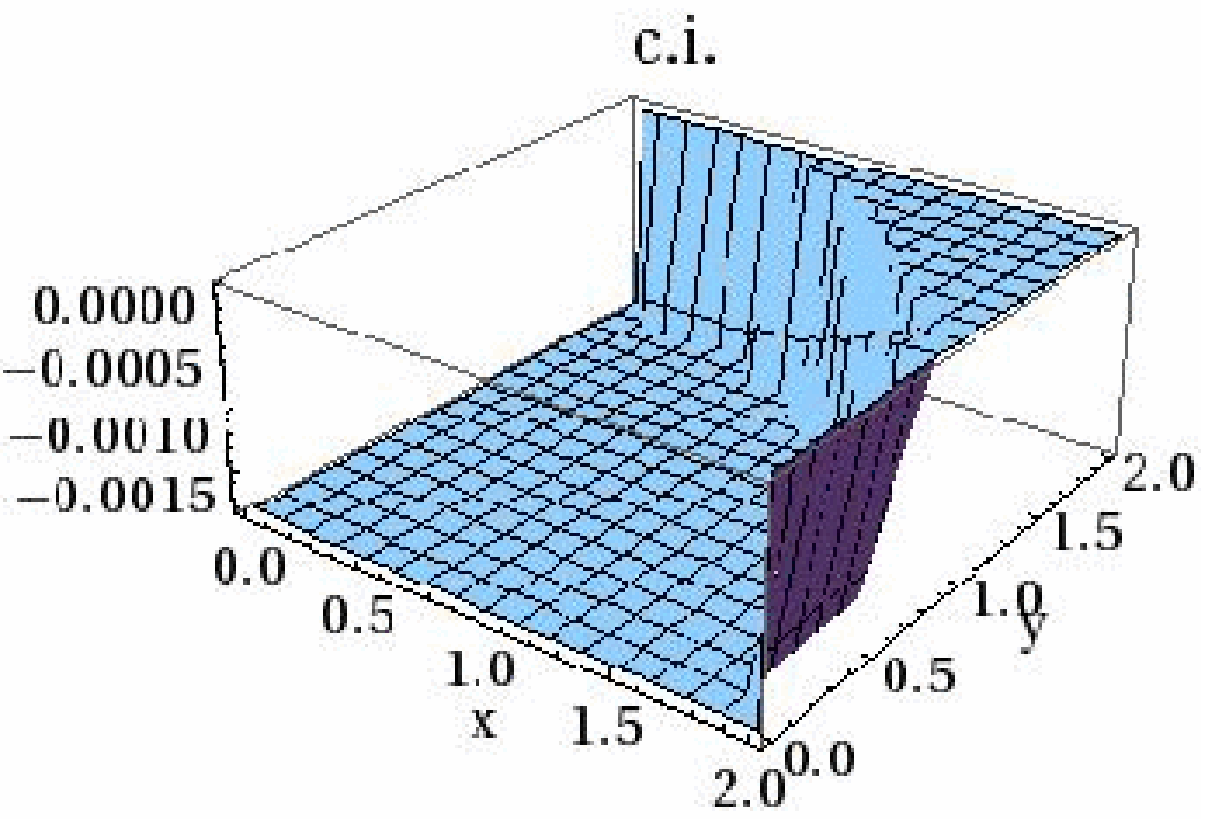}
\hspace{0.1\textwidth}
 \includegraphics[width=0.4\textwidth]{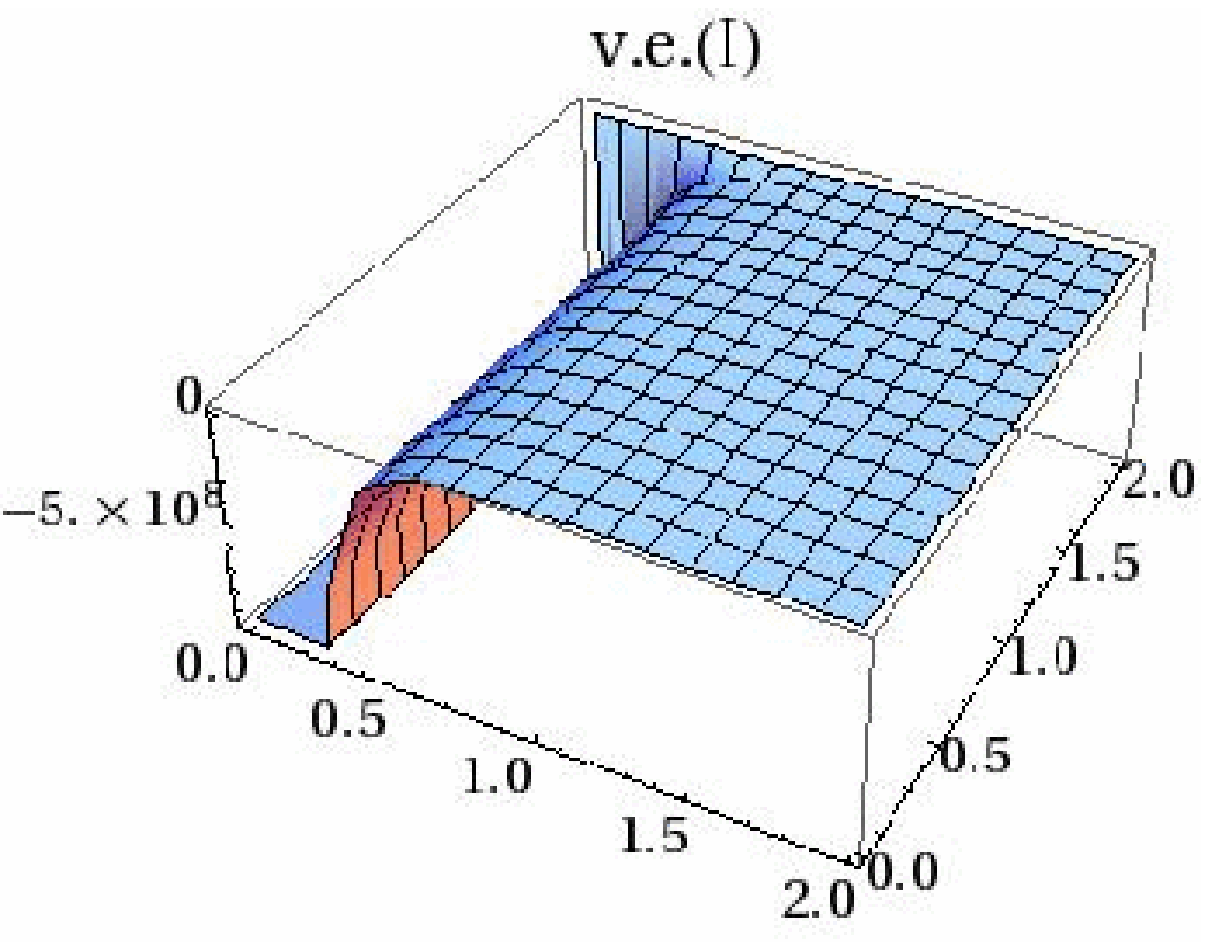}
\vspace{0.02\textwidth}
 \includegraphics[width=0.4\textwidth]{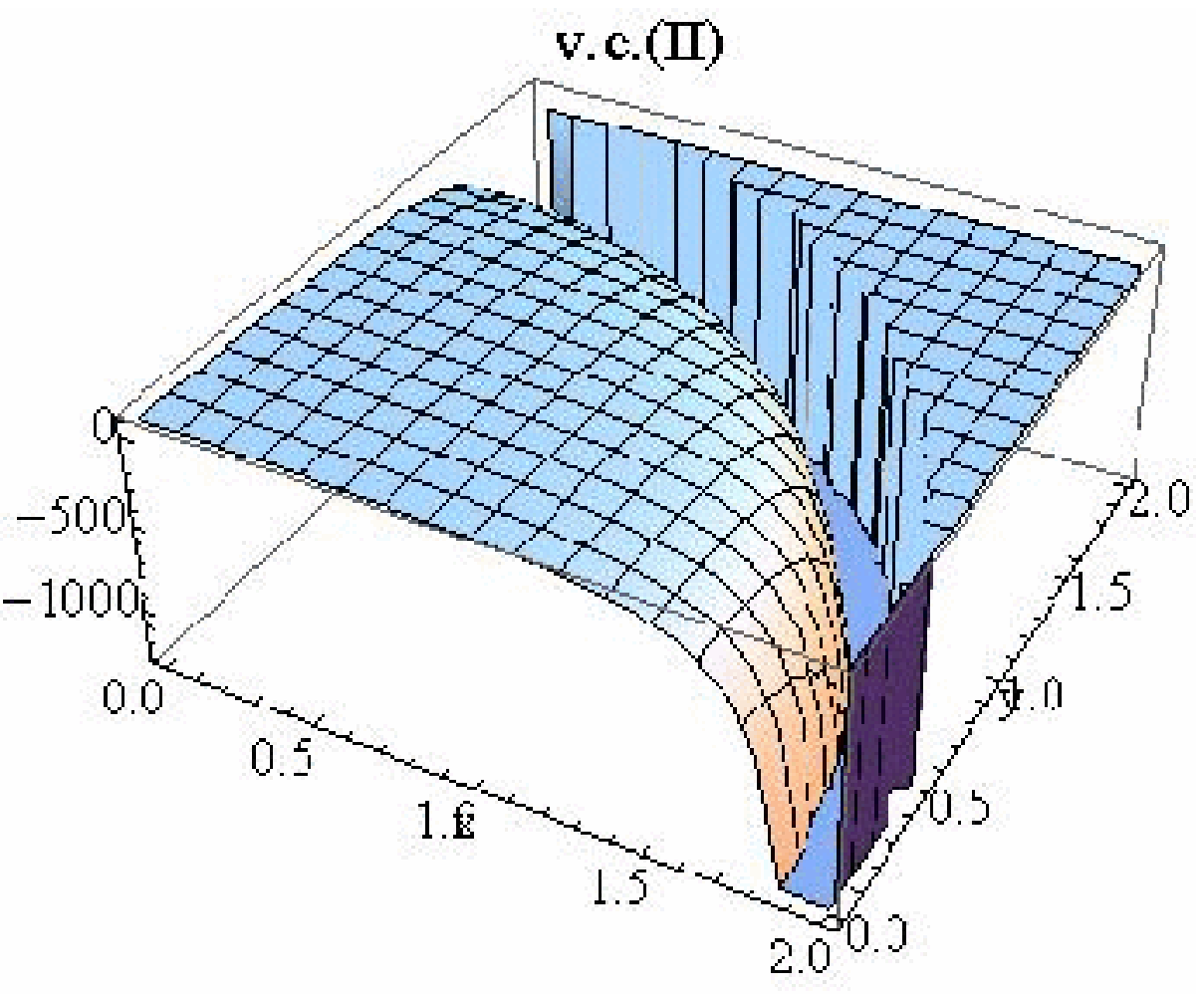}
\hspace{0.1\textwidth} \includegraphics[width=0.4\textwidth]{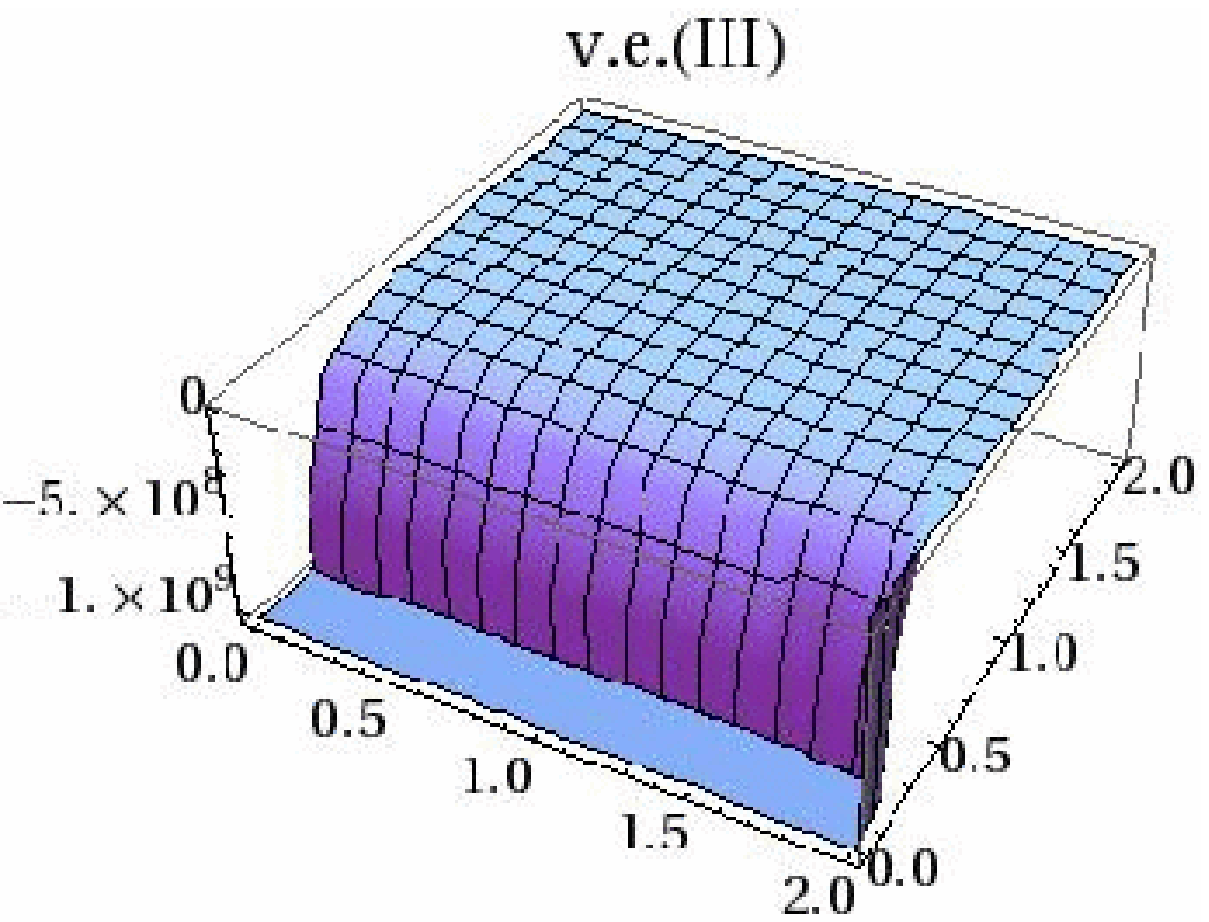}
\caption{ \label{Fig2} Plots of the isotropic functions appearing in the vector fields trispectrum (from Eq.~(\ref{fire1})): $c.i.$ is the contribution from contact-interaction diagrams, $v.e.(I)$, $v.e.(II)$ and $v.e.(III)$ are the contributions from the vector-exchange diagrams. The equilateral configuration has been considered in this figure.}
\end{figure}

\begin{figure}[t]\centering
\includegraphics[width=0.4\textwidth]{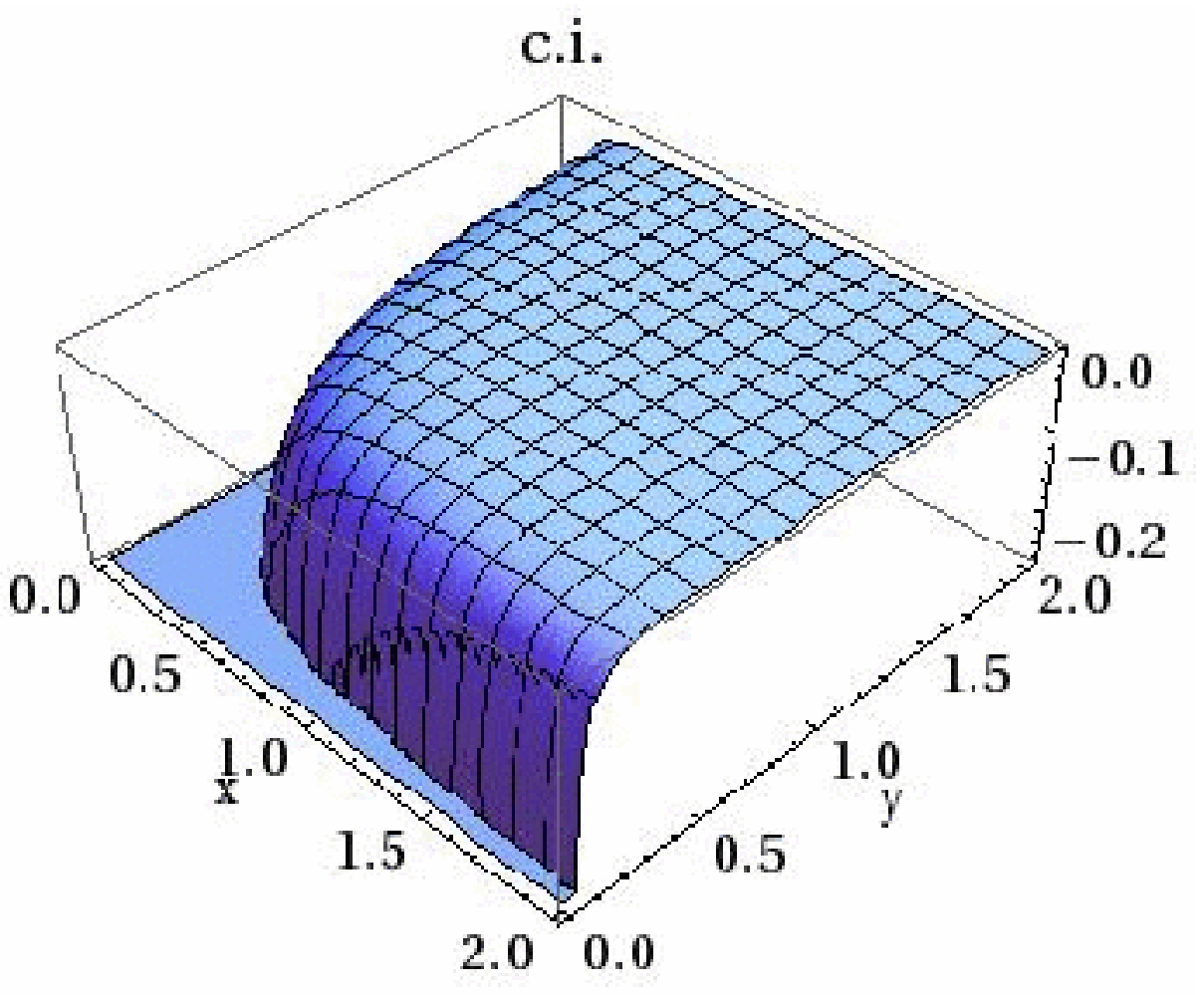}
\hspace{0.1\textwidth}
 \includegraphics[width=0.4\textwidth]{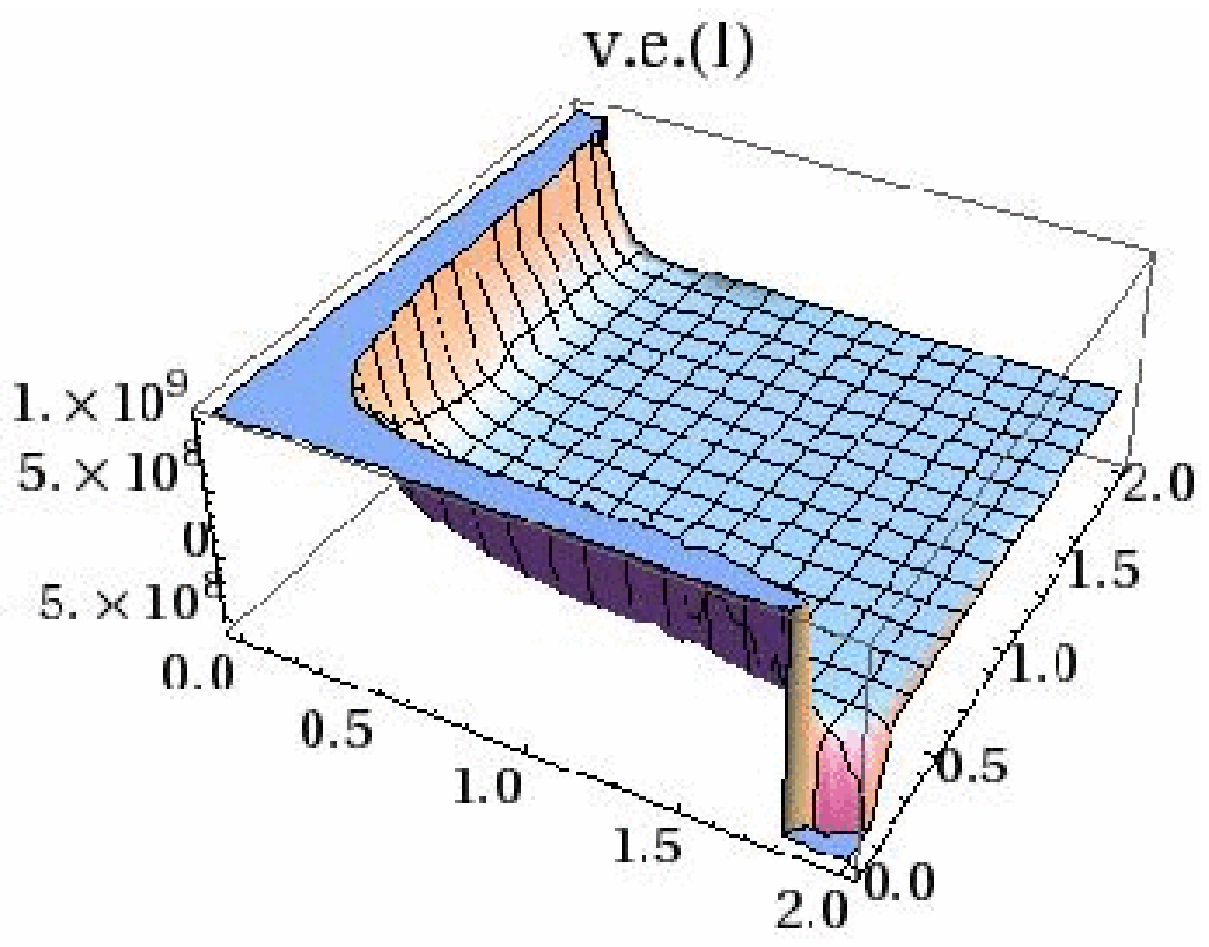}
\vspace{0.02\textwidth}
 \includegraphics[width=0.4\textwidth]{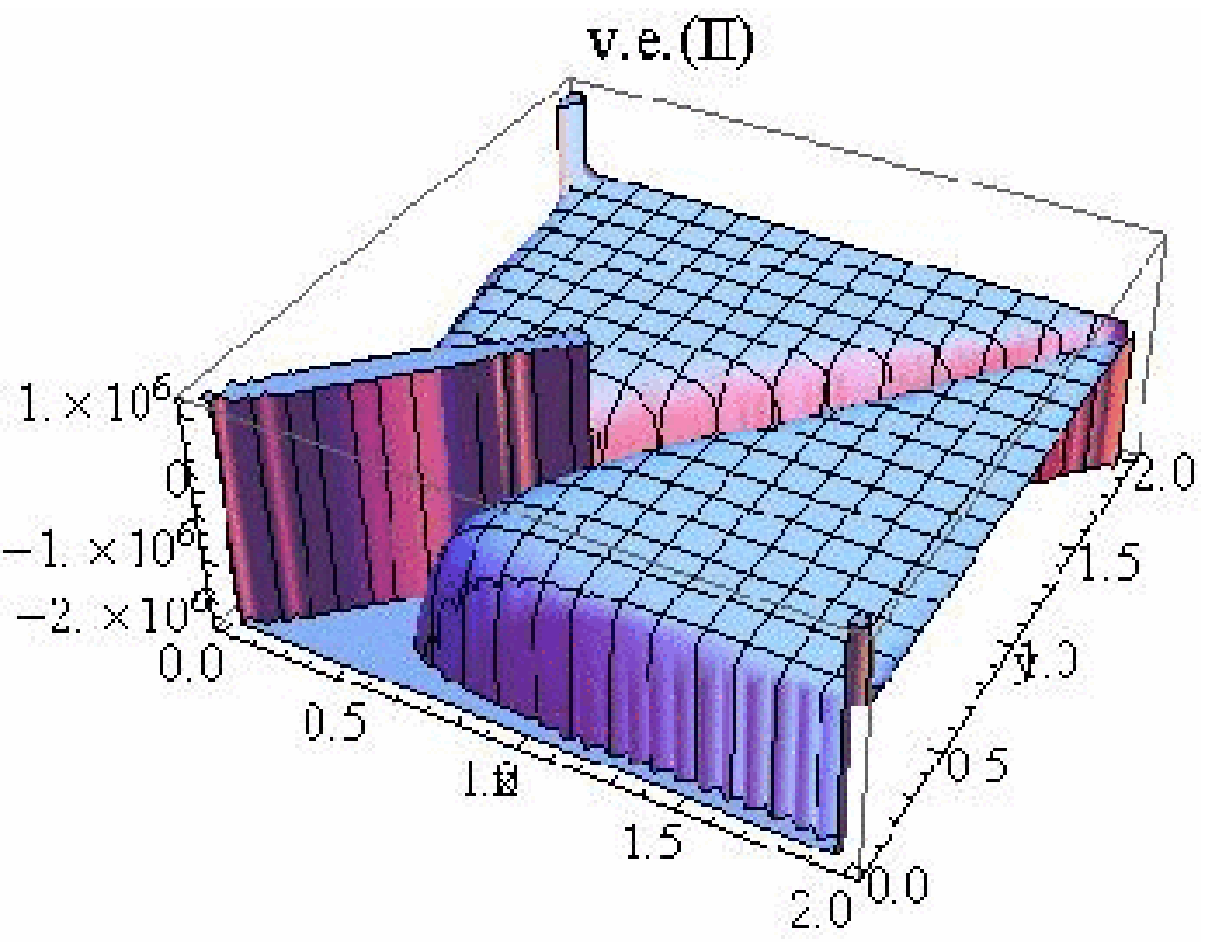}
\hspace{0.1\textwidth}
 \includegraphics[width=0.4\textwidth]{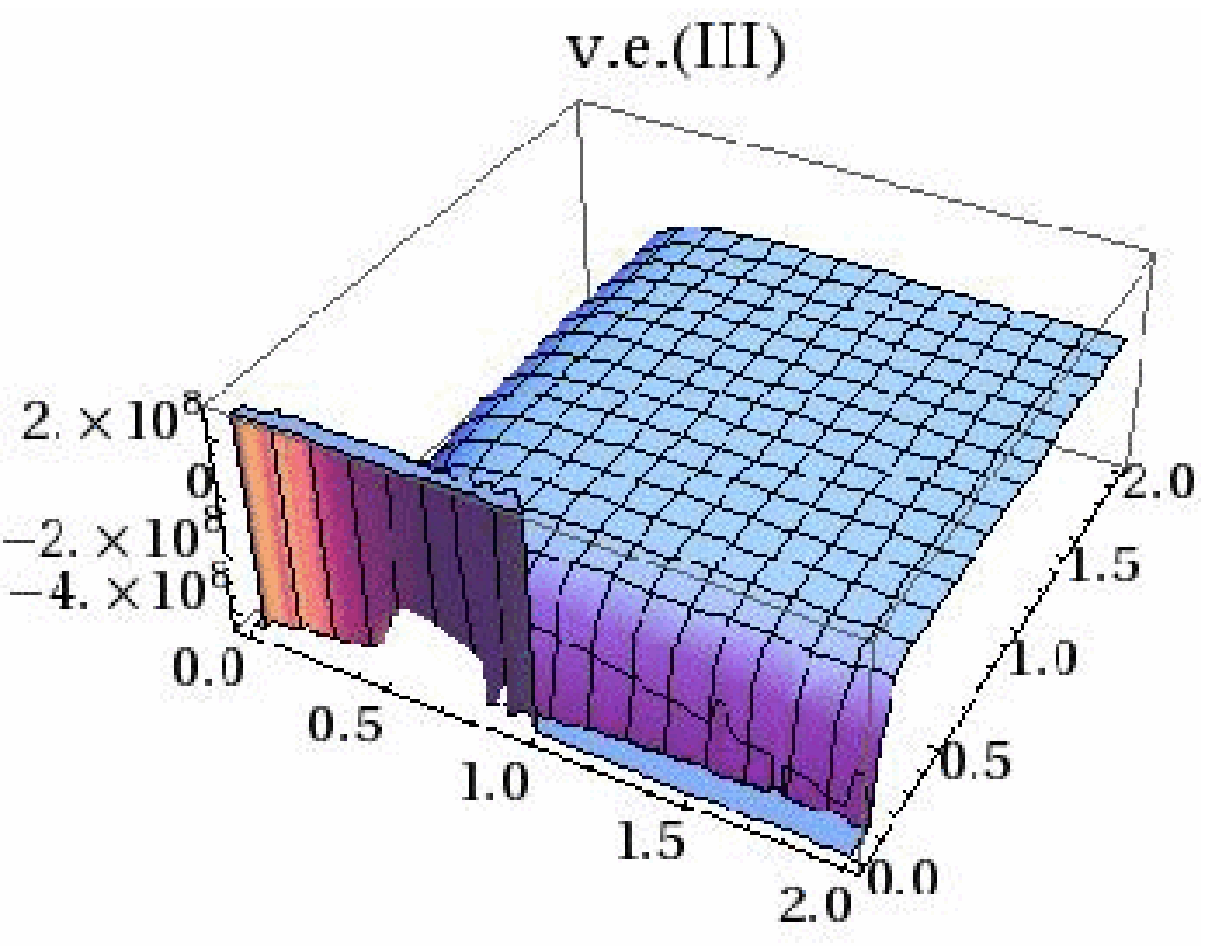}
\caption{ \label{Fig3} Plots of the contact interaction and of the vector-exchange contributions in the specialized planar configuration (plus sign).}
\end{figure}

\begin{figure}\centering
\includegraphics[width=0.4\textwidth]{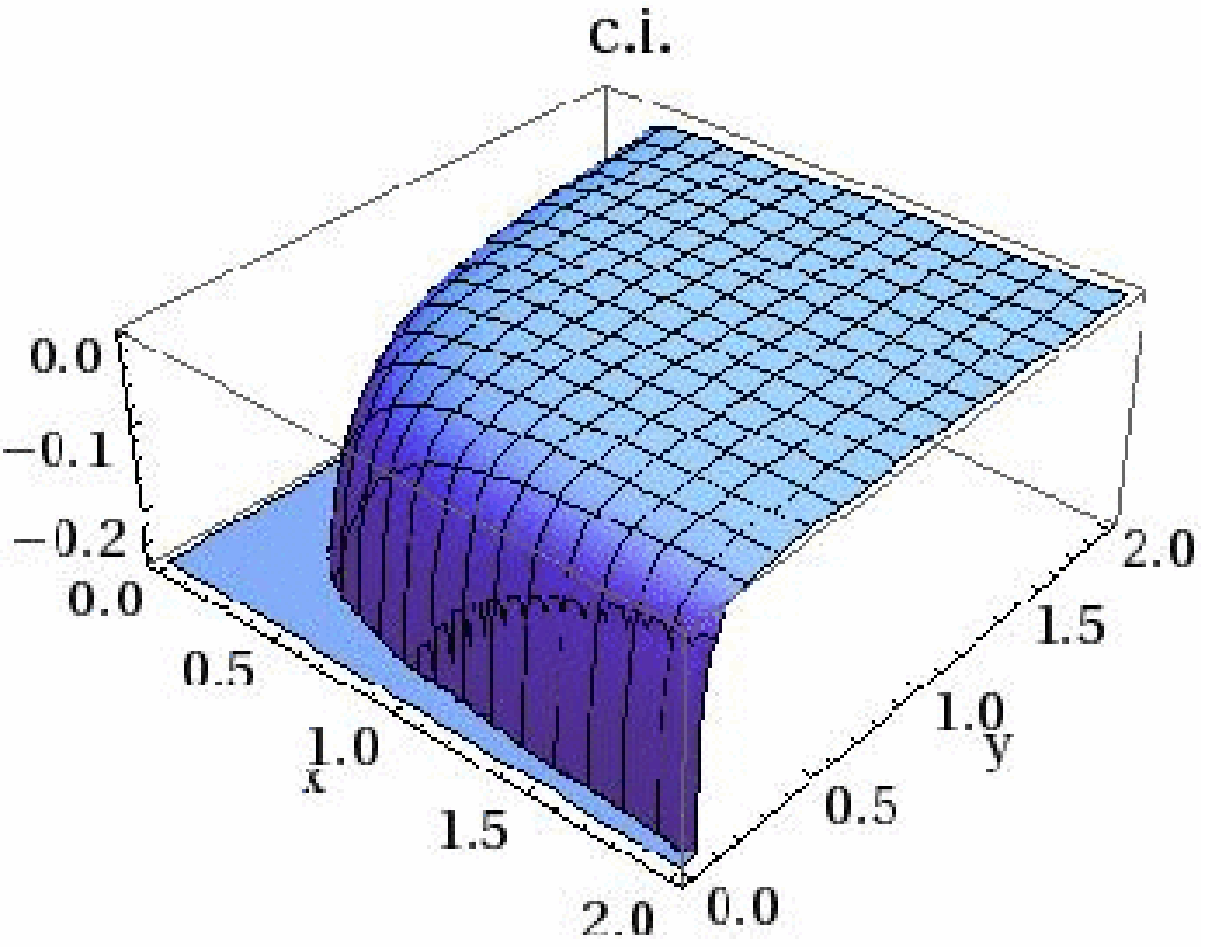}
\hspace{0.1\textwidth}
 \includegraphics[width=0.4\textwidth]{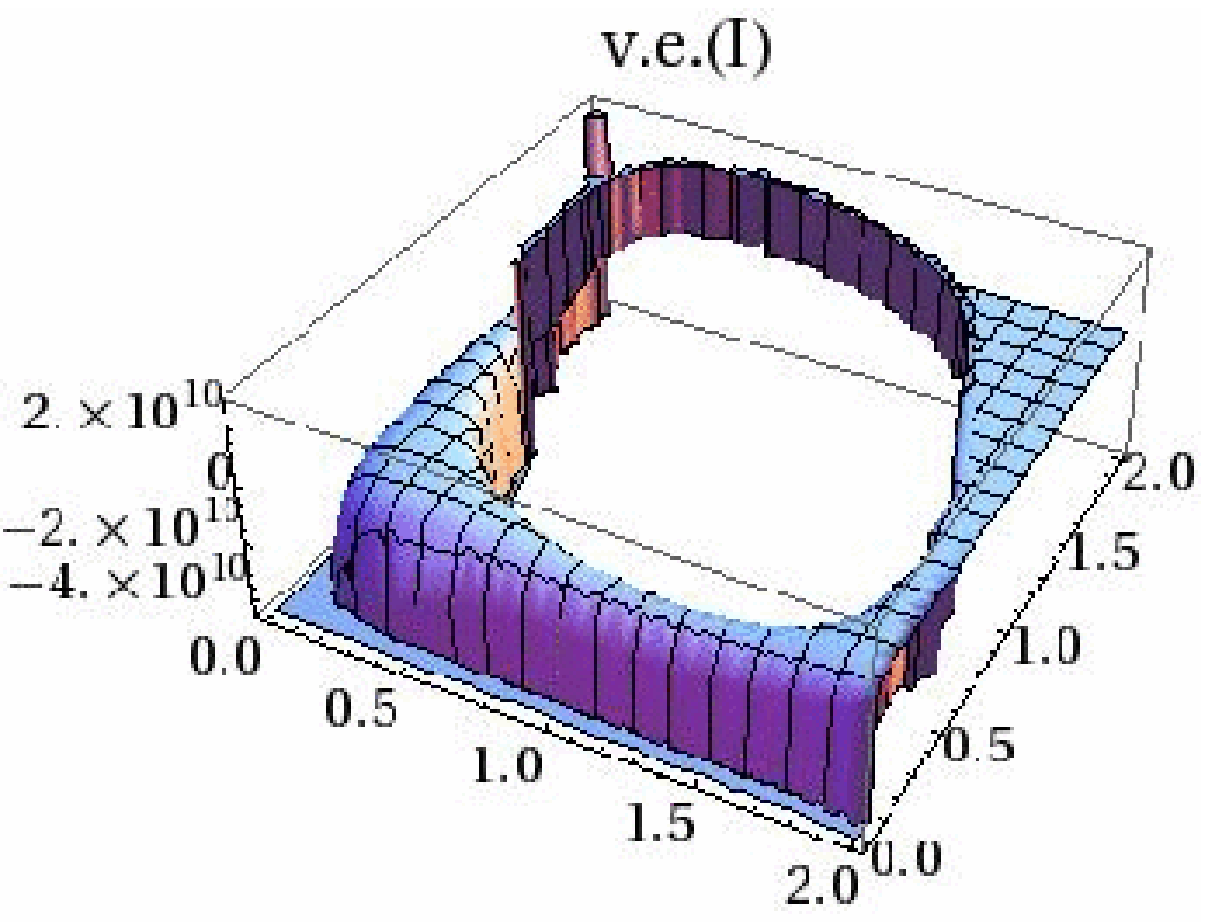}
\vspace{0.02\textwidth}
 \includegraphics[width=0.4\textwidth]{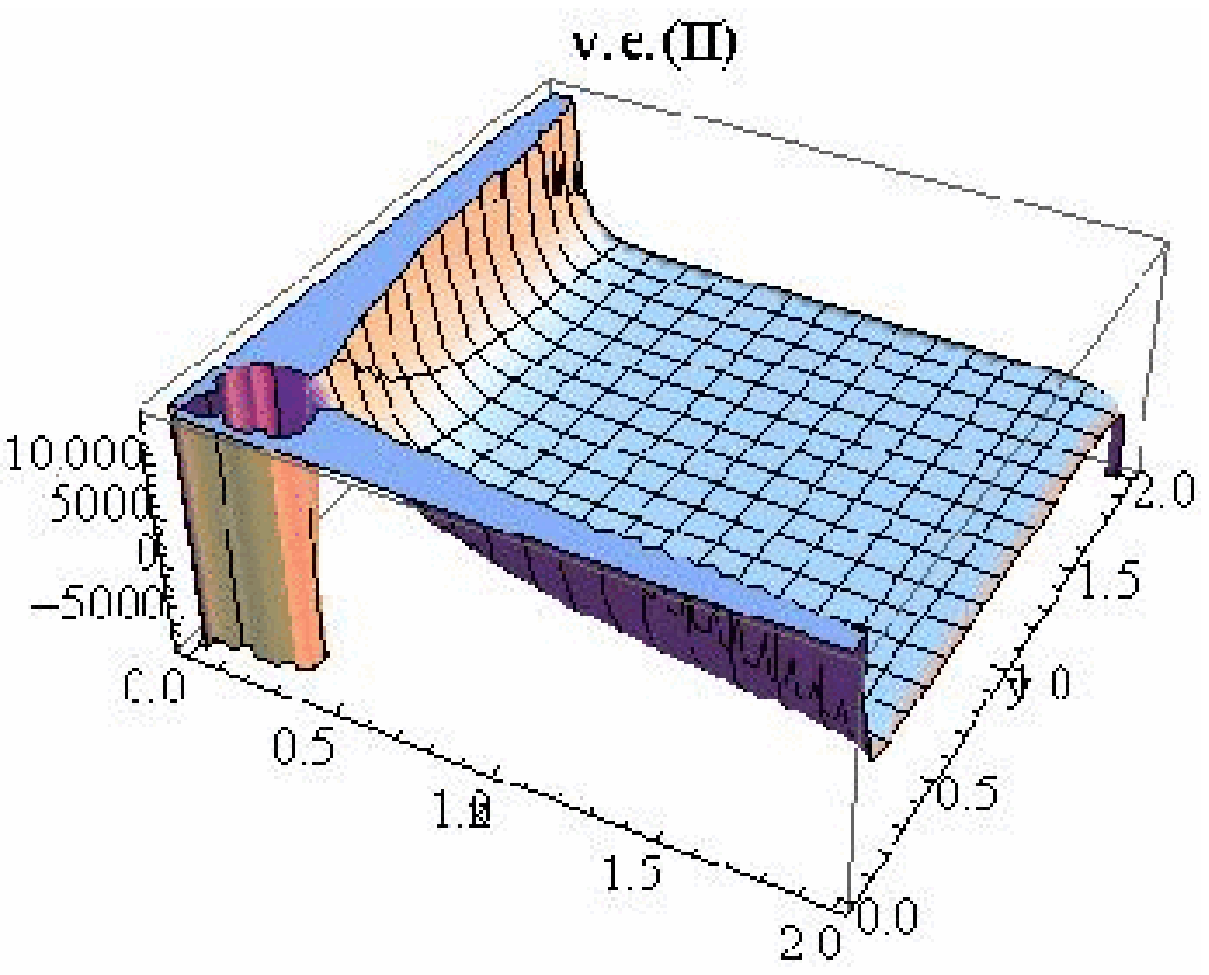}
\hspace{0.1\textwidth}
 \includegraphics[width=0.4\textwidth]{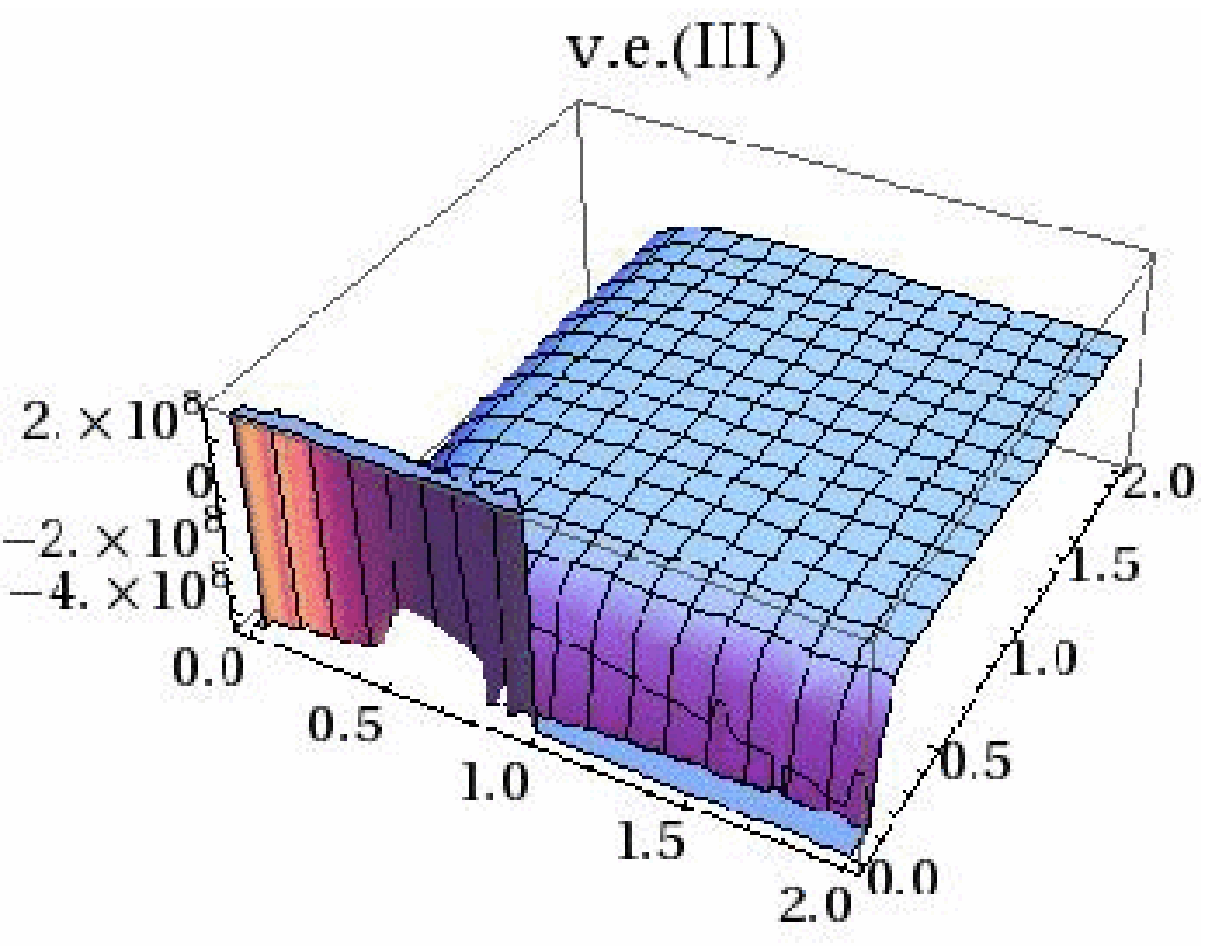}
\caption{ \label{Fig4} Plots of the contact interaction and of the vector-exchange contributions in the specialized planar configuration (minus sign).}
\end{figure}

\noindent In the equilateral configuration the four sides of the tetrahedron have the same length ($k_{1}=k_{2}=k_{3}=k_{4}$), therefore $x\equiv k_{\hat{12}}/k_{1}$ and $y\equiv k_{\hat{14}}/k_{1}$ can be chosen as variables for the plots. The plots of the isotropic functions of contact interaction and vector exchange contributions are provided in Fig.~8. The former ($c.i.$) shows a constant behaviour in this configuration, being independent of $k_{\hat{12}}$ and $k_{\hat{14}}$. The latter ($v.e.(I)$, $v.e.(II)$ and $v.e.(III)$) diverge as $k_{\hat{1i}}^{-3}$ ($i=1,2,3$ respectively for the three plots) in the limit of a flat tetrahedron, i.e. $(k_{\hat{1i}}/k_{1})\rightarrow 0$.

\noindent In the specialized planar configuration, the tetrahedron is flattened and, in addition to that, three of the six momentum variables are set equal to one another ($k_{1}=k_{3}=k_{\hat{14}}$); this leaves two independent variables, which can be $x\equiv k_{2}/k_{1}$ and $y\equiv k_{3}/k_{1}$. There is a double degeneracy in this configuration, due to the fact that the quadrangle can have internal angles larger than or smaller/equal to $\pi$, as we can see from the plus and minus signs in the expressions for $k_{\hat{12}}$ and $k_{\hat{13}}$ \cite{Chen:2009bc}
\bea\label{k12}
\frac{k_{\hat{12}}}{k_{1}}=\sqrt{1+\frac{x^2 y^2}{2}\pm\frac{xy}{2}\sqrt{(4-x^2)(4-y^2)}},\\\label{k13}
\frac{k_{\hat{13}}}{k_{1}}=\sqrt{x^2+y^2-\frac{x^2 y^2}{2}\mp\frac{xy}{2}\sqrt{(4-x^2)(4-y^2)}}.
\eea
The two cases are plotted in Figs.~9 and 10. Notice that divergences generally occur as $x,y \rightarrow 0$, as $x\rightarrow y$ and $(x,y)\rightarrow (2,2)$. 

\subsection{Features and level of anisotropy}\label{anisodiscuss}

Statistical homogeneity and isotropy are considered characterizing features of the CMB fluctuations distribution, if one ignores the issues raised by the ``anomalous'' detections we presented in the introduction. \\
Homogeneity of the correlation functions equates translational invariance and hence total momentum conservation, as enforced by the delta functions appearing on the left-hand sides of Eqs.~(\ref{ps}) through (\ref{trisp}). This invariance property can then be pictured as the three momentum vectors forming a closed triangle for the bispectrum and the four momenta arranged in a tetrahedron for the trispectrum (see Fig.~11).\\
Statistical isotropy corresponds to invariance w.r.t. rotations in space of the momentum (for the power spectrum) and of the triangle or tetrahedron made up by the momenta, respectively for the bispectrum and the trispectrum. This symmetry can be broken, as it for example happens in the $SU(2)$ case, by assuming the existence of preferred spatial directions in the early universe that might be revealed in the CMB observations. When this happens, the correlation functions are expected to be sensitive to the spatial orientation of the wave number or of the momenta triangles and tetrahedrons w.r.t. these special directions. Analitically, the bispectrum and the trispectrum will depend on the angles among the vector bosons and the wave vectors (besides the angles among the gauge bosons themselves), as shown in the coefficients $I_{n}$ and $L_{n}$ appearing in Eqs.~(\ref{fire}) and (\ref{fire1}). This implies that both the amplitude and the shape of bispectrum and trispectrum will be affected by these mutual spatial orientations. The modulation of the shapes by the directions that break statistical anisotropy was discussed with some examples both for the bispectrum and the trispectrum in our papers \cite{Bartolo:2009pa,Bartolo:2009kg}. These examples are here reported in Figs.~12 and 13. \\
In Fig.~12 we show the plot of the vector contribution to the bispectrum of $\zeta$, properly normalized in the configuration
\bea\label{instance1}
\vec{N}_{3}=N_{A}(0,0,1)\\\label{instance2}
\vec{N}_{1}=\vec{N}_{2}=N_{A}(\sin\theta\cos\phi,\sin\theta\sin\phi,\cos\theta),
\eea
where, the $(x,y,z)$ coordinate frame is chosen to be $\hat{k}_{3}=\hat{x}$ and $\hat{k}_{1}=\hat{k}_{2}=\hat{z}$ and $\delta$ is the angle between $\vec{N}_{1,2}$ and $\hat{k}_{3}$. The coefficients $I_{n}$ in this configuration become
\bea\fl
I_{EEE}&=&m^2 N^{4}_{A}\Big[-20-24\cos\delta+2\cos\theta-12\cos^2\delta+12\cos^2\theta-2\cos^3\theta+6\cos\theta\cos^2\delta\nonumber\\\fl&&-2\cos^2\theta\cos\delta+2\cos\delta\cos^3\theta\Big] ,\\\fl  
I_{lll}&=& m^2 N^{4}_{A}\Big[4\cos^2\delta\Big],\\\fl
I_{llE}&=& m^2 N^{4}_{A}\Big[4-2\cos\theta-6\cos^2\theta-4\cos^2\delta\Big],\\\fl
I_{lEl}&=&m^2 N^{4}_{A}\Big[-4\cos^2\delta \Big],\\\fl
I_{Ell}&=&m^2 N^{4}_{A}\Big[-4\cos^2\delta\Big] ,\\\fl
I_{EEl}&=&m^2 N^{4}_{A}\Big[ -2\cos^2\delta\Big],\\\fl
I_{ElE}&=& m^2 N^{4}_{A}\Big[4-4\cos^2\theta-8\cos^2\delta\Big],\\\fl
I_{lEE}&=& m^2 N^{4}_{A}\Big[2+\cos\theta-3\cos^2\theta-4\cos^2\delta\Big].
\eea
where $m^2\equiv (A)/(N_{A})$, $A$ being the background value of the $\vec{A}_{a}$'s evaluated at horizon crossing. The analytic expression of the 'non-Abelian' bispectrum normalized to the ratio $(g_{c}^{2}H^2 m^2N_{A}^{4})/(k^{6}_{1}x_{2}^{2}x_{3}^{2})$, as a function of the angles $\theta$ and $\delta$ and for fixed values of $x^{*}$, $x_{2}$ and $x_{3}$ is
\bea\label{graphs}\fl
B_{\zeta}(\theta,\delta)&\simeq &g_{c}^{2}\frac{H^2}{k_{1}^{6}}\frac{m^2 N_{A}^{4}}{10^{-1}}\Big[\cos^2\delta (8\cos\theta-1.4\times 10^3)+3\cos\delta(\cos^3\theta-\cos^2\theta-11)\nonumber\\\fl&-&11\cos 2 \delta-40-6\cos 2\theta-\cos\theta(3\cos^2\theta-30\cos\theta-10)\Big],
\eea
where we set $x^{*}=1$, while $x_{2}$ and $x_{3}$ were chosen in the 'squeezed' region, $x_{2}=0.9$ and $x_{3}=0.1$.\\
\noindent In Fig.~13 we provide a similar plot, but for the trispectrum from vector-exchange contributions and in a different configuration
\bea\label{daqui}
\hat{N}_{2}\cdot\hat{k}_{i}=0\,(i=1,...4)\nonumber\\
\hat{N}_{1}\cdot\hat{k}_{1}=\cos\delta,\quad\quad\quad\hat{N}_{1}\cdot\hat{k}_{2}=0\nonumber\\
\hat{N}_{3}\cdot\hat{k}_{2}=\cos\theta,\quad\quad\quad\hat{N}_{3}\cdot\hat{k}_{1}=0.
\eea 
In addition to that, let us assume that all the $\vec{N}_{a}$ have the same magnitude $N_{A}$. In this configuration, we have 
\bea
\Delta_{I}=\Delta_{III}=N^{A}_{4}\cos^{2}\theta\cos^{2}\delta,\quad\quad\quad\Delta_{II}=0,
\eea
therefore the the expression in Eq.~(\ref{IA}) becomes
\bea\label{zzeta}
T_{\zeta}\supset g_{c}^{2}H^{4}_{*}N_{A}^{4}\left[ISO\right]\cos^{2}\theta\cos^{2}\delta
\eea
where the expression in brackets includes an isotropic term (which is rotationally invariant) 
\bea\label{aqui}
ISO\equiv\left(\frac{x^{*}}{k}\right)^4\left(I\sum_{i=1}^{4}t_{i}+III\sum_{i=9}^{12}t_{i}\right).
\eea
Fig.~13 plots the trispectrum contribution in Eq.~(\ref{zzeta}) normalized to its isotropic part.\\

\begin{figure}\centering
 \includegraphics[width=0.37\textwidth]{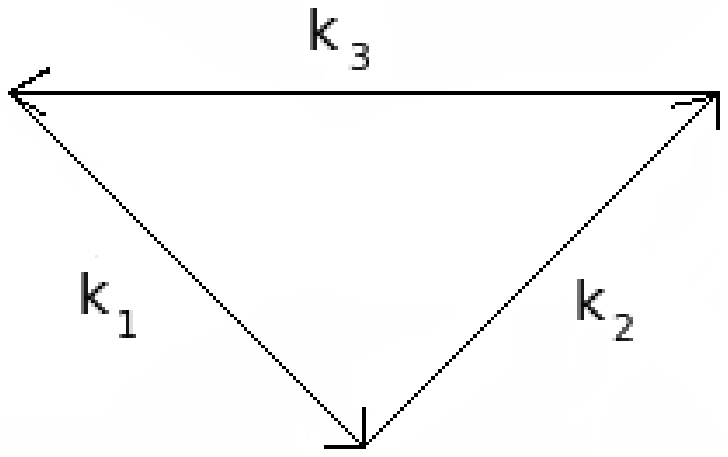}
\hspace{0.17\textwidth}
 \includegraphics[width=0.37\textwidth]{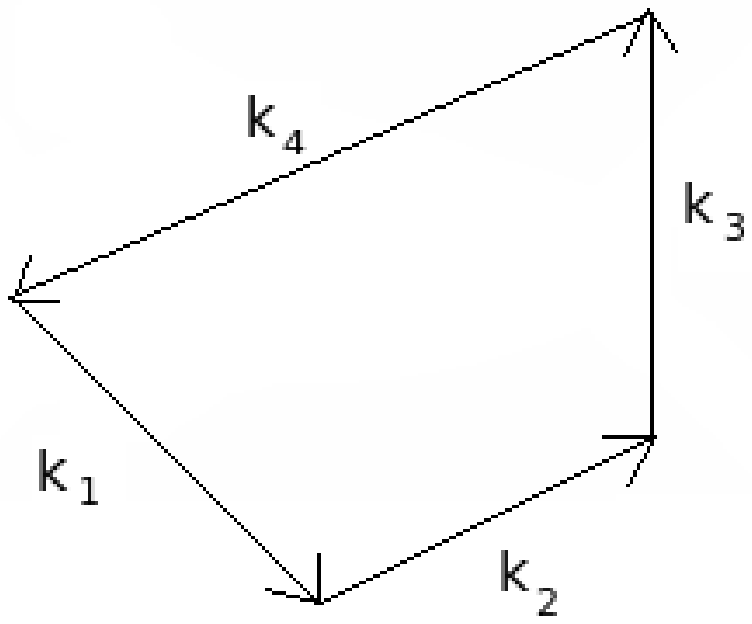}
\caption{ \label{Fig1} Representation of momentum conservation for the bispectrum (the three momenta form a closed triangle) and for the trispectrum (the momenta form a tetrahedron).}
\end{figure}

\noindent Another comment should be added concerning statistical anisotropy in the model. Notice that both the bispectrum and the trispectrum can be written as the sum of a purely isotropic and an anisotropic parts. The orders of magnitude of these two parts can, for instance, be read from Table 2 for the trispectrum: each one among $\tau_{NL}^{NA_{2}}$, $\tau_{NL}^{A_{1}}$ and $\tau_{NL}^{A_{2}}$ provide the order of magnitude of the level of both their isotropic and anisotropic contributions, which are therefore comparable; $\tau_{NL}^{NA_{1}}$ instead quantifies a purely anisotropic contribution which, as discussed in Sec.~\ref{amplitude}, can be comparable to the other three parts, if not the dominant one. A similar discussion applies to the bispectrum (see $f_{NL}^{A}$ and $f_{NL}^{NA}$ in Table 1). We can then conclude that, for the three and for the four point function, there is room in the parameter space of the theory for the anisotropic contributions to be as large as, or even larger than, the isotropic ones.

\begin{figure}\centering
 \includegraphics[width=0.4\textwidth]{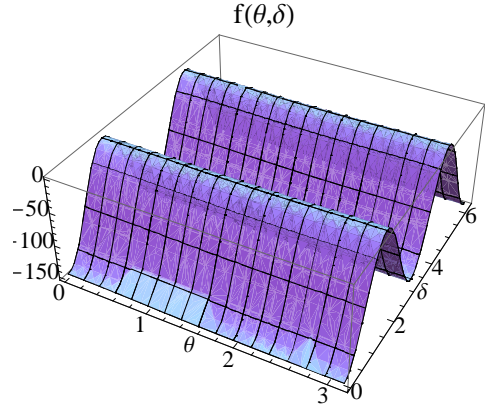}
\caption{ \label{Fig2}Plot of $f(\theta,\delta)\equiv {[(B_{\zeta}(\theta,\delta,x^{*},x_{2},x_{3})x_{2}^{2}x_{3}^{2}k_{1}^{6})/(g_{c}^{2}H^2 m^2 N_{A}^{4})]}$ evaluated at\\${(x^{*}=1,x_{2}=0.9,x_{3}=0.1)}$ in a sample angular configuration (see Eqs.~(\ref{instance1}) through (\ref{graphs})).}
\end{figure} 

\begin{figure}\centering
\includegraphics[width=0.4\textwidth]{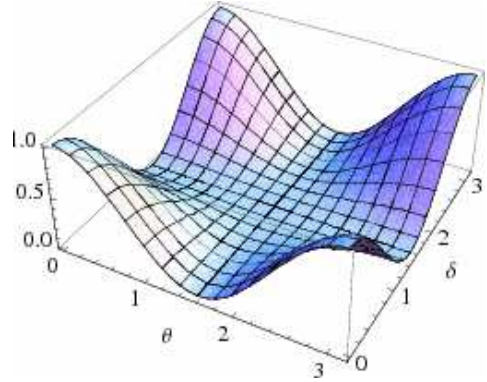}
\caption{ \label{Fig5} Plot of the anisotropic part of the trispectrum from the contribution due to vector-exchange diagrams in a sample angular configuration (see Eqs.~(\ref{daqui}) through (\ref{aqui})).}
\end{figure}

\newpage

\section{Trispectrum for $f(\tau)$ models of gauge interactions}
\label{gen}
\setcounter{equation}{0}
\def\theequation{12.\arabic{equation}}

We will now show that it is quite straightforward to extend the calculations we performed for $f=1$ to cases where $f$ is not a constant. One interesting model is the one studied in \cite{Martin:2007ue} and also recently discussed in \cite{Yokoyama:2008xw}, where the field is effectively massless ($m_{0}=\xi=0$) so the action (\ref{ac}) for the gauge field becomes
\bea\label{newL}
S= \int d^4x \sqrt{-g}\left[-\frac{1}{4}f^{2}(\phi)g^{\mu\nu}g^{\rho\sigma}F_{\mu\nu}^{a}F_{\rho\sigma}^{a}+...\right],
\eea
where again $F_{\mu\nu}^{a}\equiv\p_{\mu}B^{a}_{\nu}-\p_{\nu}B^{a}_{\mu}+g_{c}\ep^{abc}B^{b}_{\mu}B^{c}_{\nu}$.\\
Let us introduce the fields $\tilde{A}_{i}^{a}$ and $A^{a}_{i}$, related by the equations $\tilde{A}_{i}^{a}\equiv f B_{i}^{a}=a A_{i}^{a}$. The $A^{a}_{i}$ are the physical fields.\\
We can expand the perturbations of $\tilde{A}^{a}_{i}$ in terms of creation and annihilation operators in the usual way
\bea
\delta\tilde{A}^{a}_{i}(\eta,\vec{x})=\int\frac{d^3q}{(2 \pi)^3}\sum_{\lambda=R,L}\Big[e_{i}^{\lambda}(\hat{q})a^{a\lambda}_{\vec{q}}\delta\tilde{A}^{a}_{\lambda}(\eta,q)+h.c.\Big].
\eea
If $f=f_{0} a^{\alpha}$, with $\alpha$ equal either to $1$ or $-2$ ($f_{0}$ is a constant), it is possible to prove \cite{Yokoyama:2008xw} that the equation of motion for $\delta\tilde{A}^{a}_{\lambda}$ is the same as the one for $\delta B^{T}$, where by $\delta B^{T}$ we mean the transverse mode function in Eq.~(\ref{T}). This is equivalent to saying that, under the assumption $\alpha=1,-2$, the physical gauge fields are governed by the same equation of motion as a light scalar field in a de Sitter space and so they generate a scale invariant power spectrum. Let us sketch the calculation of the trispectrum in this theory. \\

\begin{figure}\centering
\includegraphics[width=0.4\textwidth]{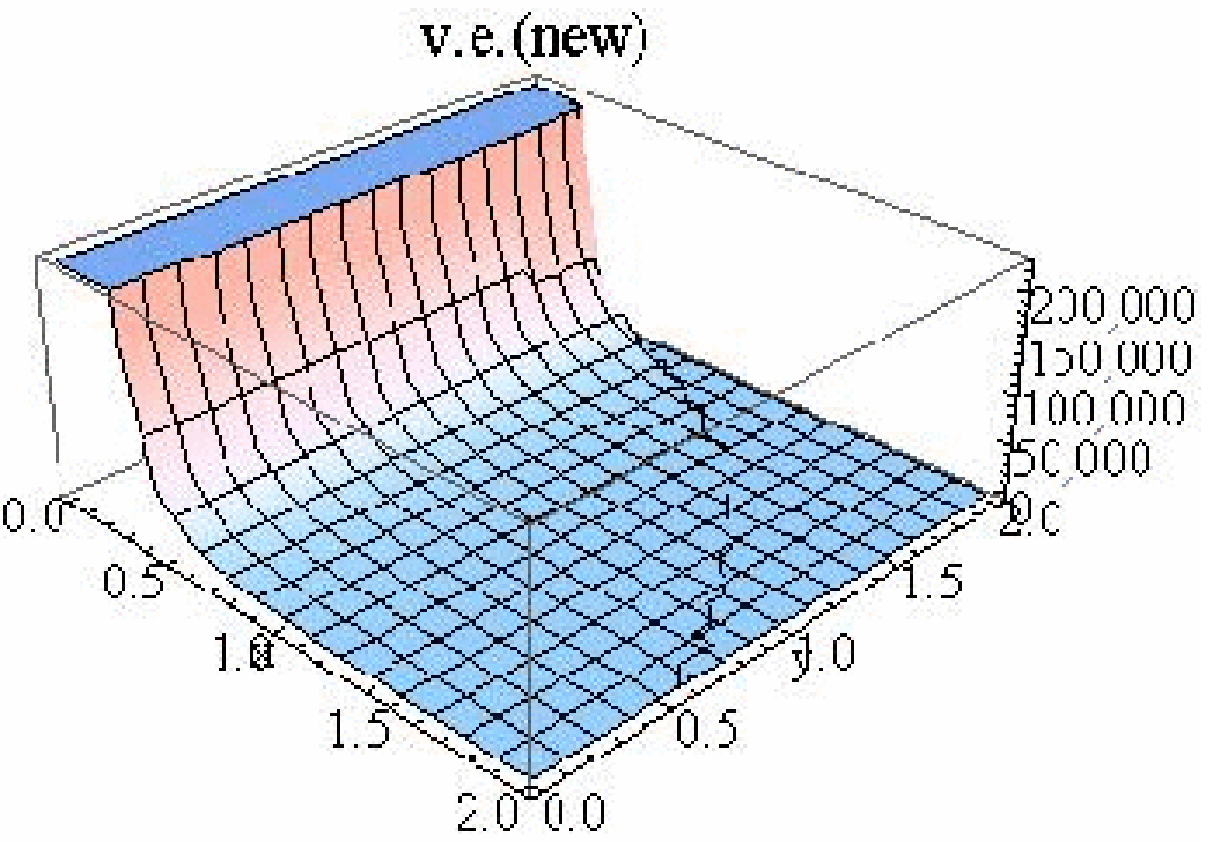}
\hspace{0.1\textwidth}
 \includegraphics[width=0.4\textwidth]{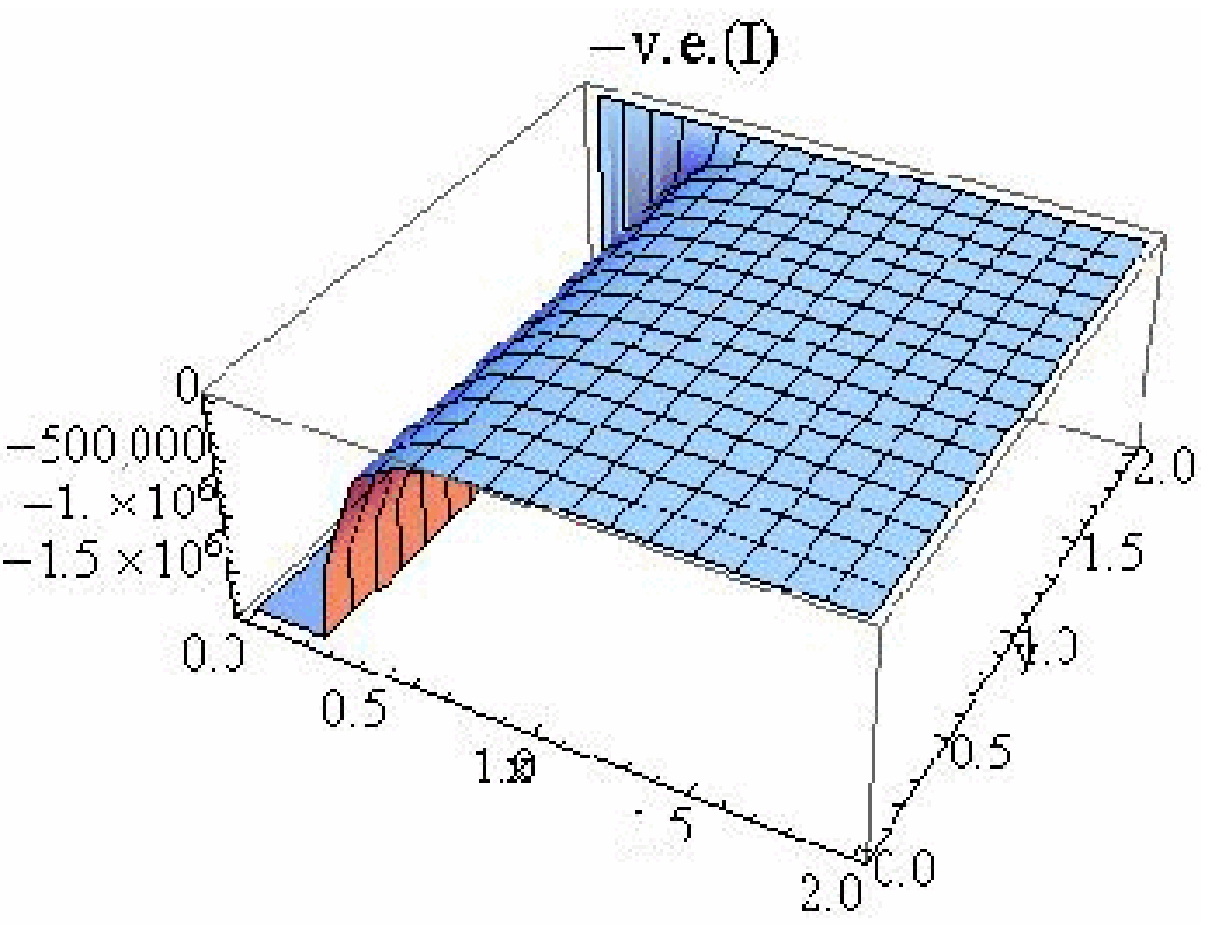}
\vspace{0.02\textwidth}
 \includegraphics[width=0.4\textwidth]{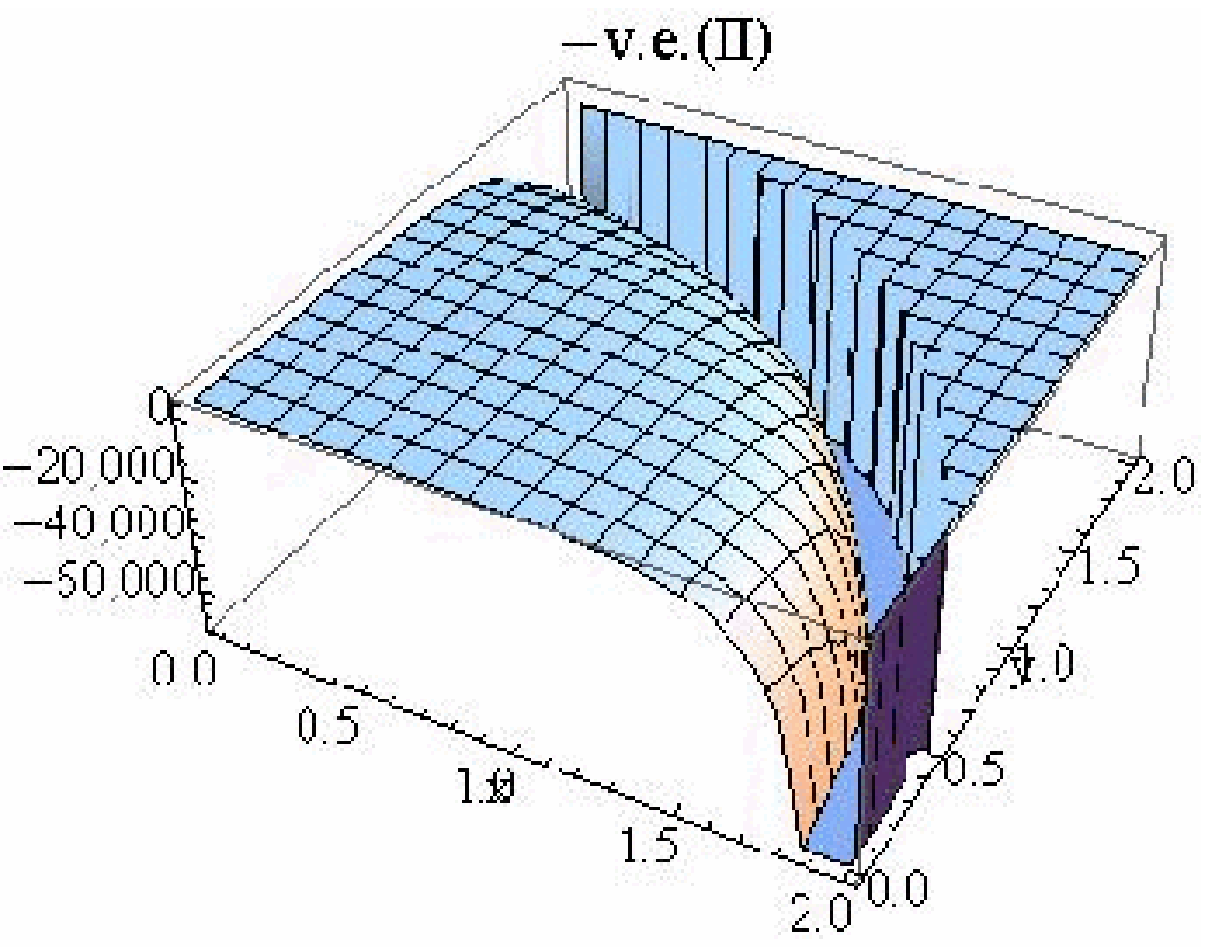}
\hspace{0.1\textwidth} \includegraphics[width=0.4\textwidth]{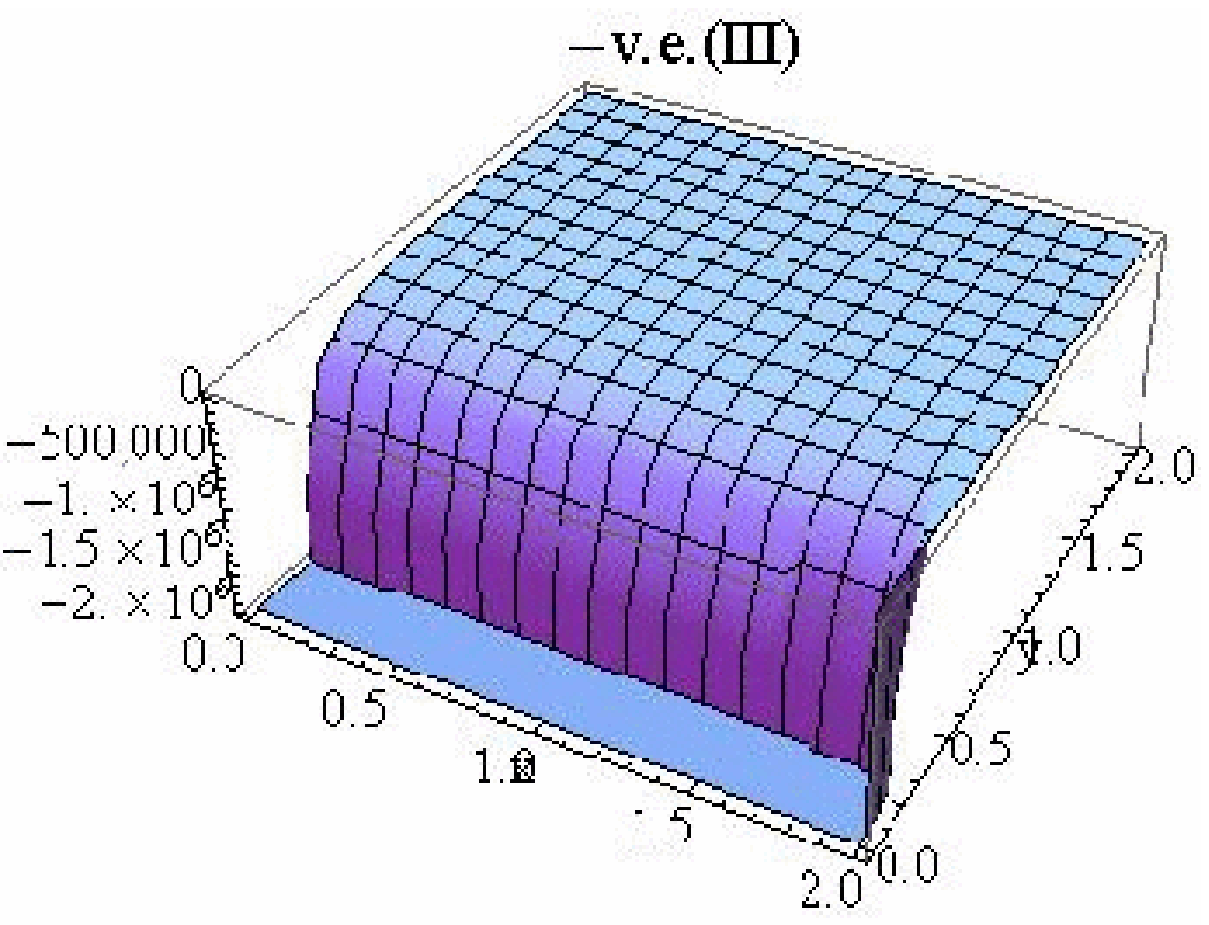}
\caption{ \label{Fig2}Plots of the isotropic functions of some of the vector-exchange contributions in the equilateral configuration, for the $f\simeq a^{-2}$ model. In this and in the next figures, ``v.e.(I,II,III)'' represent the isotropic functions associated with the very last term in square brackets in Eq.~(\ref{t}); ``v.e.(new)'' represents the isotropic function associated with the $k_{1144}$, $k_{2244}$, $k_{1133}$ and $k_{2233}$ terms in the second line of Eq.~(\ref{t}).}
\end{figure}

\begin{figure}\centering
\includegraphics[width=0.4\textwidth]{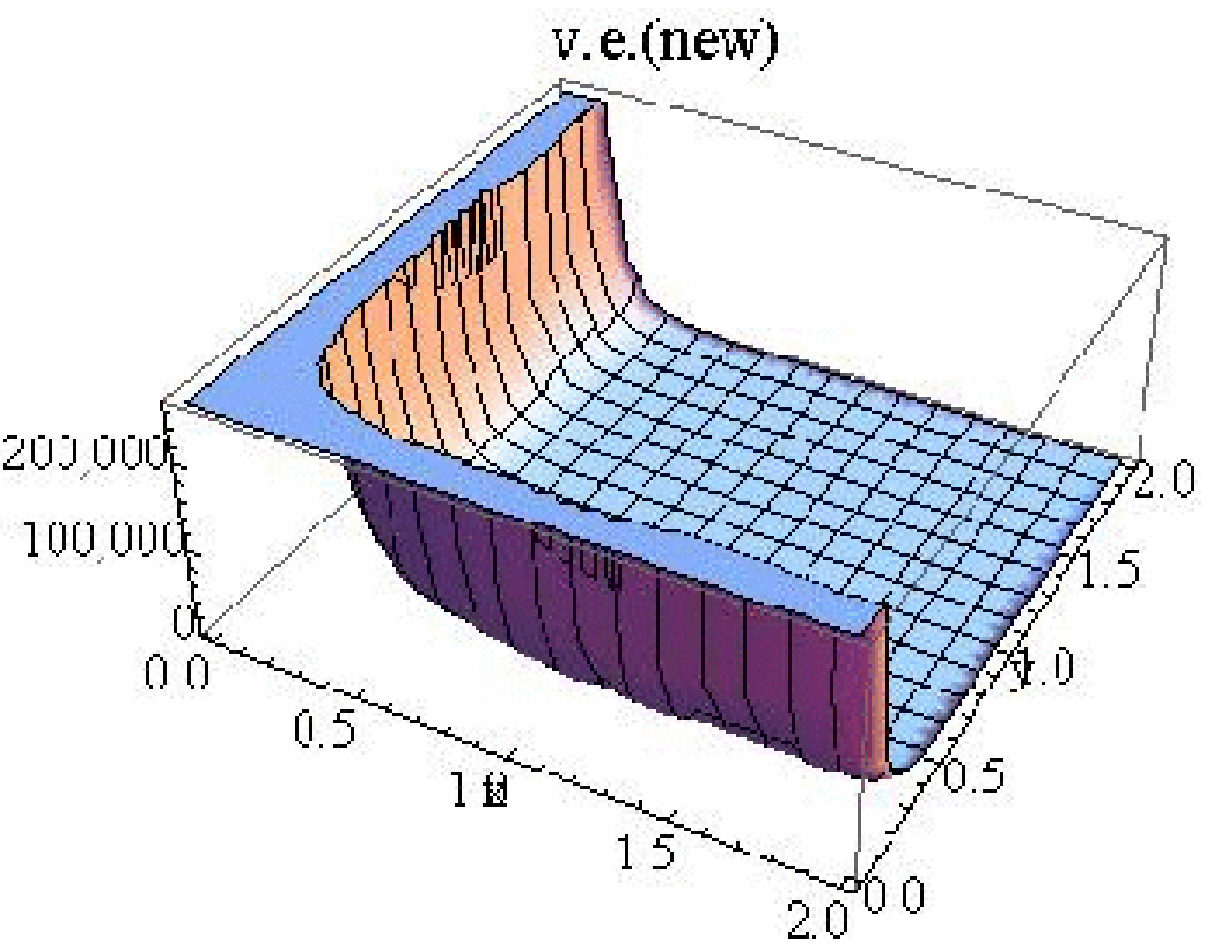}
\hspace{0.1\textwidth}
 \includegraphics[width=0.4\textwidth]{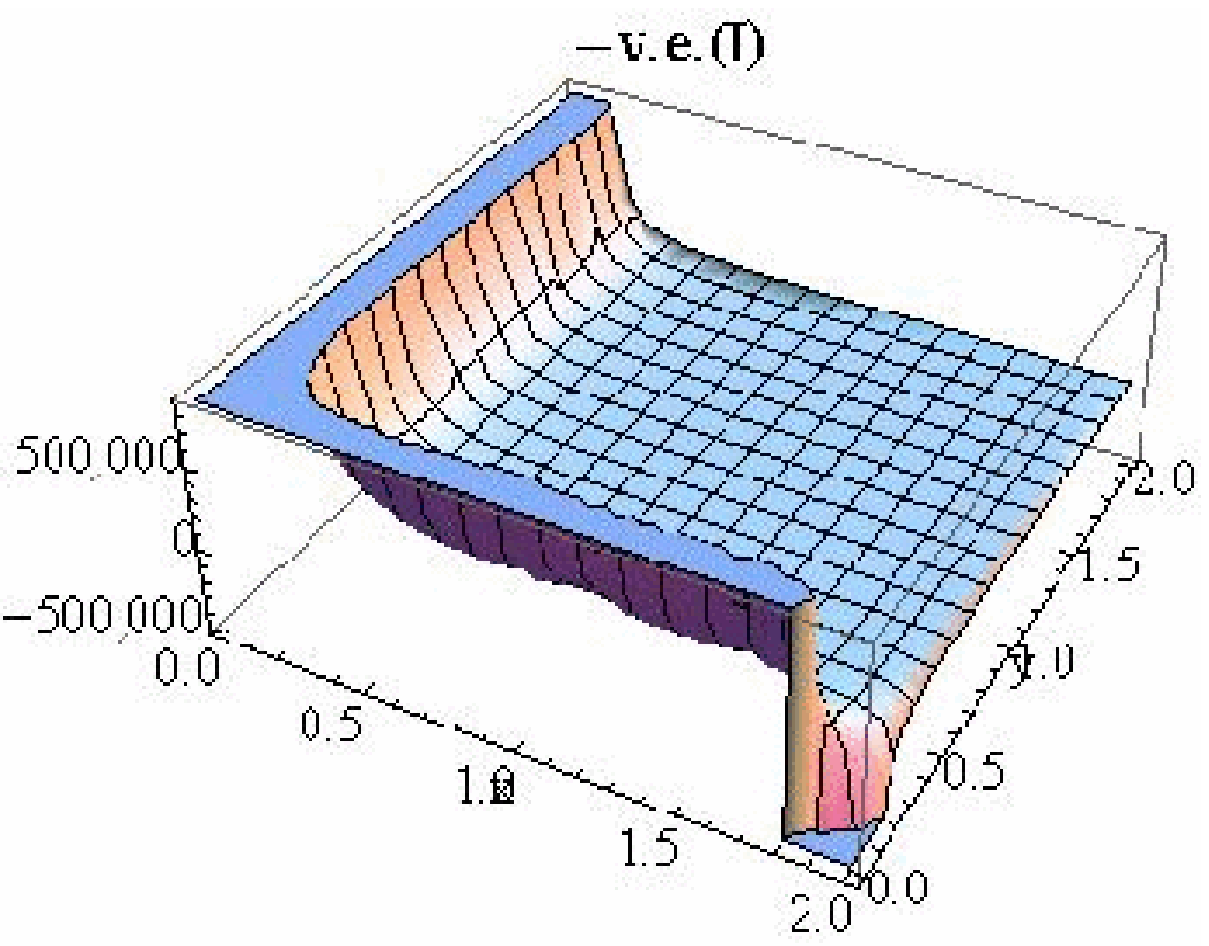}
\vspace{0.02\textwidth}
 \includegraphics[width=0.4\textwidth]{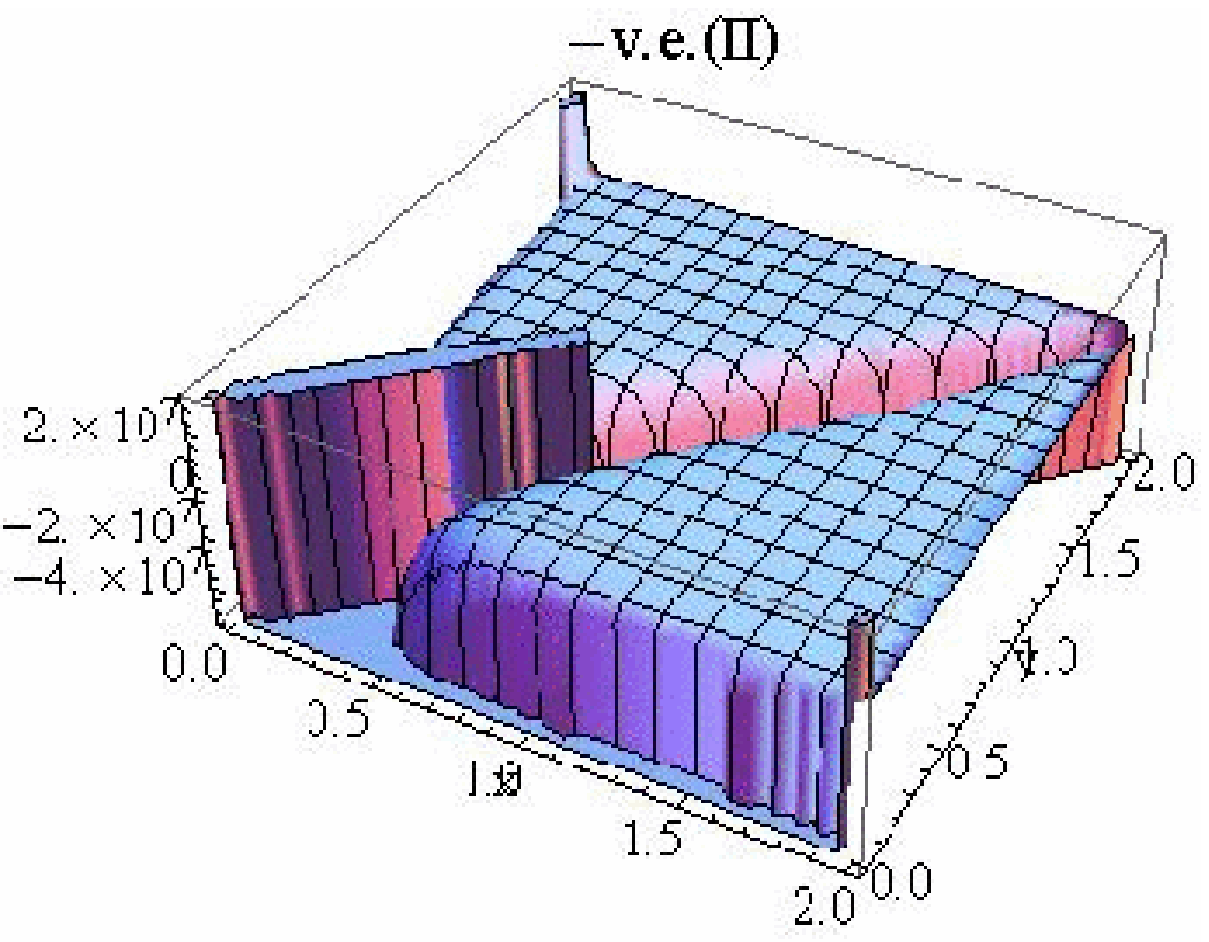}
\hspace{0.1\textwidth} \includegraphics[width=0.4\textwidth]{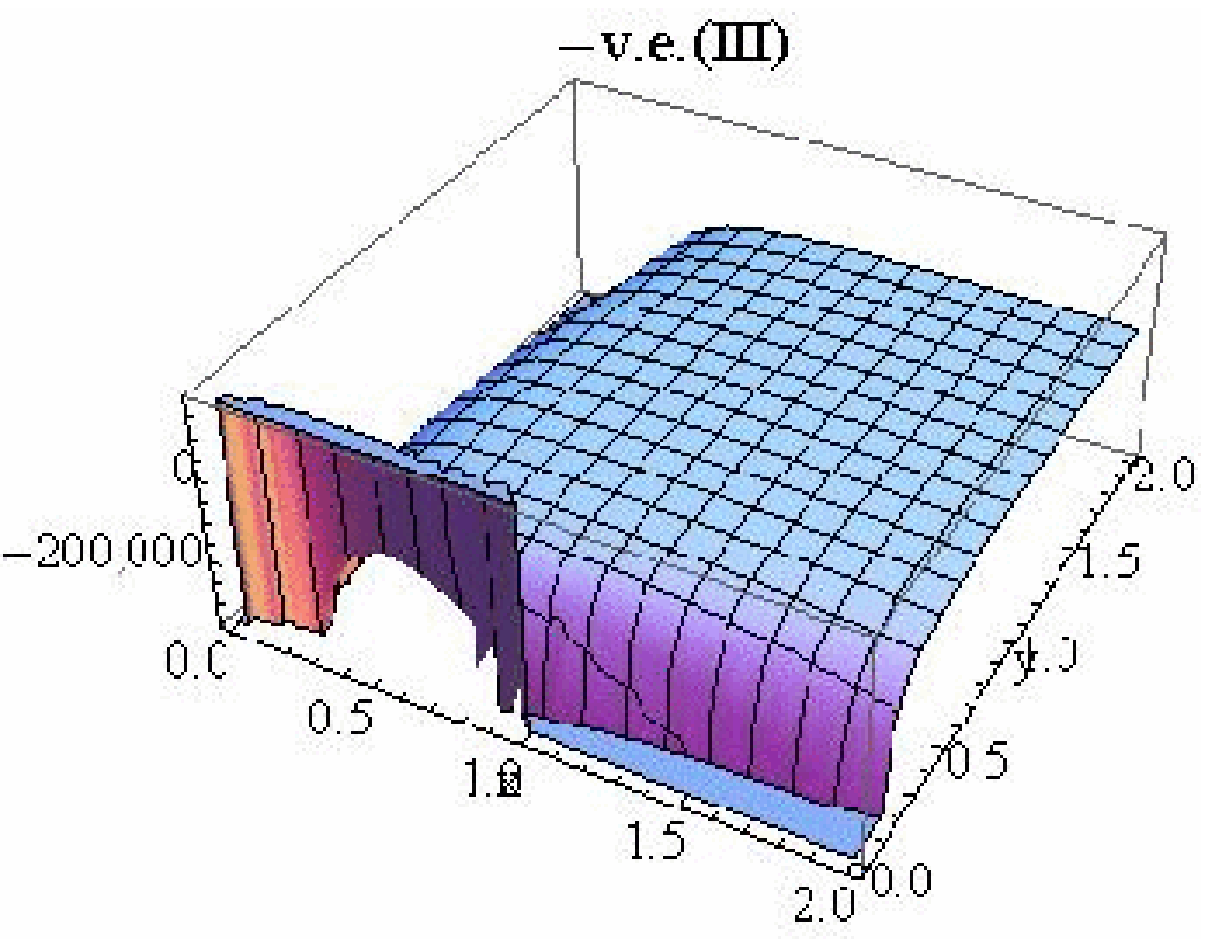}
\caption{ \label{Fig2}Plots of the isotropic functions of some of the vector-exchange contributions in the specialized planar configuration (plus sign), for the $f\simeq a^{-2}$ model.}
\end{figure}

\begin{figure}\centering
\includegraphics[width=0.4\textwidth]{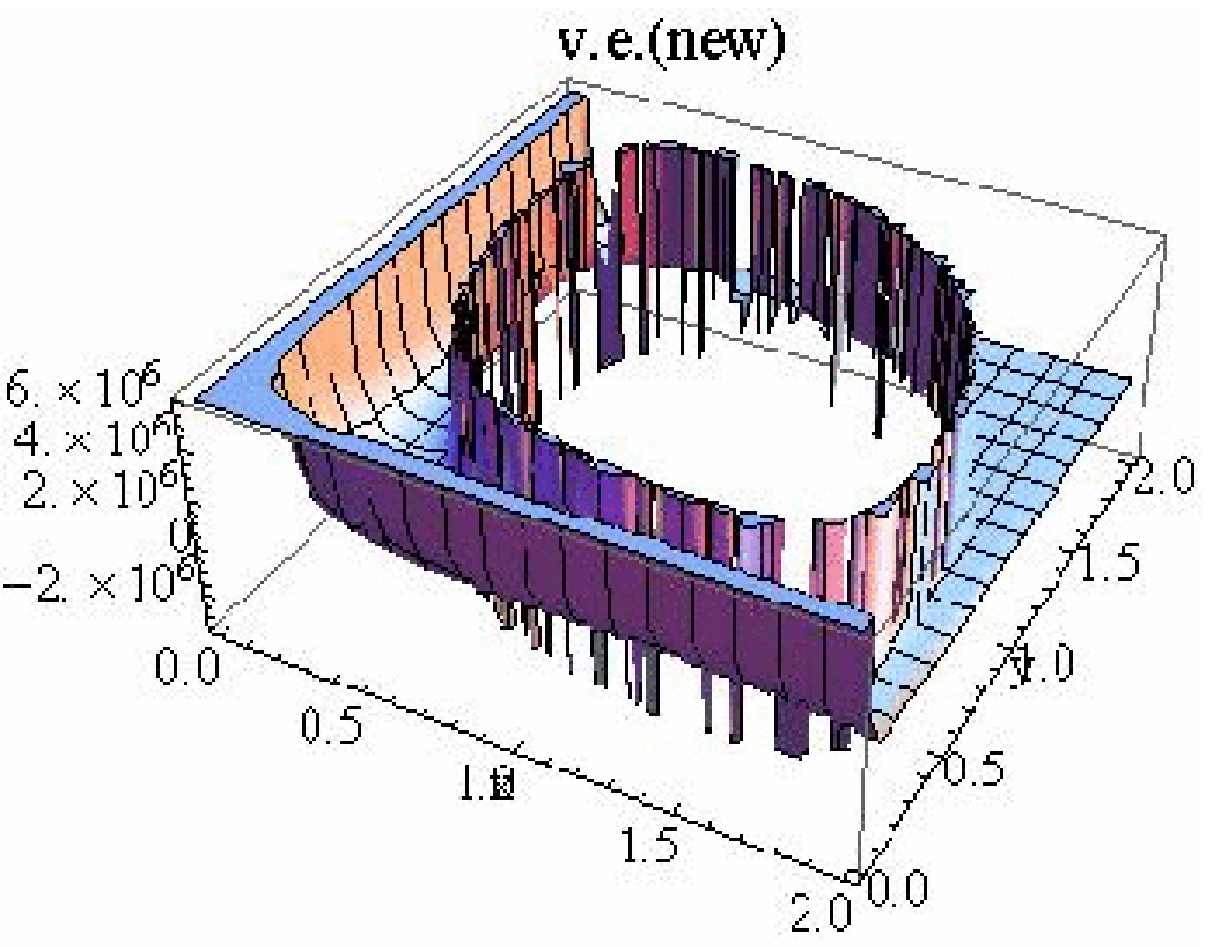}
\hspace{0.1\textwidth}
 \includegraphics[width=0.4\textwidth]{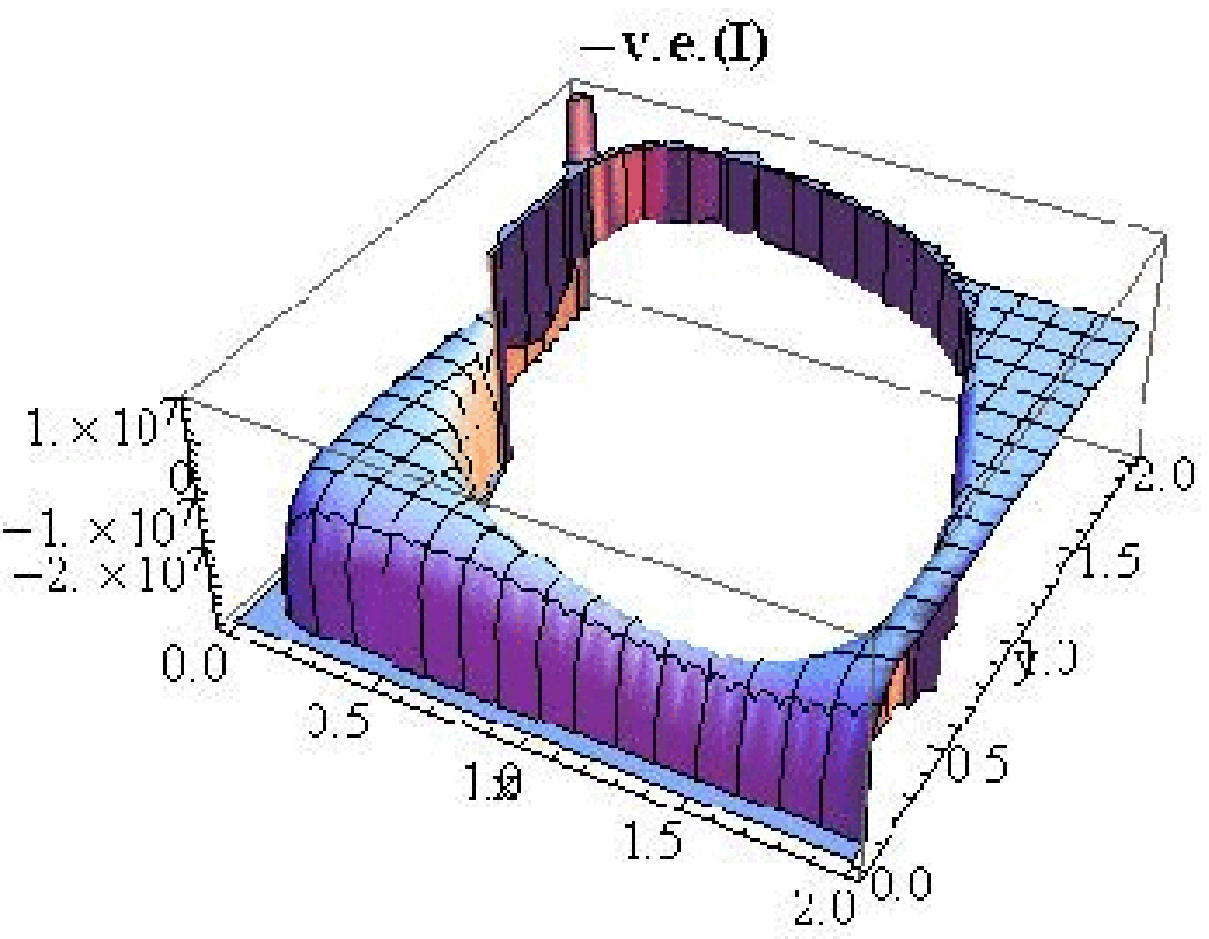}
\vspace{0.02\textwidth}
 \includegraphics[width=0.4\textwidth]{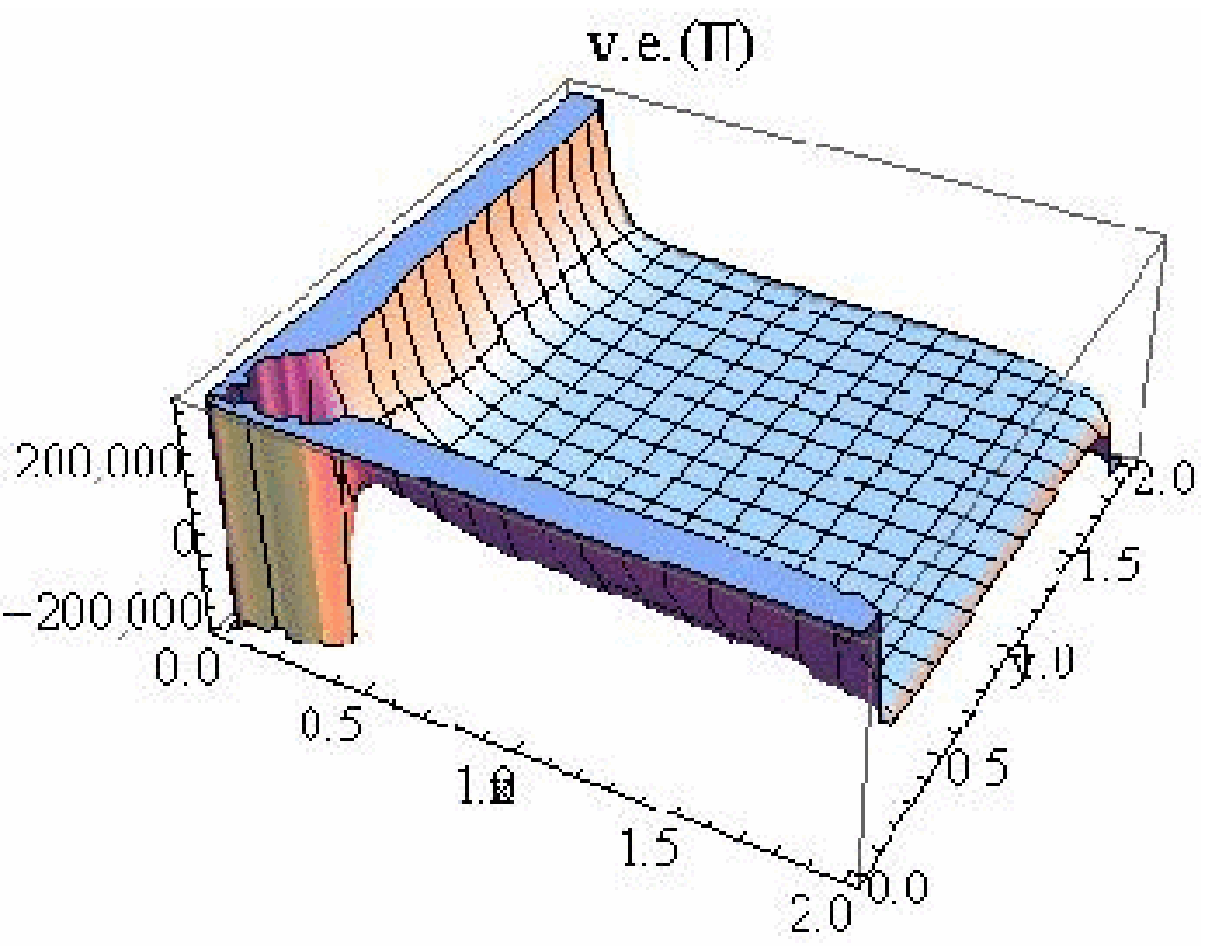}
\hspace{0.1\textwidth} \includegraphics[width=0.4\textwidth]{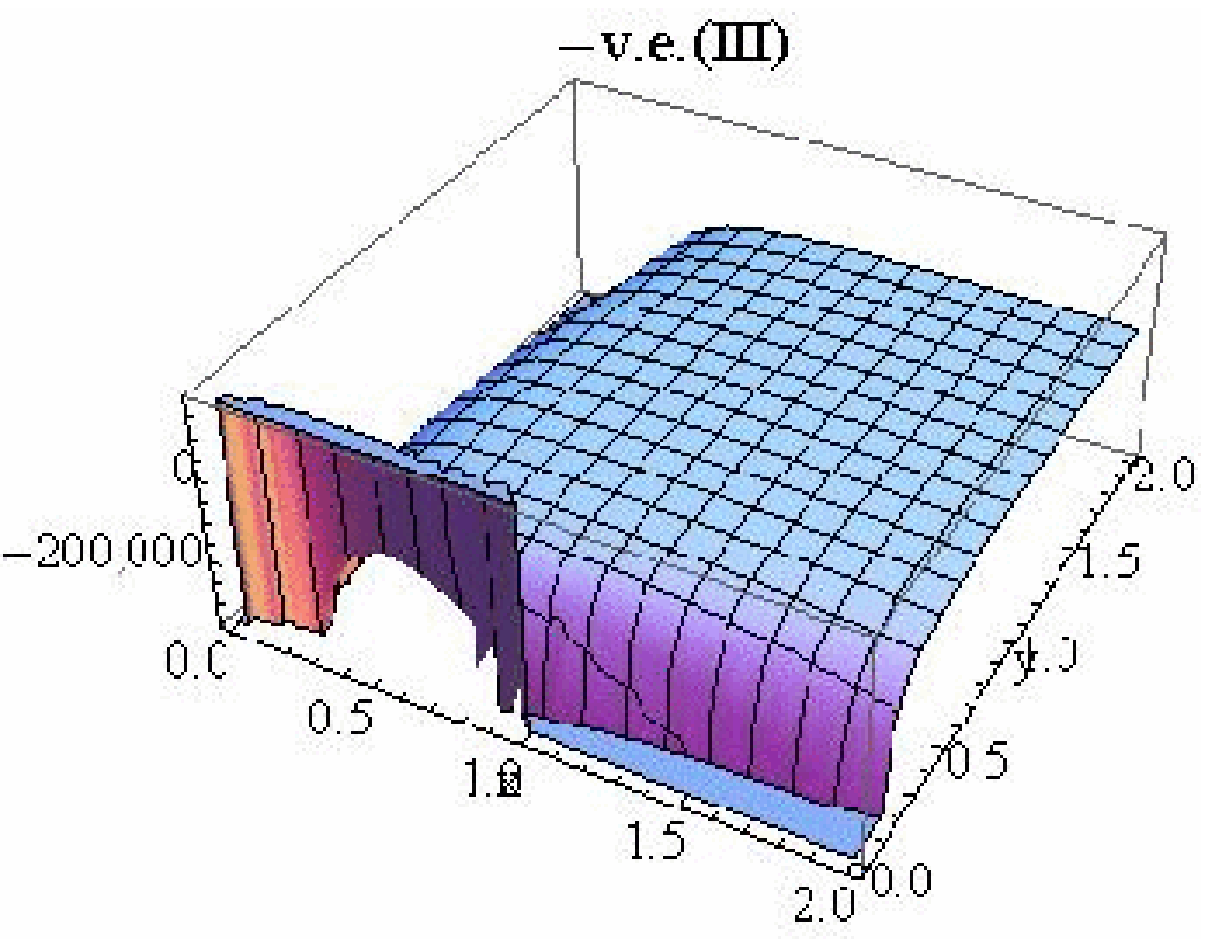}
\caption{ \label{Fig2}Plots of the isotropic functions of some of the vector-exchange contributions in the specialized planar configuration (minus sign), for the $f\simeq a^{-2}$ model.}
\end{figure}

\noindent The general expressions of the Abelian terms still hold, except that the power spectrum in Eq.~(\ref{ps}) reduces to 
\bea
P^{ab}_{ij}=T^{ab}_{ij}P_{+},
\eea
having gauged the longitudinal modes away.\\
Let us now have a look at the non-Abelian part. First of all, we need to set $n(x^{*})=0$. The interaction Hamiltonian to third and fourth order are the same as in Eqs.~(\ref{int3}) and (\ref{int4}), but with extra $f^{2}(\tau)$ factors. The anisotropy coefficients such as $I_{n}$ and $L_{n}$ that survive after setting the longitudinal mode to zero do not change. On the other hand, the wavefunctions for the gauge fields are now given by $\delta B=\delta\tilde{A}/f=\delta B^{T}/f_{0}a^{\alpha}$. The new trispectrum therefore differs because of extra scale factors inside and outside the time integrals, which in general imply a different power of $H_{*}$ in the final results and a different momentum dependence in the isotropic part of the expressions.\\
Let us take into account the trispectrum. The three non-Abelian vector contributions from Eq.~(\ref{class4}) in the $f=1$ case can be schematically written as follows 
\bea\label{ftau1}\fl
\langle \delta B^4\rangle_{line3}&\simeq& \left(\delta B^{3}\right)\int d\eta \left(\delta B\right)^{3}=\left(\delta B^{T}\right)^{3}\int d\eta \left(\delta B^{T}\right)^{3},\\\fl
\langle\delta B^4\rangle_{v.e.}&\simeq& \left(\delta B\right)^{4}\int d\eta^{'} \left(\delta B\right)^{3}\int d\eta^{''} \left(\delta B\right)^{3}\nonumber\\\fl&=&\left(\delta B^{T}\right)^{4}\int d\eta^{'} \left(\delta B^{T}\right)^{3}\int d\eta^{''} \left(\delta B^{T}\right)^{3},\\\label{ftau2}\fl
\langle \delta B^4\rangle_{c.i.}&\simeq& \left(\delta B\right)^{4}\int d\eta \left(\delta B\right)^{4}=\left(\delta B^{T}\right)^{4}\int d\eta \left(\delta B^{T}\right)^{4},
\eea
where we indicate with the subscript $line3$ the contribution from the third line of Eq.~(\ref{class4}) and with $v.e.$ and $c.e.$ respectively the vector exchange and the contact interaction contributions to the first line of (\ref{class4}). $\langle \delta B^4\rangle\equiv \langle \delta B^{a}_{i}(\vec{k}_{1})\delta B^{b}_{j}(\vec{k}_{2})\delta B^{c}_{k}(\vec{k}_{3})\delta B^{d}_{l}(\vec{k}_{4})\rangle$ and we have omitted all the gauge and vector indices, as well as complex conjugate symbols, for simplicity. Let us see now how Eqs.~(\ref{ftau1})-(\ref{ftau2}) change if $f=f_{0}a^{\alpha}$ ($\alpha=1,-2$)
\bea\label{ftau3}\fl
\langle \delta B^4\rangle_{line3}&\simeq& \left(\delta B\right)^{3}\int d\eta f^{2}\left(\delta B\right)^{3}\simeq\left(\frac{\delta B^{T}}{a^{\alpha}}\right)^{3}\int d\eta a^{2\alpha}\left(\frac{\delta B^{T}}{a^{\alpha}}\right)^{3},\\\fl
\langle \delta B^4\rangle_{v.e.}&\simeq& \left(\delta B\right)^{4}\int d\eta^{'} f^{2}\left(\delta B\right)^{3}\int d\eta^{''}f^{2} \left(\delta B \right)^{3}\nonumber\\\fl&\simeq&\left(\frac{\delta B^{T}}{a^{\alpha}}\right)^{4}\int d\eta^{'} a^{2\alpha}\left(\frac{\delta B^{T}}{a^{\alpha}}\right)^{3}\int d\eta^{''}a^{2\alpha} \left(\frac{\delta B^{T}}{a^{\alpha}} \right)^{3},\\\label{ftau4}\fl
\langle \delta B^4\rangle_{c.i.}&\simeq& \left(\delta B\right)^{4}\int d\eta f^{2} \left(\delta B \right)^{4}\simeq\left(\frac{\delta B^{T}}{a^{\alpha}}\right)^{4}\int d\eta a^{2\alpha} \left(\frac{\delta B^{T}}{a^{\alpha}} \right)^{4}.
\eea

\noindent Using $a=(-H\eta)^{-1}$ in the previous equations, we get

\bea\label{ftau5}\fl
\langle \delta B^4\rangle_{line3}&\simeq& H^{4\alpha}_{*}\left(\delta B^{T}(-\eta^{*})^{\alpha}\right)^3\int d\eta \left(\delta B^{T}\right)^{3}(-\eta)^{\alpha},\\\fl
\langle \delta B^4\rangle_{v.e.}&\simeq& H^{6\alpha}_{*}\left(\delta B^{T}(-\eta^{*})^{\alpha}\right)^{4}\int d\eta^{'} \left(\delta B^{T}\right)^{3}(-\eta^{'})^{\alpha}\nonumber\\\fl&\times&\int d\eta^{''} \left(\delta B^{T} \right)^{3}(-\eta^{''})^{\alpha},\\\label{ftau6}\fl
\langle \delta B^4\rangle_{c.i.}&\simeq& H^{6\alpha}_{*}\left(\delta B^{T}(-\eta^{*})^{\alpha}\right)^{4}\int d\eta \left(\delta B^{T}\right)^{4}(-\eta)^{2\alpha} .
\eea

\noindent Let us now consider more in details the $\alpha=-2$ case for contact-interaction and vector-exchange contributions. The expressions for the anisotropy coefficients are respectively given by

\bea\label{t}\fl
T^{EEEEE}_{ijkl}&=&k_{1}k_{3}\big(\hat{k}_{1}\cdot\hat{k}_{3}-\hat{k}_{1}\cdot\hat{k}_{\hat{12}}\hat{k}_{3}\cdot\hat{k}_{\hat{12}}\big)\big[\delta_{ij}\delta_{kl}-\delta_{ij}\hat{k}_{k4}\hat{k}_{l4}-\delta_{ij}\hat{k}_{k3}\hat{k}_{l3}-\delta_{kl}\hat{k}_{i2}\hat{k}_{j2}-\delta_{kl}\hat{k}_{i1}\hat{k}_{j1}\nonumber\\\fl&+&\delta_{ij}\hat{k}_{k3}\hat{k}_{l4}\hat{k}_{3}\cdot\hat{k}_{4}+\delta_{kl}\hat{k}_{i1}\hat{k}_{j2}\hat{k}_{1}\cdot\hat{k}_{2}+k_{1144}+k_{2244}+k_{1133}+k_{2233}-k_{2234}\hat{k}_{3}\cdot\hat{k}_{4}\nonumber\\\fl&-&k_{1134}\hat{k}_{3}\cdot\hat{k}_{4}-k_{1244}\hat{k}_{1}\cdot\hat{k}_{2}+k_{1233}\hat{k}_{1}\cdot\hat{k}_{2}+k_{1234}\hat{k}_{1}\cdot\hat{k}_{2}\hat{k}_{3}\cdot\hat{k}_{4}\big]
\eea
and by Eq.~(\ref{coeff1}), for one of the possible permutations. These expressions are more complicated w.r.t $T_{ijkl}^{lllll}$ in Eq.~(\ref{tt}) and $T_{mnop}^{llll}$ in Eq.~(\ref{coeff2}) for the longitudinal modes. As a result, when studying the shape of the trispectrum, for the isotropic functions appearing in it, several diagrams need to be taken into account, one for each term in $T^{EEEE}_{ijkl}$ and $T^{EEEEE}_{ijkl}$. For comparison with the $f=1$ case, we plotted the isotropic functions associated with the very last term in square brackets in Eq.~(\ref{t}) (see ``$-v.e(I)$'', ``$-v.e(II)$'' and ``$-v.e(III)$'' in Figs.~14, 15 and 16). By comparing these plots with the ones in Figs.~8, 9 and 10, it is evident that they have very similar shapes. On the other hand, when we consider the isotropic functions associated with terms that are not present in the $f=1$ case, several differences arise in the plots; we provide a sample in Figs.~14, 15 and 16 with the ``$v.e.(new)$'' plots, which represent isotropic functions associated with the $k_{1144}$, $k_{2244}$, $k_{1133}$ and $k_{2233}$ terms in the second line of Eq.~(\ref{t}). We verified that similar observations can be made concerning the shapes of the contact-interaction contributions.

\newpage

\section{Conclusions}

Cosmology has entered what it is called its ``precision era'': over the past few years, observations of the CMB and of other cosmological probes have been performed which have greatly improved the previous bounds on some of the fundamental parameters characterizing the early Universe cosmology. These bounds are expected to become a lot stricter with the advent of new experiments, such as the ongoing Planck satellite mission. The theoretical search for models of the early Universe has consequently been concentrated on, among other things, trying to produce more and more accurate predictions. \\ 
In this optics, alternative models to the basic single-field slow-roll inflationary scenario have been and are being investigated, with a strong focus on the computation of higher order (three and four-point) correlation functions and on higher order (beyond tree-level) corrections. All of these ``higher order'' predictions are expected to be particularly revealing of the early Universe physics, given that they are an indication of the type of fields and interactions populating the specific inflationary scenarios.   \\

\noindent In this thesis I have presented an overview of the existing results concerning some of these ``higher order'' predictions; in particular, I focused on our works on loop corrections in inflationary scalar field models and on the primordial non-Gaussianity and statistical anisotropy predictions in some inflationary models populated with non-Abelian vector fields.\\ 

\noindent Loop corrections to the power spectrum of the comoving curvature fluctuation $\zeta$ in single-field inflation arise both from the inflaton self-interactions and from the coupling of the scalar field with gravitons. We have calculated the corrections from tensor loops, previously neglected in the literature for simplicity reasons. It turnes out that one-loop corrections from tensor-scalar interactions are of the same order of magnitude as those arising from scalar self-interactions, therefore they cannot be neglected in a self-consistent calculation.\\
\noindent One loop corrections have been found to be suppressed by an $(H/m_{P})^2$ factor compared to the tree-level result; they exhibit a slightly different dependence from the external momentum, because of the presence of a logarithmic factor (which does not spoil scale invariance) and they turn out vary with time at most as fast as the logarithm of the scale factor. One-loop diagrams generally present both infrared logarithmic and ultraviolet power-law (in addition to logarithmic) divergences. The ultraviolet divergences can be treated using ordinary regularization and renormalization techniques as in flat-space quantum field theory. The infrared divergences are ``cured'' by introducing an infrared cutoff represented by the smallest observable physical mode, i.e. by considering a finite space, a sort of ``box'' of observation, with a size equal to the current horizon length. As expected, tensor modes provide, as well as the scalar modes, observably small corrections at one loop level, as long as the basic inflationary scenario is concerned. Their computation is nevertheless instructive and provides a path that can be followed in the investigation of ``non-standard'' models of inflation. More appealing results are for instance expected in inflationary scenarios described by non-canonical Lagrangians, some of which, $P(X,\phi)$ models, have been described and analysed in this thesis. \\ 

\noindent Motivated by an interest in models that combine non-Gaussianity and statistical anisotropy predictions for the CMB fluctuations, we have worked on theories of inflation where primordial vector fields effectively participate in the production of the $\zeta$ perturbations, eventually focusing on some $SU(2)$ vector fields models. The two, three and four point correlation functions in these models result as the sum of scalar and vector contributions. The latter are of two kinds, ``Abelian'' (i.e. arising from the zeroth order terms in the Schwinger-Keldysh expansion) and ``non-Abelian'' (i.e. originating from the self-interactions of the vector fields). The bispectrum and the trispectrum final results are presented as a sum of products of isotropic functions of the momenta ($F_{n}$ and $G_{n}$ in the text) multiplied by anisotropy coefficients (which we indicated by $I_{n}$ and $L_{n}$) that depend on the angles among all gauge and wave vectors. \\
The amplitude of non-Gaussianity has been evaluated through the parameters $f_{NL}$ and $\tau_{NL}$; in particular we have discussed the dependence of these functions from the non-angular parameters of the theory. We have provided the comparisons among the different (scalar versus vector, Abelian versus non-Abelian) contributions to $f_{NL}$ and $\tau_{NL}$, noticing that any one of them can be the dominant contribution depending on the selected region of parameter space. In particular, we have stressed how the anisotropic contributions to the bispectrum and the trispectrum can overcome the isotropic parts. An interesting feature of these models is that the bispectrum and the trispectrum depend on the same set of parameters and their amplitudes are therefore strictly related to one another. \\
We have presented the shapes of both the bispectrum and the trispectrum. The isotropic functions appearing in their final expressions have been analyzed separately from their anisotropy coefficients. The bispectrum isotropic functions have been found to preferably show a local shape. The trispectrum ones have been plotted selecting equilateral and specialized planar configurations. Finally, the full expressions (i.e. completed by the anisotropy coefficients) of bispectrum and trispectrum have been presented in specific momenta configuration, in order to provide a hint of the modulation of shapes and amplitudes operated by anisotropy.\\
\noindent  In our view, the most promising features of these models consists in the possibility of providing non-Gaussianity and statistical anisotropy predictions that are related to one another because of the fact that they share the same underlying theory. Models that combine both types of predictions could be more easily testable and, from non-Gaussianity measurement, more stringent statistical anisotropy predictions could be produced or viceversa.\\

\noindent Inflationary models that do not spoil the current agreement with experimental data constitute a huge variety but many of them have very distinctive features that might be confirmed or ruled out by observations. Both the the non-Gaussianity and statistical anisotropy predictions and the nature and the amount of higher corrections to the cosmological correlation functions can be ranked among these distinctive features and certainly deserve further investigations, looking forward for a confront with new and promising experimental data. \\

\noindent While completing this thesis, the Wilkinson Microwave Anisotropy Probe (WMAP) team published its seven-year data analyses \cite{Komatsu:2010fb,Bennett:2010jb}. The new bounds on non-Gaussianity from this study are given by $-10<f_{NL}^{equil}<74$ and $-214<f_{NL}^{loc}<266$, at $95\%$ CL.

\newpage

\section{Acknowledgments}

A sincere and warm thank to my supervisors Nicola Bartolo and Sabino Matarrese for letting me in the most enjoyable and fruitful working environment someone could ever hope for, for being there for me anytime I needed, for being two unique models of committment and love for physics and for being such incredibly nice people.\\
A special thank to Antonio Riotto: I have always admired him as a scientist and, in addition to this, I came to truly appreciate him as a person and as a teacher, expecially during the last year of my Ph.D, when I had the great chance to know him better and to work for him. Thanks also for always finding a spot to talk with me.


\newpage

\section{Appendices}

\vskip 1cm
\setcounter{equation}{0}
\def\theequation{15.1.\arabic{equation}}
\subsection{Computation of the interaction Hamiltonian}\label{IHAM}

The propagator of two fields $\phi_{1}$ and $\phi_{2}$ is defined by (see for example \cite{peskin})
\be
\langle\phi_{1}\phi_{2}\rangle=\int D\phi D\Pi e^{i\int d^{4}x\left(\Pi\dot{\phi}-\textit{H}\right)},
\ee
\noindent where $\Pi$ is the momentum conjugate to $\phi$ and $\textit{H}$ is the Hamiltonian density. If $H$ is quadratic in $\Pi$, as it happens for example in flat space-time for a field governed by a Lagrangian $L=\int d^{4}x \left(\frac{1}{2}\p_{\mu}\p^{\mu}\phi-V(\phi)\right)$ the square in the exponent can be completed and the integral in $\Pi$ evaluated and all is left is
\be
\langle\phi_{1}\phi_{2}\rangle=\int D\phi e^{iL}.
\ee
So, if an interaction term with time derivatives appears in the Lagrangian, $\Pi$ and $\phi$ are independent fields in the path integral. This will provide some extra vertices that need to be accounted for in the Feynman diagrams. We sketch a derivation of these extra vertices. It will turn out to be similar to what is done in \cite{seery1}, although complicated by the presence of gravitons. To keep the calculations easier we will at first ignore all spatial derivative and tensor indices, this will also make the notation simpler. Also, we will momentarily ignore all numerical real coefficients; it is instead very important to keep track of imaginary coefficients, time derivatives and powers of the scale factor $a$, so we will make sure they are all accounted for in our analysis.\\
The total action is $S=\int d\eta \left(L_{\gamma}+L_{\phi}\right)$, where
\bea
L_{\gamma}&=&a^{2}{\gap}^{2}+\Ga_{\ga}\gap {\dfp}^{2}+\Ga_{\ga}{\gap}^{3}+\Ga_{\phi}\dfp{\gap}^{2}+\Ga_{\ga \phi}\dfp\gap+\Ga_{\ga \ga}{\gap}^{2}\nonumber\\&+&\lambda_{\phi \ga \ga}\dfp
+\lambda_{\phi \phi \ga}\gap\nonumber+\lambda_{\ga \ga \ga}\gap,\\
L_{\phi}&=&a^{2}{\dfp}^{2}+\Ga_{1}{\dfp}^{2}+\Ga_{2}{\dfp}^{2}+\omega {\dfp}^{3}+\lambda \dfp,
\eea
\noindent where $f^{'} \equiv df / d\eta$ and where we define 
\bea
\Ga_{\ga}\sim a \ga\label{K1},\\
\Ga_{\phi}\sim a \df ,\\
\Ga_{1}\sim \dot{\phi} a^{2}\df,\\
\Ga_{2}\sim a^{2}{\df}^{2},\\
\Ga_{\ga \phi}\sim a^{2}\df \ga,\\
\Ga_{\ga \ga}\sim a^{2} {\ga}^{2},\\
\omega \sim a \df,\\
\lambda_{\phi \phi \phi} \sim a {\df}^{3},\\
\lambda_{\phi \ga \ga}\sim a \df {\ga}^{2},\\ 
\lambda_{\phi \phi \ga}\sim a \ga {\df}^{2},\\
\lambda_{\ga \ga \ga}\sim a {\ga}^{3}.\label{K2}
\eea
Notice that in equations (\ref{K1}) through (\ref{K2}) we use the equivalence symbol meaning that we skeep details about integrations in momenta and real coefficients.\\
The conjugate momenta are
\bea
\fl
\Pig\equiv \frac{\delta L}{\delta \gap}= a^{2}\gap+\lambda_{\ga \ga \ga}+\lambda_{\phi \phi \ga}+\Ga_{\ga \phi}\dfp+\Ga_{\ga}{\dfp}^{2}  
+\Ga_{\phi}\gap\dfp+\Ga_{\ga \ga}{\gap}^{3}+\Ga_{\ga}{\gap}^{2}\label{K3},\\
\nonumber\\
\fl
\Pif\equiv \frac{\delta L}{\delta \left(\dfp\right)}=\Ga_{\ga}\gap\dfp+\Ga_{\ga \phi}\gap+\lambda_{\phi \phi \ga}+a^{2}\dfp+\Ga_{1}\dfp+\Ga_{2}\dfp
+\lambda_{\phi \phi \phi}+\omega{\dfp}^{2}.\label{K4}
\eea
We solve perturbatively the equations (\ref{K3}) and (\ref{K4}) in order to derive $\gap$ and $\dfp$ to fourth order
\bea
\gap&=&a^{-2}\Big[\Pig+\lambda_{\ga \ga \ga}+\lambda_{\phi \phi \ga}+a^{-2}\Ga_{\ga \phi}\Pif+a^{-4}\Ga_{\ga}\Pif\Pif+a^{-4}\Ga_{\phi}\Pig\Pif\nonumber\\&+&a^{-2}\Ga_{\ga \ga}\Pig+a^{-4}\Ga_{\ga}\Pig\Pig
+a^{-4}\Ga_{\ga \phi}\Ga_{1}\Pif+a^{-4}\Ga_{\phi}\Pig\Pif\nonumber\\
&+&a^{-6}\Ga_{\phi}\Ga_{1}\Pig\Pif+a^{-2}\Ga_{\ga \ga}\Pif\Big],\label{K5}\\
\dfp&=&a^{-2}\Big[\Pif+a^{-4}\Ga_{\ga}\Pig\Pif+a^{-6}\Ga_{\ga}\Ga_{1}\Pig\Pif+a^{-2}\Ga_{\ga \phi}\Pig+\lambda_{\phi \ga \ga}\nonumber\\
&+&a^{-2}\Ga_{1}\Pif+a^{-4\Ga_{1}}\Ga_{1}\Pif+a^{-2}\Ga_{2}\Pif+\lambda_{\phi \phi \phi}+a^{-4}\omega\Pif\Pif\nonumber\\&+&a^{-4}\Ga_{\ga \phi}\Ga_{1}\Pig+a^{-2}\Ga_{1}\lambda_{\phi \ga \ga}+a^{-6}\Ga_{1}\Ga_{1}\Ga_{1}\Pif+a^{-4}\Ga_{1}\Ga_{2}\Pif\nonumber\\&+&a^{-2}\Ga_{1}\lambda_{\phi \phi \phi}+a^{-6}\Ga_{1}\omega\Pif\Pif\Big].\label{K6}
\eea
The next steps are: derive the hamiltonian $H=\Pig \gap+\Pif \dfp-L\left(\ga,\df,\gap,\dfp\right)$, where we need to plug in the solution (\ref{K5}) and (\ref{K6}) for $\gap$ and $\dfp$; construct the action as $S=S_{0}+S_{\Pi}$, where $S_{0}=\int d \eta \left(L_{\ga}+L_{\phi}\right)$ and $S_{\Pi}$ includes the terms that depend on the conjugate momenta of the fields (a change of variables similar to the one that Seery performs in \cite{seery1} over the conjugate momenta will also be necessary).\\ 
Let's consider the vertices in $S_{\Pi}$ that are involved in the corrections to the one loop point function for the scalar field
\bea\label{PI}
S_{\Pi}&\supset& \int d\epr \Big[a^{-4}\Ga_{1}\dfp\Pif+a^{-4}\Ga_{1}\Pif\Pif+a^{-2}\Ga_{1}\gap\Pif+a^{-4}\Ga_{2}\Pif\Pif\nonumber\\
&+&a^{-4}\Ga_{\phi}\dfp\Pig\Pig+a^{-4}\omega\dfp\Pif\Pif\Big].\label{K10}
\eea   
The first three vertices belong to the third order part of the action; $a^{-4}\Ga_{1}\dfp\Pif$ and $a^{-2}\Ga_{1}\gap\Pif$, provide a correction to the two point function at one loop with two vertices. Because of the presence of $\Ga_{1}$ which involves a factor of $\dot{\phi}$, it is subleading in slow roll order w.r.t. the corrections coming from fourth order vertices. We will therefore neglect these diagrams. The same applies to the second vertex, $a^{-4}\Ga_{1}\Pif\Pif$, although this may contribute to correcting the one point function 
\be\label{K8}
\langle\df_{\vec{k}}(\eta^{*})\rangle \supset C_{1} \frac{H_{*}^{2}}{k^{3}}\sqrt{\epsilon}\int d^{3}q f_{1}(\vec{q}) \int_{- \infty}^{\eta^{*}}d\epr \delta(0) \left(1-ik\epr\right)e^{ik\epr}+c.c.,
\ee
\noindent where $C_{1}$ is a numerical real coefficient and $\delta(0)$ is the Dirac delta function deriving from the propagator of the $\Pif$'s and $f_{1}$ is a scalar function of the internal momentum. The main contribution to the integral is due to times around horizon crossing since at early times the rotation to imaginary plane of the contour integral makes the exponent decrease rapidly to zero and moreover $\eta^{*}$ was chosen to be just a few e-folding after horizon crossing. Also, since the integrand function goes to zero as $\epr$ approaches zero, we get a good approximation of this integral taking the upper limit $\eta^{*}\rightarrow 0$. The result is purely imaginary and it cancels out with its complex conjugate.\\
Let us now move to the fourth order vertices. From $a^{-4}\Ga_{2}\Pif\Pif$ we have
\be\fl
\langle\df_{\vec{k_{1}}}(\eta^{*})\df_{\vec{k_{2}}}(\eta^{*})\rangle \supset \frac{H_{*}^{4}}{k^{6}} C_{2} \int d^{3}q f_{2}(\vec{q})\int^{\eta^{*}}_{-\infty}d \epr\delta(0){\left(1-ik\epr\right)}^{2}e^{2ik\epr}+c.c.
\ee   
\noindent The same consideration as in (\ref{K8}) apply to the integral above, which gives a zero contribution, as well as the following diagrams (corresponding to the last two vertices in (\ref{K10})) 
\bea
\fl
\langle\df_{\vec{k_{1}}}(\eta^{*})\df_{\vec{k_{2}}}(\eta^{*})\rangle &\supset& \frac{H_{*}^{4}}{k^{6}} C_{3} \int d^{3}q f_{3}(\vec{q})\int^{\eta^{*}}_{-\infty}d\epr\delta(0){\epr}^{2}\left(1-ik\epr\right)e^{2ik\epr}\nonumber\\
&+&\frac{H^{4}}{k^{6}} C_{4} \int d^{3}q f_{4}(\vec{q})\int^{\eta^{*}}_{-\infty}d \epr\delta(0){\epr}^{2}\left(1-ik\epr\right)e^{2ik\epr}+c.c.
\eea



\vskip 1cm
\setcounter{equation}{0}
\def\theequation{15.2.\arabic{equation}}
\subsection{Study of leading slow roll order vertices in the fourth order action}\label{Semplificazioni}

We are interested in computing the correlators just a few e-foldings after the scales we consider cross the horizon, so we can assume that the slow roll parameters remain small and can be treated as constants during this length of time. It is then correct to limit our interest to the leading order slow-roll contribution to the action .\\

\noindent Let us start from the study of the slow-roll order of the fluctuations derived as solution to the constraint equations:
\bean
\alo &=& \sqrt{\ep} \textit{Q}_{\alo}[\df],\\
\tho &=& \sqrt{\ep} \textit{Q}_{\tho}[\df],\\
\alt &=& \ep\textit{R}_{\alt}[\df^{2}]+\sqrt{\ep}\textit{S}_{\alt}[\df,\ga]+\textit{T}_{\alt}[\df^{2}],\\
\tht&=&\ep\textit{R}_{\tht}[\df^{2}]+\sqrt{\ep}\textit{S}_{\tht}[\df,\ga]+\textit{T}_{\tht}[\df^{2}]+\ep^{2}\textit{U}_{\tht}[\df^{2}]+\textit{V}_{\tht}[\ga^{2}]
\nonumber\\&+&\ep^{3/2}\textit{W}_{\tht}[\df,\ga],\\ \b_{j}&=&\ep\textit{R}_{j}[\df^{2}]+\sqrt{\ep}\textit{S}_{j}[\df,\ga]+\textit{T}_{j}[\df^{2}]+\textit{V}_{j}[\ga^{2}],
\eean
\noindent where $\textit{S}[\df,\ga]$ is a linear function of $\df$ and/or its derivatives and a linear function of $\ga$ and/or its derivatives, $\textit{R}[\df^{2}]$ is a quadratic function of $\df$ and/or its derivatives and so on.  
\noindent Notice that the first order fluctuations are subleading ($\sim \sqrt{\ep}$) w.r.t. the second order ones ($\sim \ep^{0}$). This criterium allows a suppression of a large number of terms in the 4th order action based on keeping the leading order (i.e. $\sim \ep^{0}$) terms only
\bean\fl
S_{4}&=&a^{3} \int dt d^{3}x\Big[\frac{1}{4a^{2}}\Big(\p_{i}\b_{j}+\p_{j}\b_{i}\Big)\p_{i}\b_{j}
-\frac{1}{a^{2}}\ddf\left(\p_{j}\th_{2}+\b_{j}\right)\p_{j}\df+3H^{2}\al_{2}^{2}\\\fl
&-&\frac{1}{a^{2}}\left(\frac{1}{4}\ga_{ik}\ga_{kj}\p_{j}\df\p_{i}\df+\alt\p_{i}\df\p_{i}\df-\p_{k}\tht\dot{\ga_{ab}}\p_{b}\ga_{ak}+\frac{1}{2}\dot{\ga_{ab}}\p_{k}\ga_{ab}\p_{k}\tht\right)\\\fl
&+&\frac{1}{a^{2}}\left(\b_{k}\dot{\ga_{ab}}\p_{b}\ga_{ak}-\frac{1}{2}\dot{\ga_{ab}}\b_{k}\p_{k}\ga_{ab}\right)\Big].
\eean
\noindent It can be easily shown that the terms in the action that do not contain the gravitons reproduce the ones in equation (37) of \cite{Seery:2006vu}. The contribution to the power spectrum due to these vertices has been calculated by these authors, but only for the scalar part. We will then focus on all the tensor contributions from these and from the remaining terms. Interaction vertices with both two and four tensor fluctuations will be obtained once the expressions for $\alt$, $\tht$ and $\b_{j}$ are plugged in the action. The terms in the action that we need for constructing Feynman diagrams with one loop of gravitons are
\bea\label{Z}\fl
S_{\ga^{2}}&=&a^{3}\int d^{3}x dt \Big[-\frac{1}{4a^{2}}\b_{j}\p^{2}\b_{j}-\frac{1}{a^{2}}\ddf\p_{j}\df\p_{j}\tht-\frac{1}{4 a^{2}}\p_{j}\df\p_{i}\df\ga_{ik}\ga_{kj}\\\fl&-&\frac{1}{a^{2}}\ddf\p_{j}\df\b_{j}
+\frac{1}{2a^{2}}\Big(2\dot{\ga_{ab}}\p_{a}\ga_{ak}\left(\p_{k}\tht+\b_{k}\right)-\dot{\ga_{ab}}\p_{k}\ga_{ab}\Big(\b_{k}+\p_{k}\tht\Big)\Big]\nonumber.
\eea
Let's plug the expressions for $\beta_{j}$ and $\tht$ into (\ref{Z}) considering the terms with two gravitons. The result is an ensemble of vertices which can in principle contribute to the one loop corrections to the power spectrum of the scalar field. Apart from $\p_{j}\df\p_{i}\df\ga_{ik}\ga_{kj}$, all of the other terms contain time derivatives of one, two or three of the four fields
\bea
\b_{j}\p^{2}\b_{j}&\supset&  a^{4}\Big[\p^{-4}\Big(\p_{m}\p_{j}\ddf\p_{m}\df-\p^{2}\ddf\p_{j}\df+\p_{j}\ddf\p^{2}\df\nonumber\\&-&\p_{m}\ddf\p_{m}\p_{j}\df\Big)\left(\dot{\ga_{ik}}\p_{i}\ga_{kj}-\ga_{il}\p_{i}\dot{\ga_{kj}}\right)\nonumber\\&+&\p^{-2}\Big(\p_{m}\p_{j}\ddf\p_{m}\df-\p^{2}\ddf\p_{j}\df+\p_{j}\ddf\p^{2}\df\nonumber\\&-&\p_{m}\ddf\p_{m}\p_{j}\df\Big)\p^{-2}\left(\dot{\ga_{ik}}\p_{i}\ga_{kj}-\ga_{il}\p_{i}\dot{\ga_{kj}}\right)\Big],
\eea
\bea
\ddf\p_{j}\df\p_{j}\tht &\supset& \ddf\p_{j}\df\frac{1}{16H}\p^{-2}\p_{j}\Big[\frac{1}{2a^{2}}\p_{a}\ga_{iq}\p_{a}\ga_{iq}+\dot{\ga_{lj}}\dot{\ga_{lj}}\Big]
\eea
\bea
\ddf\p_{j}\df\b_{j}&\supset& \ddf\p_{j}\df\frac{a^{2}}{2}\p^{-2}\p_{j}\Big[\dot{\ga_{ik}}\p_{i}\ga_{kj}-\ga_{ik}\p_{i}\dot{\ga_{kj}}\Big],
\eea
\bea
\dot{\ga_{ab}}\p_{a}\ga_{bk}\p_{k}\tht&\supset&\dot{\ga_{ab}}\p_{a}\ga_{bk}\frac{a^{2}}{4H}\p^{-2}\p_{k}\Big[-6H\p^{-2}\Sigma-{\ddf}^{2}-\frac{1}{a^{2}}\p_{i}\df\p_{i}\df\Big],
\eea
\bea
\dot{\ga_{ab}}\p_{a}\ga_{bk}\b_{k}&\supset&\dot{\ga_{ab}}\p_{a}\ga_{bk}2a^{2}\p^{-4}\Big[\p_{m}\p_{k}\ddf\p_{m}\df-\p^{2}\ddf\p_{k}\df\nonumber\\&+&\p_{k}\ddf\p^{2}\df-\p_{m}\ddf\p_{m}\p_{k}\df\Big],
\eea
\bea
\dot{\ga_{ab}}\p_{k}\ga_{ab}\beta_{k}&\supset& \dot{\ga_{ab}}\p_{k}\ga_{ab}\p^{-4}\Big[\p_{m}\p_{k}\ddf \p_{m}\df-\p^{2}\ddf\p_{k}\df+\p_{k}\ddf\p^{2}\df\nonumber\\&-&\p_{m}\ddf\p_{m}\p_{k}\df\Big]\label{V3ref1}
\eea
\bea
\dot{\ga_{ab}}\p_{k}\ga_{ab}\p_{k}\tht&\supset& \dot{\ga_{ab}}\p_{k}\ga_{ab}\Big[-6H\p^{-2}\Big(\p^{2}\df\ddf+\p_{j}\df\p_{j}\ddf\Big)\nonumber\\&-&{\ddf}^{2}-\frac{1}{a^{2}}\p_{i}\df\p_{i}\df\Big] \label{V3ref2}.
\eea
We will now prove that the vertices that include time derivatives do not actually contribute to the two point function. First of all notice that the tensor fields carry polarization tensors $\epsilon_{ij}$ with the property $q^{i}\epsilon_{ij}=0$ and are always contracted with other tensor fields in the calculations; this implies that, if a partial derivative index is contracted with a tensor index, that diagram will be zero. Based on this observation, we can ignore several of the vertices with time derivatives. We are eventually left with only two of them, that we will call $V_{1}$, $V_{2}$ and $V_{3}$ 
\bea
V_{1} \sim \p_{j}\left(\ddf\p_{j}\df\right)\p^{-2}\left(\p_{a}\ga_{bc}\p_{a}\ga_{bc}\right),\\
V_{2} \sim \p_{j}\left(\ddf\p_{j}\df\right)\p^{-2}\left(\dot{\ga_{ab}}\dot{\ga_{ab}}\right),\\
V_{3} \sim \dot{\ga_{ab}}\p_{k}\ga_{ab}\Big(\beta_{k}+\p_{k}\tht\Big)\label{V3}.
\eea
\noindent where (\ref{V3}) is given by the sum of (\ref{V3ref1}) and (\ref{V3ref2}). Notice that the $\ga_{ij}$ fields need to be contracted between each other and that $\sum_{\lambda,{\lambda}^{'}}\epsilon^{\lambda*}_{iq}\epsilon^{{\lambda}^{'}}_{iq}=$constant \cite{polarization}; the derivatives of $\df$ contract with derivatives of $\ga$, so this produces $\vec{k} \cdot \vec{q}$ factors. Therefore we have
\bea
\fl
\langle\df_{\vec{k_{1}}}(\eta^{*})\df_{\vec{k_{2}}}(\eta^{*})\rangle_{V_{1}+V_{2}} \sim i \delta^{(3)}(\vec{k_{1}}+\vec{k_{2}})H_{*}^{4}\int \frac{d^{3}q}{q^{3}}f_{1}(q^{2})\vec{k} \cdot \vec{q}\int^{\eta^{*}}_{- \infty}d \epr f_{2}(\epr)+c.c.,   
\eea
\noindent where $f_{1}(q^{2})$ and $f_{2}(\epr)$ are some functions of $q^{2}$ and $\epr$. This contribution is evidently zero for symmetry reasons.\\


\vskip 1cm
\setcounter{equation}{0}
\def\theequation{15.3.\arabic{equation}}
\subsection{Complete expressions of one-loop two-vertex diagrams to leading order}\label{powerloops}

\noindent In the following we provide the explicit expression for Eqs.~(\ref{appC1}) and (\ref{appC2}). 
Eq.~(\ref{appC1}) reads as
\bea
\fl
\langle\df_{\vec{k_{1}}}(\eta^{*})\df_{\vec{k_{2}}}(\eta^{*})\rangle_{(1L,2v)}^{A}=\pi\delta^{(3)}(\vec{k_{1}}+\vec{k_{2}})\frac{H_{*}^{4}}{k^{3}}
\Big(a_{1}\ln(k)+a_{2}\ln(k \ell)+a_{3}\Big),
\eea
where
\bea
a_{1} &=& -\frac{4}{15}\Big(5+5x^{*2}+2x^{*4}\Big),
\eea
\bea
a_{2} &=& \frac{8}{15x^{*2}}\Big[-2
+\Big(5-8\sigma_{c}\tilde{\sigma}_{c}+4\pi \sigma_{s}-8\sigma_{s}\tilde{\sigma}_{s}\Big)x^{*2}-8(\pi \sigma_{c}+2\tilde{\sigma}_{c}\sigma_{s}\nonumber\\&-&2\sigma_{c}\tilde{\sigma}_{c})x^{*3}
+(1+8\sigma_{c}\tilde{\sigma}_{c}-4 \pi \sigma_{s}+8\sigma_{s}\tilde{\sigma}_{s})x^{*4}\Big],
\eea
\bea
a_{3} &=& \frac{1}{1800x^{*2}}\Big[-64
-\Big(-3120+15136\sigma_{c}\tilde{\sigma}_{c}
-450{\pi}^{2}\sigma_{c}\tilde{\sigma}_{c}-7568\pi \sigma_{s}\nonumber\\&+&225{\pi}^{3}\sigma_{s}+15136\sigma_{s}\tilde{\sigma}_{s}\Big)x^{*2}
-\Big(15136\pi \sigma_{c}-450{\pi}^{3}\sigma_{c}+30272\tilde{\sigma}_{c}\sigma_{s}\nonumber\\&-&900{\pi}^{2}\tilde{\sigma}_{c}\sigma_{s}
-30272\sigma_{c}\tilde{\sigma}_{s}+900{\pi}^{2}\sigma_{c}\tilde{\sigma}_{s}\Big)x^{*3}
-\Big(-672-15136\sigma_{c}\tilde{\sigma}_{c}\nonumber\\
&+&450{\pi}^{2}\sigma_{c}\tilde{\sigma}_{c}+7568\pi \sigma_{s}
-225{\pi}^{3}\sigma_{s}
-15136\sigma_{s}\tilde{\sigma}_{c}+450{\pi}^{2}\sigma_{s}\tilde{\sigma}_{s}\Big)x^{*4}\nonumber\\
&-&208x^{*6}\Big]+\rho,
\eea
and $\rho$ is a constant left over from renormalization of ultraviolet divergences. We have defined
\bea
\sigma_{s} \equiv \sin2x^{*},\nonumber\\
\sigma_{c} \equiv \cos2x^{*},\nonumber\\ 
\tilde{\sigma}_{s} \equiv {\rm Si}(2x^{*}),\nonumber\\
\tilde{\sigma}_{c} \equiv {\rm Ci}(2x^{*}),\nonumber
\eea
\noindent where ${\rm Si}$ and ${\rm Ci}$ stand for the sine-integral and the cosine-integral functions, i.e. 
\bea
{\rm Si}(x)=\int^{x}_{0}\frac{\sin(t)}{t}dt,\nonumber\\
{\rm Ci}(x)=\int^{x}_{0}\frac{\cos(t)-1}{t}dt+\ln(x)+\gamma,\nonumber
\eea
\noindent with $\gamma$ indicating the Euler Gamma function.\\
The expression for Eq.~(\ref{appC2}) is
\bea
\langle\df_{\vec{k_{1}}}(\eta^{*})\df_{\vec{k_{2}}}(\eta^{*})\rangle_{(1L,2v)}^{C}=\pi\delta^{(3)}(\vec{k_{1}}+\vec{k_{2}})\frac{H_{*}^{4}}{k^{3}}
\left(c_{1}+c_{2}\ln(k \ell)\right),
\eea
where
\bea
c_{1}&=& \frac{1}{225}\left(\frac{8}{x^{*2}}+107+50x^{*2}\right),\\
\nonumber\\
c_{2}&=& \frac{4}{15}\left(\frac{4}{x^{*2}}+1\right).
\eea
Finally the quantity $f_{3}$ appearing in Eq.~(\ref{FR}) is given by
\be
f_{3}=a_{3}+c_{1}+{\alpha}^{'},
\ee
where ${\alpha}^{'}\equiv 2\left(1+x^{*2}/3\right)\alpha$ from Eq.~(\ref{FR1}).



\vskip 1cm
\setcounter{equation}{0}
\def\theequation{15.4.\arabic{equation}}
\subsection{Background and first order perturbation equations for the gauge fields}\label{eomotion}

The equations of motion for the gauge fields and for a Lagrangian as in Eq.~(\ref{ac}) with $f=1$ have been completely derived for the $U(1)$ case in~\cite{Dimopoulos:2006ms}. We are going to carry out a similar calculation for the $SU(2)$ case  
\bea\label{eom}
\frac{1}{\sqrt{-g}}\p_{\mu}\Big[\sqrt{-g}g^{\mu\al}g^{\nu\b}\Big(F^{(AB)a}_{\al\b}+g_{c}\ep^{abc}B^{b}_{\al}B^{c}_{\b}\Big)\Big]+M^{2}g^{\mu\nu}B_{\mu}^{a}\nonumber\\+g_{c}\ep^{abc}g^{\ga\nu}g^{\de\b}F^{(AB)b}_{\ga\de}B^{c}_{\b}+g^{2}_{c}\ep^{abc}\ep^{bb'c'}g^{\nu\al}g^{\de\b}B^{c}_{\de}B^{b'}_{\al}B^{c'}_{\b}=0
\eea
where $F_{\mu\nu}^{(AB)a}\equiv\p_{\mu}B^{a}_{\nu}-\p_{\nu}B^{a}_{\mu}$.\\
The $\nu=0$ component of the equations of motion is
\bea\label{tc}
\p_{j}\dot{B}^{a}_{j}-\p_{j}\p_{j}B^{a}_{0}+a^{2}M^{2}B^{a}_{0}+g_{c}\ep^{abc}\Big[-\Big(\p_{j}B_{j}^{b}\Big)B^{c}_{0}-2B^{b}_{j}\p_{j}B_{0}^{c}-\dot{B}^{b}_{j}B^{c}_{j}\nonumber\\+g_{c}\ep^{cb'c'}B^{b}_{j}B^{b'}_{0}B^{c'}_{j}\Big]=0
\eea
where $B^{a}_{\mu}=B^{a}_{\mu}(\vec{x},t)$.\\
Let us now move to the spatial ($\nu=i$) part of (\ref{eom})
\bea\label{sc}
\ddot{B}_{i}^{a}+H\dot{B}_{i}^{a}-\frac{1}{a^{2}}\p_{j}\p_{j}B_{i}^{a}+M^{2}B_{i}^{a}-\p_{i}\dot{B}_{0}^{a}-H\p_{i}B_{0}^{a}+\frac{1}{a^{2}}\p_{i}\p_{j}B_{j}^{a}\nonumber\\
+g_{c}\ep^{abc}\Big[HB_{0}^{b}B_{i}^{c}+\dot{B}_{0}^{b}B_{i}^{c}+B_{0}^{b}\dot{B}_{i}^{c}\Big]-g_{c}\frac{\ep^{abc}}{a^2}\Big[\Big(\p_{j}B_{j}^{b})B^{c}_{i}+B_{j}^{b}\p_{j}B^{c}_{i}\Big]\nonumber\\
+g_{c}\ep^{abc}\Big[\Big(\p_{i}B_{0}^{b}\Big)B^{c}_{0}-\dot{B}^{b}_{i}B^{c}_{0}\Big]-g_{c}\frac{\ep^{abc}}{a^2}\Big[\Big(\p_{i}B_{j}^{b})B^{c}_{j}-\Big(\p_{j}B^{b}_{i}\Big)B_{j}^{c}\Big]\nonumber\\
+g_{c}^{2}\ep^{abc}\ep^{bb'c'}\Big[B^{c}_{0}B^{b'}_{i}B^{c'}_{0}\Big]-\frac{g_{c}^{2}}{a^2}\ep^{abc}\ep^{bb'c'}\Big[B^{c}_{j}B^{b'}_{i}B^{c'}_{j}\Big]=0
\eea
If we contract Eq.(\ref{eom}) with $\p_{\nu}$, we get the integrability condition 
\bea\label{e10}\fl
(aM)^2 \dot{B^{a}_{0}}-M^2\p_{i}B_{i}^{a}+3H\Big(\p_{i}\p_{i}B_{0}^{a}-\p_{i}\dot{B_{i}^{a}}\Big)+g_{c}\epsilon^{abc}\Big[2H\Big(\p_{i}B_{i}^{b}B_{0}^{c}+B_{i}^{b}\p_{i}B_{0}^{c}+\dot{B_{j}^{b}}B_{j}^{c}\Big)-\p_{j}B_{0}^{b}\dot{B}_{j}^{c}\nonumber\\\fl
-\ddot{B_{j}^{b}}B_{j}^{c}+\p_{j}\dot{B_{0}^{b}}B_{j}^{c}-\p^{2}B_{0}^{b}B_{0}^{c}-\p_{i}\dot{B_{i}^{b}}B_{0}^{c}+\frac{1}{a^2}\Big(B_{i}^{b}\p^{2}B_{i}^{c}+\p_{i}B_{j}^{b}\p_{j}B_{i}^{c}+B_{j}^{b}\p_{j}\p_{i}B_{i}^{c}+\p^{2}B_{j}^{b}B_{j}^{c}\nonumber\\\fl-\p_{i}\p_{j}B_{i}^{b}B_{j}^{c}\Big)\Big]+g_{c}^{2}\epsilon^{abc}\epsilon^{bb^{'}c^{'}}\Big[a^{2}\Big(\dot{B}_{0}^{c}B_{0}^{b^{'}}B_{0}^{c^{'}}+B_{0}^{c}\dot{B}_{0}^{b^{'}}B_{0}^{c^{'}}+B_{0}^{c}B_{0}^{b^{'}}\dot{B}_{0}^{c^{'}}\Big)+2HB_{i}^{c}B_{0}^{b^{'}}B_{i}^{c^{'}}\nonumber\\\fl-\dot{B}_{i}^{c}B_{0}^{b^{'}}B_{i}^{c^{'}}-B_{i}^{c}\dot{B}_{0}^{b^{'}}B_{i}^{c^{'}}-B_{i}^{c}B_{0}^{b^{'}}\dot{B}_{i}^{c^{'}}-\p_{i}B_{0}^{c}B_{i}^{b^{'}}B_{0}^{c^{'}}-B_{0}^{c}\p_{i}B_{i}^{b^{'}}B_{0}^{c^{'}}-B_{0}^{c}B_{i}^{b^{'}}\p_{i}B_{0}^{c^{'}}\nonumber\\\fl+\frac{1}{a^2}\Big(\p_{i}B_{j}^{c}B_{i}^{b^{'}}B_{j}^{c^{'}}+B_{j}^{c}\p_{i}B_{i}^{b^{'}}B_{j}^{c^{'}}+B_{j}^{c}B_{i}^{b^{'}}\p_{i}B_{j}^{c^{'}}\Big)\Big]=0
\eea
which reduces to Eq.($7$) of~\cite{Dimopoulos:2006ms} in the Abelian case.\\
Combining Eq.(\ref{e10}) with Eq.(\ref{tc}) we get
\bea
\fl
(aM)^2 \dot{B^{a}_{0}}-M^2\p_{i}B_{i}^{a}+3H\Big(a^{2}M^{2}B^{a}_{0}+g_{c}\ep^{abc}\Big[-\Big(\p_{j}B_{j}^{b}\Big)B^{c}_{0}-2B^{b}_{j}\p_{j}B_{0}^{c}-\dot{B}^{b}_{j}B^{c}_{j}+g_{c}\ep^{cb'c'}B^{b}_{j}B^{b'}_{0}B^{c'}_{j}\Big] \Big)\nonumber\\\fl+g_{c}\epsilon^{abc}\Big[2H\Big(\p_{i}B_{i}^{b}B_{0}^{c}+B_{i}^{b}\p_{i}B_{0}^{c}+\dot{B_{j}^{b}}B_{j}^{c}\Big)-\p_{j}B_{0}^{b}\dot{B}_{j}^{c}\nonumber\\\fl
-\ddot{B_{j}^{b}}B_{j}^{c}+\p_{j}\dot{B_{0}^{b}}B_{j}^{c}-\p^{2}B_{0}^{b}B_{0}^{c}-\p_{i}\dot{B_{i}^{b}}B_{0}^{c}+\frac{1}{a^2}\Big(B_{i}^{b}\p^{2}B_{i}^{c}+\p_{i}B_{j}^{b}\p_{j}B_{i}^{c}+B_{j}^{b}\p_{j}\p_{i}B_{i}^{c}+\p^{2}B_{j}^{b}B_{j}^{c}\nonumber\\\fl-\p_{i}\p_{j}B_{i}^{b}B_{j}^{c}\Big)\Big]+g_{c}^{2}\epsilon^{abc}\epsilon^{bb^{'}c^{'}}\Big[a^{2}\Big(\dot{B}_{0}^{c}B_{0}^{b^{'}}B_{0}^{c^{'}}+B_{0}^{c}\dot{B}_{0}^{b^{'}}B_{0}^{c^{'}}+B_{0}^{c}B_{0}^{b^{'}}\dot{B}_{0}^{c^{'}}\Big)+2HB_{i}^{c}B_{0}^{b^{'}}B_{i}^{c^{'}}\nonumber\\\fl-\dot{B_{i}^{c}}B_{0}^{b^{'}}B_{i}^{c^{'}}-B_{i}^{c}\dot{B_{0}^{b^{'}}}B_{i}^{c^{'}}-B_{i}^{c}B_{0}^{b^{'}}\dot{B}_{i}^{c^{'}}-\p_{i}B_{0}^{c}B_{i}^{b^{'}}B_{0}^{c^{'}}-B_{0}^{c}\p_{i}B_{i}^{b^{'}}B_{0}^{c^{'}}-B_{0}^{c}B_{i}^{b^{'}}\p_{i}B_{0}^{c^{'}}\nonumber\\\fl+\frac{1}{a^2}\Big(\p_{i}B_{j}^{c}B_{i}^{b^{'}}B_{j}^{c^{'}}+B_{j}^{c}\p_{i}B_{i}^{b^{'}}B_{j}^{c^{'}}+B_{j}^{c}B_{i}^{b^{'}}\p_{i}B_{j}^{c^{'}}\Big)\Big]=0
\eea
Plugging this into Eq.(\ref{sc}) we get
\bea\label{e9}\fl
\ddot{B}_{n}^{a}+H\dot{B}_{n}^{a}-\frac{1}{a^{2}}\p_{j}\p_{j}B_{n}^{a}+M^{2}B_{n}^{a}+2H\p_{n}B_{0}^{a}\nonumber\\\fl
-\frac{1}{\Big(aM\Big)^2}\p_{n}\Big[
-3H\Big(g_{c}\ep^{abc}\Big[-\Big(\p_{j}B_{j}^{b}\Big)B^{c}_{0}-2B^{b}_{j}\p_{j}B_{0}^{c}-\dot{B}^{b}_{j}B^{c}_{j}+g_{c}\ep^{cb'c'}B^{b}_{j}B^{b'}_{0}B^{c'}_{j}\Big] \Big)\nonumber\\\fl+g_{c}\epsilon^{abc}\Big[2H\Big(\p_{i}B_{i}^{b}B_{0}^{c}+B_{i}^{b}\p_{i}B_{0}^{c}+\dot{B_{j}^{b}}B_{j}^{c}\Big)-\p_{j}B_{0}^{b}\dot{B}_{j}^{c}\nonumber\\\fl
-\ddot{B_{j}^{b}}B_{j}^{c}+\p_{j}\dot{B_{0}^{b}}B_{j}^{c}-\p^{2}B_{0}^{b}B_{0}^{c}-\p_{i}\dot{B_{i}^{b}}B_{0}^{c}+\frac{1}{a^2}\Big(B_{i}^{b}\p^{2}B_{i}^{c}+\p_{i}B_{j}^{b}\p_{j}B_{i}^{c}+B_{j}^{b}\p_{j}\p_{i}B_{i}^{c}+\p^{2}B_{j}^{b}B_{j}^{c}\nonumber\\\fl-\p_{i}\p_{j}B_{i}^{b}B_{j}^{c}\Big)\Big]+g_{c}^{2}\epsilon^{abc}\epsilon^{bb^{'}c^{'}}\Big[a^{2}\Big(\dot{B_{0}^{c}}B_{0}^{b^{'}}B_{0}^{c^{'}}+B_{0}^{c}\dot{B}_{0}^{b^{'}}B_{0}^{c^{'}}+B_{0}^{c}B_{0}^{b^{'}}\dot{B}_{0}^{c^{'}}\Big)+2HB_{i}^{c}B_{0}^{b^{'}}B_{i}^{c^{'}}\nonumber\\\fl-\dot{B_{i}^{c}}B_{0}^{b^{'}}B_{i}^{c^{'}}-B_{i}^{c}\dot{B_{0}^{b^{'}}}B_{i}^{c^{'}}-B_{i}^{c}B_{0}^{b^{'}}\dot{B_{i}^{c^{'}}}-\p_{i}B_{0}^{c}B_{i}^{b^{'}}B_{0}^{c^{'}}-B_{0}^{c}\p_{i}B_{i}^{b^{'}}B_{0}^{c^{'}}-B_{0}^{c}B_{i}^{b^{'}}\p_{i}B_{0}^{c^{'}}\nonumber\\\fl+\frac{1}{a^2}\Big(\p_{i}B_{j}^{c}B_{i}^{b^{'}}B_{j}^{c^{'}}+B_{j}^{c}\p_{i}B_{i}^{b^{'}}B_{j}^{c^{'}}+B_{j}^{c}B_{i}^{b^{'}}\p_{i}B_{j}^{c^{'}}\Big)\Big]
\Big]\nonumber\\\fl
+g_{c}\ep^{abc}\Big[HB_{0}^{b}B_{n}^{c}+\dot{B}_{0}^{b}B_{n}^{c}+B_{0}^{b}\dot{B}_{n}^{c}\Big]-g_{c}\frac{\ep^{abc}}{a^2}\Big[\Big(\p_{j}B_{j}^{b})B^{c}_{n}+B_{j}^{b}\p_{j}B^{c}_{n}\Big]\nonumber\\\fl
+g_{c}\ep^{abc}\Big[\Big(\p_{n}B_{0}^{b}\Big)B^{c}_{0}-\dot{B}^{b}_{n}B^{c}_{0}\Big]-g_{c}\frac{\ep^{abc}}{a^2}\Big[\Big(\p_{n}B_{j}^{b})B^{c}_{j}-\Big(\p_{j}B^{b}_{n}\Big)B_{j}^{c}\Big]\nonumber\\\fl
+g_{c}^{2}\ep^{abc}\ep^{bb'c'}\Big[B^{c}_{0}B^{b'}_{n}B^{c'}_{0}\Big]-\frac{g_{c}^{2}}{a^2}\ep^{abc}\ep^{bb'c'}\Big[B^{c}_{j}B^{b'}_{n}B^{c'}_{j}\Big]=0.
\eea
Let us consider the background part of the vector fields, i.e. $\p_{i}B_{\mu}^{a}=0$. Then from Eq.(\ref{tc})
\bea\label{backzero}
a^{2}M^{2}B^{a}_{0}+g_{c}\ep^{abc}\Big[-\dot{B}^{b}_{j}B^{c}_{j}+g_{c}\ep^{cb'c'}B^{b}_{j}B^{b'}_{0}B^{c'}_{j}\Big]=0.
\eea
Before proceeding with the derivation of the equations of motion for the background and the field perturbations, it is necessary to make some comments about Eqs.~(\ref{e9}) and (\ref{backzero}). One approximation that we have been using in the computation of the cosmological correlators is allowing the $A_{i}^{a}$ fields to undergo slow-roll during inflation. One possible way to achieve this is by restricting the parameter space of the background gauge fields through the request that their temporal components should be much smaller than the spatial ones, $B^{b}_{0}\ll |B^{b}_{i}|/a(t)$, $b=1,2,3$, and, in addition to that, assuming $B_{0}^{b}\simeq B_{0}^{c}$, $b,c=1,2,3$. With these assumptions, the temporal component can be factored out in Eq.~(\ref{backzero}), using the approximation $\dot{B}^{a}_{i}\simeq H B^{a}_{i}$ (valid in a slow-roll regime). A solution to (\ref{backzero}) is then given by $B_{0}=0$. Adopting this solution and plugging it in Eq~(\ref{e9}), it is easy to show that a slow-roll equation of motion for the physical fields
\bea
\ddot{A}_{i}^{a}+3H\dot{A}_{i}^{a}+m_{0}^{2}A_{i}^{a}=0
\eea  
follows from (\ref{e9}) if we set $M^2=m_{0}^{2}-2H^2$ with $\dot{H}\ll m_{0}^{2}$ and if
\bea\label{cond1}\fl
\left(\frac{g_{c}A^{1}}{m_{0}}\right)^2\ll\left|\frac{\left(A^{1}\right)^2}{\left(A^{2}\right)^2+\left(A^{3}\right)^2-\left(A^{3}\right)^2\cos^2\theta_{13}-\left(A^{2}\right)^2\cos^2\theta_{12}}\right|,\\\label{cond2}\fl
\left(\frac{g_{c}A^{2}}{m_{0}}\right)^2\ll\left|\frac{\left(A^{2}\right)^2}{\left(A^{1}\right)^2+\left(A^{3}\right)^2-\left(A^{3}\right)^2\cos^2\theta_{23}-\left(A^{1}\right)^2\cos^2\theta_{12}}\right|,\\\label{cond3}\fl
\left(\frac{g_{c}A^{3}}{m_{0}}\right)^2\ll\left|\frac{\left(A^{3}\right)^2}{\left(A^{1}\right)^2+\left(A^{2}\right)^2-\left(A^{2}\right)^2\cos^2\theta_{23}-\left(A^{1}\right)^2\cos^2\theta_{13}}\right|,
\eea
are satisfied. In the equations above, we defined $A^{a}\equiv|\vec{A}^{a}|$ and $\cos\theta_{ab}\equiv\hat{A}^{a}\cdot\hat{A}^{b}$, $a$ and $b$ running over the gauge indices. The quantities appearing on the right-hand sides of Eqs.(\ref{cond1}) through (\ref{cond3}) can be either large or small w.r.t. one, depending on the specific background configuration, i.e. on the moduli of the gauge fields and the angles $\theta_{ab}$.\\
Suppose now the conditions described above are all met, then from Eq~(\ref{e9}), in terms of the comoving fields, we have
\bea\label{eomB}
\ddot{B_{i}^{a}}+H\dot{B^{a}_{i}}+M^2B_{i}^{a}=0. 
\eea
Let us now derive the equations for the perturbations. Eq.(\ref{tc}) becomes
\bea\label{sc13}\fl
\p_{j}\de \dot{B}_{j}^{a}-\p^2\de B_{0}^{a}+a^2 M^2\de B_{0}^{a}+g_{c}\ep^{abc}\Big[-\p_{j}\de B_{j}^{b}B_{0}^{c}-2B_{j}^{b}\p_{j}\de B_{0}^{c}-\de \dot{B}_{j}^{b}B_{j}^{c}-\dot{B}_{j}^{b}\de B_{j}^{c}\nonumber\\\fl+g_{c}\ep^{cb^{'}c^{'}}\Big(\de B_{j}^{b} B_{0}^{b^{'}} B_{j}^{c^{'}}+ B_{j}^{b} \de B_{0}^{b^{'}} B_{j}^{c^{'}}+ B_{j}^{b} B_{0}^{b^{'}} \de B_{j}^{c^{'}}\Big)\Big] =0
\eea
Eq.(\ref{eom}) for the field perturbations gives
\bea\label{sc1}\fl
\de \ddot{B}_{i}^{a}+H\de\dot{B}_{i}^{a}-\frac{1}{a^{2}}\p_{j}\p_{j}\de B_{i}^{a}+M^{2}\de B_{i}^{a}+\frac{1}{a^{2}}\p_{i}\p_{j}\de B_{j}^{a}-H\p_{i}\de B_{0}^{a}-H\p_{i}\de B_{0}^{a}\nonumber\\\fl
-\frac{g_{c}}{a^2}\ep^{abc}\Big[\Big(\p_{j}\de B_{j}^{b})B^{c}_{i}+B_{j}^{b}\p_{j}\de B^{c}_{i}\Big]
-\frac{g_{c}}{a^2}\ep^{abc}\Big[\Big(\p_{i}\de B_{j}^{b})B^{c}_{j}-\Big(\p_{j}\de B^{b}_{i}\Big)B_{j}^{c}\Big]\nonumber\\\fl
-\frac{g_{c}^{2}}{a^2}\ep^{abc}\ep^{bb'c'}\Big[\de B^{c}_{j}B^{b'}_{i}B^{c'}_{j}+B^{c}_{j}\de B^{b'}_{i}B^{c'}_{j}+B^{c}_{j}B^{b'}_{i}\de B^{c'}_{j}\Big]\nonumber\\\fl
+g_{c}\ep^{abc}\Big[H\left(B^{b}_{0}\DB^{c}_{i}+\DB_{0}^{b}B_{i}^{c}\right)+\dot{\DB^{b}_{0}}B_{i}^{c}+\dot{B^{b}_{0}}\DB_{i}^{c}+\DB^{b}_{0}\dot{B_{i}^{c}}+B^{b}_{0}\dot{\DB^{c}_{i}}\Big]\nonumber\\\fl
+g_{c}\ep^{abc}\Big[\p_{i}\DB^{b}_{0}\DB_{0}^{c}-\dot{\DB_{i}^{b}}B_{0}^{c}-\dot{B_{i}^{b}}\DB^{c}_{0}\Big]\nonumber\\\fl
+g_{c}^{2}\ep^{abc}\ep^{bb^{'}c^{'}}\Big[\DB_{0}^{c}B_{i}^{b^{'}}B_{0}^{c^{'}}+B_{0}^{c}\DB_{i}^{b^{'}}B_{0}^{c^{'}}+B_{0}^{c}B_{i}^{b^{'}}\DB_{0}^{c^{'}}\Big]=0
\eea
Finally from Eq.(\ref{e9}) we get
\bea\label{perturbationsEQ}\fl
\ddot{\DB^{a}_{i}}+H\dot{\DB^{a}_{i}}-\frac{1}{a^2}\p^{2}\DB^{a}_{i}+M^2\DB^{a}_{i}+2H\p_{i}\DB^{a}_{0}+(\sim g_{c}terms)=0.
\eea
When calculating n-point functions for the gauge bosons, the eigenfunctions we need are provided by free-field solutions, i.e. by solutions of Eq.(\ref{perturbationsEQ}) with $g_{c}$ being set to zero. This is exactly the Abelian limit, in fact in this case Eq.(\ref{perturbationsEQ}) corresponds to (18) of~\cite{Dimopoulos:2006ms} and can be decomposed into a transverse and a longitudinal part
\bea\label{transverse}
\left[\p_{0}^{2}+H\p_{0}+M^2+\left(\frac{k}{a}\right)^2\right]\delta\vec{B}^{T}=0 \\\label{longitudinal}
\left[\p_{0}^{2}+\left(1+\frac{2k^2}{k^2+\left(aM\right)^2}\right)H\p_{0}+M^2+\left(\frac{k}{a}\right)^2\right]\delta\vec{B}^{||}=0
\eea
where the time derivatives are intended w.r.t. cosmic time.

\vskip 1cm
\setcounter{equation}{0}
\def\theequation{15.5.\arabic{equation}}
\subsection{Calculation of the number of e-foldings of single-(scalar)field driven inflation in the presence of a vector multiplet}\label{deltader}

Let us consider the Lagrangian in Eq.(\ref{ac}) with $f=1$ and $M^2=m_{0}^{2}-2H^2$. Let us assume that the $SU(2)$ gauge multiplet undergoes slow-roll as well as the scalar field but the latter provides the dominant part of the energy density of the universe. This last hypothesis is necessary in order to produce isotropic inflation (i.e. in order for the anisotropy in the expansion that the vector fields introduce to be negligible w.r.t. the isotropic contribution from the scalar field). The expression of the number of e-foldings is 
\bea\label{N__}
N=N_{scalar}+N_{vector}=N_{scalar}+\frac{1}{4m_{P}^{2}}\sum_{a=1,2,3}\vec{A}^{a}\cdot\vec{A}^{a}.
\eea
The previous expression can be easily derived from the equations of motion of the system neglecting terms that are proportional to the $SU(2)$ coupling constant $g_{c}$ and assuming slow-roll conditions for both the scalar the gauge fields.\\
The starting point is represented by Einstein equations
\bea\label{EEdeltaN}
H^2=\frac{8 \pi G}{3}\left(\rho_{scalar}+\rho_{vector}\right).
\eea
where we split the energy density into a scalar and a vector contribution. In slow-roll approximation, $\rho_{scalar}\sim V(\phi)$. Let us  calculate $\rho_{vector}$. The energy momentum tensor for the gauge bosons
\bea
T_{\mu\nu}^{vector}=2\frac{\delta \textit{L}}{\delta g^{\mu\nu}}-g_{\mu\nu}\textit{L}
\eea
where, as a remainder, $\textit{L}=-(1/4)g^{\mu\alpha}g^{\nu\beta}F_{\mu\nu}^{a}F^{a}_{\alpha\beta}+(M^2/2)g^{\mu\nu}B^{a}_{\mu}B^{a}_{\nu}$. So we get
\bea
T_{00}^{vector}&=&\frac{\dot{B}^{a}_{i}\dot{B}^{a}_{i}}{2 a^2}+\frac{m_{0}^2}{2a^2} B^{a}_{i} B^{a}_{i}+\frac{m_{0}^2}{2} B^{a}_{0} B^{a}_{0}-\frac{H}{a^2}\dot{B}^{a}_{i}{B}^{a}_{i}+\frac{H^2}{2a^2}{B}^{a}_{i}{B}^{a}_{i}\nonumber\\&+&\frac{g_{c}}{a^2}\ep^{abc}\dot{B}^{a}_{i}B_{0}^{b}B_{i}^{c}+\frac{g_{c}^{2}}{2a^2}\ep^{abc}\ep^{ab^{'}c^{'}}B_{0}^{b}B_{i}^{c}B_{0}^{b^{'}}B_{i}^{c^{'}}\nonumber\\&+&\frac{g_{c}^{2}}{4a^4}\ep^{abc}\ep^{ab^{'}c^{'}}B^{b}_{i}B^{c}_{j}B^{b^{'}}_{i}B^{c^{'}}_{j}
\eea
where sums are taken over all repeated indices. Let us write this in terms of the physical fields 
\bea\fl
T_{00}^{vector}&=&\frac{\dot{A}_{i}^{a}\dot{A}_{i}^{a}}{2}+\frac{m_{0}^2}{2}\left({A}_{i}^{a}{A}_{i}^{a}+A_{0}^{a}A_{0}^{a}\right)+g_{c}\ep^{abc}\left(HA_{i}^{a}+\dot{A}^{a}_{i}\right)A^{b}_{0}A^{c}_{i}+\frac{g_{c}^{2}}{2}\ep^{abc}\ep^{ab^{'}c^{'}}A_{0}^{b}A_{i}^{c}A_{0}^{b^{'}}A_{i}^{c^{'}}\nonumber\\\fl&+&\frac{g_{c}^2}{4}\ep^{abc}\ep^{ab^{'}c^{'}}{A}_{i}^{b}{A}_{j}^{c}{A}_{i}^{b^{'}}{A}_{j}^{c^{'}}
\eea
If we neglect the non-Abelian contribution and we set $A^{a}_{0}=0$, we are left with the Abelian result~\cite{Golovnev:2008cf}
\bea\label{naro}\fl
T_{00}^{vector}&=&\frac{\dot{A}_{i}^{a}\dot{A}_{i}^{a}}{2}+\frac{m_{0}^2}{2}{A}_{i}^{a}{A}_{i}^{a}
\eea
The equation of motion for the background vector multiplet $\vec{A}^{a}$ can be derived from Eq.(\ref{eomB})
\be\label{eom-de-sitter}
\ddot{A_{i}^{a}}+3H\dot{A^{a}_{i}}+m_{0}^2A_{i}^{a}=0. 
\ee
which is equal to the equation of a light scalar field of mass $m_{0}$, if $m_{0} \ll H$. If the conditions for accelerated expansions are met, Eq.(\ref{eom-de-sitter}) reduces to 
\bea
3H\dot{A^{a}_{i}}+m_{0}^2A_{i}^{a}\sim 0. 
\eea
We are now ready to derive Eq.(\ref{N__}). Let us start from the definition of N and keep in mind Eq.(\ref{EEdeltaN}), where we are assuming the existence of a scalar fields $\phi$ in de-Sitter with a separable potential governed by the usual (background) equation 
\bea
\ddot{\phi}+3H\dot{\phi}+V^{'}=0
\eea
and slowly rolling down their potential. Then we have
\bea
N&=&\int_{t^{*}}^{t} Hdt^{'}=\int_{t^{*}}^{t} H^2\frac{dt^{'}}{H}=8 \pi G\int_{t^{*}}^{t}\frac{V(\phi)}{3H}dt^{'}+8 \pi G\int_{t^{*}}^{t}\frac{V(A)}{3H}dt^{'}\nonumber\\&=&8 \pi G\int_{t^{*}}^{t}\frac{V(\phi)}{3H}\frac{dt}{d\phi}d\phi+8 \pi G\sum_{a}\int_{t^{*}}^{t}\left(\frac{m_{0}^2}{2}\right)\frac{{A}_{i}^{a}{A}_{i}^{a}}{3H}\frac{dA^{a}dt^{'}}{dA^{a}}
\eea
where $A^{a}\equiv \vec{A^{a}}\cdot\vec{A^{a}}$. So
\bea\label{N___}
N&=&8 \pi G\int_{\phi(t^{*})}^{\phi(t)}\frac{V(\phi)}{3H\dot{\phi}}d\phi+8 \pi G\sum_{a}\int_{A^{a}(t^{*})}^{A^{a}(t)}\left(\frac{m_{0}^2}{4}\right)\frac{{A}_{i}^{a}{A}_{i}^{a}}{3H\dot{A}^{a}_{j}A^{a}_{j}}dA^{a}\nonumber\\&=&-\frac{1}{m_{P}^{2}}\int_{\phi(t^{*})}^{\phi(t)}\frac{V}{V^{'}}d\phi+\frac{1}{m_{P}^{2}}\sum_{a}\int_{A^{a}(t^{*})}^{A^{a}(t)}\left(\frac{m_{0}^2}{4}\right)\frac{{A}_{i}^{a}{A}_{i}^{a}}{(-m_{0}^2){A}^{a}_{j}A^{a}_{j}}dA^{a}\nonumber\\&=&-\frac{1}{m_{P}^{2}}\int_{\phi(t^{*})}^{\phi(t)}\frac{V}{V^{'}}d\phi-\left(\frac{1}{4 m_{P}^{2}}\right)\sum_{a}\int_{A^{a}(t^{*})}^{A^{a}(t)} dA^{a}
\eea
after using the slow-roll conditions. Eq.(\ref{N__}) is thus recovered. \\
In the final expression for the bispectrum then we can substitute
\bea\label{derN}
N^{i}_{a}\equiv\frac{dN}{dA^{a}_{i}}=\left(\frac{1}{2m_{P}^{2}}\right)A^{a}_{i}
\eea
where the derivatives are as usual calculated at the initial time $\eta^{*}$.\\
The upper limits in integrals such as the ones in Eq.(\ref{N___}) depend on the chosen path in field space and so they also depend on the initial field configuration. It is important to notice though, as also stated in~\cite{Vernizzi:2006ve}, that if the final time is chosen to be approaching (or later than) the end of inflation, the fields are supposed to have reached their equilibrium values and so $N$ becomes independent of the field values at the final time $t$. Eq.(\ref{derN}) is thus recovered.

\vskip 1cm
\setcounter{equation}{0}
\def\theequation{15.6.\arabic{equation}}
\subsection{Complete expressions for the functions appearing in the bispectrum from quartic interactions}\label{bispquartic}

The anisotropy coefficients $I_{n}$ in Eq.~(\ref{fire}) are listed below
\bea\label{anisotropic-coefficients1}\fl
I_{EEE}&\equiv&\ep^{aa'b'}\ep^{ac'e}\Big[6\left(\vec{N}^{a'}\cdot\vec{N}^{c'}\right)\left(\vec{N}^{b'}\cdot\vec{A}^{e}\right)\nonumber\\\fl&+&\left(\vec{N}^{b'}\cdot\vec{A}^{e}\right)\Big[\Big(-2\left(\hat{k}_{3}\cdot\vec{N}^{a'}\right)\left(\hat{k}_{3}\cdot\vec{N}^{c'}\right)-2\left(\hat{k}_{1}\cdot\vec{N}^{a'}\right)\left(\hat{k}_{1}\cdot\vec{N}^{c'}\right)\nonumber\\\fl&+&\left(\hat{k}_{1}\cdot\vec{N}^{a'}\right)\left(\hat{k}_{3}\cdot\vec{N}^{c'}\right)\hat{k}_{1}\cdot\hat{k}_{3}+\left(\hat{k}_{3}\cdot\vec{N}^{a'}\right)\left(\hat{k}_{1}\cdot\vec{N}^{c'}\right)\hat{k}_{1}\cdot\hat{k}_{3}\Big)+(1\rightarrow 2)+(3\rightarrow 2)\Big]\nonumber\\\fl&-&\left[\left(2\left(\vec{N}^{a'}\cdot\vec{N}^{c'}\right)\left(\hat{k}_{2}\cdot\vec{N}^{b'}\right)\left(\hat{k}_{2}\cdot\vec{A}^{e}\right)\right)+(2\rightarrow 1)+(2\rightarrow 3)\right]\nonumber\\\fl&+&\Big[\Big(\hat{k}_{2}\cdot\vec{A}^{e}\Big[2\left(\hat{k}_{3}\cdot\vec{N}^{a'}\right)\left(\hat{k}_{2}\cdot\vec{N}^{b'}\right)\left(\hat{k}_{3}\cdot\vec{N}^{c'}\right)+2\left(\hat{k}_{1}\cdot\vec{N}^{a'}\right)\left(\hat{k}_{2}\cdot\vec{N}^{b'}\right)\left(\hat{k}_{1}\cdot\vec{N}^{c'}\right)\nonumber\\\fl&-&\left(\hat{k}_{1}\cdot\vec{N}^{a'}\right)\left(\hat{k}_{2}\cdot\vec{N}^{b'}\right)\left(\hat{k}_{3}\cdot\vec{N}^{c'}\right)\hat{k}_{1}\cdot\hat{k}_{3}-\left(\hat{k}_{1}\cdot\vec{N}^{a'}\right)\left(\hat{k}_{2}\cdot\vec{N}^{b'}\right)\left(\hat{k}_{3}\cdot\vec{N}^{c'}\right)\hat{k}_{1}\cdot\hat{k}_{3}\Big]\Big)\nonumber\\\fl&+&(2\leftrightarrow 1)+(3\leftrightarrow 2)\Big]\Big] \\\fl
I_{lll}&\equiv&\ep^{aa'b'}\ep^{ac'e}\Big[\Big(\left(\hat{k}_{1}\cdot\vec{N}^{a'}\right)\left(\hat{k}_{3}\cdot\vec{N}^{b'}\right)\left(\hat{k}_{2}\cdot\vec{N}^{c'}\right)\left(\hat{k}_{1}\cdot\hat{k}_{2}\right)\left(\hat{k}_{3}\cdot\vec{A}^{e}\right)\nonumber\\\fl&-&\left(\hat{k}_{3}\cdot\vec{N}^{a'}\right)\left(\hat{k}_{2}\cdot\vec{N}^{b'}\right)\left(\hat{k}_{1}\cdot\vec{N}^{c'}\right)\left(\hat{k}_{1}\cdot\hat{k}_{2}\right)\left(\hat{k}_{3}\cdot\vec{A}^{e}\right)\Big)+(1\leftrightarrow 3)+(2\leftrightarrow 3) \Big]\\\fl
I_{llE}&\equiv& \ep^{aa'b'}\ep^{ac'e}\Big[\left(\vec{N}^{b'}\cdot\vec{A}^{e}\right)\Big(\left(\hat{k}_{1}\cdot\vec{N}^{a'}\right)\left(\hat{k}_{2}\cdot\vec{N}^{c'}\right)+\left(\hat{k}_{2}\cdot\vec{N}^{a'}\right)\left(\hat{k}_{1}\cdot\vec{N}^{c'}\right)\Big)\hat{k}_{1}\cdot\hat{k}_{2}\nonumber\\\fl&+&\left[\Big(2\left(\hat{k}_{2}\cdot\vec{A}^{e}\right)\left(\hat{k}_{1}\cdot\vec{N}^{a'}\right)\left(\hat{k}_{2}\cdot\vec{N}^{b'}\right)\left(\hat{k}_{1}\cdot\vec{N}^{c'}\right)\Big)+(1\leftrightarrow 2)\right]\nonumber\\\fl&-&\Big[\Big(\left(\left(\hat{k}_{1}\cdot\vec{N}^{a'}\right)\left(\hat{k}_{2}\cdot\vec{N}^{c'}\right)+\left(\hat{k}_{2}\cdot\vec{N}^{a'}\right)\left(\hat{k}_{1}\cdot\vec{N}^{c'}\right)\right)\left(\hat{k}_{3}\cdot\vec{N}^{b'}\right)\left(\hat{k}_{3}\cdot\vec{A}^{e}\right)\hat{k}_{1}\cdot\hat{k}_{2}\Big)\nonumber\\\fl&+&(1\leftrightarrow 3)+(2\leftrightarrow 3)\Big]\Big]\\\label{anisotropic-coefficients2}\fl
I_{EEl}&\equiv& \ep^{aa'b'}\ep^{ac'e}\Big[4\left(\vec{N}^{b'}\cdot\vec{A}^{e}\right)\left(\hat{k}_{3}\cdot\vec{N}^{a'}\right)\left(\hat{k}_{3}\cdot\vec{N}^{c'}\right)\nonumber\\\fl&+&\Big[\Big(\left(\hat{k}_{2}\cdot\vec{N}^{b'}\right)\left(\hat{k}_{2}\cdot\vec{A}^{e}\right)\left(\left(\hat{k}_{1}\cdot\vec{N}^{a'}\right)\left(\hat{k}_{3}\cdot\vec{N}^{c'}\right)+\left(\hat{k}_{3}\cdot\vec{N}^{a'}\right)\left(\hat{k}_{1}\cdot\vec{N}^{c'}\right)\right)\hat{k}_{1}\cdot\hat{k}_{3}\Big)\nonumber\\\fl&+&(2\leftrightarrow 1)+(2\leftrightarrow 3)\Big]\nonumber\\\fl&-&\left[\left(2\left(\hat{k}_{2}\cdot\vec{A}^{e}\right)\left(\hat{k}_{2}\cdot\vec{N}^{a'}\right)\left(\hat{k}_{3}\cdot\vec{N}^{b'}\right)\left(\hat{k}_{2}\cdot\vec{N}^{c'}\right)\right)+(1\leftrightarrow 2)+(2\leftrightarrow 3)+(1\leftrightarrow 3)\right]\nonumber\\\fl&-&\left[\left(\left(\vec{N}^{b'}\cdot\vec{A}^{e}\right)\hat{k}_{1}\cdot\hat{k}_{3}\left(\left(\hat{k}_{1}\cdot\vec{N}^{a'}\right)\left(\hat{k}_{3}\cdot\vec{N}^{c'}\right)+\left(\hat{k}_{3}\cdot\vec{N}^{a'}\right)\left(\hat{k}_{1}\cdot\vec{N}^{c'}\right)\right)\right)+(1\leftrightarrow 2)\right] \nonumber\\\fl&+&\left[\left(\vec{N}^{a'}\cdot\vec{N}^{b'}\right)\left(\hat{k}_{3}\cdot\vec{N}^{c'}\right)\left(\hat{k}_{3}\cdot\vec{A}^{e}\right)+\left(\vec{N}^{c'}\cdot\vec{N}^{b'}\right)\left(\hat{k}_{3}\cdot\vec{N}^{a'}\right)\left(\hat{k}_{3}\cdot\vec{A}^{e}\right)\right]\Big]
\eea
where $i \rightarrow j$ means `replace $\hat{k}_{i}$ with $\hat{k}_{j}$', whereas $i \leftrightarrow j$ means `exchange $\hat{k}_{i}$ with $\hat{k}_{j}$'. \\

\noindent The isotropic functions $F_{n}$ in (\ref{fire}) are given by
\bea\fl
F_{EEE}= -\frac{1}{24k^{6}k^{2}_{1}k^{2}_{2}k^{2}_{3}x^{*2}}\left[A_{EEE}+\left(B_{EEE}\cos x^{*}+C_{EEE}\sin x^{*}\right)E_{i} x^{*}\right],\\\fl
F_{lll}=n^{6}(x^{*})F_{EEE} ,\\\fl
F_{llE}= n^{4}(x^{*})F_{EEE},\\\fl
F_{EEl}= n^{2}(x^{*})F_{EEE},
\eea
where
\bea\fl
A_{EEE}&\equiv& kx^{*2}\Big(-k^2(k^3_1+k^3_2+k^3_3-4k_1k_2k_3)-k^3(k_2k_3+k_1 k_2+k_1 k_3)\nonumber\\\fl&&+k_1k_2k_3(k^2_1+k^2_2+k^2_3-k_2k_3-k_1 k_2-k_1 k_3)x^{*2}\Big) \\
\fl
B_{EEE}&\equiv& \Big(k^3_1+k^3_2+k^3_3\Big)x^{*3}\Big(-k^3+k_1k_2k_3 x^{*2}\Big)\\
\fl
C_{EEE}&\equiv& -k\Big(k^3_1+k^3_2+k^3_3\Big)x^{*2}\Big(-k^2+(k_2k_3+k_1 k_2+k_1 k_3)x^{*2}\Big)
\eea
In the previous equations we set $k\equiv k_{1}+k_{2}+k_{3}$.


\vskip 1cm
\setcounter{equation}{0}
\def\theequation{15.7.\arabic{equation}}
\subsection{More details on computing the analytic expressions of vector-exchange diagrams}\label{exchangedetails}

We report the expressions of the functions $A$, $B$...$P$ introduced in Eq.~(\ref{1a})
\bea\label{eeeee1}\fl
A&\equiv& \left(-16k^{2}+k_{1}k_{2}x^{*2}\right)\cos\left[\frac{\left(k_{1}+k_{2}\right)x^{*}}{4k}\right]-4k\left(k_{1}+k_{2}\right)x^{*}\sin\left[\frac{\left(k_{1}+k_{2}\right)x^{*}}{4k}\right],\\\fl
B&\equiv& A\left[k_{1}\rightarrow k_{3},k_{2}\rightarrow k_{4}\right],\\\fl
C&\equiv& 4k\left(k_{1}+k_{2}\right)x^{*}\cos\left[\frac{\left(k_{1}+k_{2}\right)x^{*}}{4k}\right]+\left(-16k^{2}+k_{1}k_{2}x^{*2}\right)\sin\left[\frac{\left(k_{1}+k_{2}\right)x^{*}}{4k}\right],\\\fl
D&\equiv& C\left[k_{1}\rightarrow k_{3},k_{2}\rightarrow k_{4}\right],\\\fl
E&\equiv& \left(8k^{2}\left(k_{\hat{12}}+k_{3}+k_{4}\right)-k_{\hat{12}}k_{3}k_{4}x^{*2}\right)\cos\left[\frac{\left(k_{\hat{12}}+k_{3}+k_{4}\right)x^{*}}{4k}\right]\nonumber\\\fl&+&2k\left(k_{\hat{12}}+k_{3}+k_{4}\right)^2 x^{*}\sin\left[\frac{\left(k_{\hat{12}}+k_{3}+k_{4}\right)x^{*}}{4k}\right],\\\fl
F&\equiv& E\left[k_{3}\rightarrow k_{1},k_{4}\rightarrow k_{2}\right],\\\fl
G&\equiv&2k\left(k_{\hat{12}}+k_{1}+k_{2}\right)^2 x^{*}\cos\left[\frac{\left(k_{\hat{12}}+k_{1}+k_{2}\right)x^{*}}{4k}\right] \nonumber\\\fl&+&\left(-8k^{2}\left(k_{\hat{12}}+k_{1}+k_{2}\right)-k_{\hat{12}}k_{1}k_{2}x^{*2}\right)\sin\left[\frac{\left(k_{\hat{12}}+k_{1}+k_{2}\right)x^{*}}{4k}\right],\\\fl
L&\equiv& \left(k_{1}^{3}+k_{\hat{12}}^{3}+k_{\hat{12}}^{2}k_{2}+k_{\hat{12}}k_{2}^{2}+k_{2}^{3}(k_{\hat{12}}+k_{2})+k_{1}(k_{\hat{12}}^{2}+k_{2}^{2})\right)x^{*2}si\left[\frac{\left(k_{\hat{12}}+k_{1}+k_{2}\right)x^{*}}{4k}\right],\nonumber
\\\fl
M&\equiv& \left(k_{1}^{3}+k_{\hat{12}}^{3}+k_{\hat{12}}^{2}k_{2}+k_{\hat{12}}k_{2}^{2}+k_{2}^{3}(k_{\hat{12}}+k_{2})+k_{1}(k_{\hat{12}}^{2}+k_{2}^{2})\right)x^{*2}ci\left[\frac{\left(k_{\hat{12}}+k_{1}+k_{2}\right)x^{*}}{4k}\right],\nonumber\\\fl
N&\equiv& M\left[k_{3}\rightarrow k_{1},k_{4}\rightarrow k_{2}\right],\\\fl\label{eeeee2}
P&\equiv& L\left[k_{1}\rightarrow k_{3},k_{2}\rightarrow k_{4}\right].
\eea

\noindent The anisotropy coefficients introduced in Eq.~(\ref{final-2v}) have the following expressions
\bea\label{t1}\fl
t_{1}\equiv k_{1}k_{3}\left(\hat{k}_{1}\cdot\hat{k}_{\hat{12}}\right)\left(\hat{k}_{1}\cdot\hat{k}_{2}\right)\left(\hat{k}_{3}\cdot\hat{k}_{4}\right)\left(\hat{k}_{3}\cdot\hat{k}_{\hat{12}}\right)\\\fl
t_{2}\equiv  k_{1}k_{4}\left(\hat{k}_{1}\cdot\hat{k}_{\hat{12}}\right)\left(\hat{k}_{1}\cdot\hat{k}_{2}\right)\left(\hat{k}_{3}\cdot\hat{k}_{4}\right)\left(\hat{k}_{4}\cdot\hat{k}_{\hat{12}}\right),\\\fl
t_{3}\equiv  k_{2}k_{3}\left(\hat{k}_{2}\cdot\hat{k}_{\hat{12}}\right)\left(\hat{k}_{1}\cdot\hat{k}_{2}\right)\left(\hat{k}_{3}\cdot\hat{k}_{4}\right)\left(\hat{k}_{3}\cdot\hat{k}_{\hat{12}}\right),\\\fl
t_{4}\equiv  k_{2}k_{4}\left(\hat{k}_{2}\cdot\hat{k}_{\hat{12}}\right)\left(\hat{k}_{1}\cdot\hat{k}_{2}\right)\left(\hat{k}_{3}\cdot\hat{k}_{4}\right)\left(\hat{k}_{4}\cdot\hat{k}_{\hat{12}}\right),\\\fl
t_{5}\equiv  k_{1}k_{2}\left(\hat{k}_{1}\cdot\hat{k}_{\hat{13}}\right)\left(\hat{k}_{1}\cdot\hat{k}_{3}\right)\left(\hat{k}_{2}\cdot\hat{k}_{4}\right)\left(\hat{k}_{2}\cdot\hat{k}_{\hat{13}}\right),\\\fl
t_{6}\equiv  k_{1}k_{4}\left(\hat{k}_{1}\cdot\hat{k}_{\hat{13}}\right)\left(\hat{k}_{1}\cdot\hat{k}_{3}\right)\left(\hat{k}_{2}\cdot\hat{k}_{4}\right)\left(\hat{k}_{4}\cdot\hat{k}_{\hat{13}}\right),\\\fl
t_{7}\equiv  k_{2}k_{3}\left(\hat{k}_{3}\cdot\hat{k}_{\hat{13}}\right)\left(\hat{k}_{1}\cdot\hat{k}_{3}\right)\left(\hat{k}_{2}\cdot\hat{k}_{4}\right)\left(\hat{k}_{2}\cdot\hat{k}_{\hat{13}}\right),\\\fl
t_{8}\equiv  k_{3}k_{4}\left(\hat{k}_{3}\cdot\hat{k}_{\hat{13}}\right)\left(\hat{k}_{1}\cdot\hat{k}_{3}\right)\left(\hat{k}_{2}\cdot\hat{k}_{4}\right)\left(\hat{k}_{4}\cdot\hat{k}_{\hat{13}}\right),\\\fl
t_{9}\equiv  k_{1}k_{2}\left(\hat{k}_{1}\cdot\hat{k}_{\hat{14}}\right)\left(\hat{k}_{1}\cdot\hat{k}_{4}\right)\left(\hat{k}_{2}\cdot\hat{k}_{3}\right)\left(\hat{k}_{2}\cdot\hat{k}_{\hat{14}}\right),\\\fl
t_{10}\equiv k_{1}k_{3}\left(\hat{k}_{1}\cdot\hat{k}_{\hat{14}}\right)\left(\hat{k}_{1}\cdot\hat{k}_{4}\right)\left(\hat{k}_{2}\cdot\hat{k}_{3}\right)\left(\hat{k}_{3}\cdot\hat{k}_{\hat{14}}\right),\\\fl
t_{11}\equiv k_{1}k_{3}\left(\hat{k}_{1}\cdot\hat{k}_{\hat{12}}\right)\left(\hat{k}_{1}\cdot\hat{k}_{2}\right)\left(\hat{k}_{3}\cdot\hat{k}_{4}\right)\left(\hat{k}_{3}\cdot\hat{k}_{\hat{12}}\right),\\\label{t12}\fl
t_{12}\equiv k_{1}k_{3}\left(\hat{k}_{1}\cdot\hat{k}_{\hat{12}}\right)\left(\hat{k}_{1}\cdot\hat{k}_{2}\right)\left(\hat{k}_{3}\cdot\hat{k}_{4}\right)\left(\hat{k}_{3}\cdot\hat{k}_{\hat{12}}\right).
\eea

\noindent Let us now list all the scalar products appearing in the equations above\\

\begin{tabular}{|cc|c|cc|}\hline
$\hat{k}_{1}\cdot\hat{k}_{\hat{12}}=\frac{k_{\hat{12}}^2+k^{2}_{1}+k_{2}^{2}}{2k_{1}k_{\hat{12}}}$ & $$ & $\hat{k}_{1}\cdot\hat{k}_{\hat{13}}=\frac{2k_{1}^2+k^{2}_{4}-k_{\hat{12}}^{2}-k_{\hat{14}}^{2}}{2k_{1}k_{\hat{13}}}$ &$$ & $\hat{k}_{1}\cdot\hat{k}_{\hat{14}}=\frac{k_{\hat{14}}^2+k^{2}_{1}-k_{4}^{2}}{2k_{1}k_{\hat{14}}}$  \\
\hline
$\hat{k}_{2}\cdot\hat{k}_{\hat{12}}=\frac{k_{\hat{12}}^2+k^{2}_{1}-k_{1}^{2}}{2k_{2}k_{\hat{12}}}$ & $$ & $\hat{k}_{3}\cdot\hat{k}_{\hat{13}}=\frac{2k_{3}^{2}+k_{4}^2-k_{\hat{12}}^{2}-k_{\hat{14}}^{2}+k_{2}^{2}}{2k_{3}k_{\hat{13}}}$ & $$ & $\hat{k}_{4}\cdot\hat{k}_{\hat{14}}=\frac{k_{\hat{14}}^2-k^{2}_{1}+k_{4}^{2}}{2k_{4}k_{\hat{14}}}$ \\
\hline
$\hat{k}_{3}\cdot\hat{k}_{\hat{12}}=\frac{k_{\hat{12}}^2+k_{\hat{14}}^{2}-2k_{2}^{2}-k^{2}_{1}-k_{3}^{2}}{2k_{2}k_{\hat{13}}}$ & $$ & $\hat{k}_{2}\cdot\hat{k}_{\hat{13}}=\frac{k_{\hat{12}}^2+k_{\hat{14}}^{2}-k_{1}^{2}-2k
_{2}^{2}-k_{3}^{2}}{2k_{1}k_{3}}$ & $$ & $\hat{k}_{2}\cdot\hat{k}_{4}=\frac{k_{1}^{2}+k_{3}^{2}-k_{\hat{12}}^{2}-k_{\hat{14}}^{2}}{2k_{2}k_{4}}$\\
\hline
$\hat{k}_{4}\cdot\hat{k}_{\hat{12}}=\frac{k_{\hat{14}}^2+k_{\hat{12}}^2-k^{2}_{1}-k_{3}^{2}-2k_{4}^{2}}{2k_{4}k_{\hat{13}}}$ & $$ & $\hat{k}_{4}\cdot\hat{k}_{\hat{13}}=\frac{k_{\hat{14}}^{2}+k_{\hat{12}}^2-k^{2}_{1}-k^{2}_{3}-2k^{2}_{4}}{2k_{4}k_{\hat{13}}}$ & $$ &$\hat{k}_{2}\cdot\hat{k}_{\hat{14}}=\frac{k_{3}^{2}-k_{\hat{14}}^2-k^{2}_{2}}{2k_{2}k_{\hat{14}}}$\\
\hline
$\hat{k}_{1}\cdot\hat{k}_{2}=\frac{k_{1}^{2}+k_{2}^{2}-k_{\hat{12}}^{2}}{2k_{1}k_{2}}$ & $$ & $\hat{k}_{1}\cdot\hat{k}_{3}=\frac{k_{4}^{2}+k_{2}^{2}-k_{\hat{12}}^{2}-k_{\hat{14}}^{2}}{2k_{1}k_{3}}$ & $$ & $\hat{k}_{1}\cdot\hat{k}_{4}=\frac{k_{\hat{14}}^{2}-k_{1}^{2}-k_{4}^{2}}{2k_{1}k_{4}}$ \\
\hline
$\hat{k}_{3}\cdot\hat{k}_{4}=\frac{k_{\hat{12}}^{2}-k_{4}^{2}-k_{3}^{2}}{2 k_{4}k_{3}}$ & $$  & $\hat{k}_{3}\cdot\hat{k}_{\hat{14}}=\frac{k_{2}^2-k^{2}_{3}-k_{\hat{14}}^{2}}{2k_{3}k_{\hat{14}}}$ & $$ & $\hat{k}_{2}\cdot\hat{k}_{3}=\frac{k_{\hat{14}}^{2}-k_{2}^{2}-k_{3}^{2}}{2k_{2}k_{3}}$ \\
\hline
\end{tabular}
\\

\noindent where $k_{i}\equiv |\vec{k}_{i}|$, $\hat{k}_{i}\equiv\vec{k}_{i}/k_{i}$, $\vec{k}_{\hat{ij}}\equiv \vec{k}_{i}+\vec{k}_{j}$, $k_{ij}\equiv |\vec{k}_{i}+\vec{k}_{j}|$, $i$ and $j$ running over the four external wave vectors.

\vskip 1cm
\setcounter{equation}{0}
\def\theequation{15.8.\arabic{equation}}
\subsection{Complete expressions for functions appearing in point-interation diagrams}\label{pointdetails}

We provide here the expressions for the coefficients appearing in Eq.~(\ref{p-i})
\bea\fl
Q_{EEEE}&\equiv& x^{*3}\big[-k\big(k^{3}k^{4}_{1}+k^{3}k^{4}_{2}-k^{3}_{3}+k^{3}k_{3}^{4}+k^{5}k_{3}k_{4}-3k^{3}k^{2}_{2}k_{3}k_{4}+k^{3}k^{3}_{3}k_{4}\nonumber\\\fl&-&k^{3}_{4}+k^{3}k_{3}k_{4}^{3}+k^{3}k_{4}^{4}+k^{3}k_{2}\big(k_{3}^{3}-kk_{3}k_{4}-3k_{3}^{2}k_{4}-3k_{3}k_{4}^{2}+k_{4}^{3}\nonumber\\\fl&+&k^{2}\big(k_{3}+k_{4}\big)\big)+k_{2}^{3}\big(-1+k^{3}\big(k_{3}+k_{4}\big)\big)-3k^{3}k^{2}_{1}\big(k_{3}k_{4}+k_{2}\big(k_{3}+k_{4}\big)\big)\nonumber\\\fl&+&k^{3}_{1}\big(-1+k^{3}\big(k_{2}+k_{3}+k_{4}\big)\big)+k^{3}k_{1}\big(k_{2}^{3}+k_{3}^{3}-3k_{3}^{2}k_{4}-3k_{3}k_{4}^{2}+k_{4}^{3}\nonumber\\\fl&-&3k^{2}_{2}\big(k_{3}+k_{4}\big)+k^{2}\big(k_{2}+k_{3}+k_{4}\big)-3k_{2}\big(k_{3}^{2}+3k_{3}k_{4}+k_{4}^{2}\big)-k\big(k_{3}k_{4}\nonumber\\\fl&+&k_{2}\big(k_{3}+k_{4}\big)\big)\big)\big)\big]+x^{*5}\big[k^2\big(k^{3}_{1}k_{2}k_{3}k_{4}+k_{2}k_{3}k_{4}\big(k_{2}^{3}+k_{3}^{3}-3k_{2}k_{3}k_{4}+k_{4}^{3}\big)\nonumber\\\fl&+&k^{4}_{1}\big(k_{3}k_{4}+k_{2}\big(k_{3}+k_{4}\big)\big)-3k_{1}^{2}\big(k_{3}^{2}k_{4}^{2}+k_{2}k_{3}k_{4}\big(k_{3}+k_{4}\big)+k^{2}_{2}\big(k_{3}^{2}+k_{3}k_{4}\nonumber\\\fl&+&k^{2}_{4}\big)\big)+k_{1}\big(k_{2}^{3}k_{3}k_{4}+k_{2}^{4}\big(k_{3}+k_{4}\big)-3k_{2}^{2}k_{3}k_{4}\big(k_{3}+k_{4}\big)+k_{3}k_{4}\big(k_{3}^{3}+k_{4}^{3}\big)\nonumber\\\fl&+&k_{2}\big(k_{3}^{4}+k_{3}^{3}k_{4}-3k_{3}^{2}k_{4}^{2}+k_{4}^{4}+k_{3}k_{4}\big(k^{2}+k_{4}^{2}\big)\big)\big)\big)\big]-3x^{*7}k_{1}^{2}k_{2}^{2}k_{3}^{2}k_{4}^{2} ,\\ \fl
A_{EEEE}&\equiv&  k\left(k_{1}^{3}+k_{2}^{3}+k_{3}^{3}+k_{4}^{3}\right)x^{*3},\\ \fl
B_{EEEE}&\equiv&  k^{4}-k^{2}\left(k_{3}k_{4}+k_{2}\left(k_{3}+k_{4}+k_{1}\left(k_{2}+k_{3}+k_{4}\right)x^{*2}+k_{1}k_{2}k_{3}k_{4}x^{*4}\right)\right),\\ \fl
C_{EEEE}&\equiv& kx^{*}\left(k^{3}-\left(k_{2}k_{3}k_{4}+k_{1}\left(k_{3}k_{4}+k_{2}\left(k_{3}+k_{4}\right)\right)\right)x^{*2}\right) ,\\ \fl
D_{EEEE}&\equiv&  -kx^{*}\left(k^{3}-\left(k_{2}k_{3}k_{4}+k_{1}\left(k_{3}k_{4}+k_{2}\left(k_{3}k_{4}\right)\right)\right)x^{*2}\right),\\ \fl
E_{EEEE}&\equiv& k^{4}-k^{2}\left(k_{3}k_{4}+k_{2}\left(k_{3}+k_{4}\right)+k_{1}\left(k_{2}+k_{3}+k_{4}\right)\right)x^{*2}+k_{1}k_{2}k_{3}k_{4}x^{*4} ,\\ \fl
F_{EEEE}&\equiv&  k\left(k_{1}^{3}+k_{2}^{3}+k_{3}^{3}+k_{4}^{3}\right)x^{*3}.
\eea



\newpage

\section{References}

\begin{thebibliography}{99}

\bibitem{lrreview} For a review, see
D.~H.~Lyth and A.~Riotto,
Phys.\ Rept.\  {\bf 314}, 1 (1999). 


\bibitem{smoot92} 
G.~F.~Smoot {\it et al.},
Astrophys.\ J.\  {\bf 396}, L1 (1992)

\bibitem{bennett96} 
C.~L.~Bennett {\it et al.},
Astrophys.\ J.\  {\bf 464}, L1 (1996). 

\bibitem{gorski96} 
K.~M.~Gorski, A.~J.~Banday, C.~L.~Bennett, G.~Hinshaw, A.~Kogut, 
G.~F.~Smoot and E.~L.~Wright,
Astrophys.\ J.\  {\bf 464} (1996) L11. 


\bibitem{wmap3}
 D.~N.~Spergel {\it et al.}  [WMAP Collaboration],
  Astrophys.\ J.\ Suppl.\  {\bf 170}, 377 (2007)
  [arXiv:astro-ph/0603449].

\bibitem{Creminelli:2005hu}
  P.~Creminelli, A.~Nicolis, L.~Senatore, M.~Tegmark and M.~Zaldarriaga,
  JCAP {\bf 0605}, 004 (2006)
  [arXiv:astro-ph/0509029].

\bibitem{wmap5} 
  E.~Komatsu {\it et al.}  [WMAP Collaboration],
  arXiv:0803.0547 [astro-ph].


\bibitem{Smith:2009jr}
  K.~M.~Smith, L.~Senatore and M.~Zaldarriaga,
  JCAP {\bf 0909}, 006 (2009)
  [arXiv:0901.2572 [astro-ph]].

\bibitem{Senatore:2009gt}
  L.~Senatore, K.~M.~Smith and M.~Zaldarriaga,
  arXiv:0905.3746 [astro-ph.CO].



\bibitem{review} 
N.~Bartolo, E.~Komatsu, S.~Matarrese and A.~Riotto,
Phys.\ Rept.\  {\bf 402}, 103 (2004). 





\bibitem{prop}
 E.~Komatsu {\it et al.},
  arXiv:0902.4759 [astro-ph.CO].


\bibitem{Acqua}
  V.~Acquaviva, N.~Bartolo, S.~Matarrese and A.~Riotto,
  Nucl.\ Phys.\  B {\bf 667}, 119 (2003)
  [arXiv:astro-ph/0209156].

\bibitem{Maldacena:2002vr}
  J.~M.~Maldacena,
  JHEP {\bf 0305}, 013 (2003)
  [arXiv:astro-ph/0210603].

\bibitem{Seery:2006vu}
  D.~Seery, J.~E.~Lidsey and M.~S.~Sloth,
  JCAP {\bf 0701}, 027 (2007)
  [arXiv:astro-ph/0610210].



\bibitem{Linde:1985yf}
  A.~D.~Linde,
  Phys.\ Lett.\  B {\bf 158}, 375 (1985).

\bibitem{Kofman:1985zx}
  L.~A.~Kofman,
  Phys.\ Lett.\  B {\bf 173}, 400 (1986).

\bibitem{Polarski:1994rz}
  D.~Polarski and A.~A.~Starobinsky,
  Phys.\ Rev.\  D {\bf 50}, 6123 (1994)
  [arXiv:astro-ph/9404061].

\bibitem{GarciaBellido:1995qq}
  J.~Garcia-Bellido and D.~Wands,
  Phys.\ Rev.\  D {\bf 53}, 5437 (1996)
  [arXiv:astro-ph/9511029].

\bibitem{Mukhanov:1997fw}
  V.~F.~Mukhanov and P.~J.~Steinhardt,
  Phys.\ Lett.\  B {\bf 422}, 52 (1998)
  [arXiv:astro-ph/9710038].

\bibitem{Langlois:1999dw}
  D.~Langlois,
  Phys.\ Rev.\  D {\bf 59}, 123512 (1999)
  [arXiv:astro-ph/9906080].

\bibitem{Gordon:2000hv}
  C.~Gordon, D.~Wands, B.~A.~Bassett and R.~Maartens,
  Phys.\ Rev.\  D {\bf 63}, 023506 (2001)
  [arXiv:astro-ph/0009131].

\bibitem{Mollerach}
  S.~Mollerach,
  Phys.\ Rev.\  D {\bf 42}, 313 (1990).

\bibitem{Enqvist:2001zp}
  K.~Enqvist and M.~S.~Sloth,
  Nucl.\ Phys.\  B {\bf 626}, 395 (2002)
  [arXiv:hep-ph/0109214].


\bibitem{Lyth:2001nq}
  D.~H.~Lyth and D.~Wands,
  Phys.\ Lett.\  B {\bf 524}, 5 (2002)
  [arXiv:hep-ph/0110002].



\bibitem{Lyth:2002my}
  D.~H.~Lyth, C.~Ungarelli and D.~Wands,
  Phys.\ Rev.\  D {\bf 67}, 023503 (2003)
  [arXiv:astro-ph/0208055].

\bibitem{Moroi:2001ct}
  T.~Moroi and T.~Takahashi,
  Phys.\ Lett.\  B {\bf 522}, 215 (2001)
  [Erratum-ibid.\  B {\bf 539}, 303 (2002)]
  [arXiv:hep-ph/0110096].




\bibitem{Bartolo:2003jx}
  N.~Bartolo, S.~Matarrese and A.~Riotto,
  Phys.\ Rev.\  D {\bf 69}, 043503 (2004)
  [arXiv:hep-ph/0309033].


\bibitem{Garriga:1999vw}
  J.~Garriga and V.~F.~Mukhanov,
  Phys.\ Lett.\  B {\bf 458}, 219 (1999)
  [arXiv:hep-th/9904176].

\bibitem{ArmendarizPicon:1999rj}
  C.~Armendariz-Picon, T.~Damour and V.~F.~Mukhanov,
  Phys.\ Lett.\  B {\bf 458}, 209 (1999)
  [arXiv:hep-th/9904075].

\bibitem{Ali}
  M.~Alishahiha, E.~Silverstein and D.~Tong,
  Phys.\ Rev.\  D {\bf 70} (2004) 123505
  [arXiv:hep-th/0404084].

\bibitem{Chen:2006nt}
  X.~Chen, M.~x.~Huang, S.~Kachru and G.~Shiu,
  JCAP {\bf 0701}, 002 (2007)
  [arXiv:hep-th/0605045].



\bibitem{ghostinfl}
  N.~Arkani-Hamed, P.~Creminelli, S.~Mukohyama and M.~Zaldarriaga,
  JCAP {\bf 0404}, 001 (2004)
  [arXiv:hep-th/0312100].




\bibitem{Bartolomulti}
  N.~Bartolo, S.~Matarrese and A.~Riotto,
  Phys.\ Rev.\  D {\bf 65}, 103505 (2002)
  [arXiv:hep-ph/0112261].



\bibitem{Barttrisp}
  N.~Bartolo, S.~Matarrese and A.~Riotto,
  JCAP {\bf 0508}, 010 (2005)
  [arXiv:astro-ph/0506410].


\bibitem{SL3}
  D.~Seery and J.~E.~Lidsey,
  JCAP {\bf 0701}, 008 (2007)
  [arXiv:astro-ph/0611034].


\bibitem{VernizziWands}
  F.~Vernizzi and D.~Wands,
  JCAP {\bf 0605}, 019 (2006)
  [arXiv:astro-ph/0603799].

\bibitem{ChenWang}
  X.~Chen and Y.~Wang,
  arXiv:0911.3380 [hep-th].


\bibitem{Byrnes}
  C.~T.~Byrnes, M.~Sasaki and D.~Wands,
  Phys.\ Rev.\  D {\bf 74}, 123519 (2006)
  [arXiv:astro-ph/0611075].




\bibitem{Seery:2008ax}
  D.~Seery, M.~S.~Sloth and F.~Vernizzi,
  JCAP {\bf 0903}, 018 (2009)
  [arXiv:0811.3934 [astro-ph]].


\bibitem{SVW}
  M.~Sasaki, J.~Valiviita and D.~Wands,
  Phys.\ Rev.\  D {\bf 74}, 103003 (2006)
  [arXiv:astro-ph/0607627].




\bibitem{Huang:2006eha}
  X.~Chen, M.~x.~Huang and G.~Shiu,
  Phys.\ Rev.\  D {\bf 74}, 121301 (2006)
  [arXiv:hep-th/0610235].

\bibitem{Arroja:2008ga}
  F.~Arroja and K.~Koyama,
  Phys.\ Rev.\  D {\bf 77}, 083517 (2008)
  [arXiv:0802.1167 [hep-th]].

\bibitem{Arroja:2009pd}
  F.~Arroja, S.~Mizuno, K.~Koyama and T.~Tanaka,
  Phys.\ Rev.\  D {\bf 80}, 043527 (2009)
  [arXiv:0905.3641 [hep-th]].

\bibitem{Chen:2009bc}
  X.~Chen, B.~Hu, M.~x.~Huang, G.~Shiu and Y.~Wang,
  JCAP {\bf 0908}, 008 (2009)
  [arXiv:0905.3494 [astro-ph.CO]].




\bibitem{Gao:2009gd}
  X.~Gao and B.~Hu,
  JCAP {\bf 0908}, 012 (2009)
  [arXiv:0903.1920 [astro-ph.CO]].


\bibitem{Mizuno:2009mv}
  S.~Mizuno, F.~Arroja and K.~Koyama,
  arXiv:0907.2439 [hep-th].

\bibitem{Okamoto:2002ik}
  T.~Okamoto and W.~Hu,
  Phys.\ Rev.\  D {\bf 66}, 063008 (2002)
  [arXiv:astro-ph/0206155].


\bibitem{Kogo:2006kh}
  N.~Kogo and E.~Komatsu,
  Phys.\ Rev.\  D {\bf 73}, 083007 (2006)
  [arXiv:astro-ph/0602099].


\bibitem{Schwinger:1960qe}
  J.~S.~Schwinger,
  J.\ Math.\ Phys.\  {\bf 2}, 407 (1961).

\bibitem{Calzetta:1986ey}
  E.~Calzetta and B.~L.~Hu,
  Phys.\ Rev.\  D {\bf 35}, 495 (1987).

\bibitem{Jordan:1986ug}
  R.~D.~Jordan,
  Phys.\ Rev.\  D {\bf 33}, 444 (1986).

\bibitem{Weinberg:2005vy}
  S.~Weinberg,
  Phys.\ Rev.\  D {\bf 72}, 043514 (2005)
  [arXiv:hep-th/0506236].

\bibitem{Weinberg:2006ac}
  S.~Weinberg,
  Phys.\ Rev.\  D {\bf 74}, 023508 (2006)
  [arXiv:hep-th/0605244].

\bibitem{Seery:2007we}
  D.~Seery,
  JCAP {\bf 0711}, 025 (2007)
  [arXiv:0707.3377 [astro-ph]].

\bibitem{Seery:2007wf}
  D.~Seery,
  JCAP {\bf 0802}, 006 (2008)
  [arXiv:0707.3378 [astro-ph]].

\bibitem{Dimastrogiovanni:2008af}
  E.~Dimastrogiovanni and N.~Bartolo,
  JCAP {\bf 0811}, 016 (2008)
  [arXiv:0807.2790 [astro-ph]].











\bibitem{Vielva:2009jz}
  P.~Vielva and J.~L.~Sanz,
  arXiv:0910.3196 [astro-ph.CO].

\bibitem{Desjacques:2009jb}
  V.~Desjacques and U.~Seljak,
  arXiv:0907.2257 [astro-ph.CO].


\bibitem{http://planck.esa.int/}
See http://planck.esa.int/.




\bibitem{de OliveiraCosta:2003pu}
  A.~de Oliveira-Costa, M.~Tegmark, M.~Zaldarriaga and A.~Hamilton,
  Phys.\ Rev.\  D {\bf 69}, 063516 (2004)
  [arXiv:astro-ph/0307282].

\bibitem{Vielva:2003et}
  P.~Vielva, E.~Martinez-Gonzalez, R.~B.~Barreiro, J.~L.~Sanz and L.~Cayon,
  Astrophys.\ J.\  {\bf 609}, 22 (2004)
  [arXiv:astro-ph/0310273].


\bibitem{Eriksen:2003db}
  H.~K.~Eriksen, F.~K.~Hansen, A.~J.~Banday, K.~M.~Gorski and P.~B.~Lilje,
  Astrophys.\ J.\  {\bf 605}, 14 (2004)
  [Erratum-ibid.\  {\bf 609}, 1198 (2004)]
  [arXiv:astro-ph/0307507].



\bibitem{Bennett:1996ce}
  C.~L.~Bennett {\it et al.},
  Astrophys.\ J.\  {\bf 464}, L1 (1996)
  [arXiv:astro-ph/9601067].

\bibitem{Spergel:2003cb}
  D.~N.~Spergel {\it et al.}  [WMAP Collaboration],
  Astrophys.\ J.\ Suppl.\  {\bf 148}, 175 (2003)
  [arXiv:astro-ph/0302209].



\bibitem{Efstathiou:2003tv}
  G.~Efstathiou,
  Mon.\ Not.\ Roy.\ Astron.\ Soc.\  {\bf 348}, 885 (2004)
  [arXiv:astro-ph/0310207].


\bibitem{Land:2005ad}
  K.~Land and J.~Magueijo,
  Phys.\ Rev.\ Lett.\  {\bf 95}, 071301 (2005)
  [arXiv:astro-ph/0502237].




\bibitem{Cruz:2006fy}
  M.~Cruz, L.~Cayon, E.~Martinez-Gonzalez, P.~Vielva and J.~Jin,
  Astrophys.\ J.\  {\bf 655}, 11 (2007)
  [arXiv:astro-ph/0603859].



\bibitem{Hansen:2004mj}
  F.~K.~Hansen, P.~Cabella, D.~Marinucci and N.~Vittorio,
  Astrophys.\ J.\  {\bf 607}, L67 (2004)
  [arXiv:astro-ph/0402396].


\bibitem{Hansen:2004vq}
  F.~K.~Hansen, A.~J.~Banday and K.~M.~Gorski,
  Mon.\ Not.\ Roy.\ Astron.\ Soc.\  {\bf 354}, 641 (2004)
  [arXiv:astro-ph/0404206].




\bibitem{Groeneboom:2009cb}
  N.~E.~Groeneboom, L.~Ackerman, I.~K.~Wehus and H.~K.~Eriksen,
  arXiv:0911.0150 [astro-ph.CO].


\bibitem{Hanson:2009gu}
  D.~Hanson and A.~Lewis,
  Phys.\ Rev.\  D {\bf 80}, 063004 (2009)
  [arXiv:0908.0963 [astro-ph.CO]].

\bibitem{Wald:1983ky}
  R.~M.~Wald,
  Phys.\ Rev.\  D {\bf 28}, 2118 (1983).

\bibitem{Barrow:2005qv}
  J.~D.~Barrow and S.~Hervik,
  Phys.\ Rev.\  D {\bf 73}, 023007 (2006)
  [arXiv:gr-qc/0511127].

\bibitem{Barrow:2006xb}
  J.~D.~Barrow and S.~Hervik,
  Phys.\ Rev.\  D {\bf 74}, 124017 (2006)
  [arXiv:gr-qc/0610013].


\bibitem{DiGrezia:2003ug}
  E.~Di Grezia, G.~Esposito, A.~Funel, G.~Mangano and G.~Miele,
  Phys.\ Rev.\  D {\bf 68}, 105012 (2003)
  [arXiv:gr-qc/0305050].

\bibitem{Pereira:2007yy}
  T.~S.~Pereira, C.~Pitrou and J.~P.~Uzan,
  JCAP {\bf 0709}, 006 (2007)
  [arXiv:0707.0736 [astro-ph]].


\bibitem{Pitrou:2008gk}
  C.~Pitrou, T.~S.~Pereira and J.~P.~Uzan,
  JCAP {\bf 0804}, 004 (2008)
  [arXiv:0801.3596 [astro-ph]].

\bibitem{Dimastrogiovanni:2008ua}
  E.~Dimastrogiovanni, W.~Fischler and S.~Paban,
  JHEP {\bf 0807}, 045 (2008)
  [arXiv:0803.2490 [hep-th]].



\bibitem{Gumrukcuoglu:2008gi}
  A.~E.~Gumrukcuoglu, L.~Kofman and M.~Peloso,
  Phys.\ Rev.\  D {\bf 78}, 103525 (2008)
  [arXiv:0807.1335 [astro-ph]].






\bibitem{ArmendarizPicon:2003qk}
  C.~Armendariz-Picon and P.~B.~Greene,
  Gen.\ Rel.\ Grav.\  {\bf 35}, 1637 (2003)
  [arXiv:hep-th/0301129].

\bibitem{Boehmer:2007ut}
  C.~G.~Boehmer and D.~F.~Mota,
  Phys.\ Lett.\  B {\bf 663}, 168 (2008)
  [arXiv:0710.2003 [astro-ph]].


\bibitem{Watanabe:2009nc}
  T.~Watanabe,
  arXiv:0902.1392 [astro-ph.CO].


\bibitem{deBerredoPeixoto:2009sb}
  G.~de Berredo-Peixoto and E.~A.~de Freitas,
  Class.\ Quant.\ Grav.\  {\bf 26}, 175015 (2009)
  [arXiv:0902.4025 [gr-qc]].

\bibitem{Germani:2009iq}
  C.~Germani and A.~Kehagias,
  JCAP {\bf 0903}, 028 (2009)
  [arXiv:0902.3667 [astro-ph.CO]].


\bibitem{Kobayashi:2009hj}
  T.~Kobayashi and S.~Yokoyama,
  JCAP {\bf 0905}, 004 (2009)
  [arXiv:0903.2769 [astro-ph.CO]].


\bibitem{Koivisto:2009ew}
  T.~S.~Koivisto and N.~J.~Nunes,
  arXiv:0907.3883 [astro-ph.CO].

\bibitem{Germani:2009gg}
  C.~Germani and A.~Kehagias,
  JCAP {\bf 0911}, 005 (2009)
  [arXiv:0908.0001 [astro-ph.CO]].

\bibitem{Koivisto:2009fb}
  T.~S.~Koivisto and N.~J.~Nunes,
  Phys.\ Rev.\  D {\bf 80}, 103509 (2009)
  [arXiv:0908.0920 [astro-ph.CO]].

\bibitem{Koivisto:2009sd}
  T.~S.~Koivisto, D.~F.~Mota and C.~Pitrou,
  JHEP {\bf 0909}, 092 (2009)
  [arXiv:0903.4158 [astro-ph.CO]].












\bibitem{Yokoyama:2008xw}
  S.~Yokoyama and J.~Soda,
  JCAP {\bf 0808}, 005 (2008)
  [arXiv:0805.4265 [astro-ph]].

\bibitem{ValenzuelaToledo:2009af}
  C.~A.~Valenzuela-Toledo, Y.~Rodriguez and D.~H.~Lyth,
  Phys.\ Rev.\  D {\bf 80}, 103519 (2009)
  [arXiv:0909.4064 [astro-ph.CO]].

\bibitem{Dimopoulos:2008yv}
  K.~Dimopoulos, M.~Karciauskas, D.~H.~Lyth and Y.~Rodriguez,
  JCAP {\bf 0905}, 013 (2009)
  [arXiv:0809.1055 [astro-ph]].

\bibitem{ValenzuelaToledo:2009nq}
  C.~A.~Valenzuela-Toledo and Y.~Rodriguez,
  arXiv:0910.4208 [astro-ph.CO].

\bibitem{Karciauskas:2008bc}
  M.~Karciauskas, K.~Dimopoulos and D.~H.~Lyth,
  Phys.\ Rev.\  D {\bf 80}, 023509 (2009)
  [arXiv:0812.0264 [astro-ph]].


\bibitem{Dimopoulos:2009am}
  K.~Dimopoulos, M.~Karciauskas and J.~M.~Wagstaff,
  arXiv:0907.1838 [hep-ph].


\bibitem{Dimopoulos:2009vu}
  K.~Dimopoulos, M.~Karciauskas and J.~M.~Wagstaff,
  arXiv:0909.0475 [hep-ph].


\bibitem{Bartolo:2009pa}
  N.~Bartolo, E.~Dimastrogiovanni, S.~Matarrese and A.~Riotto,
  JCAP {\bf 0910}, 015 (2009)
  [arXiv:0906.4944 [astro-ph.CO]].

\bibitem{Bartolo:2009kg}
  N.~Bartolo, E.~Dimastrogiovanni, S.~Matarrese and A.~Riotto,
  JCAP {\bf 0911}, 028 (2009)
  [arXiv:0909.5621 [astro-ph.CO]].


















\bibitem{Mukhanov:1996ak}
V.~F. Mukhanov, L.~R.~W. Abramo, and R.~H. Brandenberger, Phys. Rev. Lett.
  {\bf 78}, 1624 (1997).


\bibitem{Abramo:1997hu}
L.~R.~W. Abramo, R.~H. Brandenberger, and V.~F. Mukhanov, Phys. Rev.
  {\bf D56}, 3248  (1997).

\bibitem{boy1} D.~Boyanovsky, H.~J.~de Vega and N.~G.~Sanchez,
  Phys.\ Rev.\  D {\bf 71}, 023509 (2005).

\bibitem{boy2} D.~Boyanovsky, H.~J.~de Vega and N.~G.~Sanchez,
    Nucl.\ Phys.\  B {\bf 747}, 25 (2006). 

\bibitem{olandesi}
  M.~van der Meulen and J.~Smit,
 JCAP {\bf 0711}, 023 (2007)
  [arXiv:0707.0842 [hep-th]]. 



\bibitem{weinberg1}
S.~Weinberg,
  Phys.\ Rev.\  D {\bf 72}, 043514 (2005)
  [arXiv:hep-th/0506236].

\bibitem{weinberg2}
S.~Weinberg,
  Phys.\ Rev.\  D {\bf 74}, 023508 (2006)
  [arXiv:hep-th/0605244].





\bibitem{sloth} M.~S.~Sloth,
  Nucl.\ Phys.\  B {\bf 748}, 149 (2006).

\bibitem{sloth2}
M.~S.~Sloth,
  Nucl.\ Phys.\  B {\bf 775}, 78 (2007).


\bibitem{seery1}
 D.~Seery,
  JCAP {\bf 0711} (2007) 025
  [arXiv:0707.3377 [astro-ph]].

\bibitem{seery2}
D.~Seery,
  JCAP {\bf 0802} (2008) 006
  [arXiv:0707.3378 [astro-ph]].


\bibitem{box4}
 D.~H.~Lyth,
  JCAP {\bf 0712}, 016 (2007)
  [arXiv:0707.0361 [astro-ph]].

 


\bibitem{box5}
N.~Bartolo, S.~Matarrese, M.~Pietroni, A.~Riotto and D.~Seery,
  JCAP {\bf 0801} (2008) 015
  [arXiv:0711.4263 [astro-ph]].






\bibitem{Enqvistetal}
  K.~Enqvist, S.~Nurmi, D.~Podolsky and G.~I.~Rigopoulos,
  JCAP {\bf 0804}, 025 (2008)
  [arXiv:0802.0395 [astro-ph]].

\bibitem{nuovo}
  N.~Bartolo, E.~Dimastrogiovanni and A.~Vallinotto,
  (to appear).

\bibitem{Leblond:2008gg}
  L.~Leblond and S.~Shandera,
  JCAP {\bf 0808}, 007 (2008)
  [arXiv:0802.2290 [hep-th]].

\bibitem{Shandera:2008ai}
  S.~Shandera,
  Phys.\ Rev.\  D {\bf 79}, 123518 (2009)
  [arXiv:0812.0818 [astro-ph]].


\bibitem{Byrnes:2007tm}
  C.~T.~Byrnes, K.~Koyama, M.~Sasaki and D.~Wands,
  JCAP {\bf 0711}, 027 (2007)
  [arXiv:0705.4096 [hep-th]].


\bibitem{deltaN4}
D.~H.~Lyth and Y.~Rodriguez,
  Phys.\ Rev.\ Lett.\  {\bf 95}, 121302 (2005)
  [arXiv:astro-ph/0504045].

\bibitem{box2}
L.~Boubekeur and D.~H.~Lyth,
  Phys.\ Rev.\  D {\bf 73}, 021301 (2006)
  [arXiv:astro-ph/0504046].

\bibitem{box3}
I.~Zaballa, Y.~Rodriguez and D.~H.~Lyth,
  JCAP {\bf 0606} (2006) 013
  [arXiv:astro-ph/0603534].

\bibitem{polarization}
S. Weinberg, ``Cosmology'' (Oxford University Press, 2008): Sec. 5.2 

\bibitem{Burgess:2009bs}
  C.~P.~Burgess, L.~Leblond, R.~Holman and S.~Shandera,
  arXiv:0912.1608 [hep-th].




\bibitem{Ford:1989me}
  L.~H.~Ford,
  Phys.\ Rev.\  D {\bf 40}, 967 (1989).

\bibitem{Golovnev:2008cf}
  A.~Golovnev, V.~Mukhanov and V.~Vanchurin,
  JCAP {\bf 0806}, 009 (2008)
  [arXiv:0802.2068 [astro-ph]].

\bibitem{Golovnev:2008hv}
  A.~Golovnev, V.~Mukhanov and V.~Vanchurin,
  JCAP {\bf 0811}, 018 (2008)
  [arXiv:0810.4304 [astro-ph]].

\bibitem{Golovnev:2009ks}
  A.~Golovnev and V.~Vanchurin,
  Phys.\ Rev.\  D {\bf 79}, 103524 (2009)
  [arXiv:0903.2977 [astro-ph.CO]].



\bibitem{ArmendarizPicon:2004pm}
  C.~Armendariz-Picon,
  JCAP {\bf 0407}, 007 (2004)
  [arXiv:astro-ph/0405267].

\bibitem{Boehmer:2007qa}
  C.~G.~Boehmer and T.~Harko,
  Eur.\ Phys.\ J.\  C {\bf 50}, 423 (2007)
  [arXiv:gr-qc/0701029].

\bibitem{Koivisto:2007bp}
  T.~Koivisto and D.~F.~Mota,
  Astrophys.\ J.\  {\bf 679}, 1 (2008)
  [arXiv:0707.0279 [astro-ph]].


\bibitem{Koivisto:2008xf}
  T.~S.~Koivisto and D.~F.~Mota,
  JCAP {\bf 0808}, 021 (2008)
  [arXiv:0805.4229 [astro-ph]].

\bibitem{Jimenez:2009py}
  J.~B.~Jimenez, R.~Lazkoz and A.~L.~Maroto,
  arXiv:0904.0433 [astro-ph.CO].


\bibitem{Jimenez:2009zza}
  J.~B.~Jimenez, R.~Lazkoz and A.~L.~Maroto,
  Phys.\ Rev.\  D {\bf 80}, 023004 (2009).




\bibitem{Himmetoglu:2008zp}
  B.~Himmetoglu, C.~R.~Contaldi and M.~Peloso,
  Phys.\ Rev.\ Lett.\  {\bf 102}, 111301 (2009)
  [arXiv:0809.2779 [astro-ph]].


\bibitem{Himmetoglu:2008hx}
  B.~Himmetoglu, C.~R.~Contaldi and M.~Peloso,
  Phys.\ Rev.\  D {\bf 79}, 063517 (2009)
  [arXiv:0812.1231 [astro-ph]].


\bibitem{Himmetoglu:2009qi}
  B.~Himmetoglu, C.~R.~Contaldi and M.~Peloso,
  Phys.\ Rev.\  D {\bf 80}, 123530 (2009)
  [arXiv:0909.3524 [astro-ph.CO]].


\bibitem{Dimopoulos:2006ms}
  K.~Dimopoulos,
  Phys.\ Rev.\  D {\bf 74}, 083502 (2006)
  [arXiv:hep-ph/0607229].


\bibitem{Dimopoulos:2008rf}
  K.~Dimopoulos and M.~Karciauskas,
  JHEP {\bf 0807}, 119 (2008)
  [arXiv:0803.3041 [hep-th]].


\bibitem{Kanno:2008gn}
  S.~Kanno, M.~Kimura, J.~Soda and S.~Yokoyama,
  JCAP {\bf 0808}, 034 (2008)
  [arXiv:0806.2422 [hep-ph]].


\bibitem{Ackerman:2007nb}
  L.~Ackerman, S.~M.~Carroll and M.~B.~Wise,
  Phys.\ Rev.\  D {\bf 75}, 083502 (2007)
  [Erratum-ibid.\  D {\bf 80}, 069901 (2009)]
  [arXiv:astro-ph/0701357].


\bibitem{Carroll:2009em}
  S.~M.~Carroll, T.~R.~Dulaney, M.~I.~Gresham and H.~Tam,
  Phys.\ Rev.\  D {\bf 79}, 065011 (2009)
  [arXiv:0812.1049 [hep-th]].


\bibitem{Lyth:2005qk}
  D.~H.~Lyth,
  JCAP {\bf 0511}, 006 (2005)
  [arXiv:astro-ph/0510443].

\bibitem{Alabidi:2006wa}
  L.~Alabidi and D.~Lyth,
  JCAP {\bf 0608}, 006 (2006)
  [arXiv:astro-ph/0604569].

\bibitem{Salem:2005nd}
  M.~P.~Salem,
  Phys.\ Rev.\  D {\bf 72}, 123516 (2005)
  [arXiv:astro-ph/0511146].


\bibitem{Bernardeau:2004zz}
  F.~Bernardeau, L.~Kofman and J.~P.~Uzan,
  Phys.\ Rev.\  D {\bf 70}, 083004 (2004)
  [arXiv:astro-ph/0403315].


\bibitem{Watanabe:2009ct}
  M.~a.~Watanabe, S.~Kanno and J.~Soda,
  Phys.\ Rev.\ Lett.\  {\bf 102}, 191302 (2009)
  [arXiv:0902.2833 [hep-th]].

\bibitem{Himmetoglu:2009mk}
  B.~Himmetoglu,
  arXiv:0910.3235 [astro-ph.CO].




\bibitem{DG}
  T.~R.~Dulaney and M.~I.~Gresham,
  arXiv:1001.2301 [astro-ph.CO].


\bibitem{Seery:2005wm}
  D.~Seery and J.~E.~Lidsey,
  JCAP {\bf 0506}, 003 (2005)
  [arXiv:astro-ph/0503692].

\bibitem{Vernizzi:2006ve}
  F.~Vernizzi and D.~Wands,
  JCAP {\bf 0605}, 019 (2006)
  [arXiv:astro-ph/0603799].

\bibitem{Martin:2007ue}
  J.~Martin and J.~Yokoyama,
  JCAP {\bf 0801}, 025 (2008)
  [arXiv:0711.4307 [astro-ph]].

\bibitem{Babich:2004gb}
  D.~Babich, P.~Creminelli and M.~Zaldarriaga,
  JCAP {\bf 0408}, 009 (2004)
  [arXiv:astro-ph/0405356].

\bibitem{peskin}
M.E. Peskin and D.V. Schroeder, ``An introduction to quantum field theory'' (1995): Sec. 9.2

\bibitem{Seery:2006vu}
  D.~Seery, J.~E.~Lidsey and M.~S.~Sloth,
  JCAP {\bf 0701}, 027 (2007)
  [arXiv:astro-ph/0610210].


\bibitem{Komatsu:2010fb}
  E.~Komatsu {\it et al.},
  arXiv:1001.4538 [astro-ph.CO].


\bibitem{Bennett:2010jb}
  C.~L.~Bennett {\it et al.},
  arXiv:1001.4758 [astro-ph.CO].









\end{thebibliography}
\end{document}